\documentclass[twoside,12pt,a4paper]{article}
\usepackage{epsfig}
\usepackage{graphicx,cite,xspace}
\usepackage{multirow}

\def\Journal#1#2#3#4{{#1} {#2} (#4) #3 }
\def\NPA{{\em Nucl. Phys.} A}

\def\PRO{{\em Prog. Theor. Phys.}}
\def\NPB{{\em Nucl. Phys.} B}
\def\PLB{{\em Phys. Lett.} B}

\def\PRL{{\em Phys. Rev. Lett.}}

\def\PREP{\em Phys. Rep.}
\def\PRD{{\em Phys. Rev.} D}
\def\PRC{{\em Phys. Rev.} C}
\def\ZPC{{\em Z. Phys.} C}

\def\RMP{{\em Rev. Mod. Phys.}}
\def\INT{{\em Int. J. Mod. Phys.} E}
\def\ARNP{{\em Ann. Rev. Nucl. Part. Sci.}}
\def\JHEP{{\em JHEP}}
\def\NAT{{\em Nature}}
\def\PPNP{{\em Prog. Part. Nucl. Phys.}}
\def\EPJC{{\em Eur. Phys. J.} C}
\def\JPG{{\em J. Phys. G: Nucl. Part. Phys.}}
\def\ARNPS{{\em Ann. Rev. Nucl. Part. Sci.}}
\def\NIMA{{\em Nucl. Instrum. Methods} A}
\def\ASTRJ{{\em Astrophys. J.}}

\providecommand*\eg{e.g.\xspace}
\providecommand*\ie{i.e.\xspace}
\providecommand*\etal{et al.\xspace}
\providecommand*\cf{\emph{cf.\xspace}}

\providecommand{\pp}{pp\xspace}
\providecommand{\ppbar}{p$\bar{\rm p}$\xspace}
\providecommand*\dau{d+Au\xspace}
\providecommand*\ppb{p+Pb\xspace}
\providecommand*\auau{Au+Au\xspace}
\providecommand*\cucu{Cu+Cu\xspace}
\providecommand*\pbpb{Pb+Pb\xspace}

\providecommand*\tc{T_{\rm C}\xspace}
\providecommand*\mub{\mu_{\rm B}\xspace}
\providecommand*\snn{\sqrt{s_{\rm NN}}\xspace}
\providecommand*\raa{R_{\rm AA}\xspace}
\providecommand*\rda{R_{\rm dA}\xspace}
\providecommand*\gevc{GeV/$c$\xspace}
\providecommand*\gevcsquare{GeV/$c^2$\xspace}

\providecommand*\jpsi{${\rm J}/\psi$\xspace}
\providecommand*\ups{$\Upsilon$\xspace}
\providecommand*\dzero{D$^0$\xspace}
\providecommand*\bzero{B$^0$\xspace}
\providecommand*\bszero{B$_{\rm s}^{0}$\xspace}

\providecommand*\dplus{D$^+$\xspace}
\providecommand*\bplus{B$^+$\xspace}

\providecommand*\bzero{B$^0$\xspace}
\providecommand*\dstar{D$^{*}$\xspace}
\providecommand*\dstarplus{D$^{*+}$\xspace}
\providecommand*\dsplus{D$_{\rm s}^{+}$\xspace}
\providecommand*\lambdac{$\Lambda_{\rm c}$\xspace}

\providecommand*\pte{p_t^e\xspace}

\providecommand*\snn{\sqrt{s_{\rm NN}}\xspace}
\providecommand*\npart{N_{\rm part}\xspace}
\providecommand*\ncol{N_{\rm coll}\xspace}

\providecommand*\ccbar{c\bar{c}\xspace}
\providecommand*\bbbar{b\bar{b}\xspace}

\providecommand{\s}{$\sqrt{s}$\xspace}
\providecommand{\pt}{\ensuremath{p_{\rm t}}\xspace}
\providecommand{\mt}{\ensuremath{m_{\rm t}}\xspace}
\providecommand{\dedx}{d$E$/d$x$\xspace}

\begin{document}
\title{Heavy-flavor production in heavy-ion collisions\\
       and implications for the properties of hot QCD matter}
\author{R.\ Averbeck
\\
ExtreMe Matter Institute EMMI and Research Division,\\ 
GSI Helmholtzzentrum f\"ur Schwerionenforschung, Darmstadt, Germany}
\date{}
\maketitle
\begin{abstract}
Hadrons carrying open heavy flavor, \ie single charm or bottom quarks, are 
among the key diagnostic tools available today for the hot and dense state of 
strongly interacting matter which is produced in collisions of heavy atomic 
nuclei at ultra-relativistic energies. First systematic heavy-flavor 
measurements in nucleus-nucleus collisions and the reference proton-proton 
system at Brookhaven National Laboratory's (BNL) Relativistic Heavy Ion 
Collider (RHIC) have led to tantalizing results. These studies are now 
continued and extended at RHIC and at CERN's Large Hadron Collider (LHC), 
where considerably higher collision energies are available. 
This review focuses on experimental results on open heavy-flavor observables
at RHIC and the LHC published until July 2012. Yields of heavy-flavor hadrons 
and their decay products, their transverse momentum and rapidity distributions,
as well as their azimuthal distributions with respect to the reaction plane in 
heavy-ion collisions are investigated. Various theoretical approaches are 
confronted with the data and implications for the properties of the hot and 
dense medium produced in ultra-relativistic heavy-ion collisions are discussed.
\end{abstract}

\tableofcontents

\clearpage

\section{Introduction}
\label{sec:intro}
\subsection{Strongly interacting matter 
            in ultra-relativistic heavy-ion collisions}
\label{subsec:qgp}
\begin{figure}[t]
\begin{center}
\includegraphics[width=0.6\linewidth]{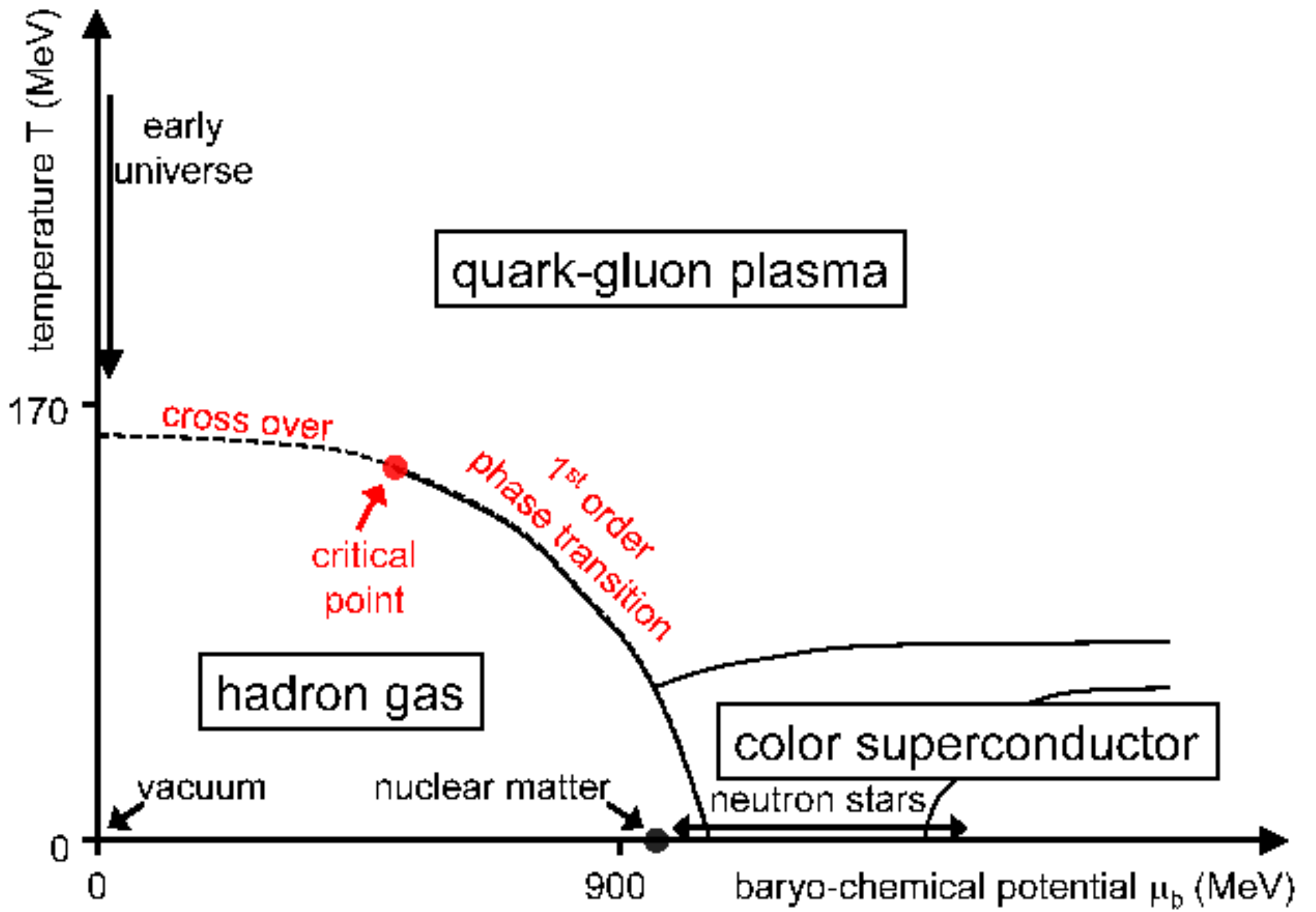}
\caption{Schematic phase diagram of strongly interacting matter for two 
         massless quarks as a function of the temperature $T$ and the 
         baryo-chemical potential $\mub$.}
\label{fig:phase}
\end{center}
\end{figure}

Quantum chromodynamics (QCD) is the underlying theory of the strong 
interaction~\cite{fritzsch73}. At sufficiently high temperature, $T$, or 
baryo-chemical potential, $\mub$, QCD inspired model calculations predict 
a phase transition of strongly interacting matter from a system of hadrons 
to a deconfined medium, dubbed a quark-gluon plasma (QGP), in which the 
relevant degrees of freedom are of partonic 
nature~\cite{itoh70,cabibbo75,collins75}. At zero baryo-chemical
potential, corresponding to zero net baryon density, QCD calculations on a 
discretized space-time lattice~\cite{wilson74,creutz77,laermann03} indicate 
that the phase transition is of the cross over 
type~\cite{fodor04,ejiri04,aoki06a}. The critical temperature, $\tc$, is still 
not known precisely even for $\mub = 0$. Values in the range 
$150 < \tc < 190$~MeV, corresponding to an energy density in the vicinity of 
1~GeV/fm$^3$, have been quoted~\cite{aoki06b,cheng06}. However, even larger
values in the range $180 < \tc < 200$~MeV can not be 
excluded~\cite{bazavov09}. For nonzero baryo-chemical potential lattice QCD 
calculations become significantly more difficult. Some calculations predict 
that the phase transition remains of the cross over type for all values of 
$\mub$. Several others indicate that towards higher $\mub$ the phase transition
will eventually be of first order. Consequently, this would imply the existence 
of a critical point in the phase diagram of strongly interacting matter. 
At very high $\mub$ the presence of more exotic phases, \eg color 
superconducting forms of strongly interacting matter, is 
expected~\cite{rischke04,buballa05,alford07}.
A schematic QCD phase diagram is shown in Fig.~\ref{fig:phase}.

Strongly interacting matter at extreme temperature and/or density is not only
of fundamental interest in the context of mapping out the QCD phase diagram, 
but it is also of astrophysical relevance. The early universe is believed to
have spent its first few microseconds after the big bang in the state of a 
quark-gluon plasma at high temperature and zero baryo-chemical potential. 
Hadronization took place when the universe expanded and cooled down below $\tc$.
QGP matter at high density and moderate temperature might be present today 
inside of neutron stars~\cite{weber}, or it might be created briefly during 
the core collapse of a supernova explosion~\cite{gentile,sagert}.

Collisions of heavy atomic nuclei at ultra-relativistic energies provide the 
unique opportunity to investigate experimentally the properties and the 
dynamics of hot and dense QCD matter in the laboratory, because only in
such collisions the necessary temperature or density can be reached.
 
First ``circumstantial evidence'' for the production of a quark-gluon plasma 
was reported~\cite{heinz00} from the fixed target heavy-ion program at the CERN 
Super Proton Synchrotron (SPS), where collisions of lead nuclei (\pbpb) were 
investigated at energies per nucleon-nucleon pair up to $\snn = 17.3$~GeV in 
the center of mass frame. 
Experiments with colliding gold nuclei (\auau) at $\snn = 200$~GeV at the 
dedicated BNL Relativistic Heavy Ion Collider (RHIC) have substantiated these 
findings and have led to a first quantitative characterization of the 
properties of the QGP~\cite{wp_brahms,wp_phenix,wp_phobos,wp_star}, revealing 
some surprises. Here, only some of the most striking observations are briefly 
summarized. 

The initially produced fireball has such a high temperature and density that 
the partons, \ie quarks and gluons, equilibrate on a time scale of less than 
1~fm/$c$. Large pressure gradients in the system lead to a hydrodynamic 
evolution of the fireball which subsequently expands, cools down, and 
undergoes hadronization. When the system is cold and dilute enough, the 
hadrons freeze out chemically and thermally and stream to the detectors. 
It is important to note that the geometrical shape of the initial equilibrated 
partonic fireball is asymmetric for collisions with non-zero impact parameter, 
exhibiting an almond-like shape averaged over many collisions.
Consequently, the pressure gradients driving the expansion are asymmetric
as well. Thus the initial spatial anisotropy is translated into an azimuthal 
anisotropy in momentum space of the produced hadrons. Originally, this 
anisotropy was quantified based on a Fourier expansion of the momentum 
distribution

\begin{equation}
E \frac{d^3N}{d^3p} = \frac{d^3N}{\pt d\phi d\pt dy}
                      \sum_{n = 0}^{\infty} 2 v_{\rm n} \cos[n(\phi - \Phi_{\rm R})]
\label{eq:fourier}
\end{equation}

where $\phi$ is the azimuthal emission angle of a particle with respect to
the orientation of the reaction plane $\Phi_{\rm R}$ in a given 
collision~\cite{voloshin}. This plane is defined by the beam direction and the 
impact parameter vector connecting the centers of the colliding nuclei in the 
transverse plane. While the short axis of the initial almond-like shape of the 
fireball is aligned with the reaction plane, the long axis is perpendicular to 
the latter. The harmonic coefficients, $v_{\rm n}$, quantify the strength of the
anisotropy. The second harmonic, $v_{2}$, is also called the elliptic flow, 
and it is believed to be sensitive to the early dynamics in the collision. 
Recently, it was recognized that fluctuations of the initial state geometry 
are important as they can generate higher $n$th-order flow components~\cite{mishra08,mishra10,takahashi09,alver10a,alver10b,teaney11,luzum11}, in 
particular odd-$n$ components, when measured with respect to the proper 
initial-state planes of symmetry $\Phi_{\rm n}$~\cite{manly06} instead of 
$\Phi_{\rm R}$.
For a given initial state configuration and equation of state of 
the expanding medium the laws of ideal or viscous relativistic hydrodynamics 
predict the momentum space azimuthal asymmetry. The observed good agreement 
between the predictions~\cite{huovinen01,kolb04,teaney02} and the measured 
$v_{2}$ of various hadron species imply that the produced medium expands 
collectively almost like a perfect fluid, \ie a fluid with a small shear 
viscosity to entropy density ratio, $\eta/s$, close to the 
conjectured~\cite{kovtun05} quantum lower bound of $\eta/s = 1/4\pi$. 
Precision measurements of $v_{3}$ or even higher orders of the azimuthal
anisotropy~\cite{vn_alice1,vn_phenix,vn_alice2,vn_atlas,vn_cms1,vn_cms2}
should improve the sensitivity to the viscosity of the medium~\cite{alver10b}.

This partonic fluid produced in \auau collisions at RHIC is so dense that 
quarks and gluons from initial partonic hard scattering processes (``jets'') 
suffer a large energy loss while they propagate through the hot and dense 
medium. This ``jet quenching'' gives rise to a strong suppression of hadrons 
originating mainly from jet fragmentation at high transverse momentum, \pt, 
relative to the expectation from a superposition of independent binary 
nucleon-nucleon collisions. To quantify this suppression the nuclear 
modification factor $\raa$ is introduced as

\begin{equation}
\raa = \frac{dN_{\rm AA}/d\pt}{\langle N_{\rm coll} \rangle \times dN_{\rm pp}/d\pt}
   = \frac{dN_{\rm AA}/d\pt}{\langle T_{\rm AA} \rangle \times d\sigma_{\rm pp}/d\pt}
\label{eq:raa}
\end{equation}

where $dN_{\rm AA}/d\pt$ is the differential invariant yield in nucleus-nucleus
collisions and $dN_{\rm pp}/d\pt$ ($d\sigma_{\rm pp}/d\pt$) is the corresponding
differential invariant yield (cross section) in \pp collisions. For a given
centrality class the average number of binary collisions is denoted by
$\langle N_{\rm coll} \rangle$ and $\langle T_{\rm AA} \rangle$ is the corresponding
nuclear thickness function, which relates $\langle N_{\rm coll} \rangle$ with
the inelastic nucleon-nucleon scattering cross section. 
$\langle N_{\rm coll} \rangle$ and $\langle T_{\rm AA} \rangle$ are obtained
via Glauber model calculations~\cite{glauber70,miller07} of the collision 
geometry taking into account the response of the detectors used for the 
centrality measurement. As expected 
for particles not participating in the strong interaction, the nuclear 
modification factor of photons from initial hard scattering processes, \eg 
quark-gluon Compton scattering, is consistent with one~\cite{phenix_direct}. 
Furthermore, this confirms that the Glauber model based calculation of the 
collision geometry is under control and that, at least at mid-rapidity at RHIC,
initial state modifications of the parton distribution functions in nuclear 
matter, \eg shadowing, anti-shadowing, or saturation effects, do not play a 
significant role. For high \pt hadrons from jet fragmentation the value 
$\raa \approx 0.2$ measured at mid-rapidity in \auau collisions at 
RHIC~\cite{raa_phenix,wp_brahms,wp_phenix,wp_phobos,wp_star} demonstrates the 
opaqueness of the produced dense fireball. When confronted with these data, 
theoretical models indicate that initial energy densities beyond 10 GeV/fm$^3$ 
are reached in the produced fireball~\cite{miklos05,vitev06}, clearly in the 
QGP regime.

The initial temperature of the fireball can not be measured directly. However,
an enhanced yield of quasi-real virtual photons\footnote{low-mass $e^+e^-$ 
pairs with high \pt ($m_{e^+e^-} < 0.3$~\gevcsquare, $1 < \pt < 5$~\gevc)} 
above known hadronic sources, was measured~\cite{phenix_thermal} in central 
\auau collisions at RHIC. Interpreting this excess as virtual thermal photon 
emission from the hot fireball, hydrodynamical models allow to infer the 
initial temperature $T_{\rm init}$ and thermalization time $\tau_0$ of the 
fireball. Values of $ T_{\rm init} \sim 300 - 600$~MeV and 
$\tau_0 \sim 0.6 - 0.15$~fm/$c$ are consistent with the data~\cite{enterria06}.

The yield of heavy quarkonia in nucleus-nucleus collisions, \ie bound states 
of charm or bottom quarks and their antiquarks, was long thought of as an 
observable directly related to deconfinement in the fireball~\cite{matsui86}. 
It was predicted that color charge screening in a deconfined QGP would 
prohibit charm quarks and antiquarks to form a bound state, the \jpsi meson, 
similar to the process of Debye screening in an electromagnetic 
plasma~\cite{matsui86}. The resulting \jpsi suppression was indeed observed 
for the first time in central nucleus-nucleus collisions at the CERN 
SPS~\cite{abreu01}. However, other mechanisms unrelated to deconfinement, \eg 
the absorption of quarkonia in cold nuclear matter or their break-up by 
interactions with other hadrons (co-movers) produced in the collision, could 
also describe the SPS data reasonably well~\cite{gavin97,spieles99,capella02}
rendering the interpretation inconclusive. Subsequently, it was realized that
in a deconfined medium a new mechanism for quarkonium production might
become important if a sufficiently large number of heavy quark antiquark
pairs is produced initially. Quarkonia could be formed in this case via
statistical hadronization at the phase boundary~\cite{pbm00,pbm01} or, earlier,
via coalescence in the QGP~\cite{thews01}. While in nucleus-nucleus collisions 
at the SPS the number of charm quark anti-quark pairs is much smaller than one,
leaving no room for a significant contribution to the \jpsi yield via 
regeneration processes, the situation is different at RHIC, where more than 
10 charm quark anti-quark pairs are produced in a central \auau collision and, 
in particular, at the LHC, where more than 100 such pairs are generated in a 
central \pbpb collision. Under such conditions \jpsi suppression might 
actually turn into an enhancement via regeneration processes~\cite{andronic07}.
First evidence in this direction was observed at RHIC in the centrality and, 
in particular, the rapidity dependence of the nuclear modification factor of 
\jpsi mesons, $R_{\rm AA}^{{\rm J}/\psi}$~\cite{phenix_jpsi}. 
$R_{\rm AA}^{{\rm J}/\psi}$ reaches its maximum value at mid-rapidity in central 
\auau collisions. This is at variance with the expectation from suppression 
via color screening or from destruction through the interaction with co-moving 
hadrons, which both should reach their maximum near mid-rapidity where the 
energy density reaches is the largest. In the regeneration picture, however, 
this maximum of $R_{\rm AA}^{{\rm J}/\psi}$ at mid-rapidity finds its natural 
explanation in the fact that in this region of phase space the density of 
charm quarks and anti-quarks is maximal. Still the situation is not conclusive 
yet, as other possible scenarios might explain the RHIC data. Most likely, 
the \jpsi saga will only be settled by decisive measurements from the LHC, 
where rather dramatic effects due to charmonium regeneration are expected when 
the total charm production yield is large enough~\cite{andronic08}. 

At the CERN LHC colliding beams of lead nuclei are investigated at 
unprecedented high energies ($\snn = 2.76$~TeV up to now) since
November 2010. The measurement of two-pion Bose-Einstein correlations
via Hanbury-Brown Twiss interferometry has shown that the volume of the
produced fireball is about twice as large as it was observed in \auau 
collisions at RHIC and has a longer life time as well~\cite{alice_hbt_pbpb}. 
The pseudorapidity density of charged particles produced at mid-rapidity also 
increases by a factor close to two going from RHIC to the 
LHC~\cite{alice_dndeta}, indicating a substantial increase of the initial 
energy density reached in \pbpb collisions at the LHC.
Remarkably, for various hadron species the elliptic flow strength $v_2$ as 
a function of \pt is the same within uncertainties at RHIC and at the 
LHC~\cite{alice_v2,atlas_v2,cms_v2}, which demonstrates that the produced 
fireball still resembles the properties of an almost perfect fluid, 
albeit with a large initial energy density.
At the LHC, high energy jets in the TeV range are copiously produced. Such
hard probes for the medium are not available at RHIC and, consequently, the 
measurement of the nuclear modification factor can be extended to much higher
\pt in \pbpb collisions at the LHC compared to \auau collisions at RHIC.
In central collisions, $\raa$ reaches its minimum value, corresponding to
maximum suppression of hadrons from jet fragmentation, for a hadron \pt
of about 6-7~\gevc~\cite{alice_raa,cms_raa}. $\raa$ increases 
towards higher values of \pt but stays significantly below one even for 
hadrons with a \pt of 100~\gevc~\cite{cms_raa}, highlighting
the enormous density of the produced medium. Also in the quarkonium sector
intriguing new observations have been made already. The nuclear modification
factor of \jpsi measured at forward rapidity~\cite{alice_jpsi_mu_pbpb} is 
significantly larger in central \pbpb collisions at the LHC compared to \auau 
collisions at RHIC, which might indicate regeneration of charmonia from a 
deconfined medium.
In addition, the suppression of various $\Upsilon$ states, \ie bottomonia, 
can be studied in a systematic way at the LHC. After first indications
for $\Upsilon$ suppression have been reported at RHIC~\cite{star_upsilon_raa}, 
where the different spin states have not been resolved yet, first data from 
\pbpb collisions at the LHC show a suppression of the $\Upsilon{\rm (1S)}$ 
ground state in central collisions at low \pt~\cite{cms_upsilon_raa}.

For the coming years the most important scientific challenge for the field
is the quantitative characterization of the state of QCD matter produced in 
ultra-relativistic heavy-ion collisions at RHIC and at the LHC. 
In this context the measurement of observables related to the production of 
hadrons carrying open heavy flavor can play a unique role as is argued in 
Section~\ref{subsec:openHF}, where the basic concepts and related theoretical 
approaches are briefly introduced. Results, relevant for the current discussion,
originating from open heavy-flavor measurements in hadronic collisions from 
the pre-RHIC era are briefly summarized in Section~\ref{sec:preRHIC}. 
The main focus of this article is a review of the 
results on heavy-flavor yields, transverse momentum and rapidity distributions,
as well as azimuthal distributions with respect to the reaction plane in 
heavy-ion collisions at RHIC (Section~\ref{sec:rhic}) and at the LHC 
(Section~\ref{sec:lhc}). Various theoretical approaches are confronted with 
the data and implications for the properties of the hot and dense QCD matter 
produced in such collisions are discussed. The article closes with a summary 
of the current status and an outlook to the future open heavy-flavor program 
at RHIC, LHC, and elsewhere in Section~\ref{sec:summary}.

\subsection{Open heavy flavor in \pp and nucleus-nucleus collisions}
\label{subsec:openHF}
Open heavy-flavor production is experimentally accessible through the
measurement of hadrons carrying charm or bottom quarks or through the 
measurement of their decay products. The most important properties of 
the relevant hadrons are summarized in Tab.~\ref{tab:HFhadrons}~\cite{pdg}.

\begin{table}[t]
\begin{center}
\caption{Properties of hadrons carrying open heavy flavor with charm or bottom
quantum numbers C~=~+1 or B~=~+1, respectively~\cite{pdg}. Given are the 
valence quark content, the isospin, spin, and parity (I(J$^{\rm P}$)), the 
hadron mass, the most important decay modes together with their branching 
ratios (B.R.), and the decay length $c\tau$. Antihadrons with negative C or B 
quantum numbers are not listed explicitly. They decay into equivalent channels 
with charge conjugate decay products.}
\label{tab:HFhadrons}
\begin{tabular}{lllllll}
\\
\hline
Particle & Quark & I(J$^{\rm P}$) & 
Mass (\gevcsquare) & Decay mode & B.R. (\%) & $c\tau$ ($\mu$m) \\
 & content & & & & & \\
\hline
\dplus & c$\bar{\rm d}$ & $\frac{1}{2}(0^-)$ & $1.8696\pm0.0002$ & 
${\rm K}^- \pi^+ \pi^+$ & $9.13 \pm 0.19$ & $312 \pm 2$ \\
 & & & & $e^+$ anything & $16.07 \pm 0.30$ & \\
 & & & & $\mu^+$ anything & $17.6 \pm 3.2$ & \\
\dzero & c$\bar{\rm u}$ & $\frac{1}{2}(0^-)$ & $1.8648\pm0.0001$ & 
${\rm K}^- \pi^+$ & $3.87 \pm 0.05$ & $123 \pm 1$ \\
 & & & & $e^+$ anything & $6.49 \pm 0.11$ & \\
 & & & & $\mu^+$ anything & $6.7 \pm 0.6$ & \\
\dsplus & c$\bar{\rm s}$ & $0(0^-)$ & $1.9685\pm0.0003$ & 
$\phi \pi^+$ & $4.5 \pm 0.4$ & $150 \pm 2$ \\
 & & & & $e^+$ anything & $6.5 \pm 0.4$ & \\
\dstarplus & c$\bar{\rm d}$ & $\frac{1}{2}(0^-)$ & $2.0102\pm0.0001$ & 
\dzero $\pi^+$ & $67.7 \pm 0.5$ & $(2.1 \pm 0.5) \times 10^{-6}$ \\
\lambdac & udc & 0($\frac{1}{2}^+)$ & $2.2865\pm0.0001$ & 
${\rm p} {\rm K}^- \pi^+$ & $5.0 \pm 1.3$ & $60 \pm 2$ \\
\hline
\bplus & u$\bar{\rm b}$ & $\frac{1}{2}(0^-)$ & $5.2792\pm0.0003$ & 
${\rm J}/\psi K^+$ & $0.1013 \pm 0.0034$ & $492 \pm 2$ \\
 & & & & $l^+ \nu_l$ anything & $10.99 \pm 0.28$ & \\
\bzero & d$\bar{\rm b}$ & $\frac{1}{2}(0^-)$ & $5.2795\pm0.0003$ & 
${\rm J}/\psi K_{\rm S}^0$ & $0.0436 \pm 0.0016$ & $455 \pm 2$ \\
 & & & & $l^+ \nu_l$ anything & $10.33 \pm 0.28$ & \\
\bszero & s$\bar{\rm b}$ & $0(0^-)$ & $5.3663 \pm 0.0006$ & 
${\rm J}/\psi \phi$ & $0.14 \pm 0.05$ & $441 \pm 8$ \\
b hadrons & & & & ${\rm J}/\psi$ anything & $1.16 \pm 0.10$ & \\
\hline
\end{tabular}
\end{center}
\end{table}

In contrast to the light u, d, and s quarks, both the charm and the bottom 
quark are ``heavy'' with (bare) quark masses significantly exceeding the 
QCD scale parameter $\Lambda_{QCD} \simeq 0.2$~GeV 
($m_{\rm c} = 1.29^{+0.05}_{-0.11}$~GeV and 
$m_{\rm b} = 4.19^{+0.18}_{-0.06}$~GeV~\cite{pdg}).
Thus, the production of a pair of a heavy quark and its antiquark in
ultra-relativistic \pp collisions proceeds exclusively through initial 
hard partonic scattering processes and, therefore, can be treated
theoretically within the framework of perturbative QCD. The unique feature of
heavy-flavor production is that such a perturbative treatment is warranted
for all momenta since the large quark mass introduces a hard scale even at 
zero momentum. This is in distinct contrast to gluon and light quark jets 
which can be treated perturbatively only at high \pt. 

Consequently, the measurement of heavy-flavor production in \pp 
collisions provides a crucial testing ground for pQCD. In the following,
only hadro-production of heavy-flavor hadrons is addressed. A more detailed
introduction, which also discusses photoproduction or the production of
heavy quarks in $e^+e^-$ collisions, can be found 
elsewhere~\cite{appel92,ellis96,mangano97}. In this pQCD approach the basic 
production process of a hadron carrying heavy flavor can be factorized into 
three components: 
\begin{itemize}
\item the (non-perturbative) initial conditions. These are mainly determined
      by the fractional momenta, $x$, the interacting partons carry from the
      colliding hadrons. The $x$ distributions of various partons inside
      hadrons have been studied mainly in deep inelastic scattering experiments.
      They depend on the squared energy-momentum transfer, $Q^2$, between the 
      two partons and they have been parametrized in the form of parton 
      distribution functions (PDF).
\item the partonic (perturbative) scattering cross section. This can be 
      calculated in perturbative QCD. At leading order (LO), the only
      processes contributing to heavy-flavor production are gluon fusion
      and quark anti-quark annihilation. At next-to-leading order (NLO)
      processes such as gluon splitting or flavor excitation have to be
      considered in addition. State of the art perturbative calculations
      of heavy-flavor production go even further in the sense that they
      calculate cross sections at fixed order with a next-to-leading-log
      resummation of higher orders in $\alpha_{\rm s}$.
\item the (non-perturbative) fragmentation of the heavy quarks into 
      heavy-flavor hadrons. Here, two cases have to be distinguished:
\begin{enumerate}
\item formation of hadrons carrying open heavy flavor: in this case the
      heavy quark and antiquark individually fragment into hadrons. The energy 
      of a heavy-flavor hadron with respect to the energy of the initially
      produced heavy quark is given by the fragmentation function, which is 
      measured in $e^+e^-$ reactions and is assumed to be universal, \ie the 
      fragmentation of a heavy quark does not depend on the mechanism by which 
      this quark was produced.
\item formation of quarkonium states: typically 1-2\% of the produced heavy
      quark-anti\-quark pairs form a bound quarkonium state instead of a pair
      of hadrons with open heavy flavor. While the physics of quarkonia is not 
      discussed further in this article (see 
      references~\cite{vogt99,brambilla05,lourenco05,rapp10} for  
      reviews of this topic instead) it is important to mention that the 
      measurement of the total charm and bottom yields would provide a 
      natural reference for the investigation of charmonium and bottomonium
      yields. 
\end{enumerate}
\end{itemize}

In the context of this review results from two modern NLO pQCD calculation
approaches will be confronted with available open heavy-flavor hadro-production
data: Fixed Order calculations with Next-to-Leading-Log resummation 
(FONLL)~\cite{fonll1,fonll2,fonll3} and calculations within the General-Mass 
Variable-Flavor-Numbering Scheme (GM-VFNS)~\cite{gmvfns,kniehl06}.

In addition to serving as a precision test of QCD, open heavy-flavor 
measurements in \pp collisions provide a baseline for heavy-flavor studies
in heavy-ion collisions, where the heavy quarks propagate through and
interact with the produced hot and dense QCD matter.

In the absence of nuclear effects, the heavy-flavor yields in nucleus-nucleus
collisions would scale with the number of binary collisions as it is the
case for all hard probes. Therefore, departures from binary scaling would 
indicate nuclear modifications of heavy-flavor observables. Two classes of
such modifications have to be distinguished from each other, \ie initial state
effects and final state effects.

Initial state effects result from the fact that the distributions of partons 
embedded in nuclei are different from the parton distribution functions in 
nucleons. Depending on $x$ and $Q^2$, different nuclear modifications of the 
parton distribution functions are observed. In the anti-shadowing region at 
medium $x$ (around $x \sim 0.1$), the parton density in nuclei is larger than 
in nucleons. The shadowing region is characterized by a depletion of the 
parton density at low $x$ ($x < 10^{-2}$) in nuclei relative to nucleons. 
Other effects, such as gluon saturation, may lead to reduced gluon densities 
at very low $x$ as well. Reviews of nuclear parton distribution functions, on 
gluon saturation, and on related issues can be found 
elsewhere~\cite{accardi03,gelis07,forte10}.

Final state effects are in-medium modifications of heavy-flavor observables 
due to the presence of hot and dense matter in ultra-relativistic heavy-ion 
collisions. The main focus of the open heavy-flavor programs at RHIC and 
at the LHC is the investigation of such modifications of open heavy-flavor 
observables, aiming to shed light on the properties of the hot QCD medium 
and the nature of parton-medium interactions. In turn, this requires a solid 
understanding of the initial state effects such that these and the final
state effects can be disentangled. A number of final state effect has been 
predicted and is investigated experimentally. 

The total open heavy-flavor yield is sensitive to modifications of the 
initial parton densities but it is not expected to be changed significantly 
via final state effects. Because the charm and bottom quark masses are large 
enough that secondary, thermal heavy-flavor production mechanisms in the early,
hot stages of a heavy-ion collision do not contribute significantly to the 
total heavy-flavor yield at RHIC. It is an interesting question whether thermal
production of heavy quark-antiquark pairs can be observed in \pbpb collisions 
at the LHC but, most likely, this mechanism will not provide a large 
contribution to the total heavy-flavor yield even at LHC energies.

A number of final state effects are expected to leave their footprint on 
the phase-space distributions of heavy-flavor hadrons. In particular, 
the interaction of partons with the medium~\cite{plumer90} is expected to be 
sensitive to the medium's energy density via the mechanism of parton energy 
loss (see~\cite{majumder11} for a recent review). This energy
loss will reflect itself in a softening of the \pt distributions of 
heavy-flavor hadrons or their decay products with respect to the same spectra
in \pp collisions. Prior to systematic heavy-flavor 
measurements at RHIC, it was generally expected that the dominant energy loss 
mechanism of heavy quarks would be radiative energy loss via medium induced 
gluon radiation.
Models incorporating this energy loss mechanism provide a reasonable
description of the high-\pt suppression of light-flavor hadrons at 
RHIC~\cite{dainese05,loizides07,glv00,wicks07,zhang07,baier97,salgado03}.
However, in QCD the radiative energy loss is not the same for all partons. 
Gluons exhibit a larger color coupling factor than quarks such that the energy 
loss of quarks is expected to be smaller than that of gluons. Furthermore, with
increasing quark mass induced gluon radiation at small angles with respect 
to the quark momentum vector is reduced owing to the so-called 
``dead cone effect''~\cite{dead_cone}. Radiative energy loss should therefore 
lead to a distinctive mass hierarchy of high-\pt hadron suppression as 
quantified by the nuclear modification factor $\raa$. In the \pt range up to 
about 10~\gevc, where the heavy quark masses are sizable with respect to their 
momenta, the energy loss is expected to decrease when going from light-flavor
hadrons, which originate mostly from gluon or light-quark jets, to hadrons 
carrying charm and hadrons carrying bottom quarks:
$\raa^\pi < \raa^{\rm c} < \raa^{\rm b}$. 

It was pointed out, however, that collisional energy loss via elastic 
scattering should be comparable in magnitude to radiative energy loss over 
a wide kinematic region~\cite{mustafa05}. This additional contribution is 
more important for heavy than for light quarks and it is included in a number 
of model calculation which are confronted with data later in this review.

Another class of models aiming at the description of heavy quarks in the
QCD medium are Langevin-based transport models~\cite{moore05,vanhees05} in 
which a heavy quark is placed into a thermal medium. In such calculations 
the interaction is given by elastic collisions, in some 
approaches~\cite{vanhees05} mediated by the excitation of 
D- or B-meson like resonant states in the medium. Heavy-quark diffusion 
and the damping of the initial non-equilibrium dynamics of heavy quarks in
the medium are at the heart of these models, which gives indirect access
to the viscosity to entropy density ratio of the QCD medium.

In another approach~\cite{adil07,sharma09}, heavy-flavor energy loss via 
repeated sequential formation of heavy-flavor hadrons and their dissociation 
in the medium is investigated. In this context, the interplay between the 
formation time of heavy-flavor hadrons and the expansion dynamics of hot QCD 
matter plays a key role.

A completely different Ansatz~\cite{horowitz11} makes use of the AdS/CFT (anti 
de-Sitter-space/conformal-field theory) conjecture, which claims a 
correspondence between certain strongly coupled string theories and
semiclassical gravitational physics. In this framework, heavy-quark energy
loss is described in terms of string drag energy loss. 

In most of the approaches listed above it is assumed that the chemical 
composition of heavy-flavor hadrons is the same in \pp and nucleus-nucleus 
collisions. However, in the light-quark sector an enhancement of baryons
relative to mesons has been observed at intermediate \pt in central heavy-ion
collisions at RHIC~\cite{adler03,arsene05,abelev06}. This becomes relevant as 
soon as heavy-flavor production is studied via the measurement of decay 
products of heavy-flavor hadrons, in particular electrons or muons from 
heavy-flavor semileptonic decays. For example, owing to a smaller semileptonic 
decay branching ratio and a softer decay lepton spectrum of 
\lambdac baryon compared to D-meson decays, the $\raa$ of electrons 
from charm hadrons decays would be smaller than one even without energy loss 
of charm in the medium~\cite{sorensen06,crochet08}.
Consequently, not only heavy-flavor mesons but also baryons 
have to be measured in exclusive decay channels in order to allow for 
unambiguous conclusions concerning in-medium modifications of heavy-flavor 
observables. Also for precision measurements of the total heavy-flavor 
production yields in \pp and heavy-ion collisions it is mandatory to directly 
determine the contribution from heavy-flavor baryons. 
Unfortunately, until today heavy-flavor baryon yields and spectra could not be
measured in heavy-ion collisions.

Further insight into the interaction between heavy quarks and the hot QCD 
medium can be gained from the measurement of the azimuthal anisotropy
of heavy-flavor hadron spectra with respect to the orientation of the reaction 
plane in heavy-ion collisions. In particular, the elliptic flow parameter $v_2$
is expected to be sensitive to the degree of thermalization of heavy quarks 
within the partonic medium. The measurement of a non-zero $v_2$ would 
certainly be indicative of a strong interaction of heavy quarks with the 
medium. In addition, heavy-flavor elliptic flow data are useful to benchmark 
models in which heavy-flavor hadrons are formed via the quark coalescence 
mechanism. Most of the theoretical approaches introduced above in the context 
of heavy-quark energy loss in the medium also provide predictions for $v_2$ 
such that a simultaneous measurement of $v_2$ and $\raa$ provides a more 
stringent test of the underlying physics. 

In summary, the value of heavy-flavor observables in hadronic collisions is
twofold. 
First, in \pp collisions perturbative QCD can be tested in a unique
way and, thus, a solid experimental and theoretical reference can be prepared
which is necessary for corresponding heavy-flavor measurements in nuclear 
collisions.
Second, in collisions of the latter type the interaction of heavy charm and 
bottom quarks with hot and dense QCD matter is expected to be not only 
quantitatively but even qualitatively different from the interaction of 
light quark and gluon jets with this medium. Consequently, heavy-flavor 
measurements allow a complementary characterization of the parton-medium
interactions and of the properties of strongly interacting matter, as it is
produced in ultra-relativistic heavy-ion collisions, which can not be achieved
otherwise.

\section{Open heavy-flavor hadro-production measurements in the pre-RHIC era}
\label{sec:preRHIC}

\begin{figure}[t]
\begin{center}
\includegraphics[width=0.49\textwidth]{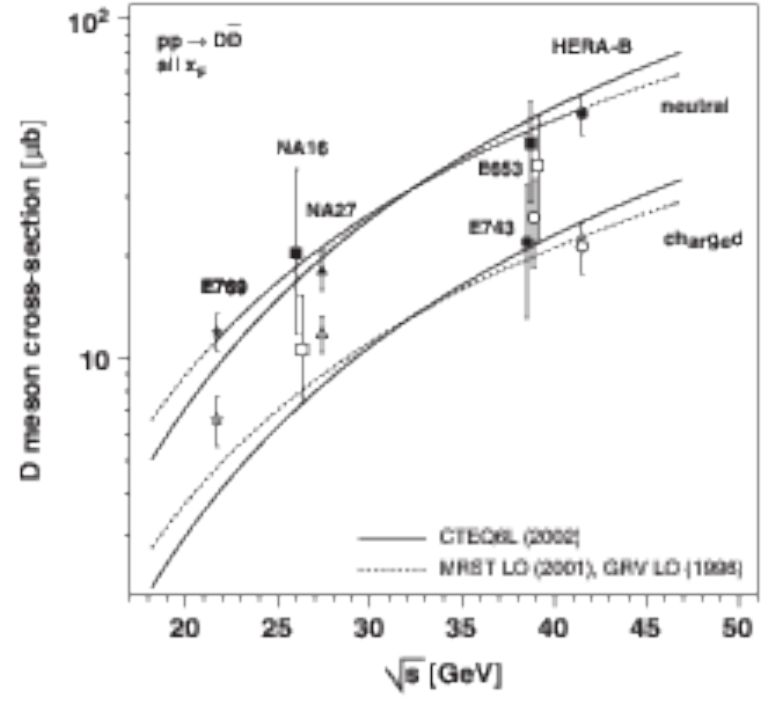}
\includegraphics[width=0.49\textwidth]{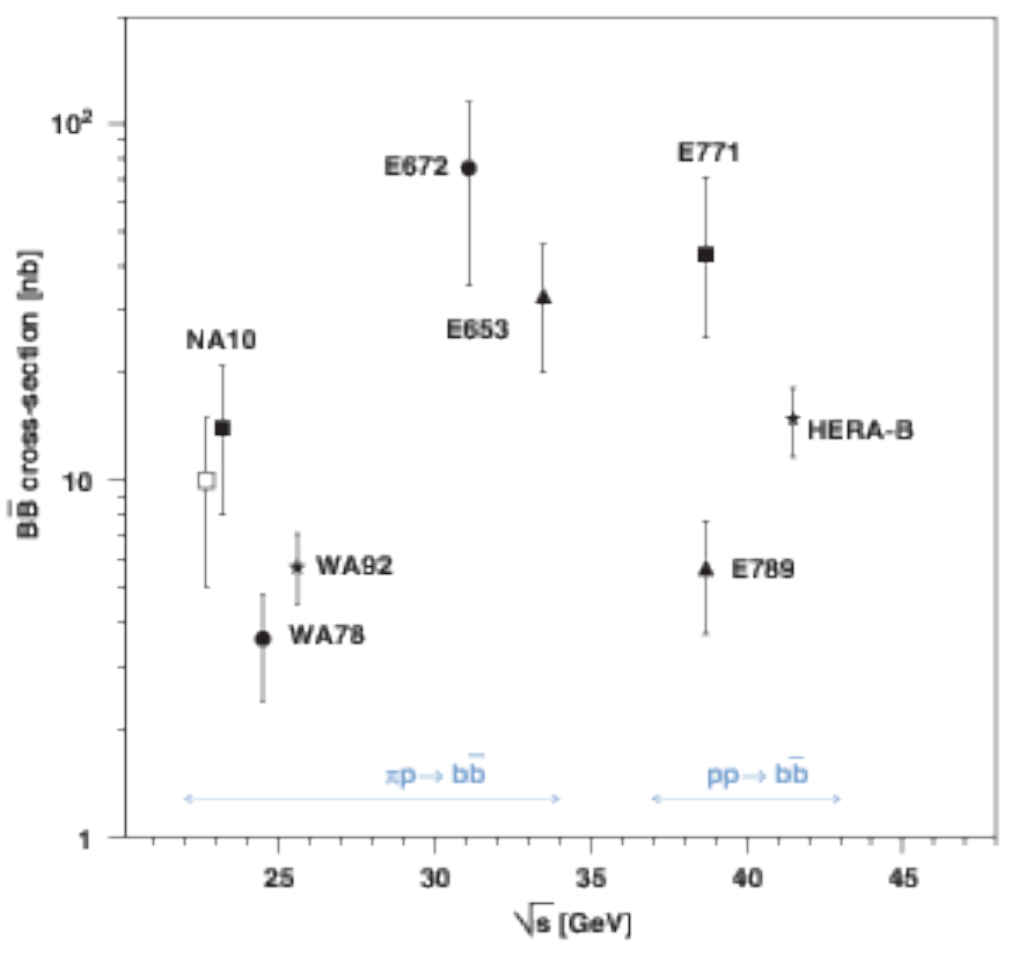}
\end{center}
\caption{Equivalent \pp production cross sections of neutral and charged 
         D-mesons as functions of $\sqrt{s}$ in fixed-target experiments
         LO pQCD calculation with the PYTHIA event generator using
         different parton distribution functions are compared with the
         experimental data (left panel). Corresponding B-meson production
         cross sections from fixed target $\pi p$ and \pp collisions
         (right panel) (reprinted from Ref.~\cite{lourenco06}).}
\label{fig:hf_pre_rhic}
\end{figure}

\subsection{Open heavy flavor from fixed target experiments}
\label{subsec:preRHIC_fixed}
The discovery of hadrons carrying open charm or bottom was preceded in both
cases by the discovery of charmonium or bottomonium, respectively. The \jpsi
charmonium state was discovered in the so-called ``November revolution'' 
simultaneously in p-Be collisions at the Brookhaven AGS~\cite{jpsi_ting} and 
in $e^+e^-$ collisions at Stanford's storage ring SPEAR~\cite{jpsi_richter}. 
The $\Upsilon$ bottomonium state was discovered in p-nucleus collisions at
FNAL~\cite{lederman}. Experimentally, the advantage of the \jpsi and the 
$\Upsilon$ compared to open heavy-flavor hadrons is related to the fact that 
the quantum numbers of these quarkonia coincide with those of the photon. 
Therefore, they can be produced resonantly in $e^+e^-$ colliders for precision 
measurements of their properties. More important in the context of heavy-ion 
collisions, the \jpsi and $\Upsilon$ can decay into dilepton pairs giving rise 
to the unique experimental signature of a high \pt dilepton pair, which can be 
triggered on.
Both neutral~\cite{goldhaber76} and charged~\cite{peruzzi76} open-charm D mesons
were first observed at the SLAC SPEAR storage ring in purely hadronic decays. 
First evidence for open-bottom production was found with the 
CLEO~\cite{bebek81} and CUSB~\cite{spencer81} detectors at the Cornell 
Electron Storage Ring (CESR) via the measurement of electrons from 
semileptonic B-meson decays. 
 
In contrast to heavy quarkonia, in particular the \jpsi meson, which played 
a crucial rule in the fixed-target heavy-ion program at the CERN SPS, it was 
only with the advent of high energy nucleus-nucleus colliders, \ie RHIC
and LHC, that open heavy-flavor hadrons could take their role as unique probes
for hot QCD matter. For many years open charm and bottom production was in the
focus of elementary particle physics only, studied mostly via photo-production
or in $e^+e^-$ experiments.

Although single electrons from heavy-flavor hadron decays were first observed
in \pp collisions at the CERN ISR collider at 
$\sqrt{s} = 52.7$~GeV~\cite{buesser74}, actually before the discovery of charm,
subsequent studies of open heavy flavor hadro-production were conducted mainly 
in fixed-target experiments at $\sqrt{s} < 50$~GeV employing pion and proton 
beams at the CERN SPS, at FNAL, and at DESY.
Most of these measurements, which are reviewed in detail 
elsewhere~\cite{lourenco06}, were specifically designed for heavy-flavor 
measurements. As a key feature they included detectors with high spatial 
resolution in the target region in order to separate the primary collision 
vertex from secondary heavy-flavor decay vertices. Bubble chambers, emulsions, 
and silicon tracking telescopes were used for this purpose. Large statistical 
samples have been collected with pion beams only. The proton beam experiments 
accumulated much less data, with the $\approx 300$ neutral and charged D-meson 
events collected by E769 representing the largest statistics data 
sample~\cite{alves96}. It is interesting to note that not only E769 but also 
most of the other experiments used nuclear targets, notably Be, Al, Cu, and W. 
To derive equivalent production cross sections for \pp (or $\pi$p) reactions a 
linear dependence of the cross section on the mass number $A$ of the target 
nucleus was assumed, 
\ie $\sigma_{{\rm pA},\pi{\rm A}} = A \cdot \sigma_{{\rm pp},\pi{\rm p}}$. 
In Fig.~\ref{fig:hf_pre_rhic} the derived D- and B-meson pair production 
cross sections in \pp or $\pi p$ collisions, extrapolated to 4$\pi$ from 
the phase space covered in the various experiments, are shown as a function 
of $\sqrt{s}$. The leading-order pQCD inspired event generator 
PYTHIA~\cite{pythia} can be used to describe describe heavy-flavor production 
in the fixed-target energy regime as was demonstrated in Ref.~\cite{pbm98}. 
More detailed and systematic studies have substantiated these findings as 
summarized in Ref.~\cite{lourenco06}.
Calculations with PYTHIA employing different parton distribution functions are 
compared with charged and neutral D-meson production cross sections from \pp 
collisions in the left panel of Fig.~\ref{fig:hf_pre_rhic}. The fact that 
higher-order diagrams are absent in PYTHIA is compensated by scaling the 
calculated cross sections up by an empirical factor, the so-called K-factor. 
This approach is justified only under the assumption that the kinematical 
distributions of heavy quarks originating from leading order and higher order 
processes are similar. Depending on the choice of other model parameters in 
PYTHIA, K-factors in the range 2.5 - 4.5 are necessary to obtain a satisfactory 
agreement of the calculation with charm production data.
The bottom production cross section is smaller than the charm production
cross section by almost three orders of magnitude in this $\sqrt{s}$ range
as shown in the right panel of Fig.~\ref{fig:hf_pre_rhic}. The experimental
uncertainties are large and some of the measurements, \eg the E771 and E789
results obtained at the same energy, are inconsistent with each
other. It is, however, remarkable that PYTHIA calculations, using a K-factor
of 2 for bottom production, do not only describe the low energy bottom 
production cross sections shown in Fig.~\ref{fig:hf_pre_rhic} within a 
factor of 2, but they are also in reasonable agreement with bottom
production data measured in \ppbar collisions at the Tevatron Collider at
$\sqrt{s}$ close to 2~TeV as discussed in Ref.~\cite{lourenco06}.  
The same systematic analysis of heavy-flavor hadro-production data has 
demonstrated that for $\sqrt{s}$ values accessible at fixed-target machines 
not only the open heavy-flavor absolute production cross sections but also the 
D-meson kinematical distributions and D$\bar{\rm D}$ pair correlations can be 
described reasonably well with PYTHIA.

\begin{figure}[t]
\begin{center}
\includegraphics[width=0.6\textwidth]{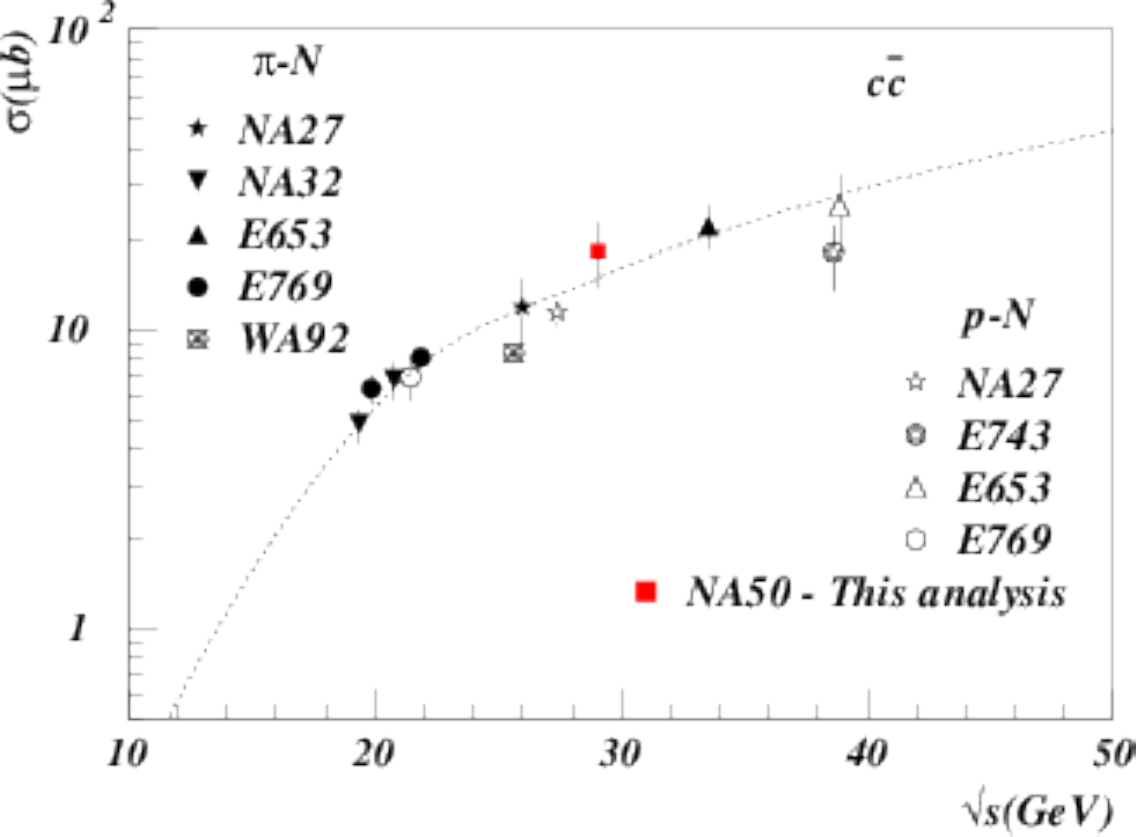}
\end{center}
\caption{Charm hadro-production cross section in the forward hemisphere 
         ($x_F > 0$) as a function of $\sqrt{s}$ in comparison with the
         $\sqrt{s}$ dependence of a corresponding PYTHIA calculation. The
         absolute normalization of the PYTHIA curve was fitted to the 
         experimental data (reprinted from Ref.~\cite{abreu00} with kind
         permission from Springer Science and Business Media).}
\label{fig:charm_na50}
\end{figure}

While in the heavy-ion program at the CERN SPS open heavy-flavor production was 
never measured directly, the contribution of simultaneous semimuonic decays
of correlated ${\rm D}\bar{\rm D}$ pairs to the dimuon continuum was carefully
studied. Particular attention was given to the intermediate mass region (IMR) 
of the dimuon mass distribution, \ie the region between the $\phi$ and the 
$J/\psi$. In addition to open charm decays, the only relevant sources for
dimuons in the IMR are the Drell-Yan process and, potentially, thermal dimuon
production in the hot QCD matter in nucleus-nucleus collisions. 
Intermediate mass dimuon production was first measured at the SPS with 
spectrometers that did not allow for a separation of prompt dimuons from the 
primary collision vertex and dimuons from displaced, secondary decay vertices 
characteristic for open charm decays. The NA38~\cite{lourenco94} and HELIOS-3 
Collaborations~\cite{angelis00} measured dimuons in p-W, S-U, and S-W 
collisions at a beam energy of 200~GeV per nucleon. Later, NA38/NA50 
investigated p-A collisions, using Al, Cu, Ag, and W as nuclear targets at a 
beam energy of 450~GeV, and \pbpb collisions at 158~GeV per 
nucleon~\cite{abreu00}. 

The correlated, opposite-sign IMR dimuon spectra measured by the three 
experiments in p-A collisions are in good agreement with the sum of the 
expected Drell-Yan and open charm contributions. In fact, NA50 determined
the ratio of dimuons from open charm decays to Drell-Yan dimuons via fits
to the mass spectra, and used PYTHIA to infer indirectly the total charm
production cross section in \pp collisions, assuming that in p-A collisions
the total open charm cross section scales with the nuclear target mass number.
The resulting charm cross section is shown in Fig.~\ref{fig:charm_na50} 
in comparison with direct measurements of the charm production cross section
and as a function of $\sqrt{s}$ together with a PYTHIA calculation similar to
the one discussed above. The cross section obtained indirectly from the IMR
dimuon spectra in p-A collisions agrees reasonably well with the $\sqrt{s}$ 
dependence obtained from direct measurements. 

\begin{figure}[t]
\begin{center}
\includegraphics[width=0.67\textwidth]{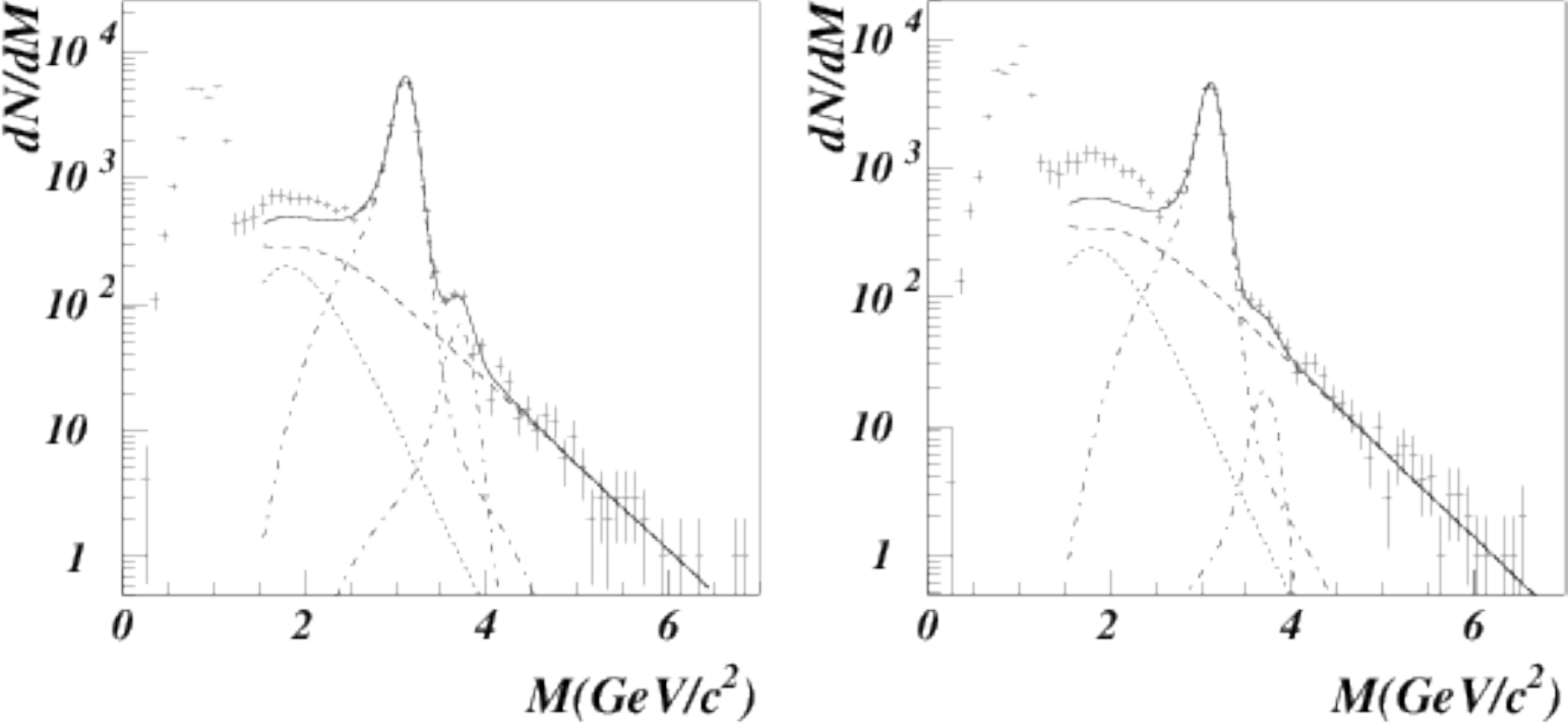}
\includegraphics[width=0.32\textwidth]{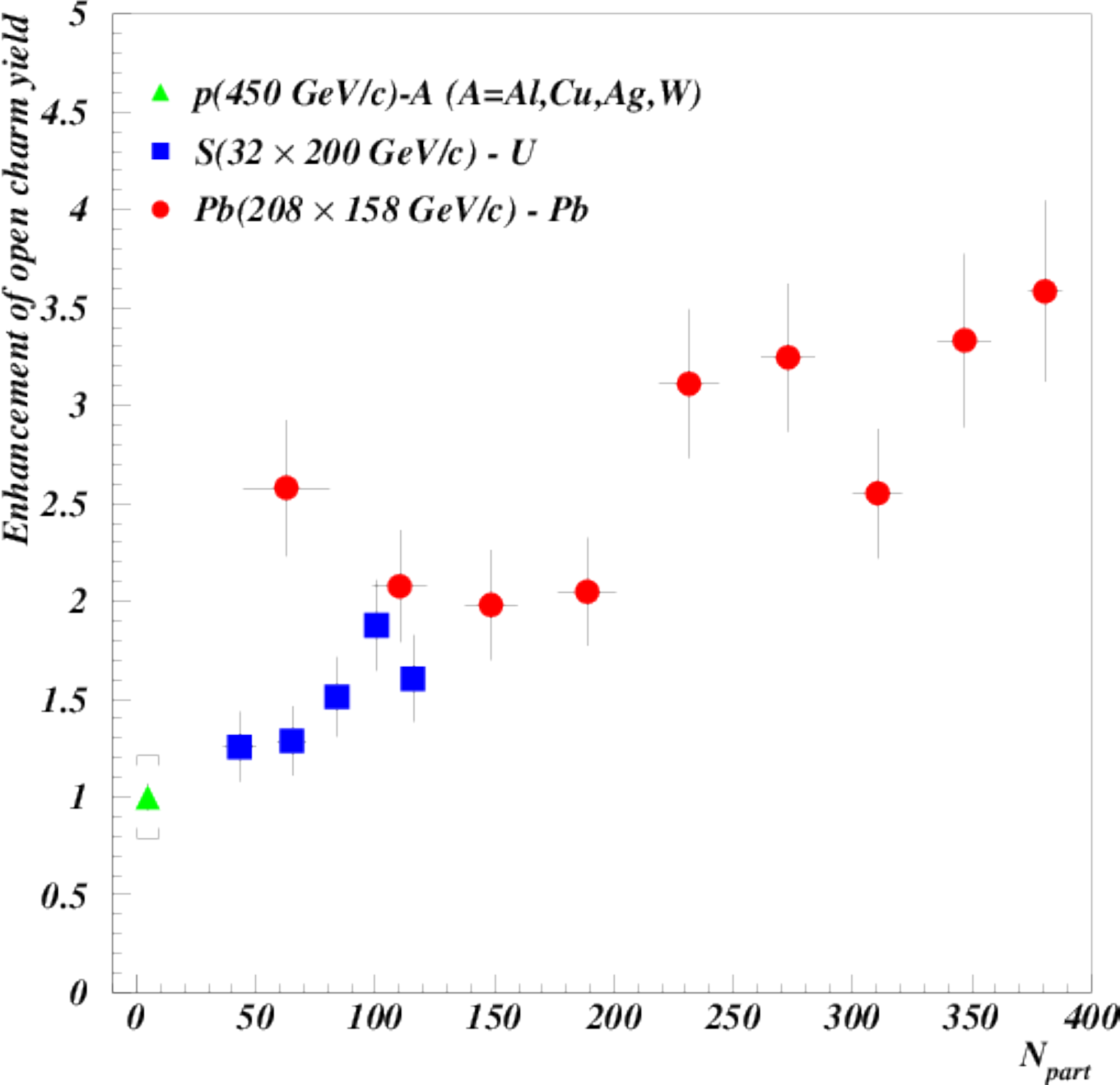}
\end{center}
\caption{Dimuon yield measured in peripheral ($\langle N_{\rm part}\rangle = 110$,
         left panel) and central ($\langle N_{\rm part}\rangle = 381$, middle 
         panel) \pbpb collisions at a beam energy of 158~GeV per nucleon in
         comparison with the contributions from expected sources. The sum 
         (solid) of Drell Yan dimuons (dashed), open charm (dotted), and the 
         dimuon decay of charmonia (dashed-dotted) is not sufficient to
         saturate the measured IMR dimuon yields. Excess relative to the
         expected open charm contribution as a function of the number of 
         participants for various collision systems (right panel) 
         (reprinted from Ref.~\cite{abreu00} with kind permission from 
         Springer Science and Business Media).}
\label{fig:na50_imr_pbpb}
\end{figure}

In contrast, in heavy-ion collisions the measured yields of IMR dimuons 
exceeded the expectation from Drell-Yan and open charm decay contributions
as demonstrated in the two left panels of Fig.~\ref{fig:na50_imr_pbpb}.
The excess, which is clearly visible in the IMR, is more pronounced in
central than in peripheral \pbpb collisions. The centrality dependence
is quantified in the right panel of Fig.~\ref{fig:na50_imr_pbpb} which
shows the excess relative to the expected dimuon yield from open charm
decays as a function of the number of participants as measured by NA38 and
NA50 in various collision systems.

Among several other potential explanations of the observed IMR dimuon excess
two interpretations were discussed as the most likely scenarios, \ie thermal
production as an additional source of IMR dimuons or an increase of the open
charm production cross section per nucleon in heavy-ion collisions compared
to p-nucleus collisions. The clarification of the origin of the IMR dimuon
excess was one of the main goals of the NA60 experiment. The NA60 setup
combined the muon spectrometer (MS) previously used by the NA38 and NA50 
experiments~\cite{abreu97} with a radiation hard silicon vertex tracker (VT) 
located in a dipole magnet providing a magnetic field of up to 
2.5~T~\cite{keil05}. The VT was positioned in between the interaction targets 
and the upstream hadron absorber in front of the MS. 
Compared to the previous experiments, the unique feature of this 
setup was the capability to match tracks of charged particles reconstructed 
in the VT and the MS both in position AND in momentum space, thus combining
the high-rate capability and selective dimuon trigger of the MS with the
excellent position and improved momentum resolution provided by the VT.
This made it possible to investigate dimuon production in In-In collisions
at a beam energy of 158~GeV per nucleon with an unprecedented precision. 
The presence of an IMR dimuon excess consistent with the earlier observations 
in \pbpb collisions was confirmed with the NA60 experiment in the In-In 
collision system~\cite{arnaldi09a,arnaldi09b}. Furthermore, it was proven that 
the excess yield does not originate from a displaced, secondary vertex but is 
consistent with prompt dimuon emission from the primary collision 
vertex~\cite{arnaldi09a,arnaldi09b}, thus ruling out an enhanced charm 
production cross section as source responsible for the excess.
Instead, the measured properties of the excess dimuons suggest that they
are related to thermal radiation from the hot QCD medium.

In summary, open heavy-flavor hadro-production in fixed target experiments
was observed to be consistent with expectations from perturbative QCD.
This includes data from fixed target heavy-ion collisions, which to 
date are available only from the SPS. Since open heavy-flavor hadrons were
never directly measured in fixed-target heavy-ion collisions it is not
precluded that such measurements can provide useful and unique information 
about the properties of strongly interacting matter in the future. In fact,
precision measurements of charm production in nucleus-nucleus collisions close 
to the production threshold is one of the key goals of the Compressed Baryonic 
Matter (CBM) experiment~\cite{cbm_book} at the future FAIR facility.

\subsection{Open heavy flavor from proton/antiproton colliders}
\label{subsec:preRHIC_collider}
Before the startup of RHIC, open heavy-flavor production was investigated
in elementary collisions at other hadron colliders giving access to much
higher $\sqrt{s}$ compared to previous fixed target experiments. For the 
discussion of medium modifications of open heavy-flavor observables in
heavy-ion collisions it is useful to investigate whether pQCD calculations
provide a theoretical baseline for heavy flavor hadro-production at high
collision energies. 

The very first open heavy flavor hadro-production measurements at a collider
date back to the experimental program at the CERN Intersecting Storage Rings 
(ISR) which is reviewed in detail elsewhere~\cite{fabjan04}.
Electrons from semileptonic charm hadron decays were observed for the first 
time in the early 1970s in \pp collisions at $\sqrt{s} = 52.7$~GeV at the 
ISR~\cite{buesser74}. However, at the time of these experiments the origin of 
single electrons measured in the range $1.6 < \pt < 4.7$~\gevc was not known 
as the charm quark was not discovered yet. Further measurements of single 
electrons and lepton pairs ($ee$ and $e\mu$) as well as full reconstructions 
of hadronic decays of open charm hadrons indicated substantial charm production
cross sections at ISR energies.

Bottom production at hadron colliders was addressed first in the late 1980s 
at the CERN Sp$\bar{\rm p}$S. The UA1 Collaboration investigated bottom 
production at high \pt ($\pt > m_b$, where $m_b$ is the bottom quark mass) 
in p$\bar{\rm p}$ collisions at $\sqrt{s} = 546$~GeV and 
630~GeV~\cite{albajar87,albajar91}. For this purpose, events containing single 
high-\pt muons ($10 < \pt^\mu < 40$~\gevc) together with a jet or events 
containing dimuon pairs ($2 < m_{\mu\mu} < 35$~\gevcsquare) were selected. 
Applying isolation cuts on the single muons and dimuon pairs, respectively, 
the contributions from bottom hadron decays to the data could be determined 
statistically. The resulting \pt-differential bottom production cross sections 
from UA1 were compared with next-to-leading-order, \ie order $\alpha_s^3$, 
pQCD calculations which were state of the art calculations at that 
time~\cite{nason88,nason89,beenakker89,beenakker91}. 
In general, a good agreement between data and pQCD calculations was observed 
within the substantial experimental and theoretical uncertainties. 
The measurement of open-charm production was difficult at the Sp$\bar{\rm p}$S.
At low \pt, no selective triggers were available, and at high \pt it was
difficult to separate charm from bottom production. The \dstarplus meson was 
the only charmed hadron which could be directly reconstructed with the UA1 
experiment in hadronic decays at high \pt in jets~\cite{albajar90}. With the 
UA2 experiment single electrons from semileptonic open charm hadron decays 
were measured in the transverse momentum range 
$0.5 < \pt < 2.0$~\gevc~\cite{botner90}. Only ten such events were observed in 
this experiment which was the first to use a RICH detector for electron 
identification at a collider. A detailed review of heavy-quark production
at the CERN Sp$\bar{\rm p}$S can be found elsewhere~\cite{ellis90}

\begin{figure}[t]
\begin{center}
\includegraphics[width=0.50\textwidth]{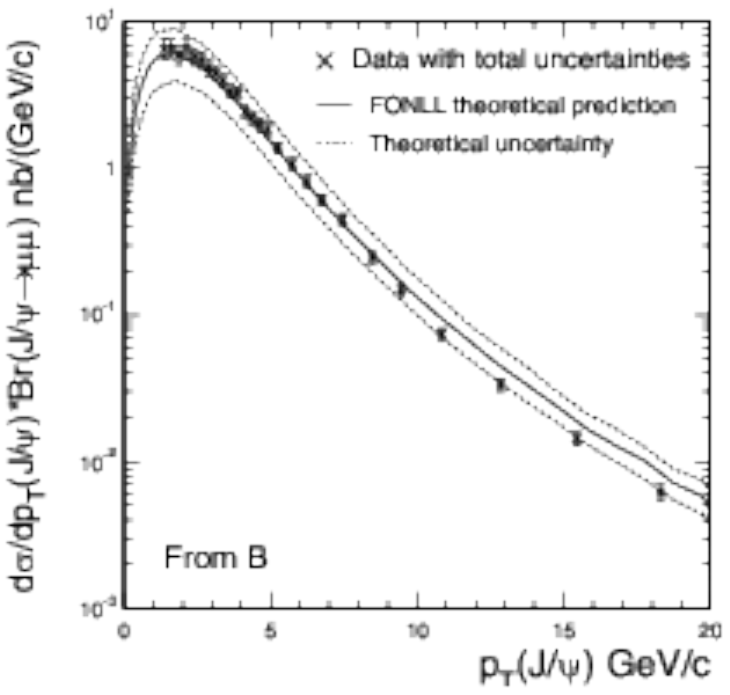}
\includegraphics[width=0.49\textwidth]{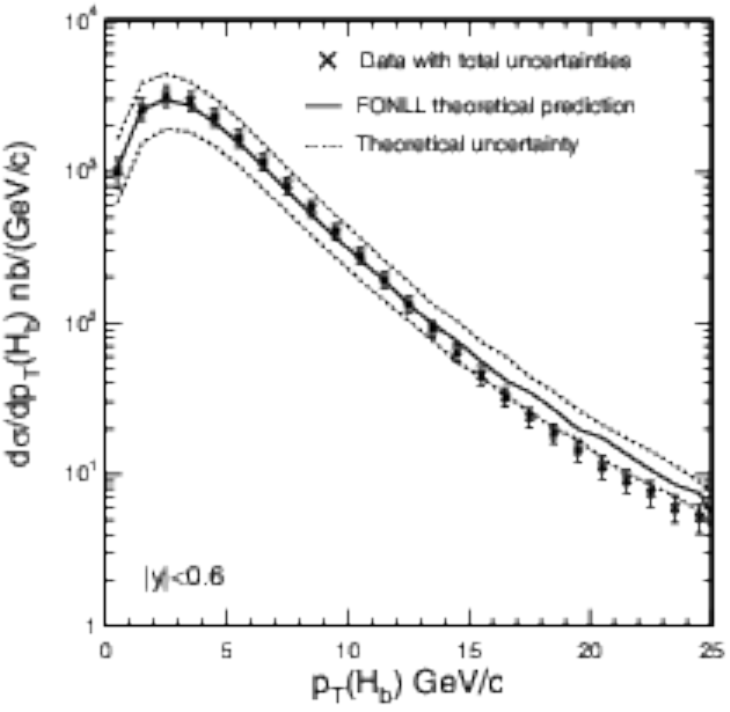}
\end{center}
\caption{\pt-differential cross section of \jpsi mesons originating from
         the decay of bottom hadrons integrated over $|y| < 0.6$ as measured
         with the CDF experiment in $\rm{p}\bar{\rm{p}}$ collisions at
         \s$ = 1.96$~TeV (left panel) and the corresponding bottom hadron
         \pt differential cross section determined from the \jpsi data 
         (right panel). The plotted \jpsi cross section includes the
         branching ratio of \jpsi mesons in the dimuon decay channel.
         FONLL pQCD calculations are compared with the data. The solid
         lines show the central theoretical values and the dashed lines
         indicate the theoretical uncertainties (reprinted with permission 
         from Ref.~\cite{acosta05}; Copyright (2005) by the American Physical 
         Society).}
\label{fig:bottom_tevatron}
\end{figure}

Open heavy-flavor production measurements, mainly focusing again on bottom
production, were conducted in the 1990s at the Fermilab Tevatron with the
CDF and D0 experiments in $\rm{p}\bar{\rm{p}}$ collisions first at 
\s$ = 1.8$~TeV (Run-I) and later at \s$ = 1.96$~TeV (Run-II). The CDF
Collaboration reported the first full reconstruction of B mesons (using
the decay ${\rm B}^\pm \rightarrow {\rm J}/\psi {\rm K}^\pm$) at high \pt (above
9~\gevc) and at midrapidity ($|y| < 1.0$) at a hadron collider~\cite{abe92}. 
Further bottom measurements with the CDF experiment employing 
electrons~\cite{abe93a} and muons~\cite{abe93b} from semileptonic bottom decays 
as well as \bzero mesons in the exclusive decay channel 
${\rm B}^0 \rightarrow {\rm J}/\psi {\rm K}^*(892)$~\cite{abe94} seemed to 
have established that bottom production cross sections measured in the 1.8~TeV
Run-I at the Tevatron exceeded the expectation from pQCD calculations.
While the first bottom-production measurements with the D0 experiment, based on 
single high-\pt muons at midrapidity, were in reasonable agreement with NLO 
pQCD calculations~\cite{abachi95}, later higher statistics measurements of 
single muons and dimuons~\cite{abbott00a,abbott00b}, and \jpsi mesons from 
displaced, secondary vertices at midrapidity~\cite{abachi96} followed the 
trend suggested by the CDF experiment that bottom was produced in excess of 
the yields expected from NLO pQCD calculations of heavy quark production. 
This apparent discrepancy between measured bottom production cross sections 
in $\rm{p}\bar{\rm{p}}$ collisions and corresponding NLO pQCD calculations 
posed a long standing puzzle. 

\begin{figure}[t]
\begin{center}
\includegraphics[width=0.99\textwidth]{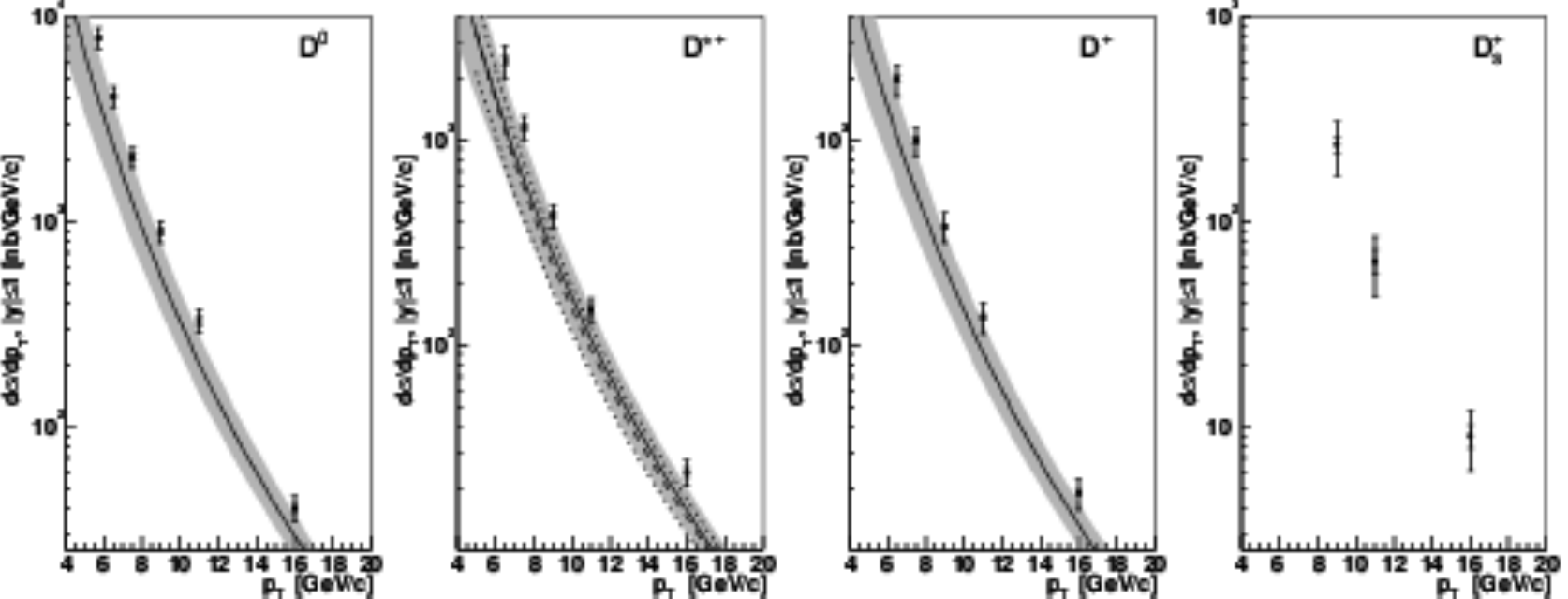}
\includegraphics[width=0.24\textwidth]{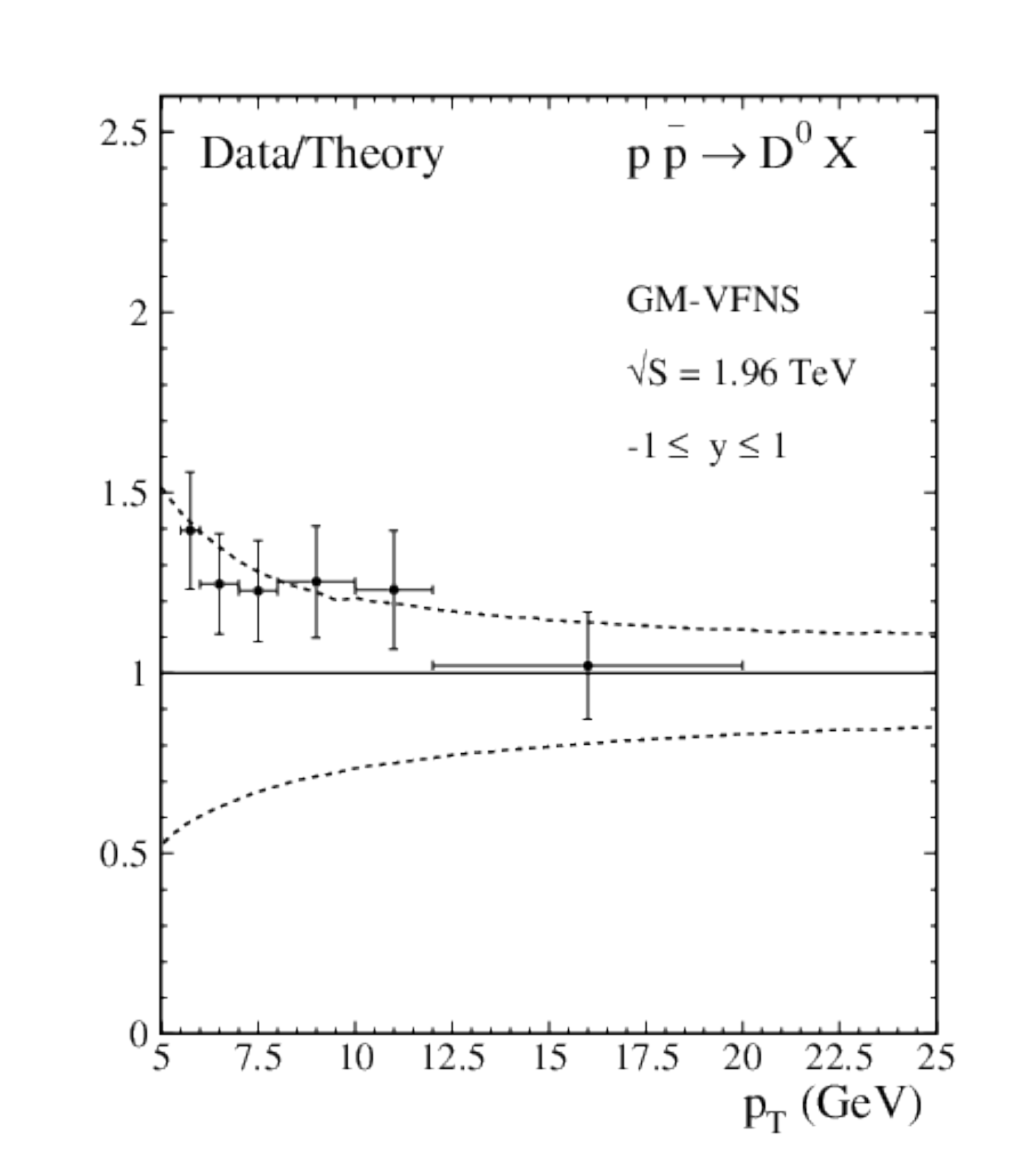}
\includegraphics[width=0.24\textwidth]{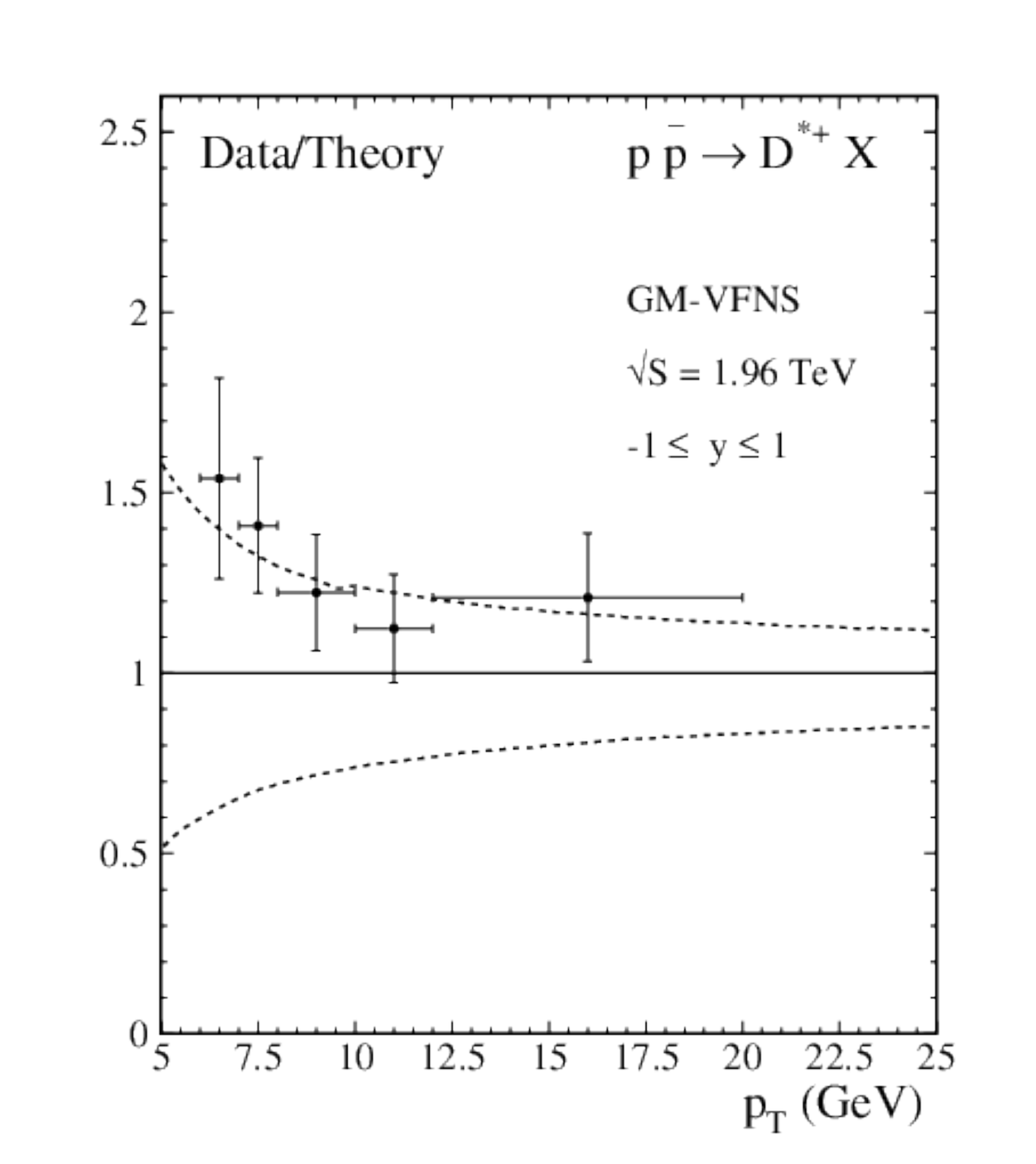}
\includegraphics[width=0.24\textwidth]{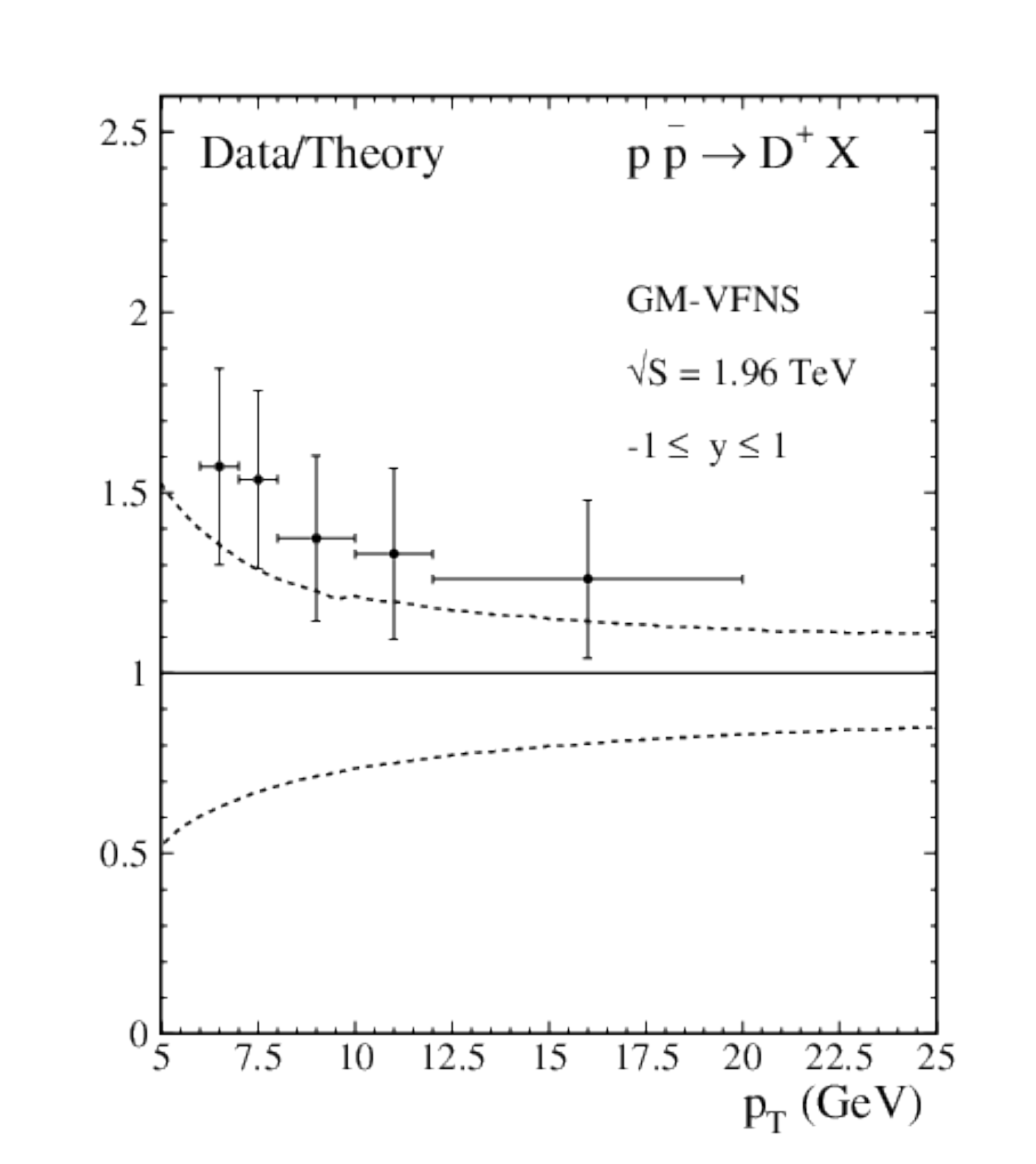}
\includegraphics[width=0.24\textwidth]{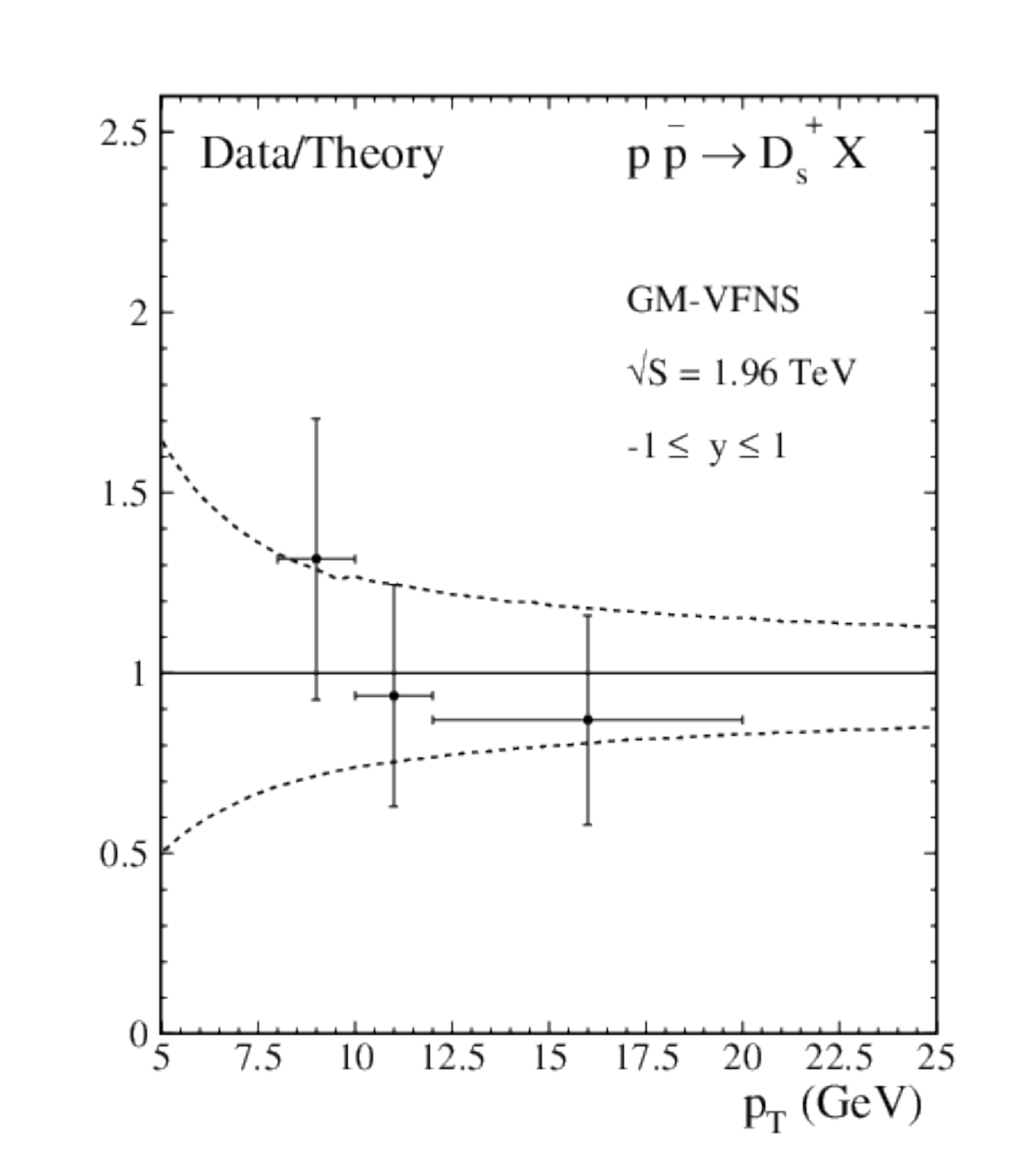}
\end{center}
\caption{\pt-differential cross sections of various D-meson species measured
         with the CDF experiment in $\rm{p}\bar{\rm{p}}$ collisions at
         \s$ = 1.96$~TeV in the rapidity range $|y| \le 1$.
         The solid curves are theoretical predictions from FONLL pQCD 
         calculations with the shaded bands indicating the theoretical 
         uncertainties. Predictions for the \dsplus cross section are not 
         available from FONLL. 
         Ratios of the measured cross sections and theoretical predictions
         from GM-VFNS pQCD calculations as functions of \pt (reprinted with 
         permission from Ref.~\cite{acosta03} (upper four panels) and 
         Ref.~\cite{kniehl06} (lower four panels); Copyrights (2003, 2006) 
         by the American Physical Society).
}
\label{fig:charm_tevatron}
\end{figure}

As it is documented in detail elsewhere~\cite{cacciari04,mangano05} the 
Tevatron bottom puzzle was finally resolved with a more careful comparison 
of experimental and theoretical observables and with the availability of much 
improved data from the Tevatron Run-II at \s$ = 1.96$~TeV. In this context, 
the most important upgrades of the CDF and D0 experiments were related to the 
trigger systems selecting the events to be recorded for further analysis. Both 
experiments included silicon vertex detectors in their trigger schemes in 
Run-II, which allowed them to trigger on the displaced secondary vertices of 
open heavy-flavor hadron decays. Due to bandwidth limitations the D0 displaced
vertex trigger could operate only on preselected events that contained an
additional muon. The high-bandwidth displaced vertex trigger of the CDF
experiment provided unique access to the measurement of fully hadronic
open heavy-flavor hadron decays. Employing this trigger, with the CDF 
experiment the \pt-differential production cross section of ${\rm J}/\psi$
from open bottom hadron decays as well as the underlying bottom hadron
production cross section was measured down to low \pt in 
$\rm{p}\bar{\rm{p}}$ collisions at \s$ = 1.96$~TeV~\cite{acosta05}. As shown in 
Fig.~\ref{fig:bottom_tevatron} this measurement is in excellent agreement 
with FONLL pQCD calculations as are all further open bottom hadron 
production results from Run-II at the Tevatron.

The CDF silicon vertex trigger did not only allow for a measurement of 
open bottom production in a unique way at the Tevatron Run-II but it 
gave also access to open charm production. The \pt-differential production 
cross sections of open charm mesons were measured~\cite{acosta03}, at high \pt 
and at midrapidity ($|y| \le 1$), in the fully reconstructed hadronic decays
\dzero$\rightarrow K^- \pi^+$, \dstarplus$\rightarrow$\dzero$\pi^+$,
\dplus$\rightarrow K^- \pi^+ \pi^+$, \dsplus$\rightarrow \phi \pi^+$,
and their charge conjugates. The measured cross sections are shown in
the upper four panels of Fig.~\ref{fig:charm_tevatron}. Results from 
FONLL pQCD calculations~\cite{fonll1,fonll2} are plotted in comparison with 
the data except for the case of the \dsplus which is not considered explicitly 
in FONLL. While the calculations agree with the data within the experimental 
and theoretical uncertainties, a general trend is observed that the measured 
cross sections lie at the upper edge of the calculated cross sections. 
This is further demonstrated in the lower four panels of 
Fig.~\ref{fig:charm_tevatron} which show the ratios of the measured 
cross sections and corresponding pQCD calculations employing the
GM-VFNS approach~\cite{kniehl06}.

In summary, open heavy-flavor hadro-production measurements conducted
at pp and $\rm{p}\bar{\rm{p}}$ colliders before the startup of RHIC can 
be reasonably well described by state of the art pQCD calculation.
Given the significant theoretical uncertainties and the fact that open
heavy-flavor measurements have been proven to be difficult, it is crucial 
to establish an experimental open heavy-flavor production baseline from 
elementary pp collisions at RHIC, at the LHC, and at future accelerators
investigating heavy-flavor hadro-production. Furthermore, such experimental
baselines have to be confronted with pQCD calculations before open heavy-flavor 
hadrons can be used with confidence as probes for the medium produced in 
ultra-relativistic heavy-ion collisions.

\section{Open heavy-flavor measurements at RHIC}
\label{sec:rhic}
\subsection{Experiments studying heavy-flavor production at RHIC}
\label{subsec:exp_rhic}
RHIC delivers heavy-ion beams since the year 2000. Two of the four heavy-ion
experiments at RHIC, PHENIX and STAR, have conducted systematic open 
heavy-flavor measurements since then, mainly in \pp and \auau collisions at 
$\snn = 200$~GeV, where substantial data sets have been recorded. In addition, 
data from \dau collisions, which give access to potential cold nuclear matter 
effects, and from \cucu collisions, which help in the investigation of the 
system size dependencies of open heavy-flavor observables, were investigated.

The results reviewed in this article originate from published measurements
with the PHENIX and STAR experiments in their configurations prior to the 
years 2008 and 2010, respectively. Since then, both PHENIX and STAR have 
embarked on ambitious detector upgrade programs. Key components in both 
upgrade projects are precision vertex spectrometers which will allow the 
separation of tracks originating from displaced, secondary open heavy-flavor 
hadron decay vertices from primary tracks pointing back to the collision 
vertex. These improvements of the experimental setups go hand in hand with 
an upgrade of the RHIC accelerator facility aiming at an increase of the 
original \auau luminosity by a factor of five. The open heavy-flavor physics 
potential of the detector and accelerator upgrades is discussed in detail 
elsewhere~\cite{frawley08}. In the following, the experimental status quo 
of the PHENIX and STAR experiments as of the years 2007 and 2009, respectively,
is briefly summarized.

\subsubsection{The PHENIX approach to open heavy-flavor measurements}
\label{subsubsec:phenix}
\begin{figure}[t]
\begin{center}
\includegraphics[width=0.49\textwidth]{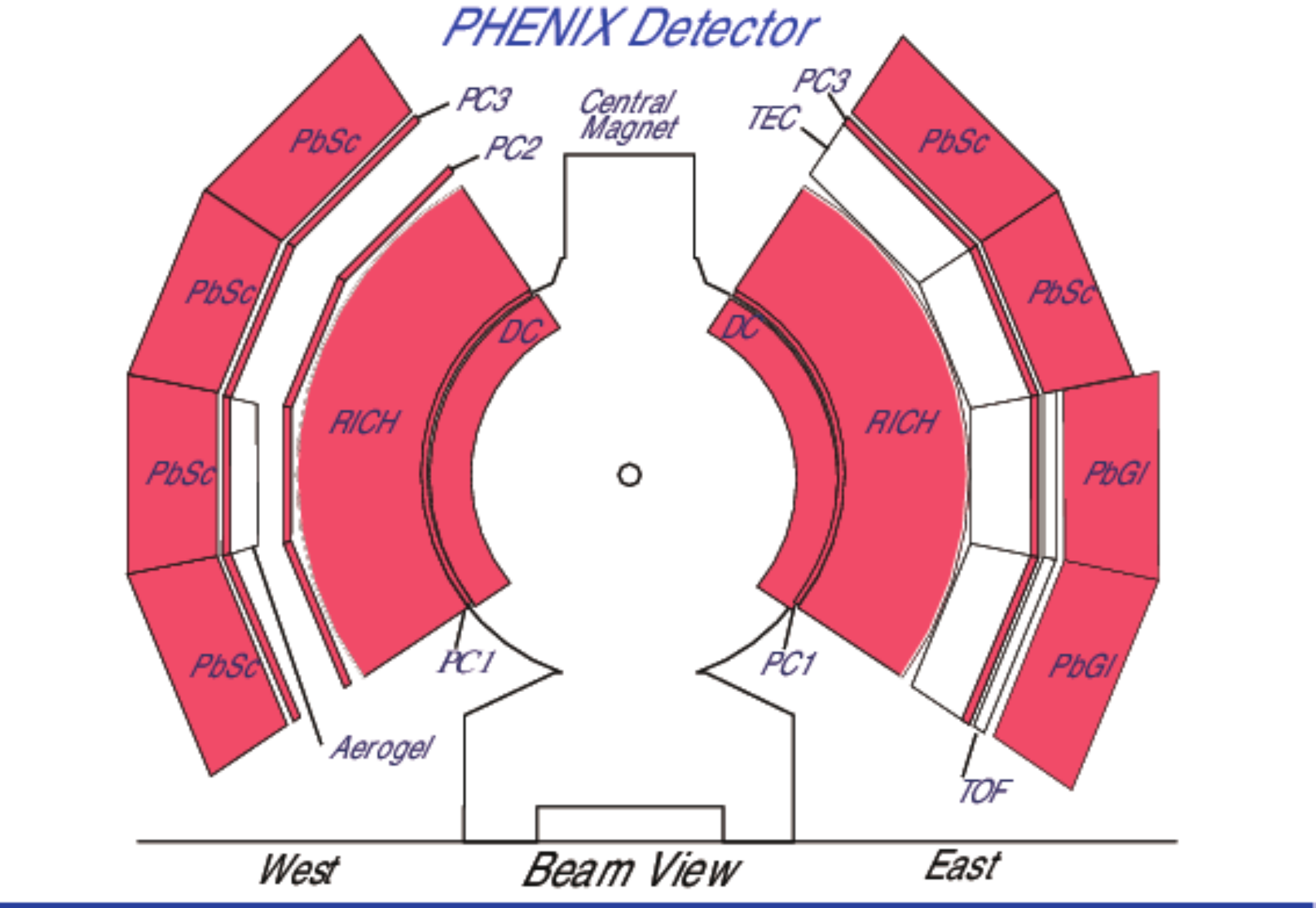}
\includegraphics[width=0.49\textwidth]{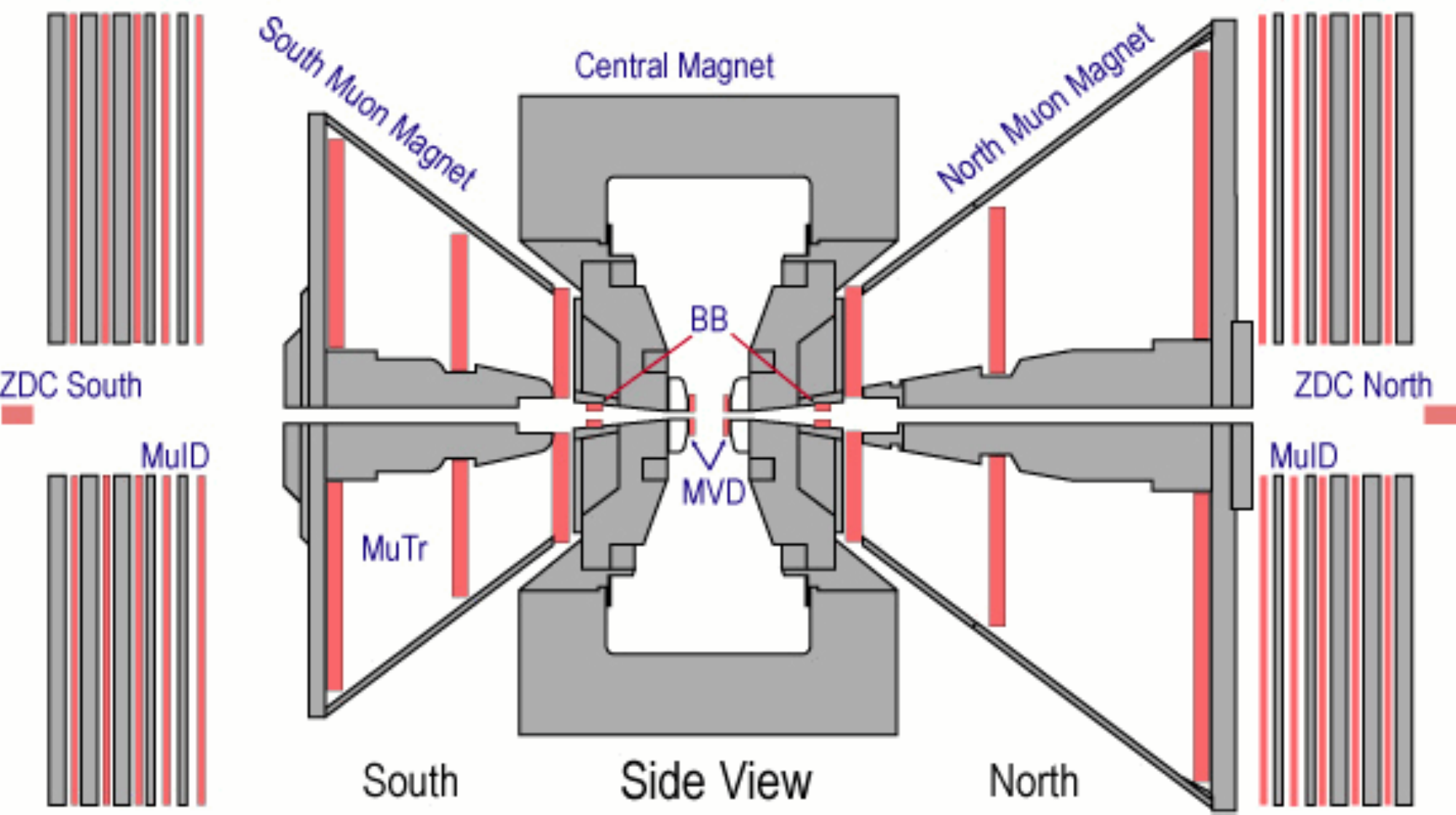}
\end{center}
\caption{Schematic beam view (left panel) and side view (right panel) of the 
         PHENIX apparatus in the year 2005. Detectors relevant for the
         heavy-flavor measurements discussed here are the drift chambers (DC)
         and the pad chambers (PC) for tracking, and the Cherenkov detectors 
         (RICH) and the electromagnetic calorimeters (PbGl, PbSc) for electron
         identification in the central arms. The relevant detectors in the
         forward and backward muon spectrometers are the muon trackers
         (MuTr) and the muon identifiers (MuID). Global beam-beam counters
         (BBC) and zero-degree calorimeters (ZDC) are used for event
         characterization (from Ref.~\cite{phenix_setup}).}
\label{fig:phenix}
\end{figure}

The PHENIX experiment comprises four spectrometers and a set of global 
detectors for event characterization.
The two central spectrometer arms, each covering $90^\circ$ in azimuth, 
allow the measurement of electrons, photons, and charged hadrons in the 
pseudorapidity range $|\eta| < 0.35$. The momenta of charged particles are 
measured in the central arms using drift chambers and pad chambers located 
in a magnetic field which is oriented in parallel to the beam axis around 
the interaction vertex. For particle identification, time-of-flight detectors, 
RICH detectors, and electromagnetic calorimeters are employed in the central 
spectrometers. In the forward and backward regions, two muon spectrometers 
cover the pseudorapidity ranges $-2.2 < \eta < -1.2$ and $1.2 < \eta < 2.4$ 
over the full azimuth. Both muon spectrometers comprise a muon tracker with 
three stations of multi-plane drift chambers placed in a radial magnetic 
field, and a muon identifier consisting of alternating layers of streamer 
tubes and steel absorbers. As global detectors, beam-beam counters and 
zero-degree calorimeters are located at forward and backward rapidity, and 
they serve a variety of purposes. With these detectors the collision time 
and vertex position along the beam direction is determined. Furthermore, the 
first level, minimum bias trigger signals for collisions are derived from 
the BBC and the ZDC. In addition, they are used to determine the collision 
centrality and the azimuthal orientation of the reaction place in 
nucleus-nucleus collisions. Schematic beam and side views of the PHENIX 
experimental setup are shown in Fig.~\ref{fig:phenix}.  A detailed description 
of the PHENIX apparatus can be found elsewhere~\cite{phenix_nim}.

In the PHENIX configuration discussed here open heavy-flavor hadrons are not
reconstructed exclusively via their hadronic decays. Instead, electrons from 
heavy-flavor hadron decays are measured with the central spectrometer arms 
around mid-rapidity in \pp and nucleus-nucleus collisions. Both the transverse 
momentum spectra of single electrons and the dielectron mass spectra are 
inspected for contributions from semileptonic heavy-flavor hadron decays. 
Correspondingly, with the forward/backward muon spectrometers single muons 
from heavy-flavor hadron decays are measured in \pp and nucleus-nucleus 
collisions. Both in the electron and in the muon measurement, the subtraction 
of background from sources other than heavy-flavor hadron decays poses the most 
important experimental challenge. 

In case of the electron measurements explained in detail in Ref.~\cite{ppg077},
remaining background from misidentified hadrons does not play a significant 
role since the combination of Cherenkov detectors and electromagnetic 
calorimetry guarantees the selection of a rather clean electron candidate 
sample. From the measured inclusive single electron \pt distributions electron 
background from non-heavy-flavor sources has to be subtracted. The most 
important components of this so-called background cocktail, which is calculated
using a MonteCarlo hadron decay generator, are Dalitz decays of the $\pi^0$ 
meson and the conversion of photons in detector material, where the majority 
of photons originates from the $\pi^0 \rightarrow \gamma\gamma$ decay. 
It is important to note that, in the PHENIX experiment, special emphasis is 
put on the minimization of the material budget in the detector aperture, 
thus minimizing the contribution from photon conversions to the measured 
electron spectra. With a relevant material budget
of less than 1\% of a radiation length, yield ratios of electrons from photon 
conversion to electrons from light meson Dalitz decays significantly 
below one are achieved in the PHENIX experiment. Similar Dalitz decay and 
conversion contributions from other light mesons, notably the $\eta$ meson, 
are also considered in the electron background cocktail. Furthermore, 
dielectron decays of light vector mesons ($\rho$, $\omega$, $\phi$) are 
included although these contributions played a negligible role only. At low
\pt, a minor contribution from weak kaon decays ($K_{\rm e3}$) has to be added 
to the background cocktail. Towards high electron \pt (above 5~\gevc) hard
quark-gluon Compton scattering processes, contributing to the electron
background via virtual photon production and via the conversion of direct 
real photons in material, play an increasingly important role. In this 
high \pt region decays of heavy quarkonia (\jpsi, \ups), and electrons from
the Drell-Yan process can not be ignored either, in fact these sources 
represent about one third of the background electron yield in the range
$5 < \pt < 10$~\gevc in \pp and \auau collisions at 
$\sqrt{s_{\rm NN}} = 200$~GeV at RHIC. In the first electron measurements with
PHENIX, which did not reach high electron \pt with good statistics, 
contributions from \jpsi and \ups decays as well as from the Drell-Yan
process were not included in the electron background 
cocktail~\cite{ppg011,ppg035,ppg037,ppg056,ppg065,ppg066}. Only with
the availability of more precise data these sources were added to the 
cocktail~\cite{ppg077}.
The background cocktail from these sources was subtracted from the inclusive
electron sample and the difference was identified with electrons from
open heavy-flavor hadron decays. A basic requirement for this cocktail 
subtraction method was the availability of the yields and phase space 
distributions of all relevant background electron sources, ideally from a 
measurement in the same apparatus. In addition, the applicability of the 
cocktail background subtraction method was in general limited to the electron 
\pt range in which the ratio of signal electrons from heavy-flavor hadron 
decays to background electrons from other sources was larger than the combined 
systematic uncertainty of the inclusive electron measurement and the 
calculated background cocktail. Since the signal-to-background ratio 
decreases with decreasing electron \pt, at low \pt (typically below 2~\gevc 
in the PHENIX case) it became advantageous to switch from the cocktail 
subtraction to the so-called converter subtraction method to determine the 
yield of electrons from heavy-flavor hadron decays. In this complementary 
approach, additional converter material with a well defined thickness was 
added to the experimental setup for a fraction of the running time of the 
experiment. The converter multiplied the yield of ``photonic'' electrons, 
\ie electrons from photon conversions and from Dalitz decays of light mesons, 
by a known factor, thus providing a measurement of this background contribution 
which was dominant at low electron \pt. A detailed description of the 
measurement of single electrons from heavy-flavor hadron decays with
the PHENIX experiment using the cocktail and converter subtraction 
techniques can be found elsewhere~\cite{ppg077}.

A disadvantage of single lepton measurements is the fact that contributions
from charm and bottom hadron decays can not be separated from each other 
without additional experimental information. Electron-hadron correlation
measurements can provide this missing information because electrons from 
heavy-flavor hadron decays are always emitted together with hadrons. Since 
the correlations are not the same in charm and bottom hadron decays they 
can be used to disentangle the corresponding relative contributions to the 
electron spectra from heavy-flavor hadron decays~\cite{ppg094}.

The contributions to the electron-positron mass spectrum from simultaneous 
semielectronic decays of correlated pairs of hadrons carrying heavy quarks 
and antiquarks, respectively, was determined with the PHENIX experiment in
a cocktail approach as well~\cite{ppg085,ppg088}. Except for photon 
conversions, the same background sources as listed above for the case of single
electrons were considered for the calculation of a background cocktail. After 
the subtraction of this cocktail background contributions from correlated 
heavy-flavor hadrons decays became accessible.

Also for the measurement of muons from semileptonic heavy-flavor hadron
decays in the forward and backward muon spectrometers of the PHENIX
experiment the subtraction of background was crucial~\cite{ppg117,ppg057}. 
In this case, three background sources were identified as the most important 
ones. First, decay muons from light hadrons decaying before reaching the
first absorber material could not be distinguished experimentally from
muons from heavy-flavor hadron decays. Second, punch-through hadrons 
produced at the collision vertex could penetrate through the MuID layers
and, therefore, were included in the inclusive muon candidate sample.
Third, muons from the decay of hadrons, which penetrated the muon arm 
front hadron absorber and decayed in the muon tracker volume, contributed
to the inclusive muon candidate sample. The overall background from
light hadron sources was quantified via MonteCarlo simulations. The
contribution from decay muons could be subtracted statistically. This was
possible because the yield of light hadron decay muons exhibited a pronounced 
vertex position dependence along the beam direction which was not the case
for the other sources. Again, a detailed description of the experimental 
techniques used to isolate the contribution from open heavy-flavor hadron 
decays to the measured single muon spectra can be found 
elsewhere~\cite{ppg117,ppg057}.

\subsubsection{The STAR approach to open heavy-flavor measurements}
\label{subsubsec:star}
\begin{figure}[t]
\begin{center}
\includegraphics[width=0.49\textwidth]{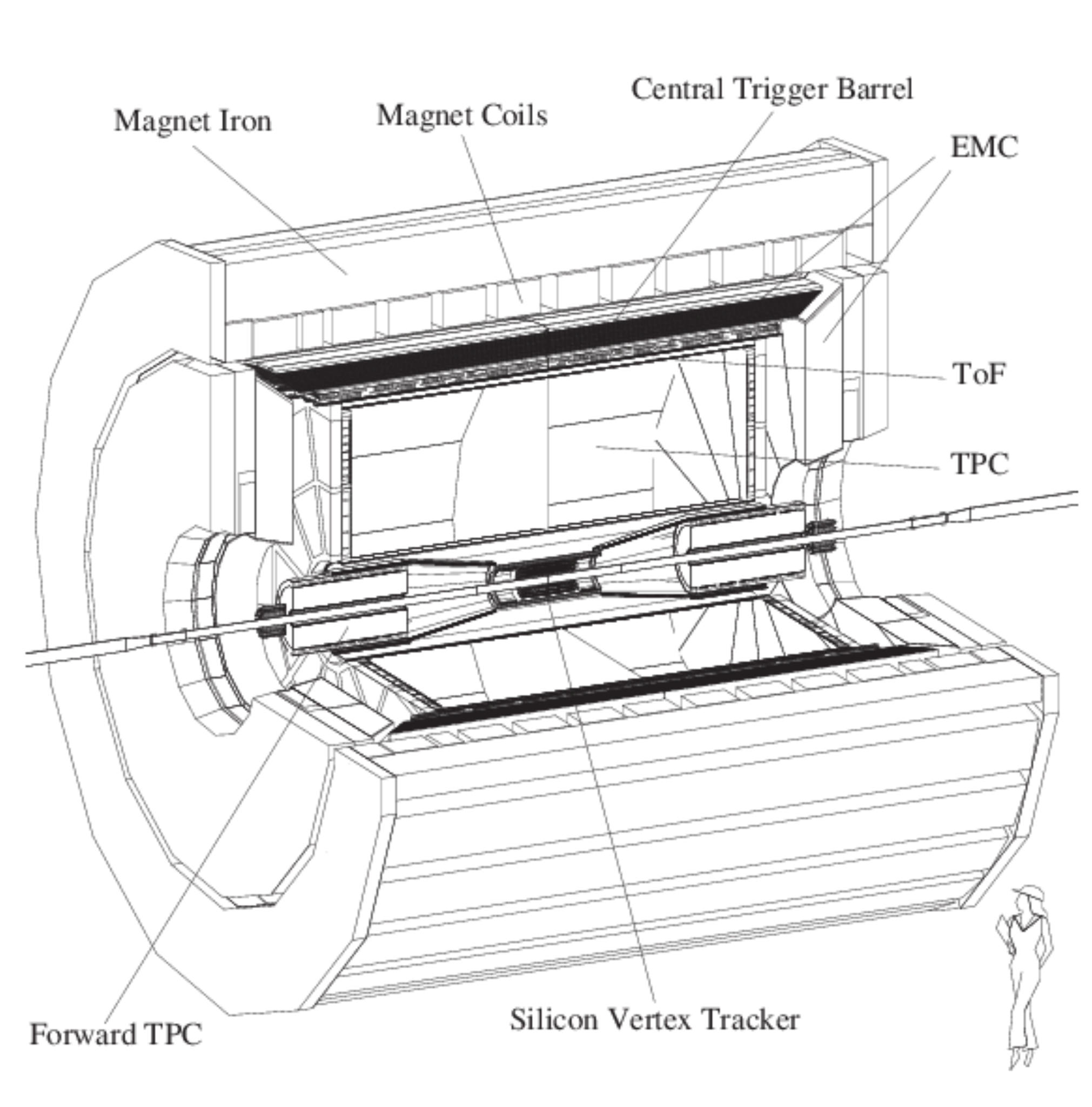}
\end{center}
\caption{Schematic view of the STAR detector. Charged particles tracks are 
         reconstructed around mid-rapidity in a large volume time projection
         chamber (TPC), which also contributes to the particle identification
         via the measurement of the charged particles' specific energy loss
         \dedx. For additional particle identification the TPC is surrounded 
         by a Time-of-Flight barrel (ToF) and an electromagnetic calorimeter
         (EMC) (from Ref.~\cite{star_setup}).}
\label{fig:star}
\end{figure}

The subdetectors of the STAR experiment relevant in this context were housed 
inside a solenoid magnet providing a field of 0.5~T. Charged particle tracking 
relied mostly on a large azimuthally symmetric time projection chamber (TPC) 
which covered the pseudorapidity range $|\eta| < 1.8$. For tracking
close to the interaction point and for vertexing the TPC originally was 
augmented by a silicon vertex tracker (SVT). This device, however, could not
be used for a clean separation of decay vertices from open heavy-flavor 
hadron decays from the primary collision vertex and, hence, it was removed
from the STAR setup prior to the RHIC run in the year 2008 to reduce
the material budget in front of the TPC inner field cage by a factor of
$\approx 10$. For particle identification, the specific energy loss 
measured in the TPC could be combined with velocity information from a 
Time-of-Flight barrel which employed multigap resistive plate technology. For 
electron identification, a barrel electromagnetic calorimeter (EMC) was 
available in addition. This device included a preshower and a shower 
maximum detector. As in the PHENIX case, beam-beam counters and zero degree 
calorimeters could be used to provide global event characteristics.
A schematic view of the STAR experimental setup is shown in 
Fig.~\ref{fig:star} and a detailed description of the STAR apparatus can 
be found elsewhere~\cite{star_nim}.

It is only with the STAR experiment that heavy-flavor hadrons have been
fully reconstructed at RHIC via an invariant mass analysis of their hadronic 
decay products. The main difficulty of such measurements is the 
substantial combinatorial background from other uncorrelated hadrons.
With increasing charged particle multiplicity the resulting 
signal-to-background ratio becomes worse such that the difficulty of such
a measurement increases from \pp to proton/deuteron-nucleus to nucleus-nucleus
collisions in which signal-to-background ratios of typically 1:1000 had to
be coped with in the STAR experiment. With STAR, \dzero mesons have been 
reconstructed in their K$\pi$ decay channel in \dau~\cite{star_d_dau} and, 
recently, in \pp collisions~\cite{star_d_pp}. A corresponding signal was 
observed in \auau collisions at low \pt as well~\cite{star_d_auau}. In addition
to the \dzero also the \dstarplus meson has been reconstructed via its 
D$^0\pi^+$ decay in \pp~\cite{star_dstar_pp} but not in nuclear collisions. 
The only possibility to increase the signal-to-background ratio is the clean 
separation of the secondary decay vertex from the primary collision vertex 
which has not been achieved with the experimental setup of the STAR experiment 
discussed here.

Heavy-flavor measurements in \auau collisions were conducted with the STAR 
experiment mainly via the semielectronic decays of heavy-flavor 
hadrons~\cite{star_e_auau}, which were also investigated for reference in the 
\pp collision system~\cite{star_e_auau,star_e_pp}. 
Electrons were identified via their specific energy loss \dedx in the STAR TPC,
their velocity as measured with the Time-of-Flight detector system, and their 
response in the electromagnetic calorimeter. A remaining contamination from
misidentified hadrons in the electron sample was quantified using the TPC \dedx
measurement, and it was subtracted statistically. As in the PHENIX case, the 
spectra of electrons from open heavy-flavor decays were obtained via the 
subtraction of electron background from the measured inclusive electron 
spectra. Originally, the electron background in the STAR experiment was 
completely dominated by photon conversions in detector material, in particular 
in the SVT. With the removal of this detector for the 2008 run at RHIC, the 
relevant material budget was reduced by a factor $\approx$10 to just above 
1\% of a radiation length, which was close to the material budget in the 
PHENIX setup. The background from photon conversions and Dalitz decays was, 
however, determined distinctively different within STAR compared to the PHENIX 
cocktail and converter subtraction approaches. Electron-positron pairs
from the relevant background sources have a very small invariant mass, below
the $\pi^0$ mass. In contrast to the PHENIX experiment, the large acceptance 
STAR apparatus provided a reasonably large efficiency ($\approx60$\%) for the 
reconstruction of these low mass pairs, such that this contribution to the 
background could be measured and subtracted from the inclusive electron 
spectrum. Not all of this so-called ``photonic background'' could be 
identified with this technique because one of the electrons might have been 
outside of the detector acceptance or had a momentum to small to be measured. 
In this case both electrons from the pair were not reconstructed as part of 
the photonic background. This inefficiency had to be determined precisely 
through simulations. Contributions from other background sources, \eg 
dielectron decays of vector mesons which were not measured directly, were 
calculated in a cocktail approach. The difference between the measured 
inclusive electron spectrum and the resulting total background from photonic 
and other sources was identified with the spectrum of electrons from open 
heavy-flavor hadron decays. 

In \pp collisions the azimuthal angular correlations of electrons from open
heavy-flavor hadron decays with hadrons or \dzero mesons was investigated
with the STAR experiment~\cite{star_e_b_d}. These procedures allowed a 
statistical separation of electrons from charm and bottom hadron decays, 
similar to the approach chosen by PHENIX.

While the STAR experiment does not include a dedicated muon spectrometer
muons could be identified at very low \pt (close to 0.2~\gevc) via their time
of flight and their ionization energy loss in the TPC~\cite{star_d_auau}. 
Muons from heavy-flavor hadron decays could be separated statistically from 
the dominant background from weak pion and kaon decays via the measurement of 
the distance of closest approach of the tracks to the primary collision vertex.
The resulting yield of low-\pt muons from heavy-flavor hadron decays provided 
some sensitivity to the total open charm cross section~\cite{star_d_auau}.

\subsection{Reference measurements in \pp collisions}
\label{subsec:rhic_pp}

\subsubsection{Overview}
The open heavy-flavor measurements with the PHENIX and STAR experiments in 
\pp collisions at $\sqrt{s} =200$~GeV at RHIC aim at two goals. First, these
data provide a crucial experimental reference for corresponding measurements
in nucleus-nucleus collisions. Such a reference is a key ingredient for the
investigation and interpretation of effects due to the presence of a hot
and dense medium produced in high-energy nucleus-nucleus collisions.
Second, these data provide an important testing ground for perturbative QCD 
calculations which, due to the large masses of the charm and bottom quarks,
should be able to predict open heavy-flavor observables even at low \pt.

As will be demonstrated in this section, predictions from NLO pQCD calculations 
within the FONLL approach~\cite{fonll1,fonll2} are in reasonable agreement 
with essentially all observables related to the production of open heavy-flavor
hadrons in \pp collisions at RHIC.

\subsubsection{Hadronic heavy-flavor hadron decays}
\label{subsubsec:rhic_d_pp}
\begin{figure}[t]
\begin{center}
\includegraphics[width=0.49\textwidth]{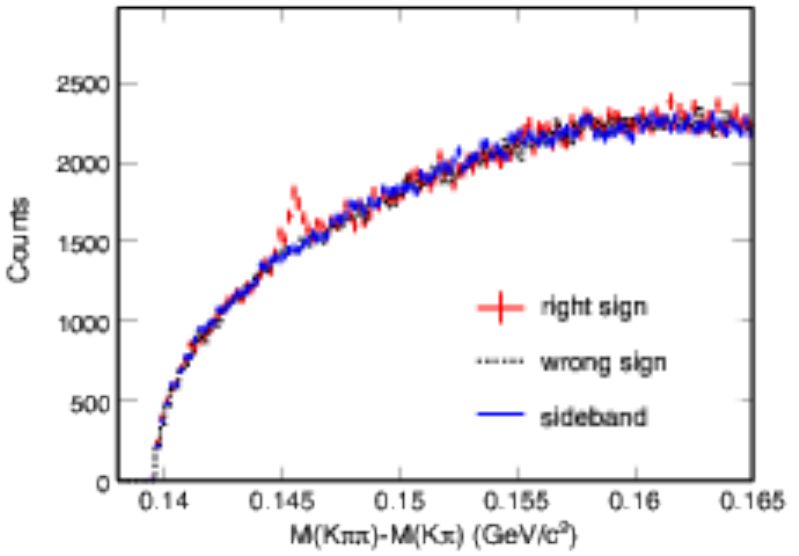}
\includegraphics[width=0.49\textwidth]{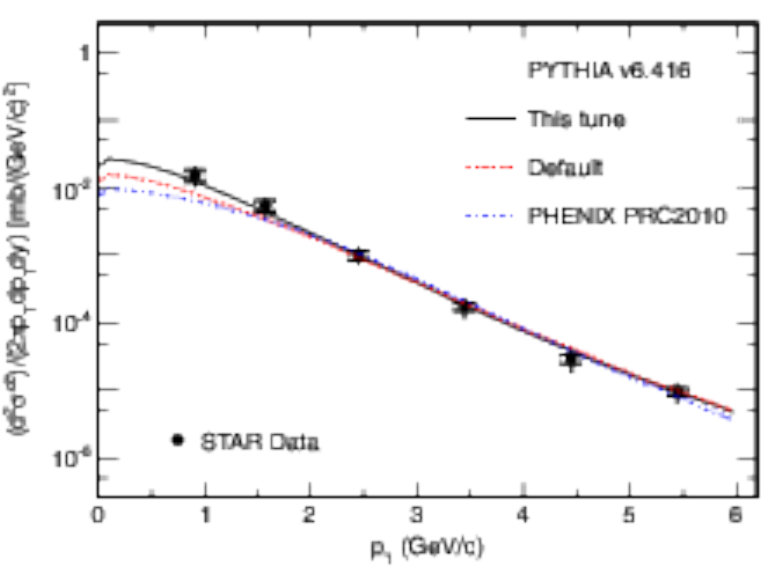}
\end{center}
\caption{Raw \dstarplus candidate signal from minimum bias \pp collisions
         at $\sqrt{s} = 200$~GeV as measured with the STAR experiment. 
         (left panel) Results from two methods to quantify the combinatorial 
         background are shown in addition (see text for details). $\ccbar$ 
         production cross section as deduced from the measured \dzero and 
         \dstarplus cross section (right panel). A power-law fit and the
         \pt differential charm cross section from a FONLL pQCD calculation 
         are shown for comparison (reprinted with permission from 
         Ref.~\cite{star_d_pp}; Copyright (2012) by the American Physical
         Society).}
\label{fig:star_d_pp}
\end{figure}

The only full reconstruction of hadronic decays of heavy-flavor hadrons in 
\pp collisions at 200~GeV at RHIC was the measurement of the 
${\rm D}^0 \rightarrow {\rm K}^- \pi^+$ decay in the range 
$0.6 < \pt < 2.0$~\gevc and the 
${\rm D}^{*+} \rightarrow {\rm D}^0 \pi^+$ decay in the range
$2.0 < \pt < 6.0$~\gevc with the STAR experiment~\cite{star_d_pp}. As an 
example, the left panel of Fig.~\ref{fig:star_d_pp} indicates the quality of 
the \dstarplus reconstruction. For minimum bias events, the distribution of 
the invariant mass difference 
$\Delta {\rm M} = {\rm M}({\rm K}\pi\pi) - {\rm M}({\rm K}\pi)$ is shown
for ``right sign'' combinations (${\rm K}^\mp \pi^\pm \pi^\pm$), which
contain the \dstarplus signal, for ``wrong sign'' combinations
(${\rm K}^\pm \pi^\mp \pi^\pm$), which are combinatorial background, and
for \dzero ``side band'' combinations, which select ${\rm K}\pi$ pairs
that are not consistent with the \dzero meson mass. Both the ``wrong sign''
and the ``side band'' method reproduce the combinatorial background well.
After corrections the measured \pt differential production cross sections 
for \dzero and \dstarplus mesons in \pp collisions at $\sqrt{s} = 200$~GeV 
were divided by the charm quark fragmentation ratios 
$0.565 \pm 0.032$ for $c \rightarrow {\rm D}^0$ and
$0.224 \pm 0.028$ for $c \rightarrow {\rm D}^{*+}$, respectively, to
convert the meson cross sections to the $\ccbar$ cross section. 
The resulting \pt differential charm cross section is shown in the
right panel of Fig.~\ref{fig:star_d_pp} in comparison with the upper
and lower limits of the corresponding cross section as predicted with 
a FONLL pQCD calculation~\cite{fonll1,fonll2}. The upper limit of the 
calculation is consistent with the measured charm hadron cross sections over 
the full \pt range. The same was observed at higher energy with the CDF 
experiment at the Tevatron (see Fig.~\ref{fig:charm_tevatron}) and with the 
ALICE experiment at the LHC (see Fig.~\ref{fig:alice_d_in_pp}).

The \pt differential cross section was parametrized with a power-law function
(indicated in the right panel of Fig.~\ref{fig:star_d_pp}) and extrapolated
to zero \pt. Furthermore, the rapidity density of the cross section at 
midrapidity was extrapolated to the full phase space based on calculations 
taking into account the rapidity dependence of charm production, resulting in
an extrapolation factor of $4.7 \pm 0.7$.
The rapidity density of the cross section and the total $\ccbar$ production 
cross sections were
\begin{eqnarray}
\frac{d\sigma_{\ccbar}}{dy}|_{y=0} & = & 170 \pm 45 {\rm (stat.)} ^{+38}_{-59} {\rm (sys.)} \; \mu{\rm b} \\
\sigma_{\ccbar} & = & 797 \pm 210 {\rm (stat.)} ^{+208}_{-295} {\rm (sys.)} \; \mu{\rm b}.
\end{eqnarray}

\subsubsection{Semielectronic heavy-flavor hadron decays}
\label{subsubsec:rhic_e_pp}
\begin{figure}[t]
\begin{center}
\includegraphics[width=0.55\textwidth]{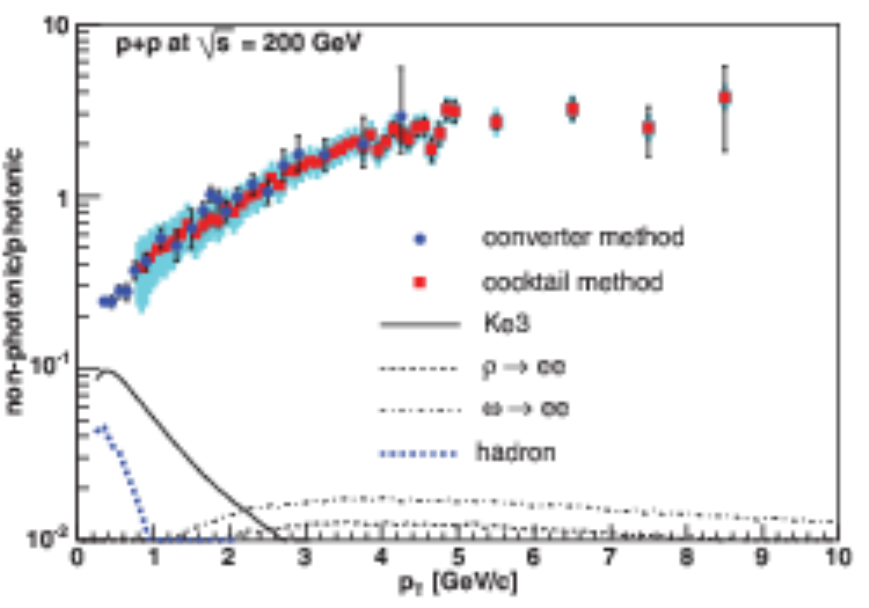}
\includegraphics[width=0.28\textwidth]{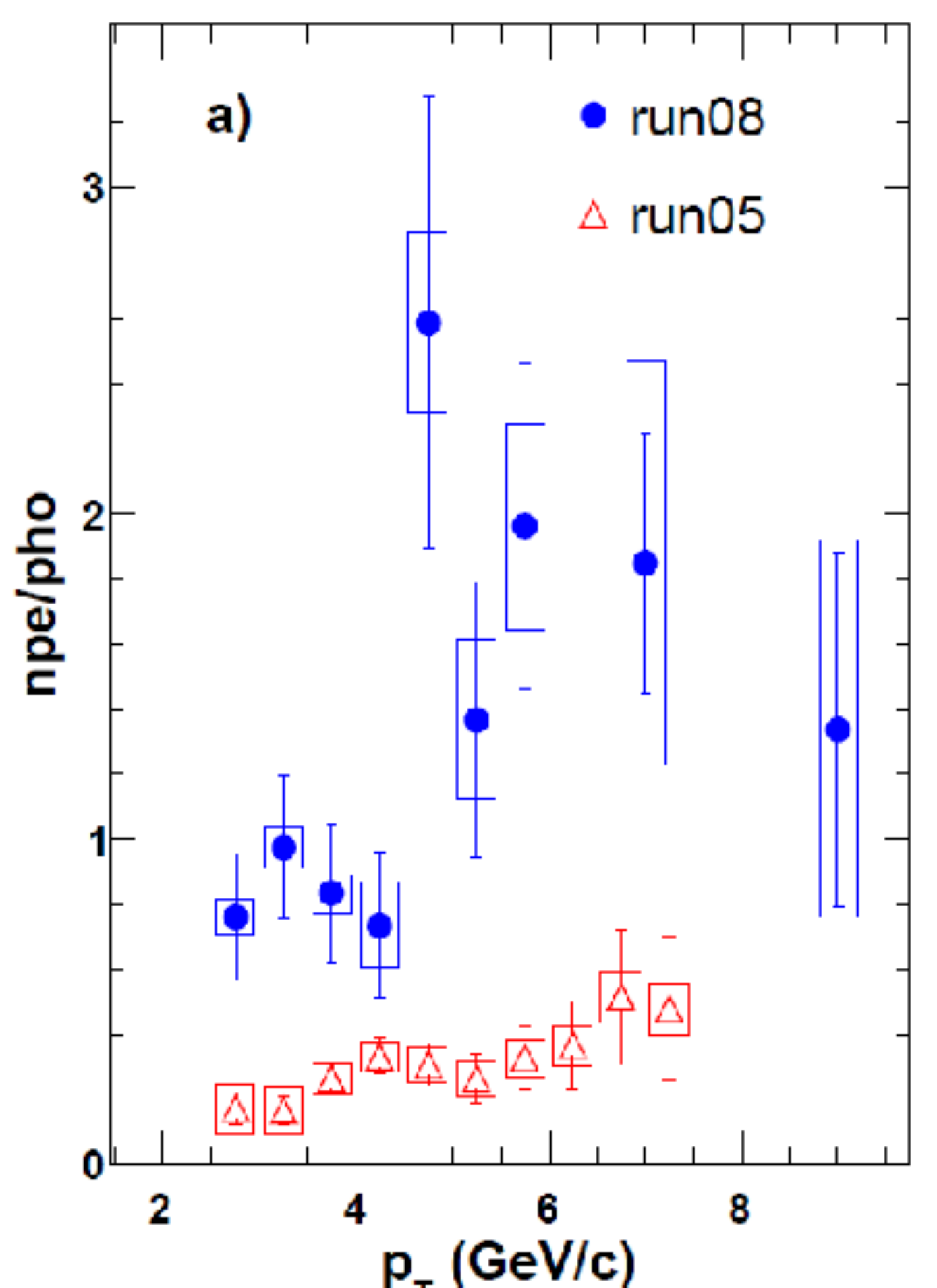}
\end{center}
\caption{Ratios of electrons from ``non-photonic'' sources to electrons from
         ``photonic'' sources measured in \pp collisions at 
         $\sqrt{s} = 200$~GeV with the PHENIX experiment (left panel) and the
         STAR experiment (right panel) at RHIC. The ``non-photonic'' electron
         spectrum is dominated by electrons from open heavy-flavor hadron 
         decays. Contributions from other sources do not play an important
         role as demonstrated for the PHENIX case in the left panel (taken 
         from Ref.~\cite{ppg077}). The important impact of the material budget 
         is shown in the right panel for the STAR experiment, in which the 
         material budget was reduced significantly in the 2008 RHIC run 
         compared to the 2005 run (see text for details) (reprinted with 
         permission from Ref.~\cite{star_e_pp}; Copyright (2011) by the 
         American Physical Society).}
\label{fig:rhic_e_sb}
\end{figure}

With both the PHENIX and STAR experiments extensive measurements of electrons
from open heavy-flavor hadron decays have been conducted at RHIC. The layout
of the PHENIX apparatus was optimized such that the material budget in the
detector aperture was kept at a minimum. The measured ratio~\cite{ppg077} of 
electrons of ``non-photonic'' origin, \ie mainly electrons from open 
heavy-flavor hadron decays, to electrons of ``photonic'' origin, \ie electrons 
from photon conversions in material in the detector aperture and from light 
meson Dalitz decays, is shown in the left panel of Fig.~\ref{fig:rhic_e_sb} 
for \pp collisions at $\sqrt{s} = 200$~GeV. Two complementary techniques to 
separate non-photonic from photonic electrons, \ie the cocktail subtraction 
method and the converter subtraction method as described in more detail above, 
gave consistent results for the non-photonic/photonic electron yield ratio. For
\pt above $\approx 2$~\gevc this ratio exceeded one. The original layout
of the STAR apparatus included the rather thick (in terms of radiation lengths)
SVT detector (see above) which was removed from the experimental setup prior
to the 2008 run of RHIC. As demonstrated in the right panel of 
Fig.~\ref{fig:rhic_e_sb} the non-photonic/photonic electron yield ratio 
measured in \pp collisions at $\sqrt{s} = 200$~GeV with the STAR experiment
was similar in magnitude to the ratio measured with PHENIX when the SVT was
not part of the setup~\cite{star_e_pp}. With the SVT in place, this ratio was 
substantially smaller, rendering the measurement of electrons from open 
heavy-flavor hadron decays very challenging, in particular at low \pt. 

\begin{figure}[t]
\begin{center}
\includegraphics[width=0.49\textwidth]{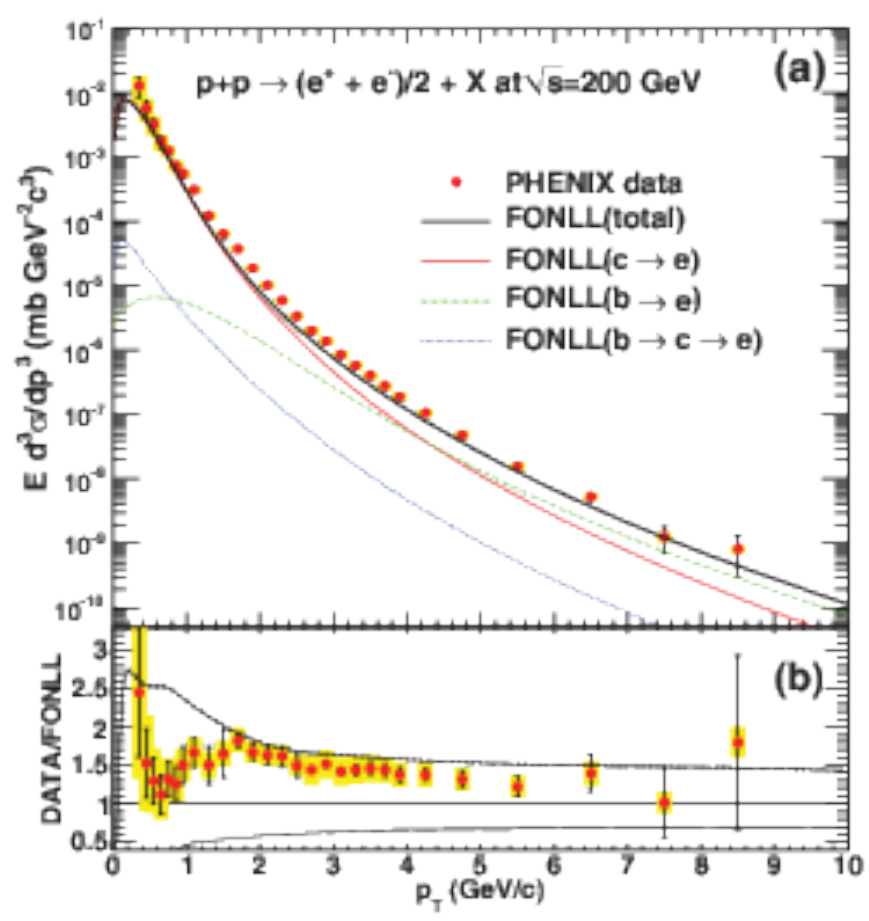}
\includegraphics[width=0.42\textwidth]{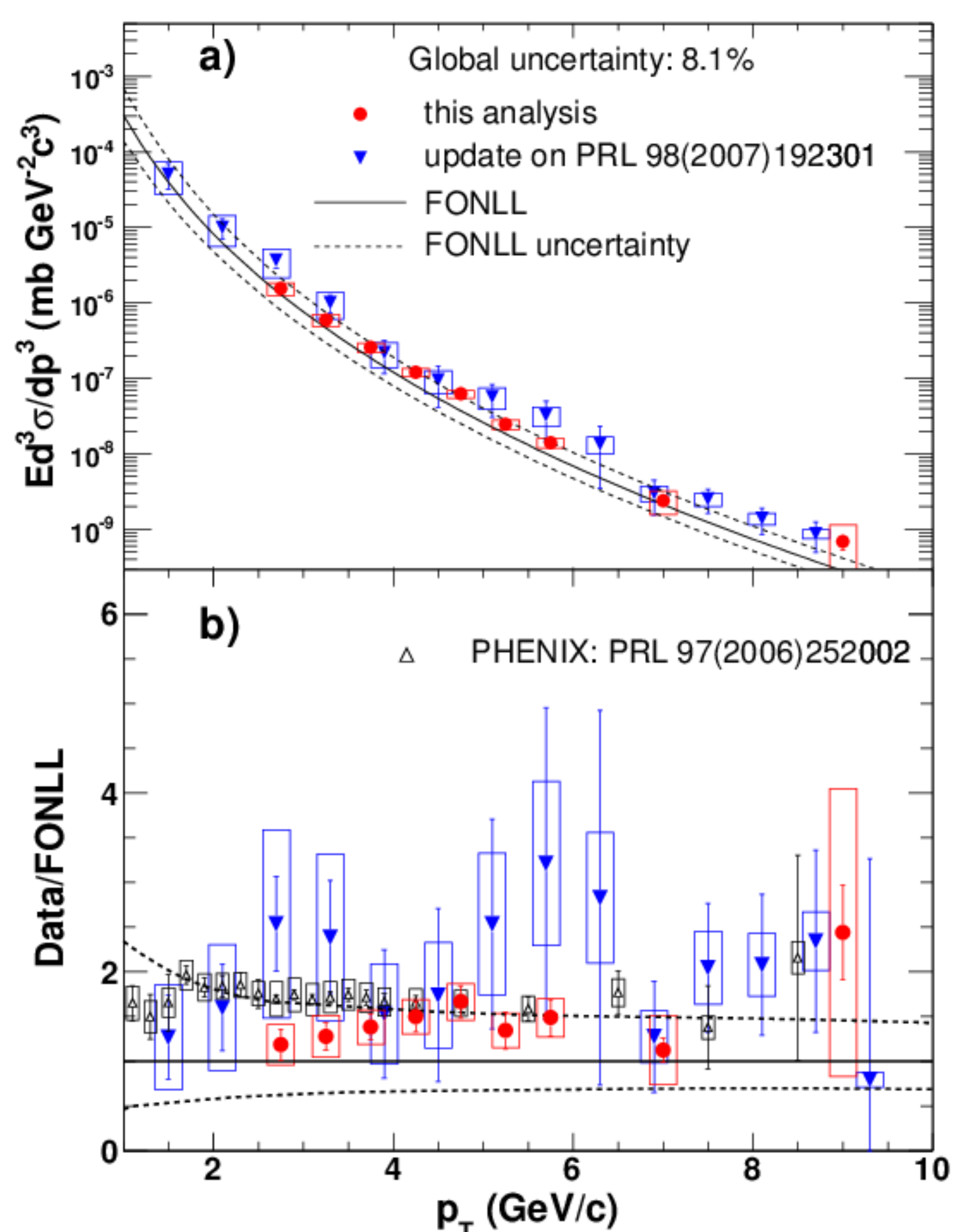}
\end{center}
\caption{Invariant \pt differential production cross section of electrons 
         from open heavy-flavor hadron decays measured in \pp collisions 
         at $\sqrt{s} = 200$~GeV with the PHENIX experiment (left panels) 
         and the STAR experiment (right panels) at RHIC. The upper panels in 
         both plots include a comparison of the data with corresponding cross 
         sections from a FONLL pQCD calculation. The lower panels show the
         ratio of the data and the calculation, where the upper (lower)
         curves indicate the upper (lower) limit of the FONLL calculation.
         More details are provided in the text (reprinted with permission from 
         Refs.~\cite{ppg077} (left) and \cite{star_e_pp} (right); 
         Copyright (2011) by the American Physical Society).}
\label{fig:rhic_e_pp}
\end{figure}

After the first measurements of electrons from open heavy-flavor hadron decays
in \pp collisions with the PHENIX and STAR experiments at RHIC, over the
following few years the precision of the data was improved significantly and 
the transverse momentum range covered was extended both to lower and higher 
\pt. The current status of these measurements is documented in 
Fig.~\ref{fig:rhic_e_pp}, which shows the invariant \pt differential cross
sections of electrons from heavy-flavor decays measured with the PHENIX 
experiment (left panel)~\cite{ppg077} in the range $0.3 < \pt < 9.0$~\gevc and 
with the STAR experiment (right panel)~\cite{star_e_pp}. 
For the STAR experiment two data sets are shown. In the 2005 RHIC run, with 
the SVT in place, electrons were measured in the range $1.2 < \pt < 10$~\gevc 
(triangle symbols). With additional data from the 2008 run, in which the SVT 
was removed from the STAR setup, the precision of the measurement could be 
significantly improved in the range $3 < \pt < 10$~\gevc (circle symbols). 
Corresponding predictions of the electron cross section from heavy-flavor 
decays from a FONLL pQCD calculation~\cite{fonll1,fonll2} are compared with 
the data in the upper panels of Fig.~\ref{fig:rhic_e_pp}. 
The ratios of the data and the FONLL calculation are shown in the lower panels 
of Fig.~\ref{fig:rhic_e_pp} on a linear scale to allow a better quantitative 
comparison. The PHENIX and STAR measurements are consistent with each other 
and, within the experimental and the theoretical uncertainties, which
are dominated by the uncertainties of the chosen normalization and 
factorization scales, the pQCD calculation is in agreement with the data. 
However, a clear tendency is observed that the measured cross sections are 
close to the upper limit of the predicted cross section, consistent with the
STAR D-meson measurement (see Fig.~\ref{fig:star_d_pp}).

According to the FONLL calculation the electron spectrum from heavy-flavor
hadron decays is completely dominated by charm decays at low \pt,
which is expected because of the much smaller total bottom production cross
section compared to charm and, in addition, because of the harder bottom decay 
electron spectrum compared to electrons from charm decays. Therefore, from the 
measured \pt differential cross section of electrons from heavy-flavor hadron 
decays the total charm production cross section could be determined. The 
necessary extrapolation of the measured electron cross section from the \pt 
and rapidity intervals covered with the PHENIX experiment to the full phase 
space was done in three steps. First, the measured invariant 
cross section was extrapolated to zero \pt to obtain the integrated electron 
cross section per unit rapidity at midrapidity over all \pt. The \pt range 
covered by the PHENIX data included more than 50\% of the total electron yield 
from charm decays at midrapidity. Second, the integrated electron cross section
was converted into a charm cross section, taking into account the effective
charm to electron branching ratio and a kinematic correction factor to account
for the different rapidity distributions of D mesons and their decay electrons.
Third, the resulting rapidity density of the cross section at midrapidity was 
extrapolated from midrapidity to the full phase space. A systematic uncertainty
of 15\% was quoted for the extrapolation factor which was $\approx 6.6$. 
Details concerning this procedure can be found elsewhere~\cite{ppg077}. 
The resulting rapidity density of the charm production cross section at 
midrapidity and the total charm production cross section from the electron 
measurement with PHENIX are

\begin{eqnarray}
\frac{d\sigma_{\ccbar}}{dy}|_{y=0} & = & 119 \pm 12 {\rm (stat.)} \pm 38 {\rm (sys.)} \; \mu{\rm b}\\
\sigma_{\ccbar} & = & 551 \pm 57 {\rm (stat.)} \pm195 {\rm (sys.)} \; \mu{\rm b}.
\end{eqnarray}

\begin{figure}
\begin{center}
\includegraphics[width=0.55\textwidth]{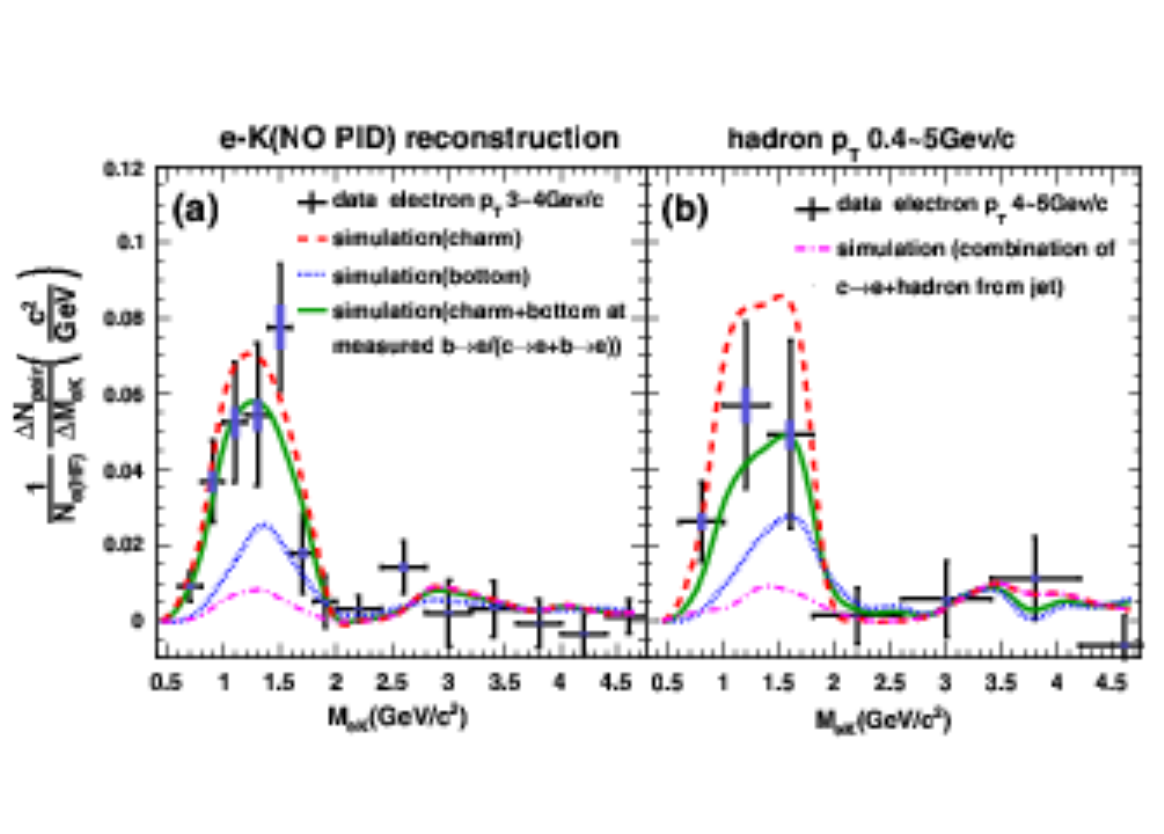}
\includegraphics[width=0.44\textwidth]{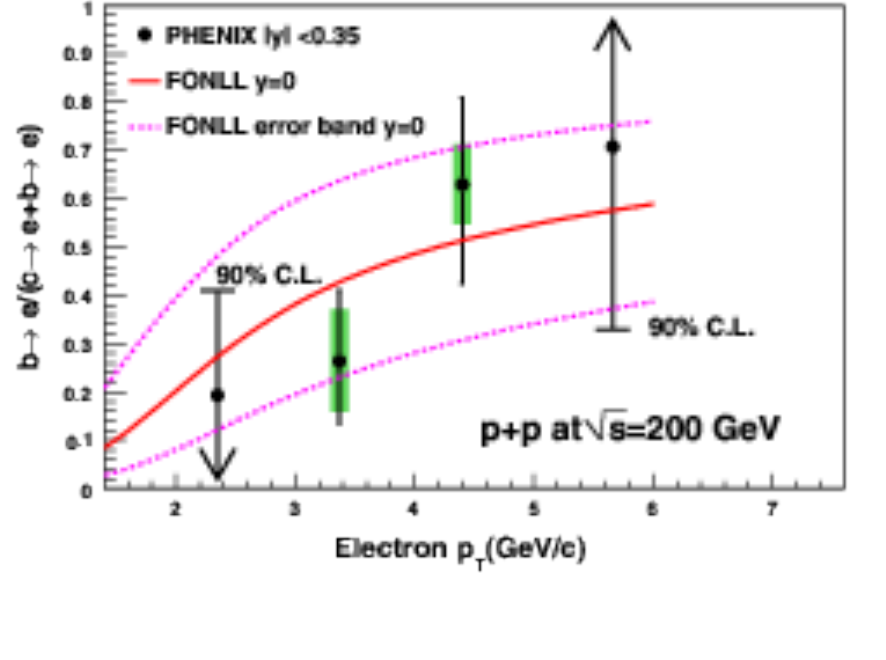}
\includegraphics[width=0.49\textwidth]{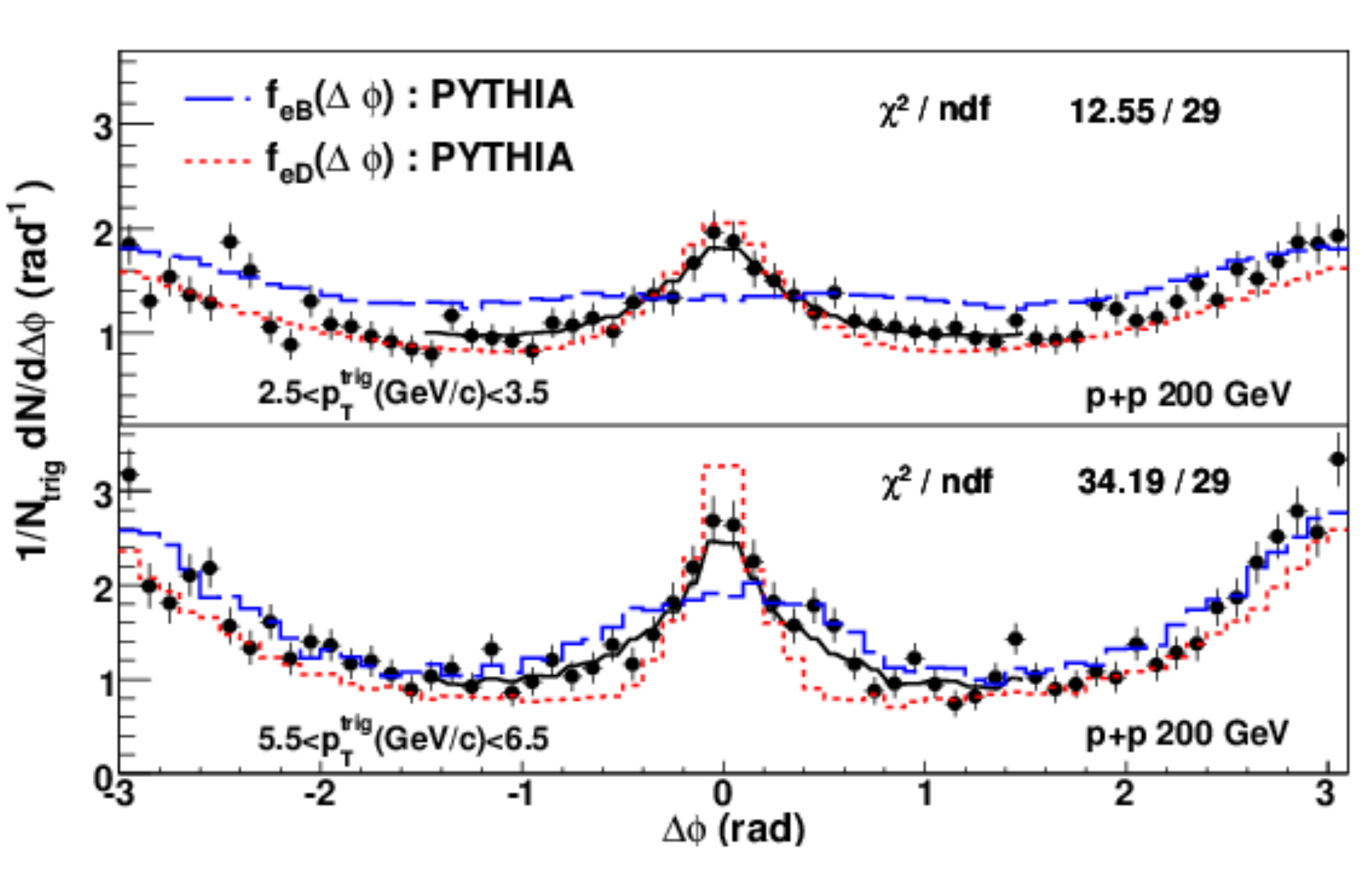}
\includegraphics[width=0.49\textwidth]{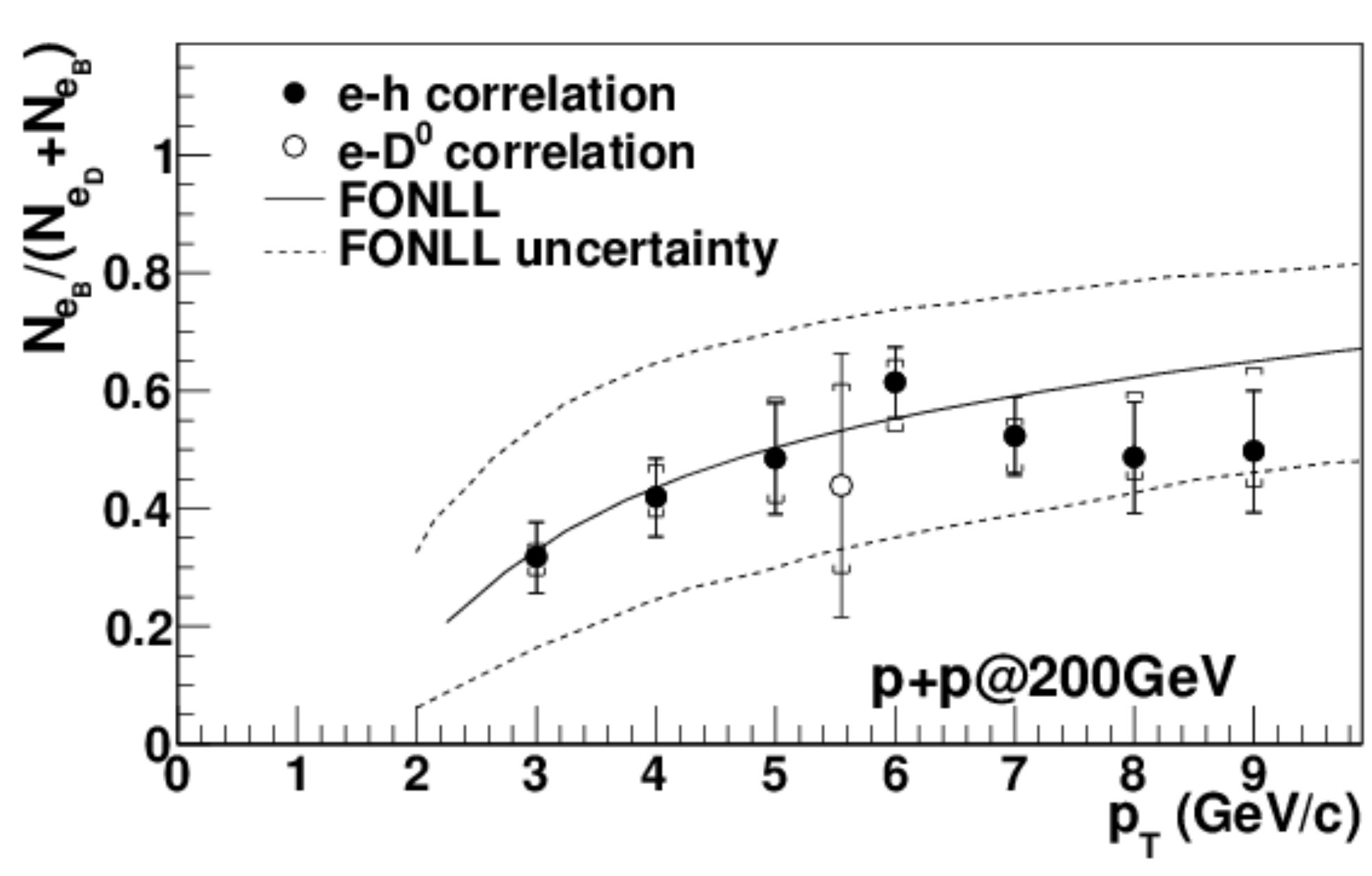}
\end{center}
\caption{Statistical separation of the charm and bottom decay contributions
         to the electron sample from heavy-flavor hadron decays in \pp
         collisions at $\sqrt{s} = 200$~GeV with the PHENIX experiment
         (upper two panels) and with the STAR experiment (lower two panels). 
         The invariant mass distribution of electron-hadron pairs measured
         with PHENIX is compared to charm and bottom decay templates obtained 
         from simulations using PYTHIA and EVTGEN for two electron \pt ranges 
         (upper left panel). The azimuthal angle between electrons from 
         heavy-flavor decays and hadrons measured with STAR is compared to 
         charm and bottom decay templates obtained with PYTHIA simulations for 
         two electron \pt ranges (lower left panel). Relative contributions of 
         bottom decays to the total yield of electrons from heavy-flavor 
         hadron decays as functions of the electron \pt are compared to 
         predictions from FONLL pQCD calculations (upper and lower left panel)
         (reprinted with permission from  Refs.~\cite{ppg094} (upper) and 
         \cite{star_e_b_d} (lower); Copyright (2009, 2010) by the American 
         Physical Society).}
\label{fig:rhic_eh}
\end{figure}

Unfortunately, the electron data measured with the STAR experiment did not 
extend to low enough electron \pt, such that an extrapolation to the full
phase space suffered from large systematic uncertainties and a sensible total 
charm production cross section from the electron measurement alone could not be 
quoted.

With increasing electron \pt the contribution from bottom decays grows. 
According to the FONLL calculation, at about 4~\gevc the contributions from 
charm and bottom decays are similar, and towards higher \pt bottom decays 
become the dominant source of electrons from heavy-flavor hadron decays. 

It is important to disentangle the charm and bottom decay contributions 
experimentally such that the FONLL pQCD predictions can be tested separately,
and to provide individual references for the decay electron spectra from the 
two heavy quark species in \auau collisions at RHIC. Both in PHENIX and STAR
such separation techniques were developed based on electron-hadron correlation
measurements. Due to the fact that the combinatorial background in two-particle
correlation measurements increases dramatically with particle multiplicity,
the statistical separation of electrons from charm and bottom decays was 
possible only in \pp collisions at RHIC up to now.

In the PHENIX case, the extraction of the relative contributions from charm
and bottom decays was based on the partial reconstruction of the decay
${\rm D} \rightarrow e^+ {\rm K}^- X$ (and charge conjugate)~\cite{ppg094}. 
The invariant-mass distribution of unlike charge-sign electron-hadron pairs
exhibits a correlated signal below the D-meson mass because of the charge
correlation in D-meson decays. The upper left panel of Fig.~\ref{fig:rhic_eh} 
shows this invariant-mass distribution normalized to the total yield of 
electrons from heavy-flavor decays in two hadron \pt bins. In this analysis, 
kaon identification was not performed but the mass of all reconstructed hadrons 
was assumed to be the kaon mass. Using a combination of the 
PYTHIA~\cite{pythia} and EVTGEN~\cite{evgen} simulation packages template 
distributions were generated that described the expected electron-hadron 
invariant-mass distributions from charm and bottom decays, shown as long-dashed
and short-dashed lines, respectively, in the upper left panel of 
Fig.~\ref{fig:rhic_eh}. 
A linear combination of these template distributions was fitted to the 
data with the relative contributions to the total distributions as the only 
fit parameter. The resulting contribution of bottom decays to the total 
electron spectrum from heavy-flavor hadron decays is shown in the upper right 
panel of Fig.~\ref{fig:rhic_eh} as a function of the electron \pt. Below
3~\gevc and above 5~\gevc only upper and lower limits, respectively, of
the relative bottom decay contribution could be measured. Nevertheless,
a clear trend is observed that the bottom decay contribution increases
with increasing electron \pt. Moreover, the measured relative contributions
are in good agreement with the expectation from a FONLL pQCD 
calculation~\cite{fonll1,fonll2}.

\begin{figure}[t]
\begin{center}
\includegraphics[width=0.49\textwidth]{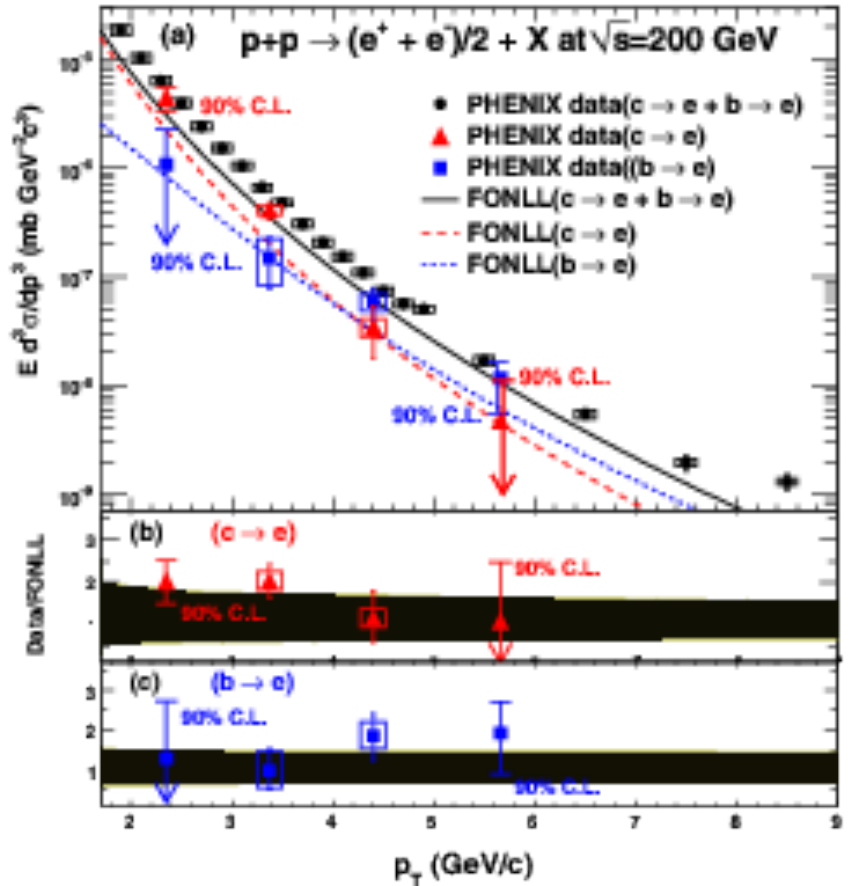}
\includegraphics[width=0.49\textwidth]{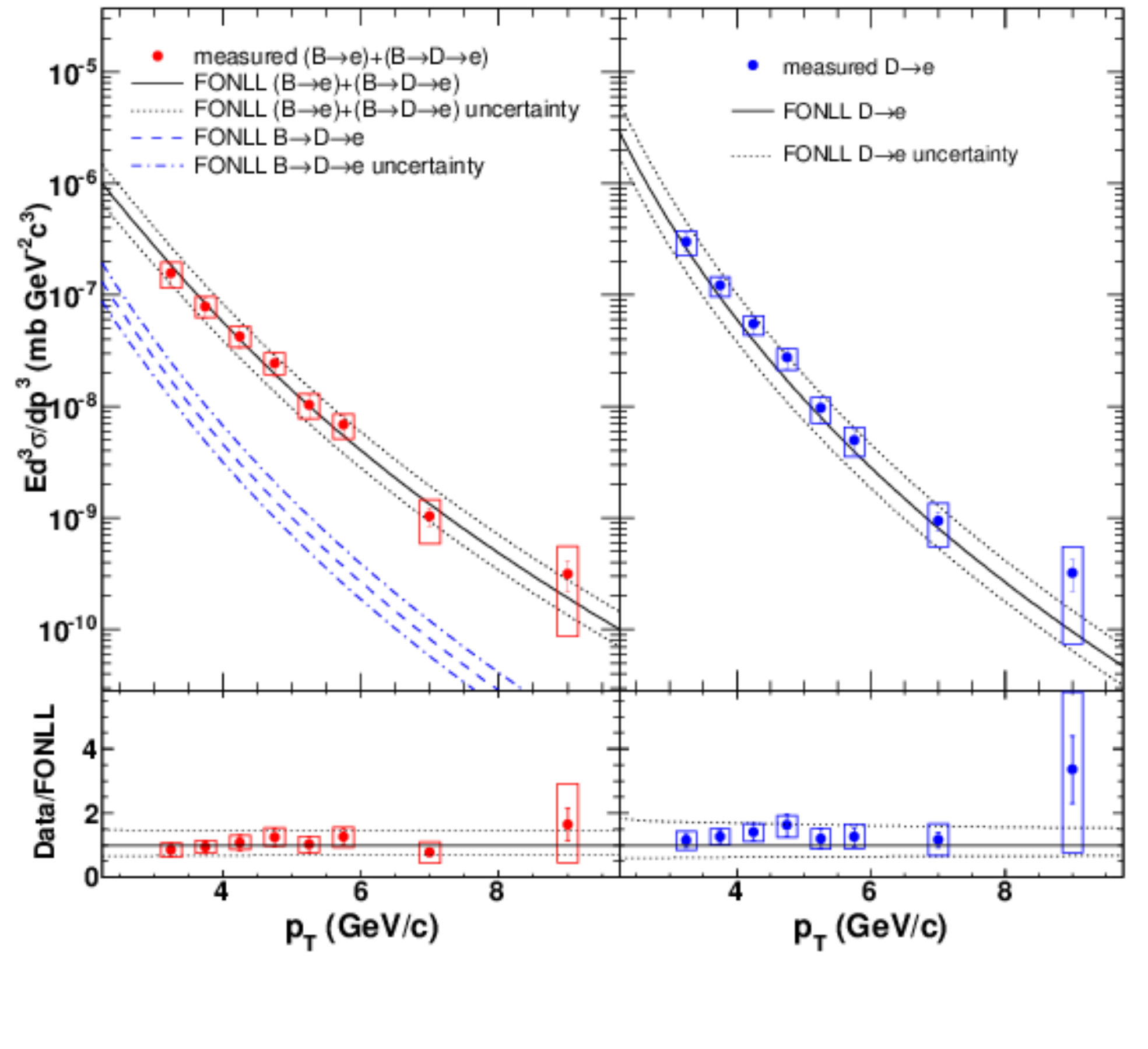}
\end{center}
\caption{Invariant \pt differential production cross sections of electrons 
         from charm and bottom hadron decays, respectively, as measured with
         the PHENIX (left panels) and STAR experiments (right panels) in \pp
         collisions at $\sqrt{s} = 200$~GeV. Predictions from a FONLL pQCD
         calculation (shown as lines) are compared with the data on an
         absolute scale (upper panels) and in form of the ratio data/FONLL
         (lower panels) (reprinted with permission from  Refs.~\cite{ppg094} 
         (left) and \cite{star_e_b_d} (right); Copyright (2009, 2010) by the 
         American Physical Society).}
\label{fig:rhic_cb_e_pp_fonll}
\end{figure}

In the STAR case, the relative contributions from charm and bottom decays
were separated statistically from each other via the measurement of angular
correlations~\cite{star_e_b_d}. In a first approach, the distribution of the 
azimuthal angle $\Delta\phi$ between electrons from heavy-flavor decays and 
charged hadrons was studied. Such distributions, normalized to the number of 
electrons from heavy-flavor hadron decays, are shown for two electron \pt bins 
in the lower left panel of Fig.~\ref{fig:rhic_eh}. The shapes of these 
distributions, in particular the width of the near side peak at 
$\Delta\phi = 0$ is largely given by the decay kinematics of the charm and 
bottom hadron decays, where the larger mass of bottom hadrons leads to a 
broader peak compared to charm hadron decays. Template distributions reflecting
the expected shapes of the electron-hadron azimuthal distributions as shown in 
the lower left panel of Fig.~\ref{fig:rhic_eh} were generated with 
PYTHIA~\cite{pythia}. The contribution of bottom decays relative to the sum of 
charm and bottom decays was obtained from a fit of the measured shape of the 
angular distribution using a linear combination of the charm and bottom 
templates with the relative normalization as fit parameter. A second, 
independent measurement of the contribution from bottom decays was conducted 
using the angular correlation of electrons from heavy-flavor hadron decays and 
\dzero mesons reconstructed via their ${\rm K} \pi$ decay. Again, template 
distributions obtained via a PYTHIA~\cite{pythia} simulation were used to 
quantify the charm and bottom decay contributions.
The observed away-side correlation signal in the electron-\dzero azimuthal
angle distribution was attributed to prompt charm hadron pair production
($\approx 75$\%) and bottom hadron decays ($\approx 25$\%), while the near
side correlation signal was mainly due to bottom hadron decays only.
The contributions of bottom decays to the total electron spectrum from 
heavy-flavor hadron decays as obtained with these two methods are shown in the 
lower right panel of Fig.~\ref{fig:rhic_eh} as functions of the electron \pt. 
Because of the much larger solid angle coverage of the STAR apparatus compared 
to PHENIX, pair correlations could be measured with a much larger geometrical
acceptance which is a clear advantage for such analyses. Results from
the two techniques used in the STAR experiment agree with each other and
with prediction from a FONLL pQCD calculation~\cite{fonll1,fonll2} as 
demonstrated in the lower right panel of Fig.~\ref{fig:rhic_eh}.

From the measured invariant cross section of electrons from heavy-flavor hadron 
decays at $\sqrt{s} = 200$~GeV (see Fig.~\ref{fig:rhic_e_pp}) and the relative 
contributions from charm and bottom decays (see Fig.~\ref{fig:rhic_eh}) it is 
straight forward to determine the electron production cross sections from charm 
and bottom hadron decays separately and compare them with the corresponding 
predictions from a FONLL pQCD calculation~\cite{fonll1,fonll2}. The results 
from the PHENIX (STAR) experiments are shown in the left (right) panel of 
Fig.~\ref{fig:rhic_cb_e_pp_fonll}. For both cases FONLL is compared with the 
data on an absolute scale in the top panels while the ratio of the data to the 
calculation is shown in the lower panels. Given that FONLL provided a decent 
prediction for both the cross section of electrons from all heavy-flavor hadron
decays and the for the relative contributions of charm and bottom decays to 
these cross sections it is not surprising that a good agreement between the 
FONLL predictions and the individual electron spectra from charm and bottom 
decays, respectively, was observed both with the PHENIX and STAR experiments.

Similar to the charm case (see above), the electron production cross sections
from bottom decays were extrapolated to the full phase to determine the
bottom production cross section. The extrapolation done for the STAR data
was based on the PYTHIA generator~\cite{pythia} in the minimum bias mode
for \pp collisions and in an alternative, exclusive bottom production mode in 
which only leading order processes were included (see Ref.~\cite{star_e_b_d} 
for details). The total bottom production cross sections from these two 
approaches were $\sigma_{\bbbar} = 1.34$~$\mu$b and 
$\sigma_{\bbbar} = 1.83$~$\mu$b, respectively~\cite{star_e_b_d}. Corresponding 
extrapolations of the PHENIX data~\cite{ppg094} using PYTHIA and FONLL resulted
in

\begin{eqnarray}
\frac{d\sigma_{\bbbar}}{dy}|_{y=0} & = & 0.92^{+0.34}_{-0.31} {\rm (stat.)}^{+0.39}_{-0.36} {\rm (sys.)} \; \mu{\rm b}\\
\sigma_{\bbbar} & = & 3.2^{+1.2}_{-1.1} {\rm (stat.)}^{+1.4}_{-1.3} {\rm (sys.)} \; \mu{\rm b}.
\end{eqnarray}

Both the measurements with the STAR and PHENIX experiments are in agreement
within uncertainties with the total bottom production cross section
$\sigma_{\bbbar} = 1.87^{+0.99}_{-0.67} \; \mu{\rm b}$ predicted with 
FONLL~\cite{fonll1,fonll2}.
To a large extent, the substantial uncertainties of the measurements were
driven by the unavailable cross section measurements for electron production
from bottom decay below an electron \pt of 3-4~\gevc, indicating that
future bottom production measurements have to include the low \pt region
in order to provide a sensitive test of pQCD calculations over the full
phase space.

\begin{figure}[t]
\begin{center}
\includegraphics[width=0.6\textwidth]{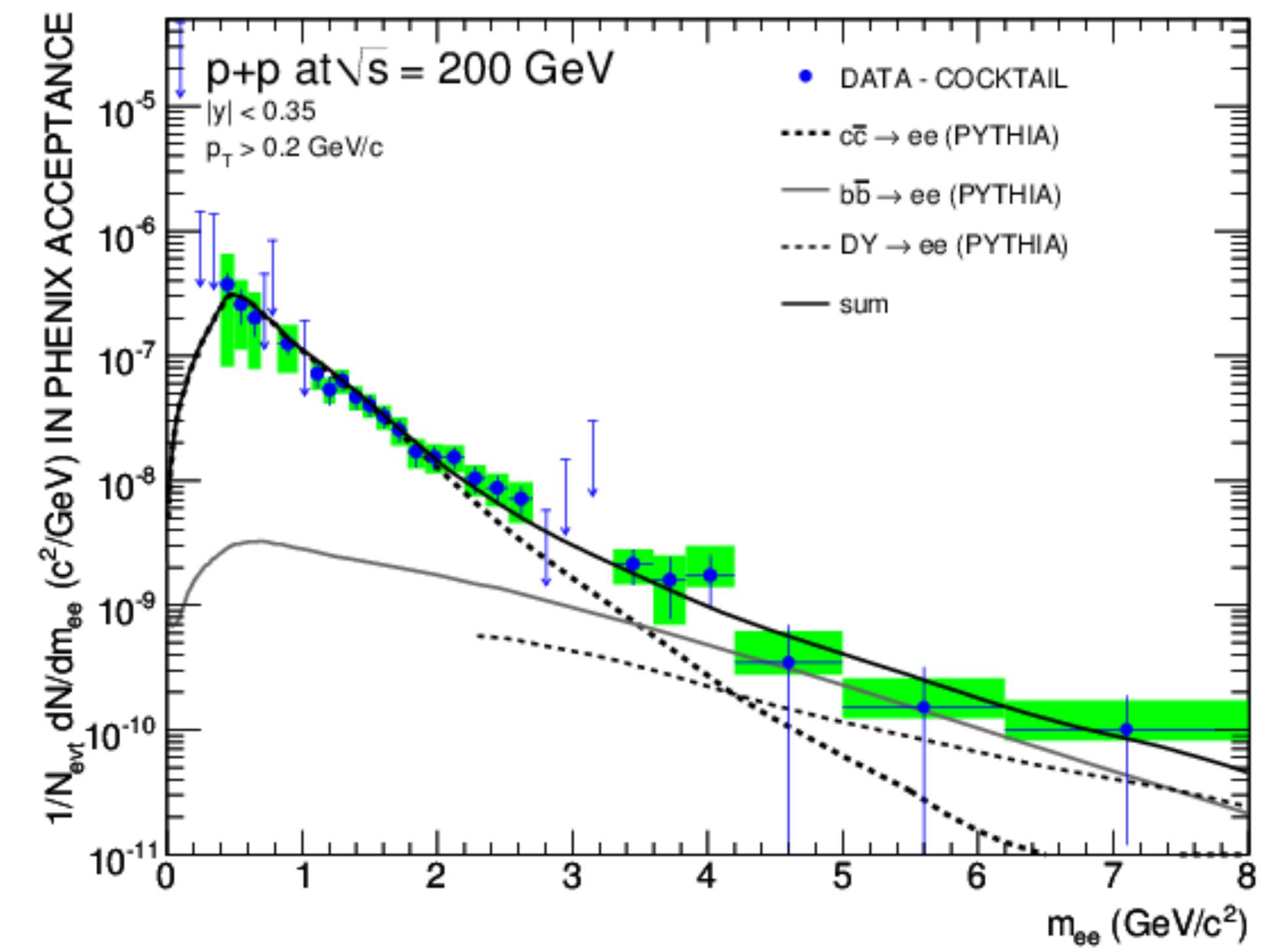}
\end{center}
\caption{Electron-positron pair mass distribution from correlated decays of
         heavy-flavor hadrons measured in \pp collisions at 
         $\sqrt{s} = 200$~GeV with the PHENIX experiment. Arrows indicate upper
         limits at 95\% confidence level where the systematic uncertainties
         were larger than the measured pair yields after background subtraction
         from other sources. Estimated contributions from charm decays (thick
         dashed line), bottom decays (thin solid line), pairs from the Drell-Yan
         process (thin dashed line), and the sum of these (thick solid line)
         are shown for comparison. See text for more details (reprinted from
         Ref.~\cite{ppg085}).}
\label{fig:phenix_pp_dielectron}
\end{figure}

In addition, with the PHENIX experiment the production of correlated 
electron-positron pairs from the decay of heavy-flavor hadron pairs
from associated production in a hard scattering process was investigated
in \pp collisions at $\sqrt{s} = 200$~GeV~\cite{ppg085,ppg088}. 
To statistically isolate such electron-positron pairs, in a first step the 
combinatorial background from uncorrelated background pairs was subtracted from
the measured electron-positron pair mass distribution. The resulting signal 
distribution was corrected for the reconstruction efficiency. A pair cocktail 
from sources other than heavy-flavor hadron decays was calculated, similar to 
the single-electron case, and it was subtracted from the mass distribution of 
correlated electron-positron pairs. The remaining mass spectrum, shown in
Fig.~\ref{fig:phenix_pp_dielectron}, was interpreted in terms of 
electron-positron pairs from correlated open heavy-flavor hadron decays. 
To statistically disentangle these contributions mass distribution templates 
were calculated with PYTHIA~\cite{pythia} which were then fitted to the data 
with the relative normalization as free parameters after the contribution from 
the Drell-Yan process was subtracted. As sources of systematic uncertainties in 
addition to the experimental uncertainties, the value of PYTHIA parameters
such as the intrinsic $k_{\rm t}$ as well as the choice of the parton 
distribution function, and the uncertainty of the relative abundance of 
charm hadron species were considered. Furthermore, the dynamical correlations
between heavy-flavor quark-antiquark pairs are not well known, which gave rise
to additional systematic uncertainties. Within this approach, the total charm
and bottom production cross sections were determined as 

\begin{eqnarray}
\sigma_{\ccbar} & = & 518 \pm 47 {\rm (stat.)} \pm 135 {\rm (sys.)} \pm 190 {\rm (model)} \; \mu{\rm b}\\
\sigma_{\bbbar} & = & 3.9 \pm 2.5 {\rm (stat.)}^{+3}_{-2} {\rm (sys.)} \; \mu{\rm b},
\end{eqnarray}

where the model dependent uncertainty of the extrapolation to the full phase
space is quoted separately for the charm cross section. Alternatively, the
rapidity density of the charm production cross section at midrapidity and the
total charm production cross section were determined after the bottom
decay contribution was subtracted based on a bottom cross section of 
3.7~$\mu$b as obtained from a systematic study~\cite{jaroschek} of the total 
bottom production cross section as a function of $\sqrt{s}$ in \pp collisions 
with PYTHIA~\cite{pythia}. The corresponding values are 

\begin{eqnarray}
\frac{d\sigma_{\ccbar}}{dy}|_{y=0} & = & 118.1 \pm 8.4 {\rm (stat.)} \pm 30.7 {\rm (sys.)} \pm 39.5 {\rm (model)} \; \mu{\rm b}\\
\sigma_{\ccbar} & = & 544 \pm 39 {\rm (stat.)} \pm 142 {\rm (sys.)} \pm 200 {\rm (model)} \; \mu{\rm b}.
\end{eqnarray}

In summary, a consistent picture emerged from the measurement of electrons 
and correlated electron-positron pairs from open heavy-flavor hadron decays 
with the PHENIX and STAR experiments in \pp collisions at $\sqrt{s} = 200$~GeV 
at RHIC. The measured \pt differential single electron invariant cross sections
are consistent with corresponding predictions from FONLL pQCD calculations.
Also, such FONLL predictions are consistent with the total charm and bottom 
production cross sections obtained from single electron and electron-positron 
pair measurements with PHENIX and STAR at RHIC. These electron data provide a 
crucial reference for corresponding measurements in nucleus-nucleus collisions.

\subsubsection{Semimuonic heavy-flavor hadron decays}
\label{subsubsec:rhic_mu_pp}
\begin{figure}[t]
\begin{center}
\includegraphics[width=0.5\textwidth]{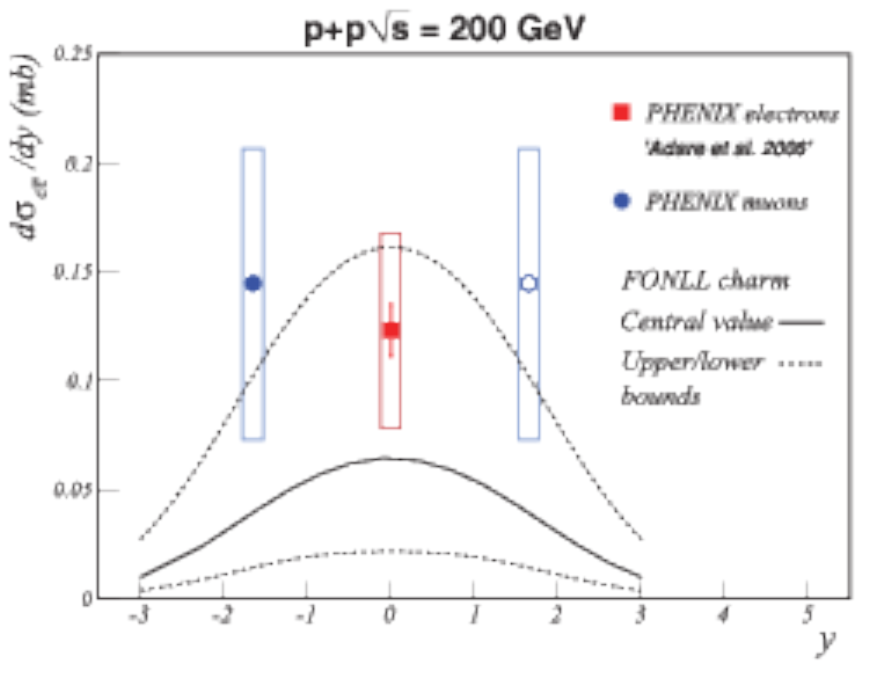}
\end{center}
\caption{Charm production cross section as a function of rapidity in \pp
         collisions at $\sqrt{s} = 200$~GeV measured with the 
         PHENIX experiment using semileptonic open heavy-flavor hadron
         decays to electrons at midrapidity (closed square) and to muons 
         at forward rapidity (closed circle). A prediction from FONLL pQCD 
         calculation is shown for comparison (reprinted with permission from 
         Ref.~\cite{ppg117}; Copyright (2012) by the American Physical 
         Society).}
\label{fig:phenix_charm_rapidity}
\end{figure}

The PHENIX experiment provides the unique opportunity at RHIC to study the 
rapidity dependence of open heavy-flavor production via the measurement
of electrons at midrapidity (see above) and muons at forward rapidity from 
semileptonic decays of heavy-flavor hadrons. For the muon measurement it
is crucial to subtract background from primary hadrons which penetrate
all absorber layers and from other muon sources, \ie from the light hadron 
decays before they reach the first hadron absorber material and from the decay 
of light hadrons that are produced inside the muon tracker volume of PHENIX.
The techniques necessary for the background subtraction procedure were 
developed for the first measurement of muon production at forward
pseudorapidity ($1.5 \le |\eta| \le 1.8$) over the transverse momentum range
$1 \le \pt \le 3$~\gevc in \pp collisions at $\sqrt{s} = 200$~GeV~\cite{ppg057}.
The background from misidentified punchthrough hadrons was significantly larger
for positively charged compared to negatively charged muon candidates, mainly
because of the relatively large nuclear absorption length for positively
charged kaons in the absorber material. Therefore, only negatively charged
muon candidates were considered for the further analysis. After the 
statistical background subtraction a muon excess remained which was
attributed to semimuonic decays of open heavy-flavor hadrons, mainly D-mesons
in the covered muon \pt range. A comparison of the \pt differential invariant
cross section of the excess muons with corresponding calculations from PYTHIA
and FONLL showed that the measured cross sections exceeded significantly both 
predictions~\cite{ppg057}. The experimental systematic uncertainties, however, 
were huge and could not be quantified very precisely because of the lack of 
statistics available from the data set recorded during the 2001/2002 RHIC run.

With new data from the 2005 RHIC run the measurement could be improved
significantly~\cite{ppg117}. The production cross section of negatively 
charged muons from open heavy-flavor hadron decays was measured over an 
extended \pt range ($1 \le \pt \le 7$~\gevc) for rapidities $1.4 < |y| < 1.9$. 
The shape of the \pt differential invariant cross section predicted by a
FONLL pQCD calculation agreed well with the measurement. Therefore, FONLL was 
used to extrapolate the measured spectrum to zero \pt and to calculate from 
the muon cross section a charm cross section at forward rapidity:

\begin{equation}
\frac{d\sigma_{\ccbar}}{dy}|_{\langle y \rangle = 1.65} 
  = 139 \pm 29 {\rm (stat.)}^{+51}_{-58} {\rm (sys.)} \; \mu{\rm b}
\end{equation}

The rapidity dependence of the charm production cross section as predicted
from a FONLL pQCD calculation is compared to the cross section measured
via decay electron at midrapidity and decay muons at forward rapidity in 
Fig.~\ref{fig:phenix_charm_rapidity}, where the combined muon result from both
muon spectrometers is shown with a closed circle symbol at negative rapidity
and reflected to forward rapidity (open circle). Within the substantial 
experimental and theoretical uncertainties the predicted cross sections are
in agreement with the data. Clearly, more precise data are needed in order
to sensibly discuss the shape of the rapidity distribution of charm in
\pp collisions at RHIC.

\subsubsection{Total heavy-flavor production cross sections at RHIC}
\label{subsubsec:rhic_hf_pp}
As discussed in this section heavy-flavor production has been studied
in detail in \pp collisions at RHIC in various channels. In addition to the
spectral shapes of heavy-flavor observables, which are well described by NLO
pQCD calculations, the rapidity densities $d\sigma_{\rm c\bar{c}}/dy$ and
$d\sigma_{\rm b\bar{b}}/dy$ and the total heavy-flavor production cross sections 
$\sigma_{\rm c\bar{c}}$ and $\sigma_{\rm b\bar{b}}$ are the most relevant observables.
The corresponding values measured in various channels are summarized in
Tab.~\ref{tab:HF_sigma_RHIC}.

\begin{table}[t]
\begin{center}
\caption{Rapidity densities $d\sigma_{\rm q\bar{q}}/dy$ and total cross sections
         $\sigma_{\rm q\bar{q}}$ for the production of heavy quark antiquark pairs 
         $\rm q\bar{q}$ (here, 'q' stands for either charm, 'c', or bottom, 'b')
         measured in various channels in \pp collisions at $\sqrt{s} = 200$~GeV
         at RHIC. In addition, the \pt and $y$ ranges covered by the 
         measurements are given. Due to the limited \pt ranges, the measured
         yields have to be extrapolated to zero and high \pt to determine
         $d\sigma_{\rm q\bar{q}}/dy$, and the measured fractions are listed.
         Errors quoted are the statistical (first values) and total systematic
         (second values) uncertainties.}
\label{tab:HF_sigma_RHIC}
\begin{tabular}{lllll}
\\
\hline
channel & kinematic coverage & 
$d\sigma_{\rm q\bar{q}}/dy$ & $\sigma_{\rm q\bar{q}}$ & ref. \\
 & in $y$ and \pt & fraction measured & & \\
& & in covered $y$ range & & \\
\hline
& & & & \\
${\rm c\bar{c}} \rightarrow$~\dzero, \dstar + X & 
$|y| < 1$ & 
$170 \pm 45 ^{+38}_{-59}$~$\mu$b & 
$797 \pm 210 ^{+208}_{-295}$~$\mu$b & \cite{star_d_pp} \\
& $0.6 < \pt < 2.0$~\gevc (\dzero) & 
\multirow{2}{*}{$\Bigg\}$$\approx$67\%} & & \\
& $2.0 < \pt < 6.0$~\gevc (\dstar) & & & \\

${\rm c\bar{c}} \rightarrow$~e$^\pm$ + X & 
$|y| < 0.35$ & 
$119 \pm 12 \pm 38$~$\mu$b & 
$551 \pm 57 \pm 195$~$\mu$b & \cite{ppg077} \\
& $0.3 < \pt < 9.0$~\gevc & $\approx$55\% & & \\

${\rm c\bar{c}} \rightarrow$~e$^+$e$^-$ + X & 
$|y| < 0.35$ & 
$118.1 \pm 8.4 \pm 50.0$~$\mu$b & 
$544 \pm 39 \pm 245$~$\mu$b & \cite{ppg085} \\
& $\pte > 0.2$~\gevc & $\approx$80\% & & \\

${\rm c\bar{c}} \rightarrow$~e$^+$e$^-$ + X & 
$|y| < 0.35$ & 
not quoted & 
$518 \pm 47 \pm 233$~$\mu$b & \cite{ppg085} \\
& $\pte > 0.2$~\gevc & $\approx$80\% & & \\

${\rm c\bar{c}} \rightarrow$~$\mu^{-}$ + X & 
$1.4 < |y| < 1.9$ & 
$139 \pm 29 ^{+51}_{-58}$~$\mu$b & 
not quoted & \cite{ppg117} \\
& $1< \pt < 7$~\gevc & $\approx$6\% & & \\

${\rm b\bar{b}} \rightarrow$~e$^\pm$ + X & 
$|y| < 1$ & 
$4.0 \pm 0.5 \pm 1.1$~nb & 
1.34 -- 1.83 $\mu$b & \cite{star_e_b_d} \\
& $3.0 < \pt < 10.0$~\gevc & $\approx$6.5\% & & \\

${\rm b\bar{b}} \rightarrow$~e$^\pm$ + X & 
$|y| < 0.35$ & 
$0.92^{+0.34}_{-0.31}{^{+0.39}_{-0.36}}$~$\mu$b & 
$3.2^{+1.2}_{-1.1}{^{+1.4}_{-1.3}}$~$\mu$b & \cite{ppg094} \\
& $3.0 < \pt < 5.0$~\gevc & $\approx$5\% & & \\

${\rm b\bar{b}} \rightarrow$~e$^+$e$^-$ + X & 
$|y| < 0.35$ & 
not quoted & 
$3.9 \pm 2.5 ^{+3}_{-2}$~$\mu$b & \cite{ppg085} \\
& $\pte > 0.2$~\gevc & $\approx$99\% & & \\
\hline
\end{tabular}
\end{center}
\end{table}

\subsection{Heavy flavor in \dau collisions}
\label{subsec:rhic_dau}

Heavy-flavor observables are in agreement with predictions from FONLL pQCD 
calculations in \pp collisions at RHIC, providing an experimental and 
theoretical baseline for corresponding measurements in collisions of heavy 
nuclei in which modifications of such observables are expected due to the 
interaction of heavy quarks with the produced hot and dense medium. 
However, in such collisions also cold nuclear matter effects play a role. 
Initial state effects, such as the modification of the parton distribution 
functions in nuclei (\eg shadowing or saturation~\cite{kharzeev03}), 
broadening of momentum distributions due to the scattering of incoming partons,
Cronin enhancement~\cite{cronin75,straub95} of particle production at 
intermediate and high \pt, and energy loss of partons in cold nuclear matter 
are expected to leave their footprint on open heavy-flavor observables. 
In collisions of heavy nuclei it is nearly impossible to cleanly disentangle 
cold matter effects from those induced by the presence of a hot and dense 
medium. Thus, it is necessary to investigate collisions of protons or deuterons
with heavy nuclei. In such collisions the production of a hot and dense medium 
is not anticipated and, therefore, cold nuclear matter effects become 
accessible. 

\begin{figure}[t]
\begin{center}
\includegraphics[width=0.49\textwidth]{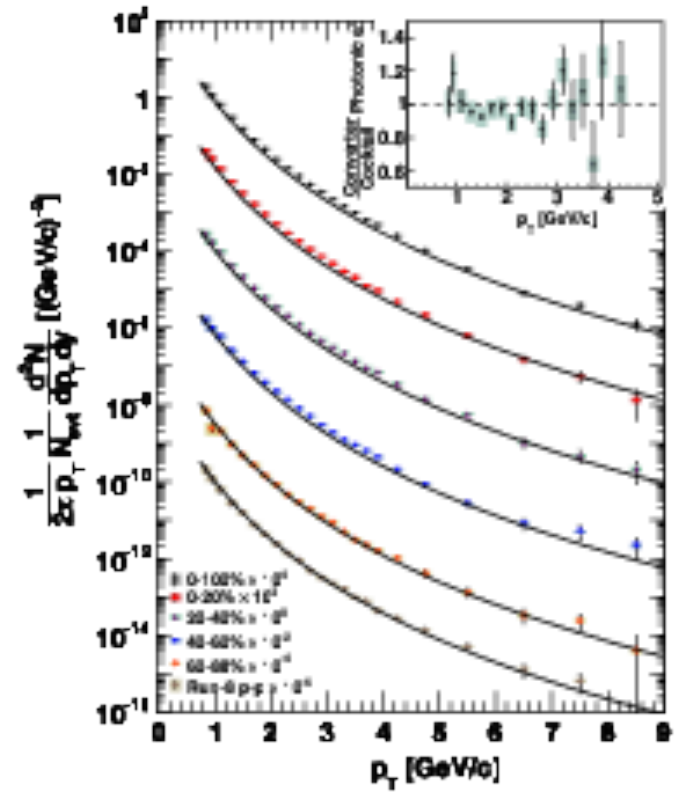}
\includegraphics[width=0.49\textwidth]{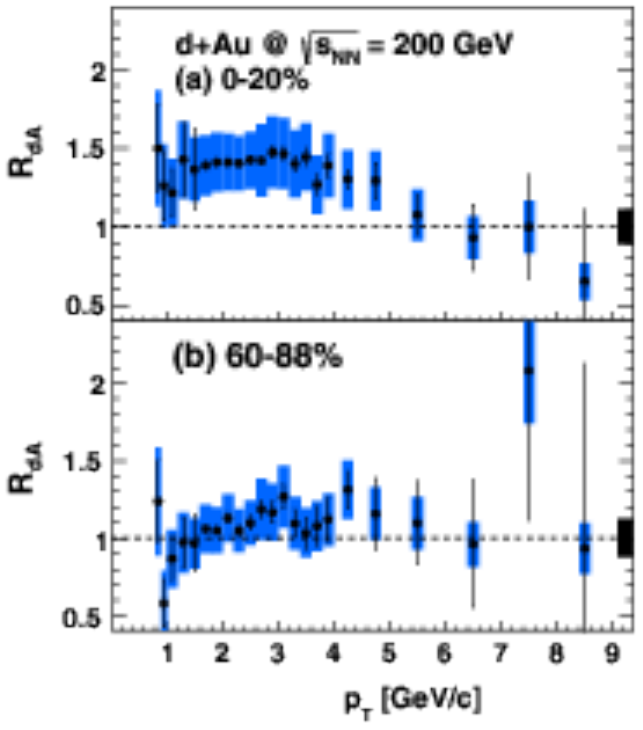}
\end{center}
\caption{Electrons from heavy-flavor hadron decays in various centrality
         classes of \dau collisions and in \pp collisions measured with the 
         PHENIX experiment at $\sqrt{s_{\rm NN}} = 200$~GeV. The inset shows 
         the ratio of electrons from heavy-flavor hadron decays measured via 
         the converter and cocktail subtraction techniques, respectively 
         (left panel). Nuclear modification factor $\rda$ of electrons from 
         heavy-flavor hadron decays in the most central (a) and most peripheral
         (b) \dau centrality classes (right panel). See text for more details
         (reprinted with permission from Ref.~\cite{ppg131}; Copyright (2012) 
         by the American Physical Society).}
\label{fig:phenix_e_dau}
\end{figure}

At RHIC, \dau collisions were investigated first in an exploratory run in
2003. Data from this run firmly 
established~\cite{brahms_dau,phenix_dau,phobos_dau,star_dau} that for light 
hadrons in central \auau collisions the strong suppression at high \pt relative
to binary scaled \pp collisions and the suppression of back-to-back azimuthal
angular correlations is due to final-state interactions with the produced
hot and dense medium and is not induced by cold nuclear matter effects.
However, the integrated luminosity sampled in this run at RHIC was not
sufficient for detailed measurements of open heavy-flavor observables.
With the STAR experiment the production of \dzero mesons was 
investigated~\cite{star_d_dau} in the transverse momentum range 
$0.1 < \pt < 3$~\gevc, and electrons from heavy-flavor hadron decays were 
measured in the range $1 < \pt < 10$~\gevc~\cite{star_e_auau}.
Within large uncertainties the \dau data were consistent with data from
\pp collisions scaled by the number of binary collisions to account for
the different collision geometry. The nuclear modification factor $\rda$
of electrons from open heavy-flavor hadron decays was observed to be 
consistent with a moderate Cronin enhancement. However, within substantial 
uncertainties the electron data were also consistent with an $\rda$ of one
for all \pt. Thus, no significant indication for cold nuclear matter effects 
on open heavy-flavor observables could be claimed based on the data from the 
2003 run at RHIC. 

The 2008 \dau run at RHIC constituted a major step forward in terms of the
statistics available for analysis. For example, with the PHENIX experiment 
the sampled luminosity of 80 nb$^{-1}$ from the 2008 run corresponds to 30 
times the sampled luminosity from the 2003 run. Using this data set the 
invariant \pt distribution of electrons from heavy-flavor hadron decays was 
measured in \dau collisions at $\sqrt{s_{\rm NN}} = 200$~GeV in several 
centrality classes~\cite{ppg131}. The collision centrality was determined 
event-by-event using the information from the PHENIX beam-beam counters. Within
a Glauber MonteCarlo calculation the mean number of participants, 
$\langle \npart \rangle$, and the mean number of binary collisions, 
$\langle \ncol \rangle$, was determined for each \dau centrality class. 
Electrons were reconstructed at midrapidity ($|\eta| < 0.35$) in the transverse
momentum range $0.85 < \pt < 8.5$~\gevc. Background from sources other than 
open heavy-flavor hadron decays was statistically subtracted using the cocktail
and converter subtraction techniques as discussed in 
section~\ref{subsubsec:rhic_e_pp}. 

The resulting electron spectra from heavy-flavor hadron decays are shown in 
the left panel of Fig.~\ref{fig:phenix_e_dau} for minimum bias \dau collisions,
for four \dau centrality classes, and for \pp data recorded in the 2008 run at 
RHIC with the identical experimental setup as it was used for the \dau 
measurement. The lines represent fits to the previously published \pp reference
as discussed in section section~\ref{subsubsec:rhic_e_pp}, scaled by the 
appropriate number of binary collisions $\ncol$ for each \dau centrality 
classes. The \pp data from the 2008 run of RHIC agree very well with the \pp 
reference. To facilitate a direct comparison of the \dau electron spectra with 
the \pp reference the nuclear modification factor $\rda$, defined as

\begin{equation}
\rda = \frac{dN^e_{\rm dA}d\pt}{\langle \ncol \rangle \times dN^e_{\rm pp}d\pt}
\end{equation}

was calculated for each \dau centrality class. $\rda$ is shown as a function 
of \pt for the most peripheral and the most central \dau centrality class
in the right panel of Fig.~\ref{fig:phenix_e_dau}. 
While in peripheral collisions $\rda$ is consistent with one within large 
systematic uncertainties, in central \dau collisions an enhancement of the 
yield of electrons from heavy-flavor hadron decays is observed up to a 
transverse momentum of 5~\gevc. 
Measurements in \pp collisions suggest that this \pt range is dominated by 
charm hadron decays (see section~\ref{subsubsec:rhic_e_pp}). Because the total 
heavy-flavor production cross section is expected to scale with the number of 
binary collisions, this enhancement at intermediate \pt can be interpreted in 
terms of a pronounced Cronin-like broadening~\cite{cronin75} of the 
heavy-flavor hadron spectra while simultaneously keeping the integrated yield 
unchanged. A similar Cronin broadening was observed in the light quark sector 
in \dau collisions at RHIC, with the enhancement growing with hadron 
mass~\cite{cronin_phenix}.

These electron data provide clear evidence that open heavy-flavor hadrons are 
subject to cold nuclear matter effects at RHIC. Consequently, such effects
are expected to be present in the initial state of nucleus-nucleus collisions
at RHIC. However, in the latter case the Cronin-like enhancement will be
convoluted with effects from the produced hot and dense medium such as partonic
energy loss leading to a suppression of particle yields at high \pt. While
further experimental input from fully reconstructed open heavy-flavor hadrons
and from a clean separation of charm and bottom observables is necessary to 
disentangle all relevant effects at RHIC, it is important to keep this first
direct evidence for cold nuclear matter modifications of an open heavy-flavor 
observable in mind when interpreting data from nucleus-nucleus collisions
at RHIC.

\subsection{Heavy flavor in \auau collisions}
\label{subsec:rhic_auau}
\subsubsection{Hadronic heavy-flavor hadron decays}
\label{subsubsec:rhic_d_auau}
The hot medium produced in heavy-ion collisions at RHIC is expected to leave
its footprint on heavy-flavor observables because heavy quarks interact with
this medium while propagating through it. In this context, detailed direct 
measurements of heavy-flavor hadrons in exclusive hadronic decay channels 
in nucleus-nucleus and the reference \pp system are highly desirable. While
\dzero and \dstarplus measurements with the STAR experiment have become
available for \pp collisions at $\sqrt{s} = 200$~GeV recently (see 
section~\ref{subsubsec:rhic_d_pp}), the situation is different for \auau
collisions at the same available energy per nucleon-nucleon pair. \dzero
mesons have been reconstructed with the STAR experiment at low \pt 
($\pt < 2$~\gevc) in minimum bias ($0 - 80\%$~central) \auau collisions 
via their ${\rm K} \pi$ decay~\cite{star_d_auau}. Due to the unfavorably small 
ratio of the \dzero signal to combinatorial background the extracted yield 
carried a systematic uncertainty of $40 - 50\%$. The direct measurement of 
invariant \pt differential heavy-flavor hadron yields over an extended \pt 
range as a function of the nucleus-nucleus collision centrality will only be 
possible once data become available from the upgraded STAR and PHENIX 
experiments, which include silicon vertex spectrometers capable of separating 
secondary heavy-flavor hadron decay vertices from the primary collision vertex.

\subsubsection{Semielectronic heavy-flavor hadron decays}
\label{subsubsec:rhic_e_auau}
\begin{figure}[t]
\begin{center}
\includegraphics[width=0.49\textwidth]{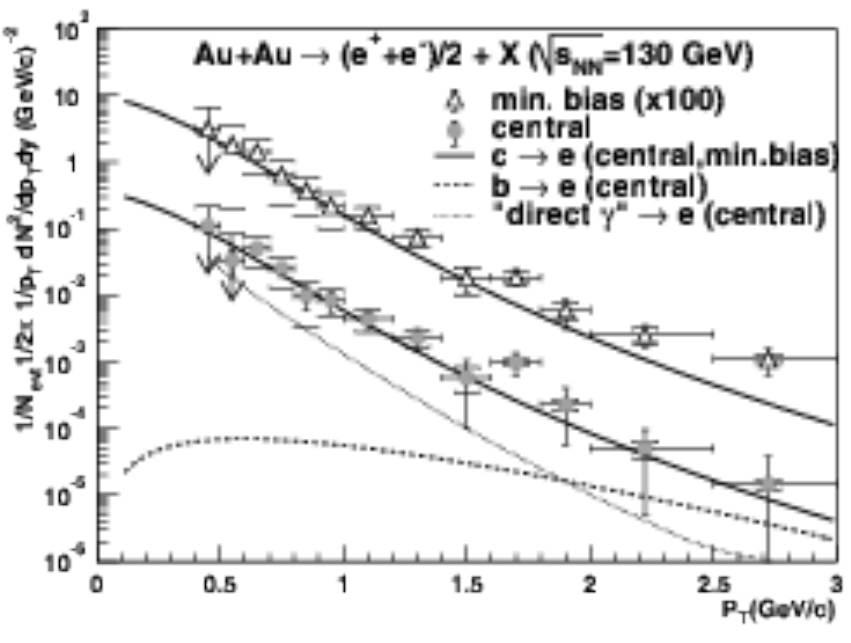}
\includegraphics[width=0.49\textwidth]{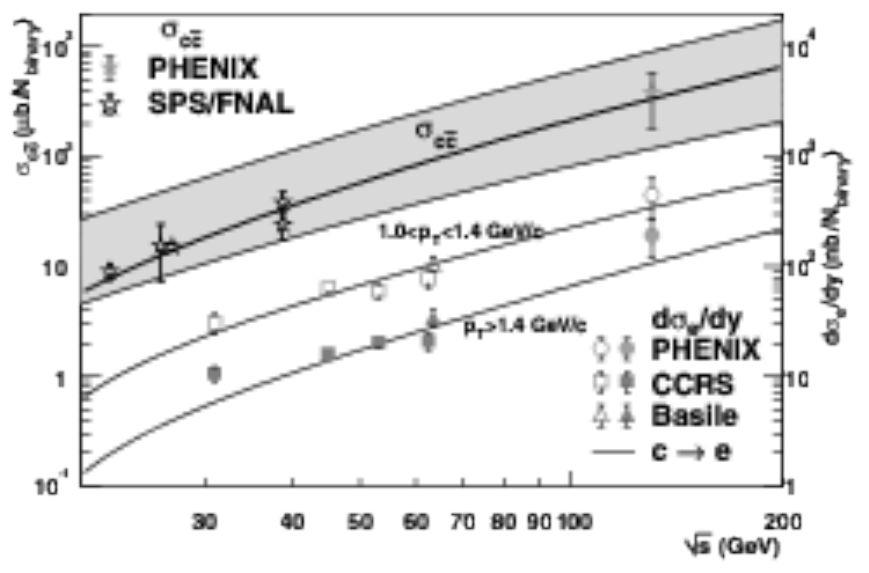}
\end{center}
\caption{\pt differential yields of electrons from open heavy-flavor hadron
         decays measured with the PHENIX experiment in minimum bias and central
         \auau collisions at $\sqrt{s_{\rm NN}} = 130$~GeV. For comparison
         corresponding PYTHIA calculations are shown (left panel). Electron 
         production cross sections from heavy-flavor hadron decays from PHENIX 
         and ISR measurements for the \pt ranges $1.0 < \pt < 1.4$~\gevc (open 
         symbols) and $\pt > 1.4$~\gevc (closed symbols), and the derived total
         charm production cross section per binary collision from PHENIX and 
         other experiments as a function of $\sqrt{s}$. The solid lines and 
         shaded band represent corresponding calculations from PYTHIA and an 
         NLO pQCD calculation respectively (right panel). See text for more 
         details (reprinted with permission from Ref.~\cite{ppg011}; Copyright 
         (2002) by the American Physical Society).}
\label{fig:phenix_e_auau130}
\end{figure}

The first measurement of open heavy-flavor production in nucleus-nucleus
collisions~\cite{ppg011} was conducted with the PHENIX experiment using data 
recorded during the very first run at RHIC. In \auau collisions at 
$\sqrt{s_{\rm NN}} = 130$~GeV, electrons from semileptonic heavy-flavor hadron 
decays were statistically separated from electron background from other 
sources (mainly Dalitz decays of light mesons and photon conversions in 
material) using a cocktail subtraction technique as described above. The
resulting electron \pt spectra from heavy-flavor hadron decays are shown
in the left panel of Fig.~\ref{fig:phenix_e_auau130} for minimum bias
($0 - 92\%$ central) and central ($0 - 10\%$ central) \auau collisions
in the range $0.5 < \pt < 3$~\gevc. A PYTHIA calculation~\cite{pythia} of the 
electron spectrum from heavy-flavor hadron decays in \pp collisions at 
$\sqrt{s} = 130$~GeV was scaled with the nuclear overlap integrals $T_{\rm AA}$
for minimum bias and central \auau collisions, respectively, and is compared
to the data in the left panel of Fig.~\ref{fig:phenix_e_auau130}. The PYTHIA 
parameters were tuned such that charm production data from fixed target 
experiments at the SPS and at FNAL as well as single electron data from the
ISR were reproduced well. Within large uncertainties the scaled PYTHIA
calculations agree with the measured electron yields, which in this rather 
low \pt range predominantly are due to charm hadron decays. This observation
suggests that the total charm yield scales geometrically with the nuclear
overlap integral or, equivalently, with the number of binary collisions when
going from \pp to \auau collisions, as it is expected for hard probes such as
heavy quarks. The measured yield of electrons with $\pt > 0.8$~\gevc at 
midrapidity corresponds to about 15\% of the yield integrated over the full
\pt range. Using the spectral shapes in \pt and $y$ as predicted from PYTHIA
the total charm production cross section per binary collision was estimated.
The latter is shown in the right panel of Fig.~\ref{fig:phenix_e_auau130}
together with corresponding results from ISR 
measurements~\cite{buesser74,basile81} as a function of $\sqrt{s}$. 
In addition, the $\sqrt{s}$ dependence of rapidity densities per 
binary collision of heavy-flavor electron yields at midrapidity in two
electron \pt intervals are depicted. For comparison, corresponding results
from PYTHIA and from a HVQMNR NLO pQCD calculation~\cite{mangano93} of the 
total charm production cross section are shown. Within substantial 
uncertainties the PYTHIA and NLO pQCD calculations are in agreement with the 
data.

While these electron data provided a first glimpse of open heavy-flavor physics
in nucleus-nucleus collisions, the limited precision of the measurement excluded
any detailed conclusion concerning the centrality dependence of open 
heavy-flavor yields, not to speak of the \pt distributions of electrons
from heavy-flavor hadron decays.

\begin{figure}[t]
\begin{center}
\includegraphics[width=0.49\textwidth]{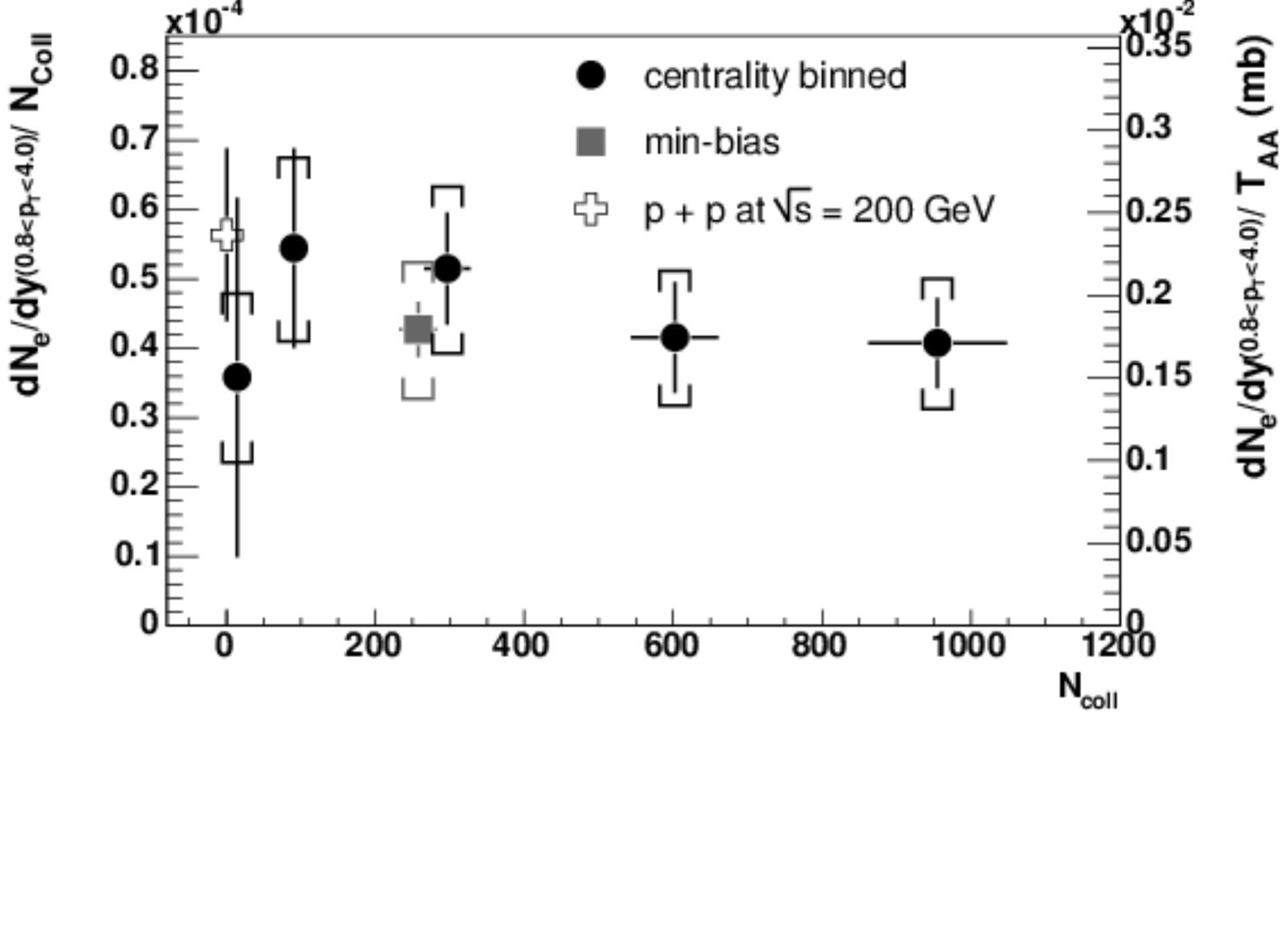}
\includegraphics[width=0.49\textwidth]{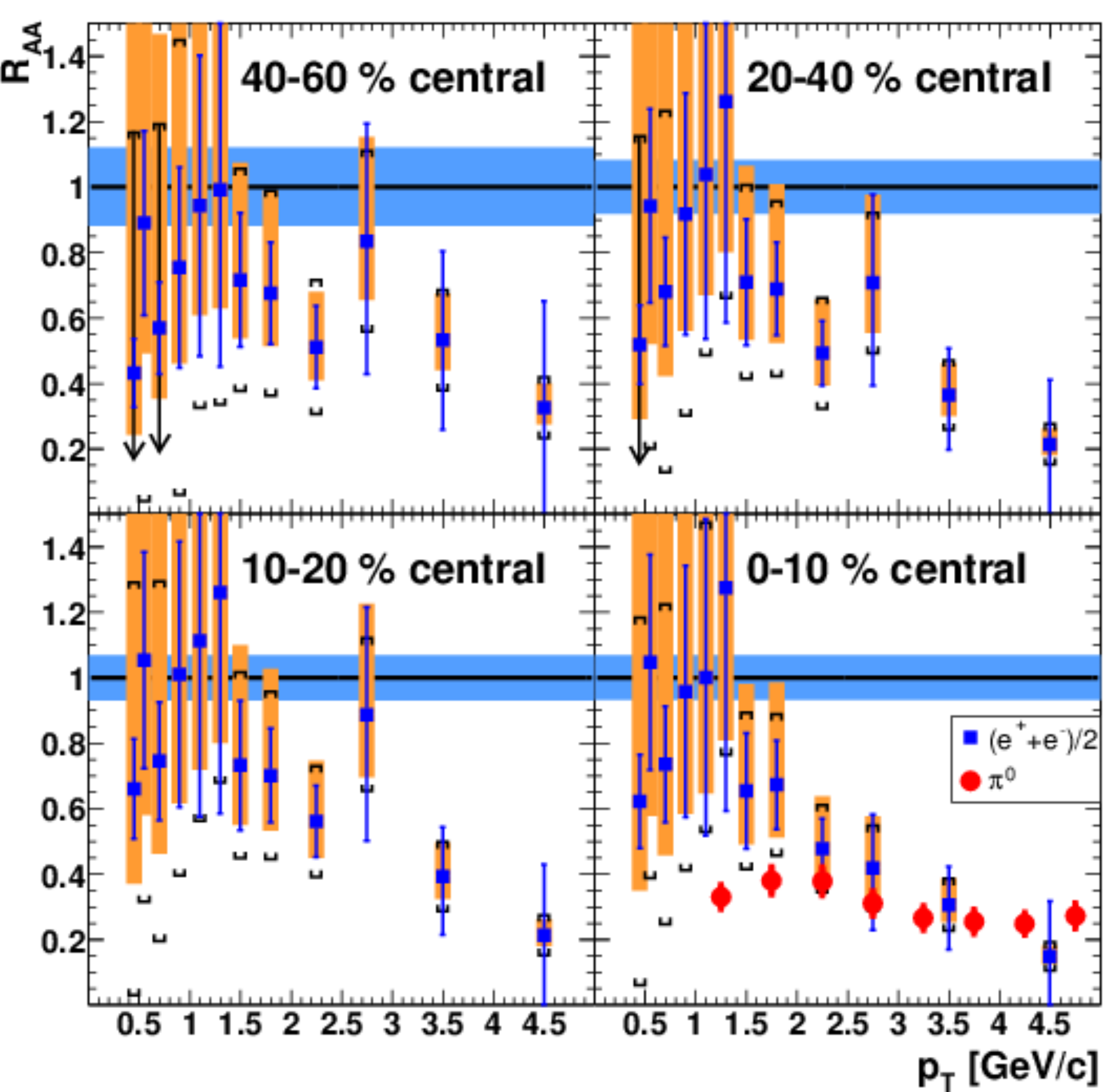}
\end{center}
\caption{Electron yield (left axis) and cross section (right axis) from 
         heavy-flavor hadron decays divided by the number of binary collisions 
         as measured with the PHENIX experiment in \auau collisions at 200~GeV 
         in the range $0.8 < \pt < 4.0$~\gevc as a function of the number of 
         binary collisions, \ie as function of collision centrality. For 
         comparison the corresponding cross section from \pp collisions
         is shown in addition (left panel). Nuclear modification factor $\raa$ 
         of electrons from heavy-flavor hadron decays in four \auau centrality 
         classes at $\sqrt{s_{\rm NN}} = 200$~GeV as a function of \pt. Error 
         bars (brackets) depict the statistical (systematic) errors from the 
         \auau measurement. Error boxes show the systematic errors from the 
         \pp reference, and the band around one indicates the relative 
         systematic uncertainties of the nuclear overlap integrals $T_{\rm AA}$.
         For the most central collisions the neutral pion $\raa$ is shown for 
         comparison (right panel) (reprinted with permission from 
         Refs.~\cite{ppg035} (left) and \cite{ppg056} (right); Copyright (2005,
         2006) by the American Physical Society).}
\label{fig:phenix_auau_binary}
\end{figure}

During the 2001 run period of RHIC a higher statistics electron data sample
was recorded with reduced systematic uncertainties at midrapidity from \auau 
collisions at $\sqrt{s_{\rm NN}} = 200$~GeV with the PHENIX experiment. 
Electrons from heavy-flavor hadron decays were measured in the \pt range 
$0.4 < \pt < 5.0$~\gevc, which is sensitive to charm production only. In a 
first step~\cite{ppg035}, these higher quality data allowed the measurement 
of the centrality dependence of the electron yields from heavy-flavor hadron 
decays in the \pt interval $0.4 < \pt < 4.0$~\gevc. The background from other 
electron sources was subtracted using the converter subtraction technique (see 
section~\ref{subsubsec:phenix} for details), which is superior to the cocktail
subtraction technique at low \pt where the signal to background ratio is small.
The resulting electron yields per binary collision are shown in the left panel
of Fig.~\ref{fig:phenix_auau_binary} as a function of the number of binary
collisions, \ie as a function of the \auau collision centrality. Within the
experimental uncertainties the electron yields, which are a measure of the 
number of produced $\ccbar$ pairs, scale with the number of binary collisions 
and in good agreement with the corresponding cross section measured in \pp 
collisions at the same $\sqrt{s}$. 

The total charm yield scales with the number of binary collisions consistent
with point-like production in a hard process. Final state interactions of 
charm quarks with the hot and dense medium produced in \auau collisions at 
RHIC are not expected to have an effect on the total open charm yield. However,
such effects will influence the momentum distribution of charm hadrons and, 
consequently, their decay products. Electron spectra from heavy-flavor hadron 
decays should be sensitive to such medium modifications for $\pt > 2.5$~\gevc. 
While the converter subtraction technique is well suited for the measurement 
of the total yield of electrons from heavy-flavor hadron decays (\ie for low 
\pt electrons), it is of advantage to use the complementary cocktail 
subtraction technique for a precision measurement of the spectral shape 
towards larger electron \pt.

Models utilizing radiative energy loss via induced gluon radiation as the 
relevant energy loss mechanism for heavy quarks in the hot and dense medium
produced in \auau collisions at RHIC predict only a moderate energy loss of 
charm and, in particular, bottom quarks at transverse momenta up to a few 
\gevc~\cite{dglv,bdmps}. In any case, due to the dead cone 
effect~\cite{dead_cone} the radiative energy loss of heavy quarks should be 
smaller than for light quarks. In view of this expectation, the 
discovery~\cite{ppg056} of a substantial suppression 
of the electron yield from heavy-flavor hadron decays at intermediate and high 
\pt in central \auau collisions relative to the binary collision scaled yields 
measured in \pp collisions came as a complete surprise.

The right panel of Fig.~\ref{fig:phenix_auau_binary} shows the nuclear 
modification factor $\raa$ as a function of \pt for electrons from heavy-flavor 
hadron decays in four \auau centrality classes at 
$\sqrt{s_{\rm NN}} = 200$~GeV~\cite{ppg056}. The \pp reference was taken from 
the measurement described in Ref.~\cite{ppg037}.
Due to small signal to background ratio at low \pt the electrons yields at low
\pt suffer from large systematic uncertainties which are reflected by the
substantial uncertainties of $\raa$ for $\pt < 1.5$~\gevc. However, for all
centrality classes $\raa$ is consistent with one in the low \pt region, which 
includes the dominant fraction of the total yields, in agreement with the 
observed scaling of the total charm yield with the number of binary collisions. 
Towards higher \pt a suppression of electrons from heavy-flavor hadron decays
develops which is most pronounced in the most central collisions at the highest
\pt. For comparison, the nuclear modification factor of neutral pions from 
\cite{ppg014} is shown for the case of the 10\% most central \auau collisions. 
For $\pt > 3$~\gevc the nuclear modification factor of electrons from 
heavy-flavor hadron decays agrees quantitatively with the $\raa$ of neutral
pions in central \auau collisions at RHIC. This discovery presents a major 
challenge for models attempting to describe the partonic energy loss of heavy 
quarks traversing the medium created at RHIC.

\begin{figure}[t]
\begin{center}
\includegraphics[width=0.54\textwidth]{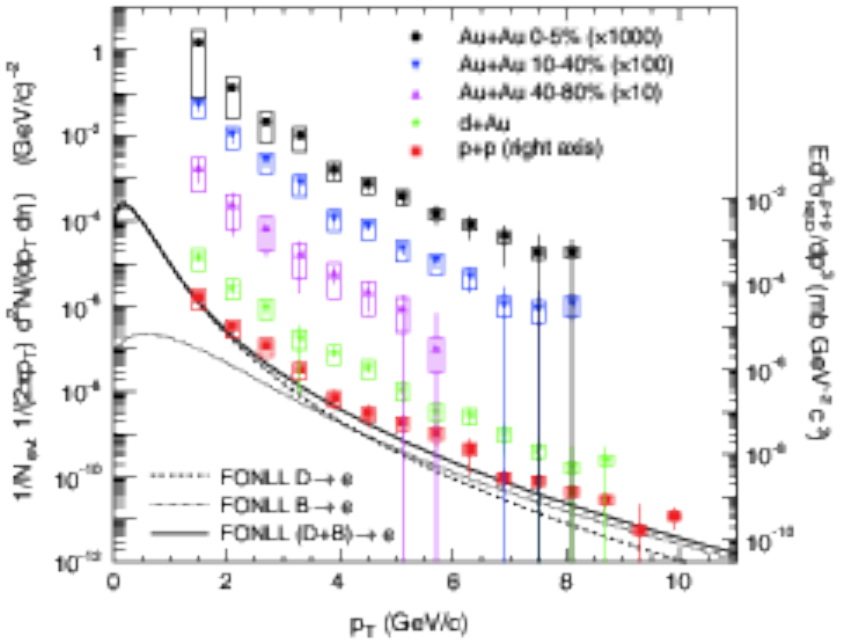}
\includegraphics[width=0.44\textwidth]{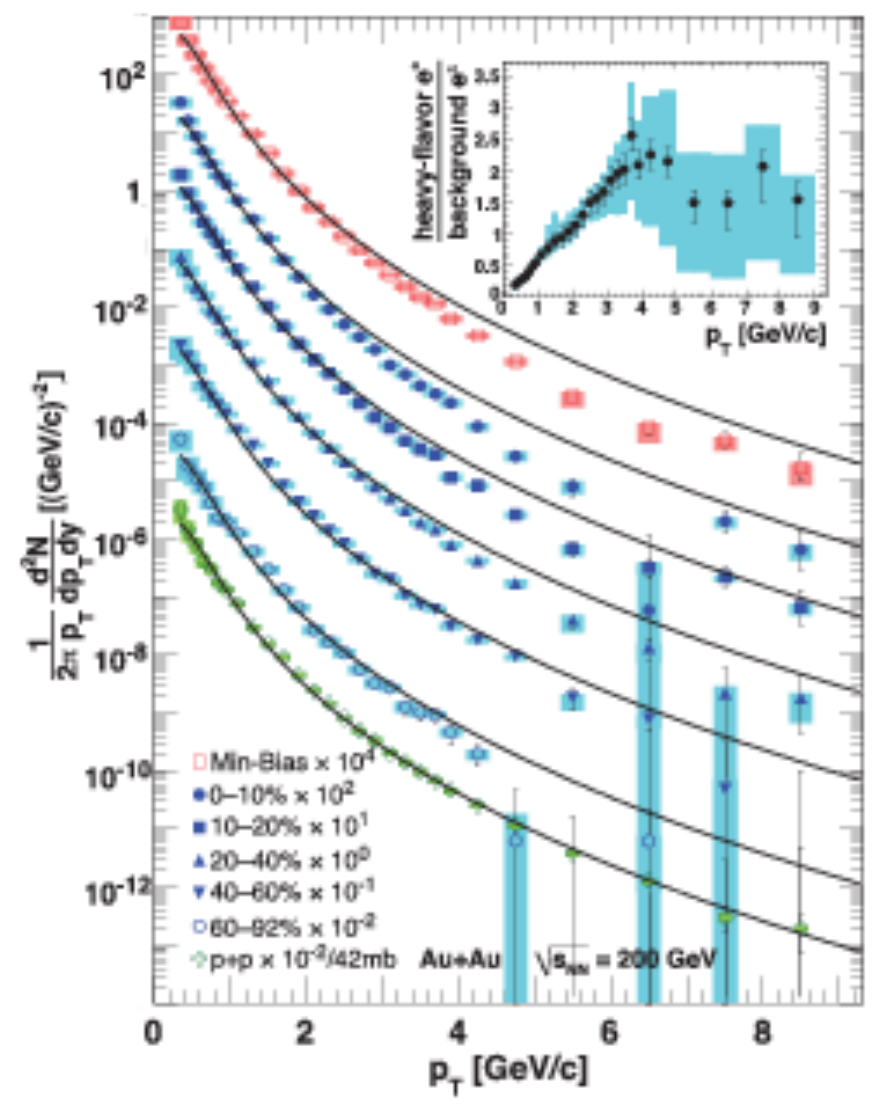}
\end{center}
\caption{Electron \pt spectra from heavy-flavor hadron decays measured with the
         STAR experiment in \pp, \dau, and various \auau centrality classes
         at $\sqrt{s_{\rm NN}} = 200$~GeV. FONLL predictions~\cite{fonll1,fonll2}
         are compared with the data for the \pp case, for which the right axis 
         indicates the corresponding cross section values (left panel). 
         Invariant \pt differential distributions of electrons from 
         heavy-flavor hadron decays measured with the PHENIX experiment for 
         various \auau centrality classes and for \pp collisions. The solid 
         lines are a from a FONLL calculation and the inset shows the signal 
         to background ratio for minimum bias \auau collisions (right panel)
         (reprinted with permission from Refs.~\cite{star_e_auau} (left) and 
         \cite{ppg077} (right); Copyright (2007, 2011) by the American Physical
         Society).}
\label{fig:rhic_e_auau200}
\end{figure}

Detailed model comparisons became possible with higher statistics electron data
from heavy-flavor hadron decays recorded in \auau collisions at 
$\sqrt{s_{\rm NN}} = 200$~GeV with the PHENIX~\cite{ppg066,ppg077} and STAR 
experiments~\cite{star_e_auau} in the 2004 run at RHIC. With the STAR 
experiment electrons were measured in this run in the transverse momentum 
range $1.2 < \pt < 10$~\gevc. Electron background from photonic sources, \ie 
light meson Dalitz decays and electrons from photon conversions in material, 
was measured via an invariant mass technique. 
Remaining background from other sources was subtracted using a cocktail 
approach. The left panel of Fig.~\ref{fig:rhic_e_auau200} depicts the 
resulting heavy-flavor electron spectra measured with the STAR experiment in 
three \auau centrality classes, \dau collisions, and the reference \pp 
system~\cite{star_e_auau}. For comparison results from FONLL pQCD 
calculations~\cite{fonll1,fonll2} scaled to the \pp 
measurement are shown, which indicate that up to 3~\gevc electrons originate
mainly from charm hadron decays while towards higher \pt bottom hadron decays
become increasingly more important. With the PHENIX experiment electrons 
have been measured in the transverse momentum range $0.3 < \pt < 9.0$~\gevc at 
midrapidity ($|y| < 0.35$). Electron background from other sources was 
subtracted statistically using the converter method for $\pt < 1.6$~\gevc
and the cocktail method for higher \pt. The resulting electron spectra from
heavy-flavor hadron decays for various \auau centrality 
classes~\cite{ppg066,ppg077} are shown in the right panel of 
Fig.~\ref{fig:rhic_e_auau200} together with the corresponding measurement from 
\pp collisions~\cite{ppg065}. The solid lines depict the
line shape from a FONLL calculation~\cite{fonll1,fonll2} normalized to the \pp 
data and scaled with the number of binary collisions for each \auau centrality 
class. 

\begin{figure}[t]
\begin{center}
\includegraphics[width=0.5\textwidth]{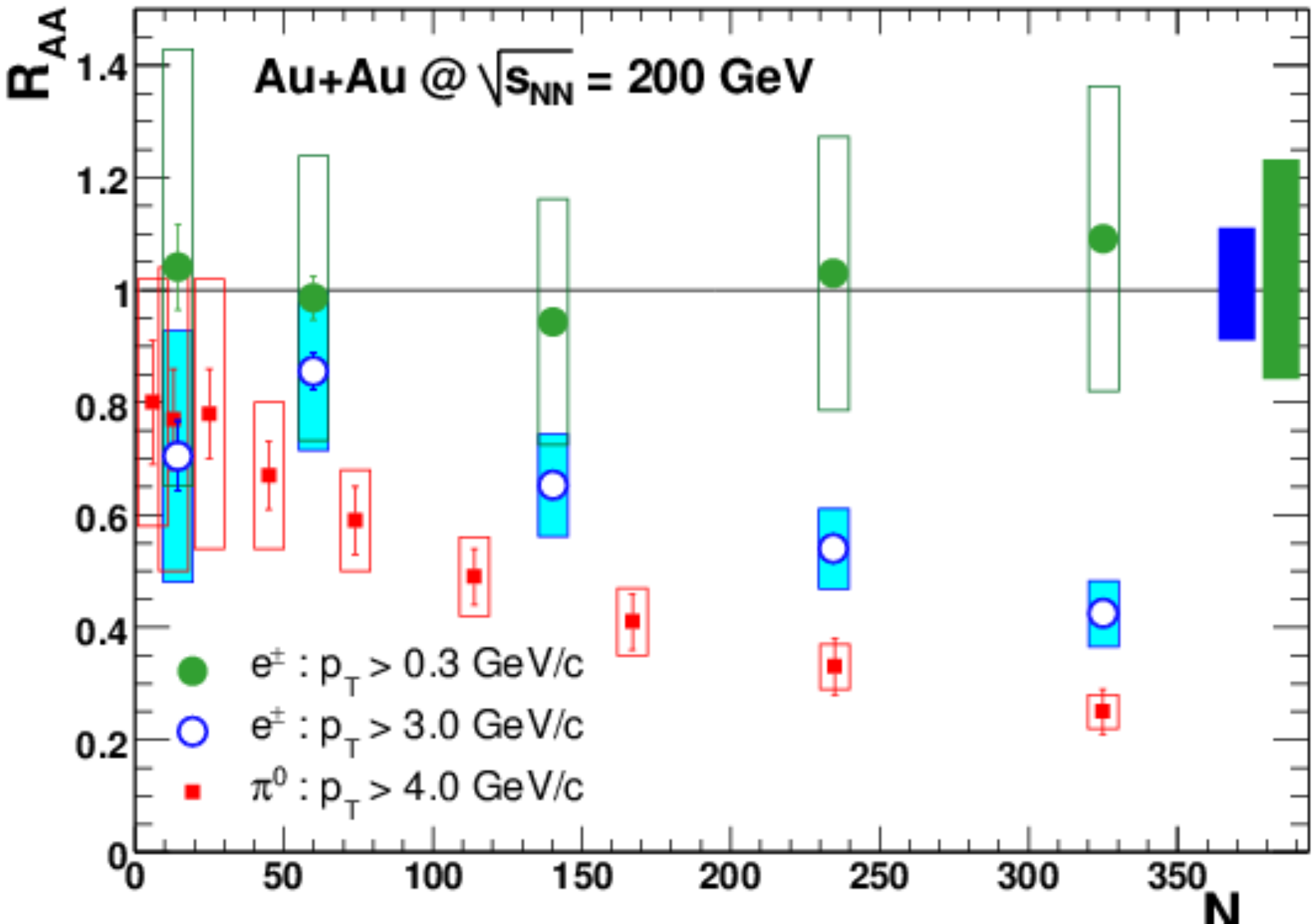}
\includegraphics[width=0.49\textwidth]{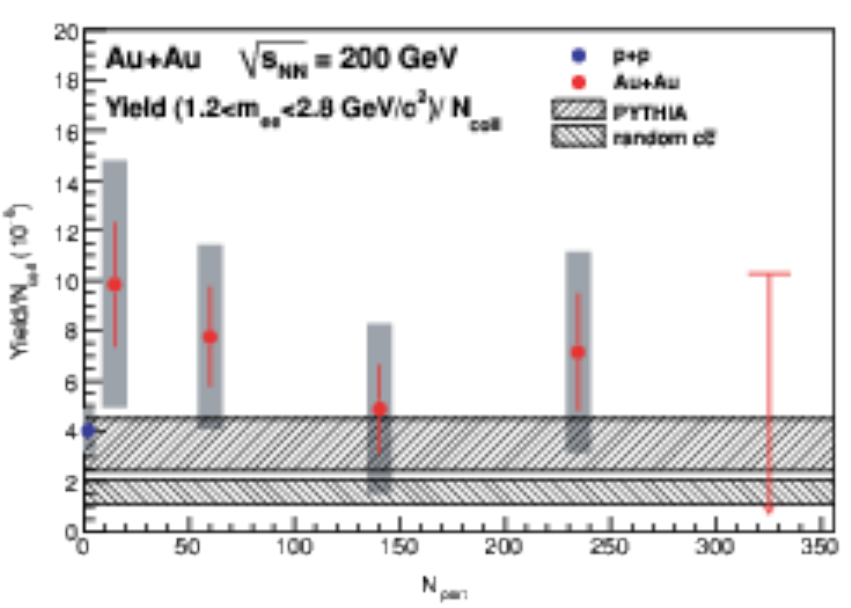}
\end{center}
\caption{Nuclear modification factor $\raa$ of electrons from heavy-flavor 
         hadron decays measured with the PHENIX experiment in \auau collisions 
         at $\sqrt{s_{\rm NN}} = 200$~GeV as a function of the collision 
         centrality (quantified by the number of participants, $\npart$) for 
         the electron transverse momentum ranges $\pt > 0.3$~\gevc and 
         $\pt > 3.0$~\gevc, respectively. 
         The boxes at $\raa = 1$ indicate the common relative 
         uncertainty from the \pp reference for $\pt > 0.3$~\gevc (right box)
         and $\pt > 3.0$~\gevc (left box). $\raa$ of neutral pions with 
         $\pt > 4.0$~\gevc~\cite{ppg014} is shown for comparison (left panel). 
         Dielectron yield in the intermediate mass range 
         $1.2 < m_{\rm ee} < 2.8$~\gevcsquare per binary collision 
         collision as a function of $\npart$ in comparison with two bands 
         corresponding to different estimates of the contribution from charm 
         hadron decays. See text for more details (right panel) (reprinted with
         permission from Refs.~\cite{ppg066} (left) and \cite{ppg088} (right); 
         Copyright (2007, 2010) by the American Physical Society).}
\label{fig:phenix_raa_vs_centrality}
\end{figure}

While the binary collision scaled \pp reference agrees with the electron yields
from heavy-flavor hadron decays at low \pt for all \auau centrality classes
this is different at high \pt where the heavy-flavor electron yields in \auau
are suppressed relative to the scaled \pp reference. The suppression is more
pronounced in central collisions compared to more peripheral centrality
classes. This behavior is quantified in the left panel of 
Fig.~\ref{fig:phenix_raa_vs_centrality} which shows the nuclear modification 
factor $\raa$ of electrons from heavy-flavor hadron decays for two \pt ranges 
as a function of the number of participating nucleons, $\npart$, in \auau 
collisions~\cite{ppg066}. For the range $\pt > 0.3$~\gevc, which includes more 
than 50\% of the total electron yield from heavy-flavor hadron decays, $\raa$ 
is consistent with one, confirming the binary collision scaling of the total 
heavy-flavor hadron yield in \auau collisions at RHIC. For the range 
$\pt > 3$~\gevc, $\raa$ decreases with increasing centrality reaching a value 
of $\approx 0.4$ for the 10\% most central \auau collisions at 
$\sqrt{s_{\rm NN}} = 200$~GeV. 
The suppression of electrons from heavy-flavor hadron decays can not be 
compared directly to light hadron suppression in the same \pt range. However, 
electrons from charm hadron decays with $\pt > 3$~\gevc originate to a large 
extent from the decay of D mesons with $\pt > 4$~\gevc~\cite{pythia}. 
Therefore, the electron $\raa$ above 3~\gevc is compared with the neutral 
pion $\raa$ above 4~\gevc~\cite{ppg014}. In this intermediate \pt range the 
trend visible in the left panel of Fig.~\ref{fig:phenix_raa_vs_centrality} 
indicates a smaller suppression of heavy-flavor hadrons relative to 
light-flavor mesons in the hot and dense medium produced in \auau collisions at
RHIC. However, for a firm conclusion it would be necessary to consider also 
cold nuclear matter effects, which might be different for light and 
heavy-flavor hadrons and which have been measured with limited precision only 
up to now~\cite{star_e_auau,ppg131}. Furthermore, the
interpretation of electron spectra from heavy-flavor hadron decays in \auau
collisions is not straight forward because of the decay kinematics involved
and, most important, because the relative contributions from charm and bottom
hadron decays can be different from those measured with substantial 
uncertainties in \pp collisions at the same energy~\cite{ppg094,star_e_b_d}.

A similar general picture emerges from the centrality dependence of dielectron 
yields per binary collision measured with the PHENIX experiment~\cite{ppg088} 
in the intermediate mass range $1.2 < m_{\rm ee} < 2.8$~\gevcsquare as shown in 
the right panel of Fig.~\ref{fig:phenix_raa_vs_centrality}. This mass region is 
dominantly populated by dielectrons from correlated charm hadron decays. The
measured yields are compared with two extreme scenarios concerning the
correlation between the original charm quark and antiquark. In the first 
approach, a PYTHIA calculation for \pp collisions~\cite{pythia} is simply 
scaled with the number of binary collisions to the \auau case, thus keeping 
the $\ccbar$ correlation intact, \ie unmodified by the interaction with the hot
and dense medium in \auau collisions. In the second approach, the kinematic
$\ccbar$ correlation is assumed to be completely destroyed by the interaction 
with the medium, \ie the electrons and positrons are sampled randomly from the 
measured electron spectra from heavy-flavor hadron decays~\cite{ppg066}. This 
approach gives rise to a considerable softening of the dielectron mass spectra
resulting in a reduced dielectron yield in the intermediate mass region
compared to the PYTHIA scaling as indicated in the right panel of 
Fig.~\ref{fig:phenix_raa_vs_centrality}. In view of the suppression of 
the electron yields from heavy-flavor hadron decays at high \pt the observed 
binary collision scaling of the intermediate mass dielectron yield for all
collisions centralities might point to the cancellation of two competing
effects. With increasing centrality the energy loss of charm increases, 
resulting in a softening of the dielectron mass spectra and, consequently, in
a reduced dielectron yield in the intermediate mass region. This might be 
compensated by an increasing yield of dielectrons from thermal radiation
with centrality, which was observed previously with the NA60 
experiment~\cite{arnaldi09a,arnaldi09b} at the CERN SPS (see 
section~\ref{subsec:preRHIC_fixed}).

\begin{figure}[t]
\begin{center}
\includegraphics[width=0.5\textwidth]{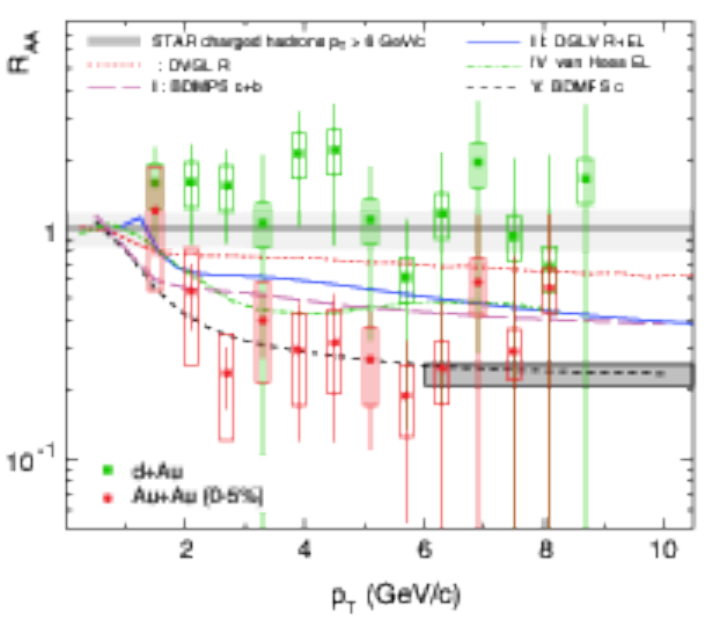}
\end{center}
\caption{Nuclear modification factor $\raa$ of electrons from heavy-flavor
         hadron decays measured as a function of \pt with the STAR experiment 
         in \dau and in the 5\% most central \auau collisions at 
         $\sqrt{s_{\rm NN}} = 200$~GeV. For comparison the corresponding $\raa$
         of charged hadrons~\cite{star_had_sup} is indicated as well as results
         from various theoretical model calculations. See text for more details
         (reprinted with permission from Ref.~\cite{star_e_auau}; Copyright 
         (2007) by the American Physical Society).}
\label{fig:star_e_raa}
\end{figure}

In order to address the suppression of electrons from heavy-flavor hadron
decays quantitatively and to relate this with the energy loss of heavy quarks
traversing the hot and dense medium produced in central \auau collisions
at RHIC, model calculations of the nuclear modification factor $\raa(\pt)$ 
from various theoretical approaches are confronted with electron data from 
the most central \auau collisions. 

Fig.~\ref{fig:star_e_raa} shows such a comparison for the case of the STAR 
experiment~\cite{star_e_auau}. The nuclear modification factor of electrons 
from heavy-flavor hadron decays in \dau collisions is consistent with a 
moderate Cronin enhancement albeit with large uncertainties. For the 5\% most 
central \auau collisions predictions of the heavy-flavor electron $\raa$ are 
shown from five different model calculations. Curve I corresponds to radiative 
energy loss via a few hard scattering processes in the medium calculated within
the DGLV formalism~\cite{dglv}, where an initial gluon density of 
$dN_{\rm g}/dy = 1000$ was assumed, consistent with hadron suppression in the 
light quark sector. Another radiative energy loss calculation, employing 
multiple soft collisions within the BDMPS formalism~\cite{bdmps} is shown as 
curve II, where the maximum transport coefficient $\hat{q} = 14$~GeV$^2$/fm was
used which is still consistent with light hadron suppression. Both radiative 
energy loss calculations predict less suppression than is observed. Adding 
collisional, \ie elastic, energy loss to a DGLV-based calculation and 
introducing path length fluctuations~\cite{wicks07} leads to
a larger, but still not sufficient suppression as demonstrated by curve III. 
In a complementary approach (curve IV) heavy-quark energy loss is implemented 
via elastic scattering involving charm and bottom resonance excitations in the 
medium as well as leading order t-channel gluon exchange~\cite{vanhees05}. In 
particular towards high \pt, where the contribution from bottom hadron decays 
is expected to become important heavy-quark energy loss models consistently 
seem to underpredict the observed suppression. However, given the relatively 
large uncertainties of the STAR electron data it is difficult to draw definite 
conclusions going beyond the general statement that in central \auau collisions
the yield of electrons from heavy-flavor hadron decays is strongly suppressed 
at high \pt, consistent with substantial energy loss of heavy quarks in the 
hot and dense medium produced at RHIC.

\begin{figure}[t]
\begin{center}
\includegraphics[width=0.6\textwidth]{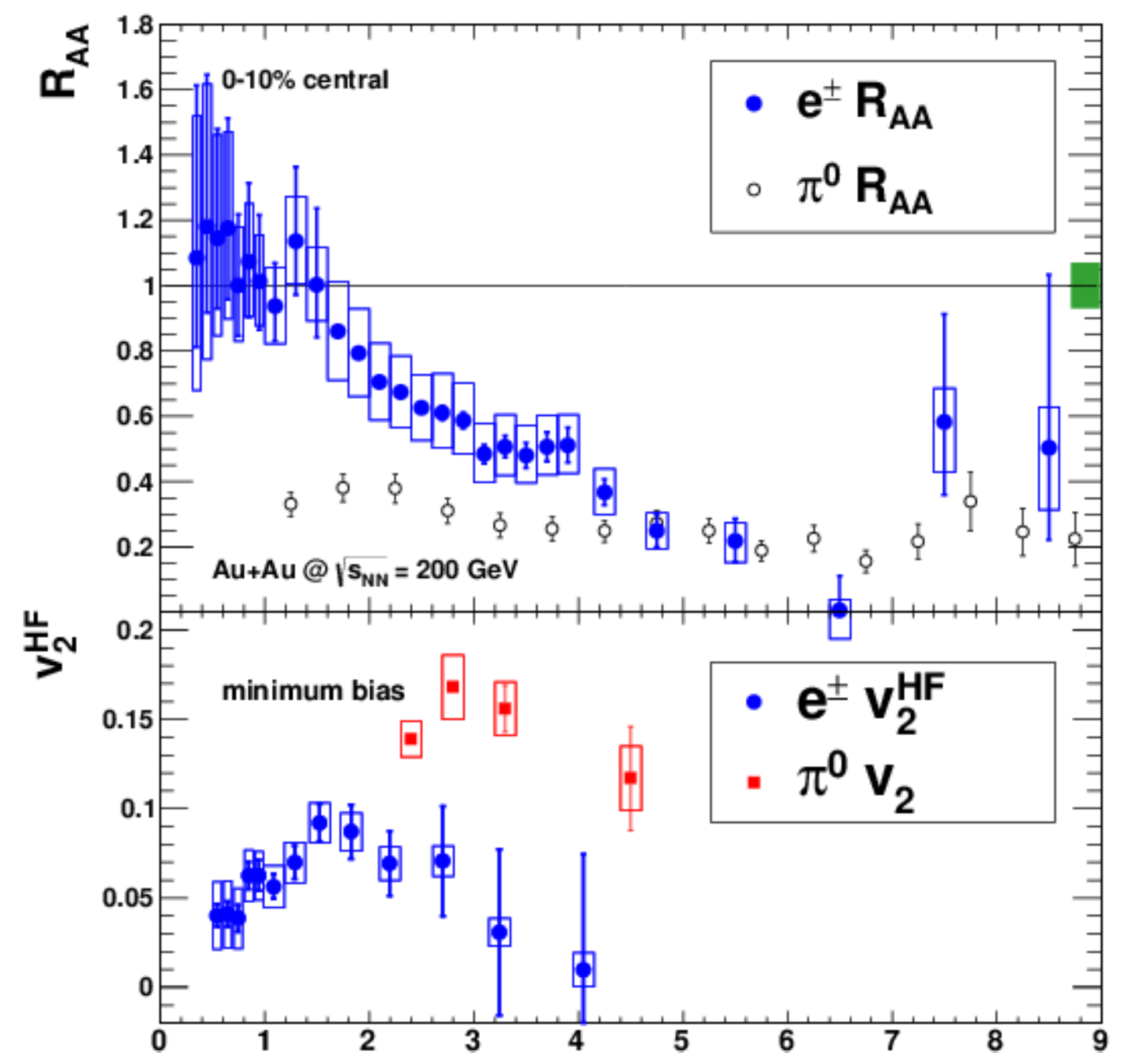}
\end{center}
\caption{$\raa$ of electrons from heavy-flavor hadron decays measured as
         a function of \pt with the PHENIX experiment in the 10\% most central 
         \auau collisions at 200~GeV~\cite{ppg077} in comparison with the 
         corresponding neutral pion $\raa$ from Ref.~\cite{ppg014} (upper 
         panel). Elliptic flow magnitude $v_2$ of electrons from heavy-flavor 
         hadron decays~\cite{ppg077} and of neutral pions~\cite{ppg046} in 
         minimum bias \auau collisions (lower panel).}
\label{fig:phenix_raa_v2}
\end{figure}

The PHENIX electron data from heavy-flavor hadron decays provide a higher 
precision and, therefore, allow a better discrimination between different 
theoretical model calculations~\cite{ppg077}. The nuclear modification factor 
$\raa$ of electrons from heavy-flavor hadron decays, shown as a function of 
\pt for the 10\% most central \auau collisions in the upper panel of 
Fig.~\ref{fig:phenix_raa_v2}, is larger compared to the neutral pion 
$\raa$~\cite{ppg014} at low and intermediate \pt but towards high \pt the 
suppression of the two particle species is of similar size. Further constraints
are provided by the magnitude $v_2$ of the elliptic flow of electrons from 
heavy-flavor hadron decays as compared for minimum bias \auau collisions to the 
corresponding neutral pion $v_2$~\cite{ppg046} in the lower panel of
Fig.~\ref{fig:phenix_raa_v2}. Although for \pt above 2~\gevc the $\pi^0$ $v_2$ 
is larger than the $v_2$ of electrons from heavy-flavor decays the latter 
is clearly larger than zero up to a \pt of $\approx 3$~\gevc. 
The suppression of electrons from 
heavy-flavor hadron decays at high \pt and the simultaneous participation 
of these hadrons in the collective dynamical expansion of the fireball, as 
indicated by the non-zero flow of decay electrons, can be interpreted in terms 
of strong interactions of heavy flavor with the hot and dense medium produced 
at RHIC. By confronting predictions from various theoretical model calculations
with the electron data one might hope to unravel the nature of the mechanisms
responsible for the observed heavy-flavor medium modifications and, at the
same time, approach a quantitative understanding of the properties of hot
QCD matter. 

A first question is whether the measured $\raa$ and $v_2$ of electrons from
heavy-flavor hadron decays require heavy-flavor interactions with the medium
on a partonic level or whether (pre)hadronic interactions are sufficient to
explain the observed medium effects. The $\raa$ and $v_2$ of D mesons was 
calculated in the framework of the hadron-string-dynamics transport model 
(HSD)~\cite{hsd} which implements heavy-flavor interactions with the medium on 
a (pre) hadronic level. Only at low \pt the predicted D-meson $\raa$ is 
consistent with the electron measurement. Towards higher \pt the calculated 
D-meson energy loss is not sufficient to account for the small electron $\raa$.
In addition, the substantial $v_2$ of electrons from heavy-flavor hadron decays
can not be reproduced in this model, indicating that strong partonic
interactions are relevant for the observed medium effect on heavy-flavor
observables in \auau collisions at RHIC.

\begin{figure}[t]
\begin{center}
\includegraphics[width=0.45\textwidth]{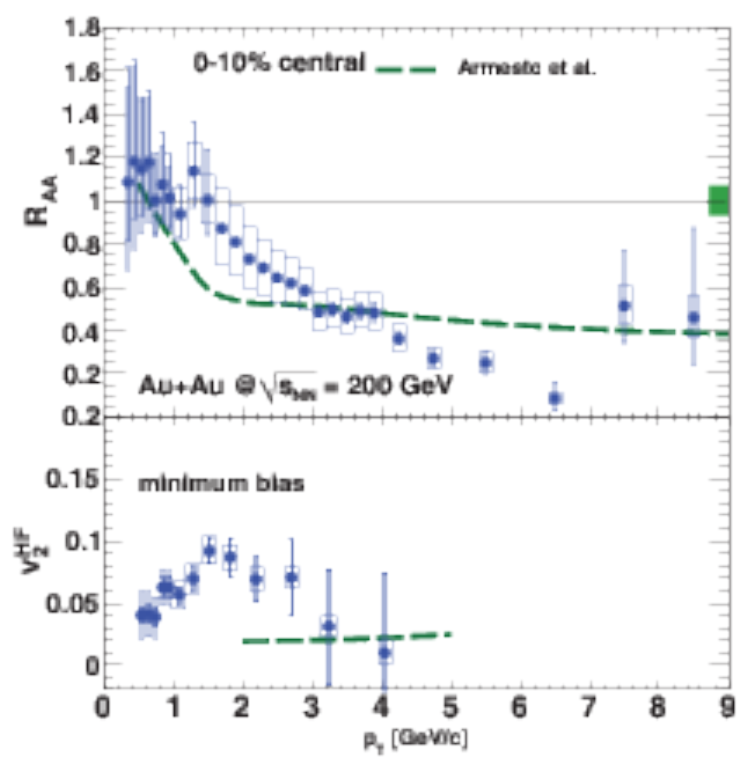}
\includegraphics[width=0.45\textwidth]{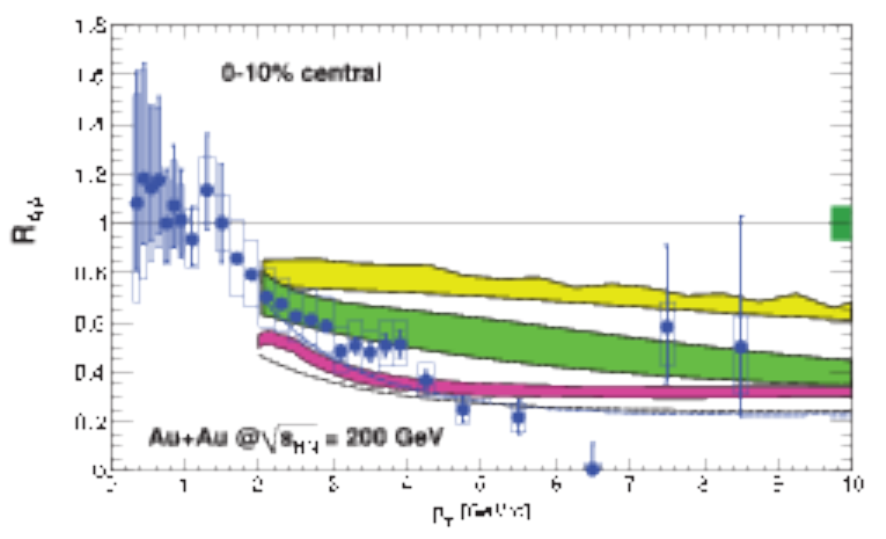}
\end{center}
\caption{Heavy-flavor electron $\raa$ in central \auau collisions and $v_2$ in
         minimum bias collisions predicted from a radiative energy loss
         calculation within the BDMPS framework~\cite{bdmps} (left panel). 
         $\raa$ in central \auau collisions from a DGLV 
         calculation~\cite{wicks07} with radiative energy loss only (uppermost 
         band) and with additional collisional energy loss (middle band) for 
         electrons from charm and bottom hadron decays. The thin dashed curves 
         correspond to charm hadron decays only. The lowest band shows the 
         prediction from a collisional dissociation model~\cite{adil07} (right 
         panel) (reprinted with permission from Ref.~\cite{ppg077}; 
         Copyright (2011) by the American Physical Society).}
\label{fig:phenix_raa_vs_models_rad}
\end{figure}

Originally it was expected that the dominant energy loss mechanism of heavy
quarks in hot QCD matter at RHIC would be induced gluon radiation, leading
to a reduced suppression of heavy quarks with respect to light quarks due to
the dead-cone effect. Predictions using the BDMPS framework~\cite{bdmps}, in 
which radiative partonic energy loss is implemented via multiple soft 
interactions of the partons with the medium, are confronted with the PHENIX 
electron data~\cite{ppg077} in the left panel of 
Fig.~\ref{fig:phenix_raa_vs_models_rad}. The relevant parameter
concerning the strength of the parton-medium interaction in this model is the
transport coefficient $\hat{q}$. For the curves shown in the left panel of 
Fig.~\ref{fig:phenix_raa_vs_models_rad} $\hat{q} = 14$~GeV$^2$/fm was used
which is close to the upper limit still consistent with the hadron suppression
observed in the light-flavor sector. In particular at high \pt, the model
prediction is difficult to reconcile with the data if bottom hadron decays
contribute significantly to the electron spectra in this region. Since this
BDMPS calculation does not include any collective effects at low \pt the 
comparison of the predicted $v_2$ with the data is of limited usefulness.
It is interesting to note that the predicted $v_2 > 0$ provides a lower limit
of $v_2$ at high \pt which is due to the energy loss in the medium in 
combination with the initial geometry of the collision zone.

An alternative radiate energy loss approach is implemented in the DGLV 
framework~\cite{wicks07,dglv}, in which partons interact with the medium via 
a few hard collisions. The predicted $\raa$ of electrons from heavy-flavor 
hadron decays, shown as the uppermost band in the right panel of 
Fig.~\ref{fig:phenix_raa_vs_models_rad}, is clearly above the measured $\raa$ 
at high \pt, consistent with the BDMPS prediction. It was pointed 
out~\cite{mustafa05} that for heavy quarks radiative and collisional energy 
loss via elastic scattering are of similar magnitude over a broad
kinematic range at RHIC. Adding collisional energy loss to the DGLV calculation
leads to a reduction of the electron $\raa$ as demonstrated by the middle band
in the right panel of Fig.~\ref{fig:phenix_raa_vs_models_rad} which is still
above the measured electron $\raa$ at high \pt.

\begin{figure}[t]
\begin{center}
\includegraphics[width=0.45\textwidth]{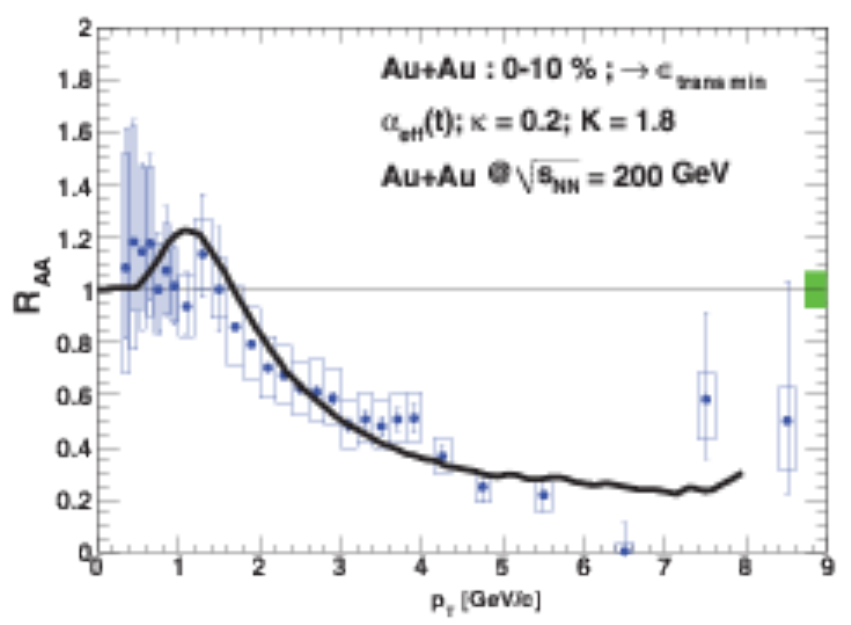}
\includegraphics[width=0.45\textwidth]{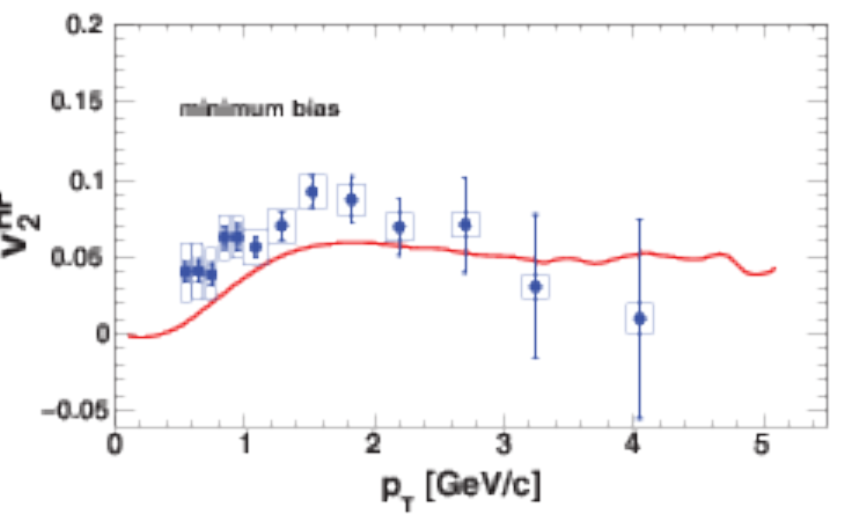}
\end{center}
\caption{Nuclear modification factor $\raa$ of electrons from heavy-flavor
         hadron decays in central \auau collisions (left panel) and 
         corresponding elliptic flow magnitude $v_2$ in minimum bias collisions
         (right panel) as predicted from a collisional energy loss 
         model~\cite{gossiaux08,gossiaux09a,gossiaux09b} in comparison with 
         the PHENIX data (reprinted with permission from Ref.~\cite{ppg077}; 
         Copyright (2011) by the American Physical Society).}
\label{fig:phenix_raa_vs_models_pol}
\end{figure}

Following a different Ansatz~\cite{gossiaux08,gossiaux09a,gossiaux09b}, 
collisional energy loss in pQCD was implemented using a running coupling 
constant $\alpha_{\rm eff}$ and replacing the Debye mass $m_{\rm D}$ by a hard 
thermal loop calculation. An important parameter in this model is a $K$ factor 
which is a scaling parameter for the collisional energy loss. $\raa$ and $v_2$ 
predicted from this model are compared with the data in 
Fig.~\ref{fig:phenix_raa_vs_models_pol}. With a $K$ factor of 1.8, the model 
prediction is in good agreement with the measured
$\raa$ of electrons from heavy-flavor hadron decays, not only for the most
central collisions but for all \auau collision centralities. Furthermore,
$K = 1.8$ leaves room for additional contributions from radiative or other
energy loss mechanisms which are not implemented explicitly. For the 
calculation of the elliptic flow magnitude $v_2$ the interaction time is 
important in this model, where a later decoupling of heavy quarks from the 
medium in general can produce larger values of $v_2$. The curve shown in the 
right panel of Fig.~\ref{fig:phenix_raa_vs_models_pol} corresponds to the 
situation in which the hadronization of heavy quarks takes place at the end 
of the mixed phase.

\begin{figure}[t]
\begin{center}
\includegraphics[width=0.45\textwidth]{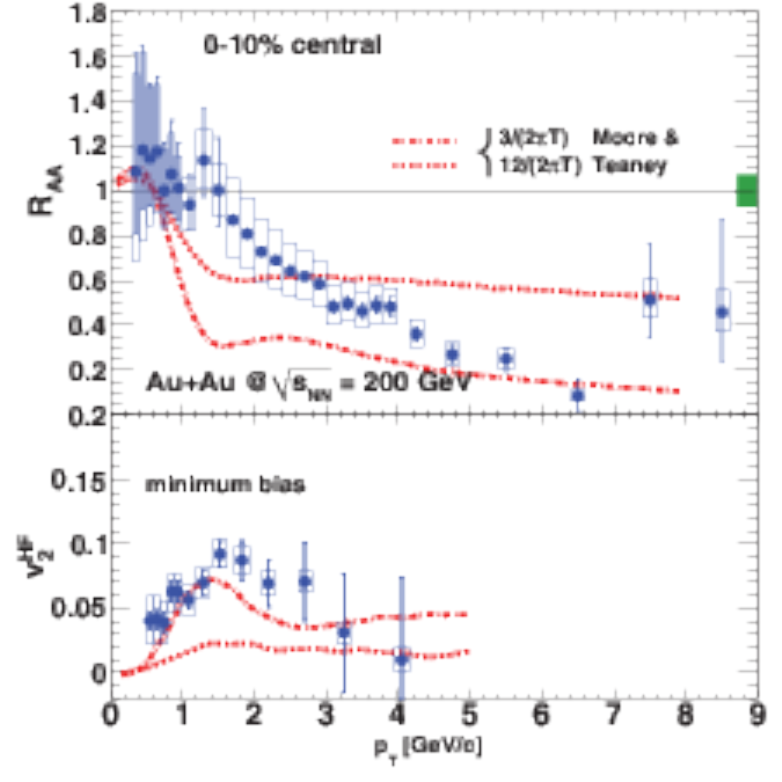}
\includegraphics[width=0.45\textwidth]{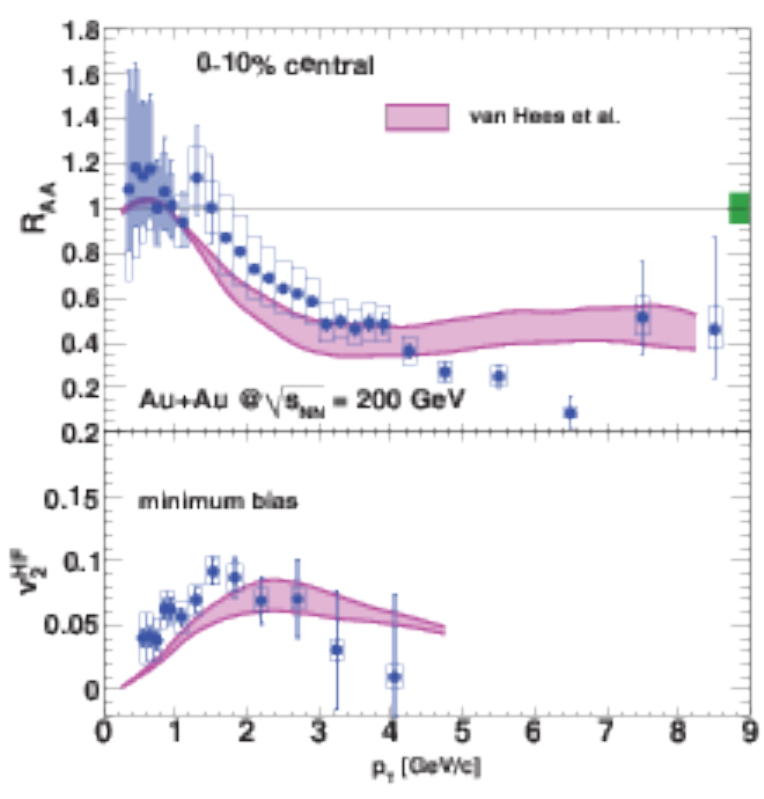}
\end{center}
\caption{Nuclear modification factor $\raa$ of electrons from heavy-flavor
         hadron decays in central \auau collisions and corresponding elliptic 
         flow magnitude $v_2$ in minimum bias collisions as predicted from 
         Langevin-based transport models without~\cite{moore05} (left panel) 
         and with~\cite{vanhees05,vanhees08} (right panel) resonance excitation
         of heavy-flavor like hadrons in the medium. See text for details 
         (reprinted with permission from Ref.~\cite{ppg077}; 
         Copyright (2011) by the American Physical Society).}
\label{fig:phenix_raa_vs_models_diff}
\end{figure}

Another class of models uses a Langevin-based transport approach to describe 
the propagation of heavy quarks in the hot medium, \ie these quarks are placed
into a thermal medium and the quark-medium interaction is implemented by 
uncorrelated elastic momentum kicks. A relevant parameter in this approach is
the heavy-quark diffusion coefficient. The predictions from one of these 
models~\cite{moore05} are compared with the data for two different values of 
the diffusion coefficient in the left panel of 
Fig.~\ref{fig:phenix_raa_vs_models_diff}. 
While the general features of the data are reproduced by this model, it can 
not simultaneously describe the measured $\raa$ and $v_2$ of single electrons 
from heavy-flavor hadron decays with a single value of the heavy-quark 
diffusion coefficient.
In a similar approach, also based on Langevin transport, the elastic 
scattering of heavy quarks in the medium is mediated by the resonant excitation
of charm and bottom states in the medium~\cite{vanhees05,vanhees08}. 
Predictions from this model are confronted with the data for two different 
choices of the resonance widths in the right panel of 
Fig.~\ref{fig:phenix_raa_vs_models_diff}. Both the measured suppression and 
the azimuthal anisotropy of electrons from heavy-flavor hadron decays are 
reproduced by the model calculation.

The diffusion constant, which is the driving parameter of both Langevin-based
transport models introduced above, can be related to the shear viscosity $\eta$
of the hot QCD matter through which the heavy quarks propagate, where the ratio
of $\eta$ to the entropy density $s$ is the quantity that determines the 
damping rate in the system. In that sense the energy loss of heavy quark in the
medium and the simultaneous participation of the heavy quarks in the medium's
collective expansion can be interpreted as a damping of the initial 
heavy-quark dynamics by the hot and dense medium. Arguments based on the 
uncertainty principle exclude zero viscosity for any thermal 
system~\cite{danielewicz85}. Furthermore, it was shown~\cite{kovtun05} that 
specific classes of conformal field theories with gravity duals in the 
anti-de-Sitter space (AdS/CFT correspondence) exhibit a ratio 
$\eta/s = 1/4\pi$. It was conjectured that this value represents a 
bound for all relativistic thermal field theories~\cite{kovtun05}. As it is 
discussed in detail in Ref.~\cite{ppg066,ppg077}, the $\raa$ and $v_2$ 
measurements of electrons from heavy-flavor hadron decays suggest a value of 
$\eta/s ~ (1.33-2)/4\pi$, close to the conjectured lower bound. In turn, 
AdS/CFT-based energy loss models~\cite{horowitz11} give a reasonable 
description of the measured $\raa$ of electrons from heavy-flavor hadron 
decays while the measured $v_2$ is severely underestimated.
$\eta/s$ can not only be inferred from measurements sensitive to the coupling
of heavy quarks to the medium as discussed here, but also from observables
related to the bulk of the medium, involving light flavor quarks only.
Such estimates based on elliptic flow measurements of light 
hadrons~\cite{lacey07,drescher07} and measurements of 
fluctuations~\cite{gavin06} and entropy production~\cite{dumitru07} are 
supported by detailed viscous relativistic hydrodynamic 
calculations~\cite{romatschke07,luzum09} and they are consistent with
the $\eta/s$ value estimated here using electrons from heavy-flavor hadron 
decays.

It is an interesting question whether heavy-quark bound states can be formed
in hot QCD matter. The relevant pQCD dynamics of open heavy-flavor production
was worked out in Ref.~\cite{adil07}, and the GLV framework was 
extended to include composite $q\bar{q}$ systems such that the medium induced 
dissociation probability of D and B mesons propagating through the hot and 
dense medium could be calculated. In this scenario the sequential fragmentation
of heavy quarks and dissociation of heavy-flavor mesons results in an effective
energy loss. Quantitatively, the formation time and the detailed expansion 
dynamics of the produced hot QCD matter as well as the in-medium formation time
of the heavy-flavor hadrons play a role. The resulting nuclear modification 
factor of electrons from heavy-flavor hadron decays, shown as the lowest band in
the right panel of Fig.~\ref{fig:phenix_raa_vs_models_rad}, is in reasonable
agreement with the measurement as are result from improved calculations which
combine sequential heavy-flavor hadron formation and dissociation with charm
and bottom energy loss on the parton level~\cite{sharma09}.

In all above model calculations, addressing the production of heavy-flavor
hadron decay electrons, the same chemical composition of charm and bottom
hadrons is assumed for both \pp and \auau collisions. However, in the 
light-flavor sector an enhancement of baryon yields at intermediate \pt
relative to meson yields has been observed in central \auau collisions at 
RHIC~\cite{abelev06}. A similar baryon anomaly in the heavy-quark sector could 
lead to an enhancement of \lambdac relative to D mesons. Given the smaller 
semileptonic decay branching ratio of charm baryons compared to mesons and 
because of a softer \pt spectrum of electrons from charm baryon decays, 
$\raa < 1$ would be a consequence of such a baryon anomaly~\cite{sorensen06} 
even without any additional heavy-flavor energy loss in the hot QCD matter. 
Assuming a charm baryon to meson ratio close to one a nuclear modification 
factor of about 0.65 was predicted~\cite{crochet08} for electrons from 
heavy-flavor hadron decays. Without additional experimental input it will be 
impossible to resolve this issue. Clearly, the reconstruction
of displaced secondary decay vertices and exclusive measurements of hadronic 
heavy-flavor decays are highly desirable as they would not suffer from such
ambiguities.

\begin{figure}[t]
\begin{center}
\includegraphics[width=0.4\textwidth]{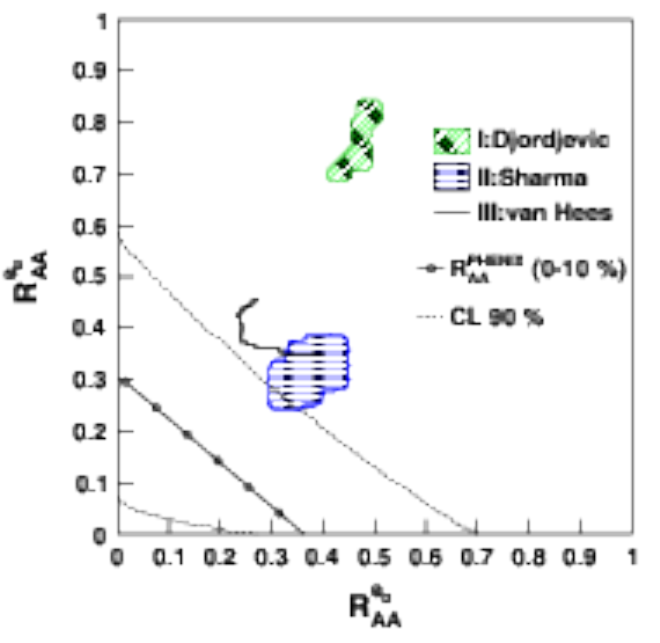}
\end{center}
\caption{90\% confidence level contours for $\raa$ of electrons from D-meson
         and B-meson decays for electron $\pt > 5$~\gevc in comparison with
         model calculations. See text for more details (reprinted with 
         permission from Ref.~\cite{star_e_b_d}; Copyright (2010) by the 
         American Physical Society).}
\label{fig:star_d_b_raa}
\end{figure}

The issue of charm versus bottom energy loss suffers from similar ambiguities.
The relative contribution, $r_{\rm B}$, from bottom hadron decays to the 
electron spectra from heavy-flavor hadron decays has been measured with 
substantial uncertainties in \pp collisions~\cite{star_e_b_d,ppg094}. For 
\auau collisions, however, experimental constraints are missing. Therefore, 
the only practical approach to separate the nuclear modification factors of 
electrons from charm hadron decays, $R_{\rm AA}^{e_{\rm D}}$, and from bottom hadron
decays, $R_{\rm AA}^{e_{\rm B}}$, is to combine the measurement of $r_{\rm B}$ in 
\pp collisions with the measurement of $R_{\rm AA}^{e_{\rm HF}}$ of electrons from 
all heavy-flavor hadron decays in \auau collisions via the equation

\begin{equation}
R_{\rm AA}^{e_{\rm HF}} = 
(1 - r_{\rm B}) R_{\rm AA}^{e_{\rm D}} + r_{\rm B} R_{\rm AA}^{e_{\rm B}}.
\end{equation}

\begin{figure}[t]
\begin{center}
\includegraphics[width=0.6\textwidth]{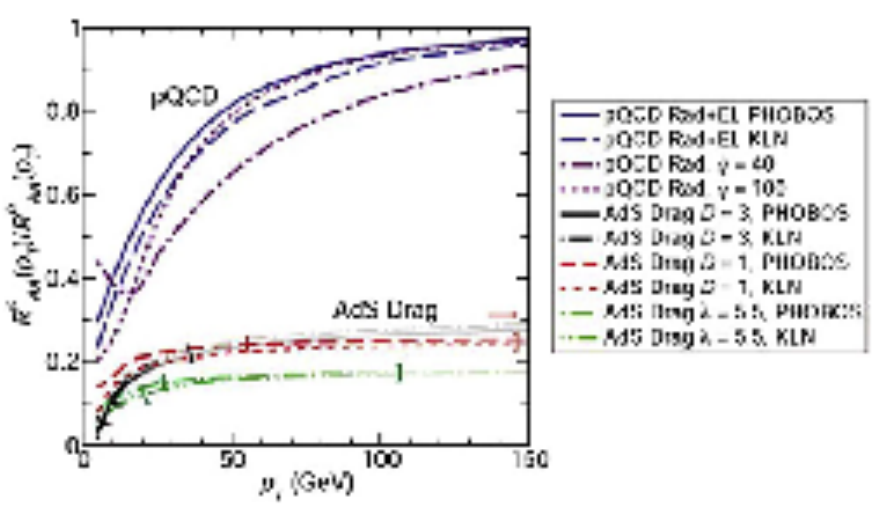}
\end{center}
\caption{Ratio of the nuclear modification factors of charm and bottom 
         calculated within the frameworks of AdS/CFT~\cite{horowitz08} and 
         WHDG pQCD~\cite{wicks07} with a wide range of input parameters
         as a function of \pt (reprinted from Ref.~\cite{horowitz08}).}
\label{fig:ratio_horowitz}
\end{figure}

Since above an electron \pt of about 5~\gevc both $r_{\rm B}$, as measured with
the STAR experiment in \pp collisions~\cite{star_e_b_d}, and 
$R_{\rm AA}^{e_{\rm HF}}$, as measured in central \auau collisions with the PHENIX 
experiment~\cite{ppg066,ppg077}, do not show a significant
\pt dependence, such an attempt to separate the contributions from charm and
bottom hadrons decays was conducted in this \pt region as discussed in detail 
in Ref.~\cite{star_e_b_d}. The results are summarized in 
Fig.~\ref{fig:star_d_b_raa}, which shows the most probable values for 
$R_{\rm AA}^{e_{\rm D}}$ and $R_{\rm AA}^{e_{\rm B}}$ obtained from the data as open 
circles and the corresponding 90\% confidence limits as dashed lines. Clearly 
the data imply a substantial suppression not only of electrons from charm but 
also from bottom hadron decays. The predictions from three models are 
confronted with the data in Fig.~\ref{fig:star_d_b_raa}. A DGLV calculation 
with radiative energy loss only (model I)~\cite{dglv} predicts a too small 
suppression for charm and, in particular, bottom energy loss. A collisional 
energy loss model including collisional dissociation, partonic energy loss, 
and cold nuclear matter effects (model II)~\cite{sharma09} is in reasonable 
agreement with the data as is a model in which collisional energy loss is 
mediated by resonant elastic scattering (model III)~\cite{vanhees05}.
Not shown are calculations within the AdS/CFT 
framework~\cite{gubser06,herzog06}, which also predict small $\raa$ for 
electrons from charm and bottom hadron decays. Even more important, the ratio 
of the nuclear modification factors of charm and bottom are 
predicted~\cite{horowitz08} to be significantly different for AdS/CFT 
calculations and pQCD-based calculations as shown for LHC energies in
Fig.~\ref{fig:ratio_horowitz}, emphasizing the importance of separate 
measurements of charm and bottom hadron energy loss in the future.

In summary, the measurements of electrons from heavy-flavor hadron decays in 
\pp and \auau collisions at RHIC as reviewed here constitute the first 
systematic application of open heavy-flavor observables as a tool to probe 
the properties of the hot QCD matter produced in high-energy nucleus-nucleus 
collisions. It was shown that the total charm yield scales proportional to
the number of binary collisions in nucleus-nucleus collisions at RHIC, \ie 
charm quark-antiquark pairs are almost exclusively produced in point-like 
hard scattering processes. Additional production processes such as thermal 
production in the hot medium do not play a significant role. The produced
heavy quarks propagate through the medium where they are subject to strong
interactions with the medium, which do not change the overall heavy-flavor
yields but modify the phase-space distributions to a large extent. Electrons
from heavy-flavor hadron decays are suppressed at high \pt, quantitatively
similar to what is observed for light hadrons. This observation is in stark
contrast with the expectation from energy-loss models based on radiative
energy-loss mechanisms. Collisional energy loss can not be ignored for heavy
quarks and/or alternative energy-loss mechanism need to be considered. At 
the same time, the measured $v_2$ of these electrons indicates that charm 
quarks participate in the collective expansion of the produced hot QCD medium.
These observations support the characterization of the hot and dense QCD
matter formed in \auau collisions at RHIC as a strongly interacting medium
with partonic degrees of freedom which features a minimal ratio of shear 
viscosity to entropy density and, therefore, was dubbed an almost perfect
fluid. However, semielectronic heavy-flavor hadron decays provide only indirect
information. In particular, a precision separation of charm from bottom is
difficult via this approach, and the chemistry of charm and bottom hadron
states is completely inaccessible. These issue can only be addressed with 
the upgrades of the PHENIX and STAR experiments.

\subsubsection{Semimuonic heavy-flavor hadron decays}
\label{subsubsec:rhic_mu_auau}
\begin{figure}[t]
\begin{center}
\includegraphics[width=0.325\textwidth]{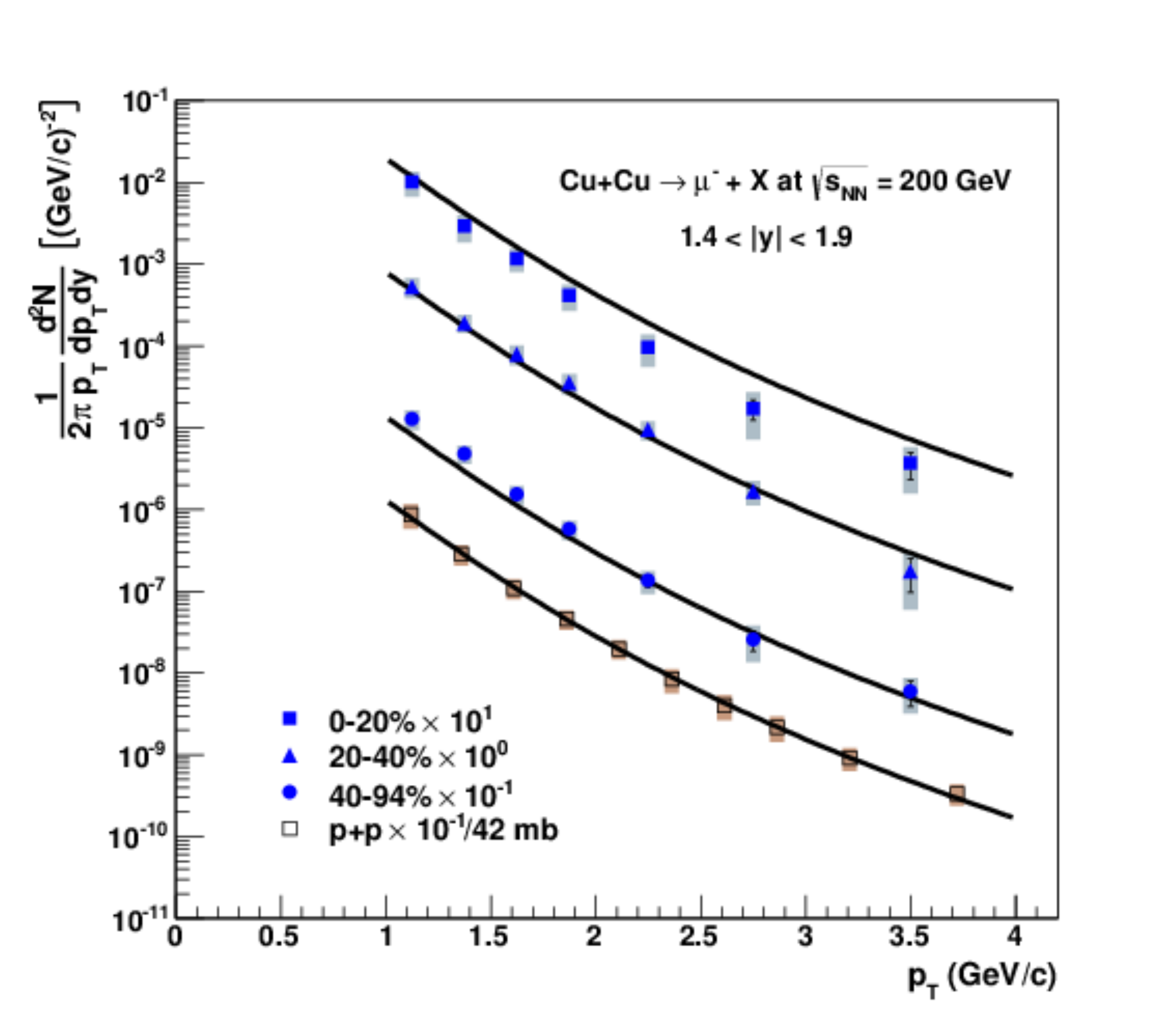}
\includegraphics[width=0.325\textwidth]{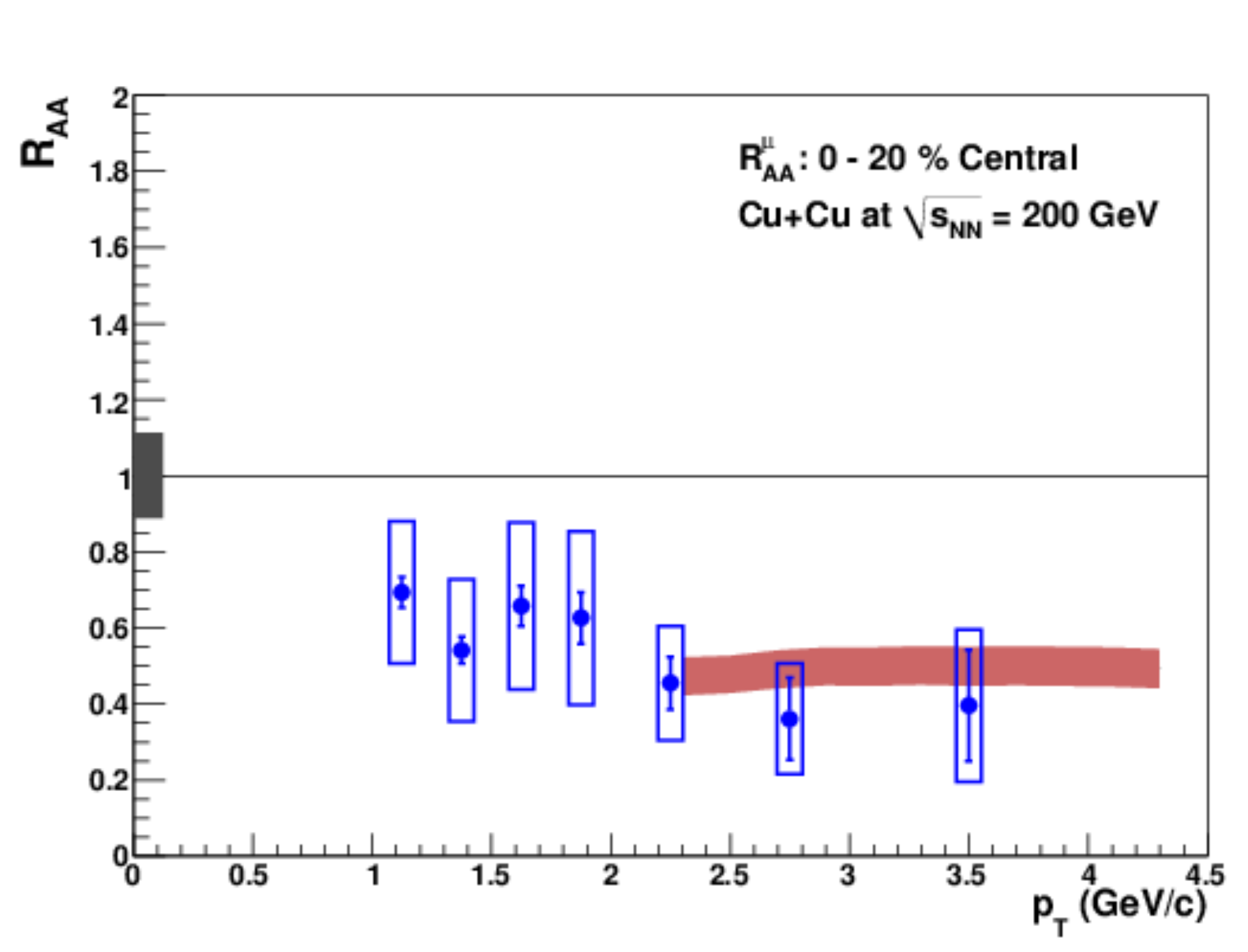}
\includegraphics[width=0.325\textwidth]{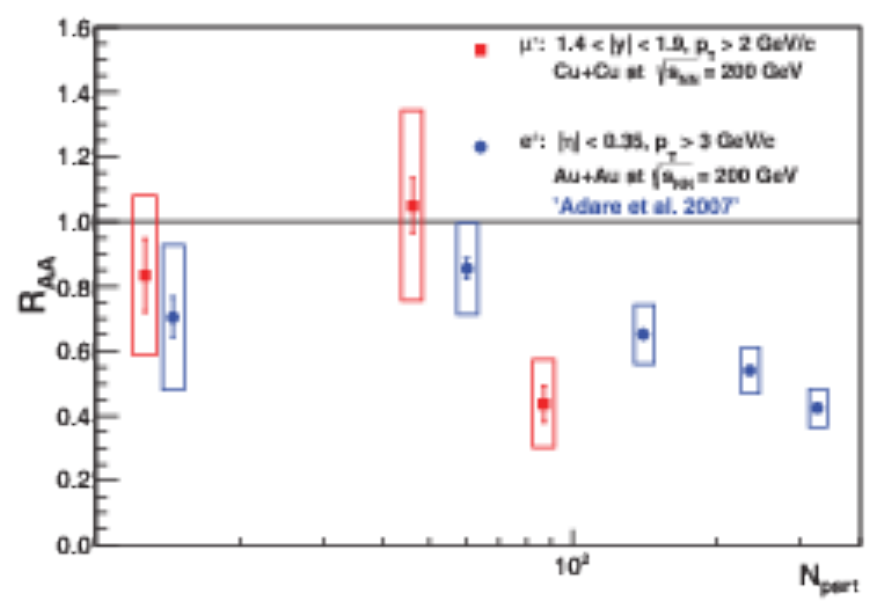}
\end{center}
\caption{\pt-differential invariant yields of negatively charged muons from 
         heavy-flavor hadron decays in \pp and three centrality classes of
         \cucu collisions measured at forward rapidity with the PHENIX 
         experiment at $\sqrt{s_{\rm NN}} = 200$~GeV. The solid lines depict
         a fit to the \pp data scaled by the appropriate number of binary
         collisions to the \cucu centrality classes (left panel). Nuclear
         modification factor $\raa$ for negative muons from heavy-flavor
         hadron decays measured as a function of \pt in the 20\% most central
         \cucu collisions. For comparison, a prediction from a collisional
         dissociation model~\cite{sharma09} is shown (middle panel). 
         Comparison of the nuclear modification factors of heavy-flavor muons 
         at forward rapidity ($1.4 < y < 1.9$) and $\pt > 2$~\gevc in \cucu 
         collisions~\cite{ppg117} and heavy-flavor electrons at mid-rapidity 
         ($|\eta| < 0.35$) and $\pt > 3$~\gevc in \auau 
         collisions~\cite{ppg066} as a function of the collision centrality, 
         expressed by the number of participating nucleons (right panel)
         (reprinted with permission from Ref.~\cite{ppg117}; Copyright (2012) 
         by the American Physical Society).}
\label{fig:mu_cucu}
\end{figure}

While extensive systematic studies of heavy-flavor production in heavy-ion 
collisions have been conducted at RHIC in the semielectronic decay channel
at mid-rapidity, corresponding measurements in the semimuonic decay channel
at forward rapidity have become available only recently. 

With the PHENIX experiment negatively charged muons from the decay of 
heavy-flavor hadrons have been measured in the rapidity range $1.4 < y < 1.9$ 
and for transverse momenta $1 < \pt < 4$~\gevc in three centrality classes of 
\cucu collisions at $\sqrt{s_{\rm NN}} = 200$~GeV~\cite{ppg117}. Background, 
mainly from light hadrons, was estimated using a MonteCarlo approach and it was
subtracted statistically from the inclusive muon candidate sample. The 
resulting invariant yields measured in \cucu collisions are compared as a
function of the muon \pt with the corresponding data from \pp collisions in 
the left panel of Fig.~\ref{fig:mu_cucu}, where the solid lines correspond to a
fit of the \pp data scaled with the appropriate numbers of binary collisions to
the \cucu centrality classes. While the muon yields from heavy-flavor hadron 
decays in peripheral and mid-central \cucu collisions are in reasonable
agreement with the binary collision scaled \pp reference, for the 20\% most
central \cucu collisions a suppression of such muons is observed at high \pt
relative to the \pp fit. 

This suppression is quantified via the nuclear modification factor $\raa$
of muons from heavy-flavor hadron decays shown as a function of \pt in the
middle panel of Fig.~\ref{fig:mu_cucu} for the 20\% most central \cucu 
collisions. A prediction from a theoretical model~\cite{sharma09} agrees well 
with the measurement at high \pt. This specific model includes elastic and 
inelastic in-medium energy loss of heavy quarks, the dissociation of heavy 
mesons in medium, and cold nuclear matter effects such as shadowing and initial
state energy loss due to multiple scattering of the incoming 
partons~\cite{sharma09}. A related model prediction~\cite{adil07} is also in 
reasonable agreement with the measured nuclear modification factor of electrons
from heavy-flavor hadron decays in \auau collisions at mid-rapidity (see 
Fig.~\ref{fig:phenix_raa_vs_models_rad}).

As demonstrated in the right panel of Fig.~\ref{fig:mu_cucu}, the suppression
of muons from heavy-flavor hadron decays at forward rapidity in central \cucu
collisions is comparable to the suppression observed in central \auau 
collisions for electrons from heavy-flavor hadron decays~\cite{ppg066,ppg077}. 
This is a surprising observation since the Bjorken energy density of hot QCD 
matter produced at mid-rapidity in central \auau collisions at RHIC is expected
to be larger by at least a factor of two compared to the energy density in the
forward rapidity region in \cucu collisions. Consequently, the substantial
suppression observed in \cucu collisions might point to significant cold
nuclear matter effects playing a role in particular at forward rapidity.

Major progress at forward rapidity requires improved heavy-flavor production 
measurements at RHIC. In particular, the PHENIX forward silicon vertex 
detector will allow the separation of charm and bottom contributions to the 
muon sample from heavy-flavor hadron decays.

\section{Open heavy-flavor measurements at the LHC}
\label{sec:lhc}
\subsection{Experiments studying heavy-flavor production at the LHC}
\label{subsec:exp_lhc}
The LHC delivered the first \pp collisions at injection energy, \ie at
$\sqrt{s} = 0.9$~TeV in November 2009. While data were recorded with all
experiments during the commissioning phase at this energy, the accumulated
statistics was not sufficient to investigate the production of heavy-flavor
hadrons. Substantial data sets, providing already detailed information on
heavy-flavor production in \pp collisions, were recorded by all four large
LHC experiments at an energy of $\sqrt{s} = 7$~TeV in the years 2010/2011. 
In particular, these data provided a reference for the first \pbpb collision
data recorded at the LHC at $\sqrt{s_{\rm NN}} = 2.76$~TeV in 
November/December 2010 and 2011 with the ALICE, ATLAS, and CMS experiments. 
The LHCb experiment does not participate in heavy-ion runs. Because of the 
difference in the available energy per nucleon-nucleon pair the 7~TeV
\pp data have to be scaled to the lower energy of 2.76~TeV to provide
a reference for measurements in \pbpb collisions. As will be demonstrated
in Section~\ref{subsec:lhc_pp}, all heavy-flavor production data from 
7~TeV \pp collisions at the LHC are in good agreement with predictions 
from FONLL pQCD calculations~\cite{fonll3}. Therefore, FONLL driven $\sqrt{s}$ 
scaling procedures could be used to determine \pp references at the proper 
energy for \pbpb collisions. In order to validate the FONLL based $\sqrt{s}$ 
scaling and to provide experimental references for observables which can not be
scaled by pQCD model calculations, the LHC delivered \pp collisions at the 
energy of $\sqrt{s} = 2.76$~TeV for a short period of a few days in March 2011.
While the LHC operates at a collision energy of $\sqrt{s} = 8$~TeV since 
April 2012 relevant results from this experimental \pp campaign are not 
yet available.

This review discusses two issues related to the LHC experiments. 
First, it is demonstrated that perturbative QCD calculations are able to 
reasonably well describe heavy-flavor hadron production in \pp collisions at 
the LHC, using data from all four large experiments in the hadronic and 
semileptonic decay channels at mid-rapidity and at forward rapidity. It is 
actually a big success of the experiments as well as of the theoretical 
calculations that this can be concluded already during the very first years 
of LHC operations. It took years to resolve discrepancies that seemingly 
existed between pQCD calculations and measurements at the Fermilab Tevatron. 
Also at RHIC, long standing discrepancies between the electron spectra from 
heavy-flavor hadron decays measured with the PHENIX and STAR experiments, 
respectively, could be resolved only after a while.
Second, the first published results on open heavy-flavor production in \pbpb 
collisions obtained with the ALICE and CMS experiments are summarized. While 
these measurements confirm the substantial suppression of heavy-flavor probes 
at high \pt with respect to binary collision scaled \pp measurements as 
observed at RHIC, they also represent a major step forward not only in terms
of the energy scale. For the first time, hot QCD matter effects on exclusive
heavy-flavor hadron observables can be investigated. In \pbpb collisions,
D mesons have been reconstructed with the ALICE experiment and with the CMS 
experiment \jpsi mesons originating from b-hadron decays have been studied.
Such measurements have only been possible because all of the four large LHC
experiments include precision silicon vertex detectors, which allow the
experimental separation of displaced secondary vertices from heavy-flavor 
hadron decays from the primary collision vertex. Augmented by inclusive 
measurements, \eg of electrons and muons from semileptonic heavy-flavor 
hadron decays, these first exclusive results are only the tip of the iceberg 
of near time future results to be expected from systematic heavy-flavor 
production studies with all experiments at the LHC.
In the following, the experimental approaches of the four large LHC experiments
to open heavy-flavor measurements are summarized briefly, with special
emphasis on the ALICE experiment.

\subsubsection{The ALICE approach to open heavy-flavor measurements}
\label{subsubsec:alice}
The ALICE experiment~\cite{alice} was designed and optimized to cope with the 
large particle multiplicities in central \pbpb collisions at the LHC. 
The ALICE apparatus consists of two main parts: a central barrel detector 
system housed inside the L3 solenoid magnet, which provides a field of 0.5~T, 
and a forward muon spectrometer ($-4 < \eta < -2.5$) including a dipole magnet,
which provides a field integral of 3~Tm. This setup is augmented by several 
smaller auxiliary detectors for event characterization and for triggering 
purposes.
Special emphasis was put on excellent tracking and particle identification 
capabilities, in particular also at low momenta. With these capabilities
excellent measurements of heavy-flavor production are possible through the 
reconstruction of heavy-flavor hadrons in their hadronic decay channels at 
mid-rapidity as well as via semileptonic decays of charm and bottom hadrons
both at mid-rapidity and forward rapidity. 

The central barrel tracking system consists of the silicon Inner Tracking 
System (ITS) and the Time Projection Chamber (TPC). In the pseudorapidity 
range $|\eta| < 0.9$ tracks are reconstructed with a momentum resolution 
better than 4\% for $\pt < 20$~\gevc. The distance of closest approach of 
tracks to the interaction vertex is measured with a resolution better than 
$75 \mu$m for $\pt > 1$~\gevc in the plane transverse to the beam 
direction~\cite{d_in_pp}.

Charged hadrons are identified via their specific energy loss \dedx in the
TPC and a time-of-flight (TOF) measurement in the ALICE TOF detector. The
same techniques allow a clean electron identification in the range
$\pt < 6$~\gevc. At higher \pt, the Transition Radiation Detector (TRD) and, 
alternatively, the Electromagnetic Calorimeter (EMCal) are necessary for the
measurement of a pure electron sample. 

The measurement of D mesons at mid-rapidity ($|y| < 0.5$) is based on the
selection of displaced-vertex topologies~\cite{d_in_pp}. Tracks are selected 
which originate from a secondary vertex consistent with a large decay length 
as expected from the D-meson lifetimes. In addition, a good alignment of the 
reconstructed D-meson momentum with the flight line from the collision vertex 
to the secondary vertex is required. Since the relevant D-meson decay channels 
(\mbox{\dzero$\rightarrow {\rm K}^- \pi^+$}, 
\mbox{\dplus$\rightarrow {\rm K}^- \pi^+ \pi^+$}, and 
\mbox{\dstarplus$\rightarrow$\dzero$\pi^+$}) involve a kaon as decay product
the identification of the charged kaon in the TPC and TOF detectors helps to
reduce the background. In an invariant mass analysis the raw D-meson yields 
are determined which, then, are corrected for geometrical acceptance and for
reconstruction and particle identification efficiency based on a detailed
simulation of the apparatus. Feed down from B-meson decays is estimated
from a FONLL pQCD calculation~\cite{fonll3} and it is subtracted. 

Bottom production can be investigated via the measurement of \jpsi mesons 
from the decay of bottom hadrons, \cf Tab.~\ref{tab:HFhadrons}. \jpsi mesons
are reconstructed via their $e^+e^-$ decays at mid-rapidity ($|y| < 0.9$).
The fraction of the total \jpsi yield originating not from the primary 
collision vertex but from a displaced bottom-hadron decay vertex is 
determined via the measurement of the pseudo-proper decay length of the 
\jpsi mesons~\cite{alice_jpsi_b}. 

Electrons from heavy-flavor hadron decays are measured at 
mid-rapidity~\cite{alice_e_pp} in a two step procedure. 
First, the inclusive electron spectrum is determined. 
Electron candidate tracks are identified with the TPC, TOF, TRD, and EMCal 
detectors. The small remaining hadron background is estimated via fits of 
the TPC \dedx distribution in momentum slices and it is subtracted from the 
electron candidate spectra. After corrections for geometrical acceptance and 
tracking and particle identification efficiencies derived from simulations the 
inclusive electron spectrum is obtained. From this inclusive spectrum a 
cocktail of electrons from sources other than heavy-flavor decays is 
subtracted. Similar to corresponding analyses at RHIC, the most important 
components of this background cocktail are Dalitz decays of the $\pi^0$ and 
$\eta$ meson as well as photon conversions in the beam pipe and in the first 
pixel layer of the ITS. In addition, decays of light vector mesons 
($\rho, \omega, \phi$), heavy quarkonia ($J/\psi, \Upsilon$), and
direct radiation from hard scattering processes are included. The cocktail 
input is based on the $\pi^0$ and $\eta$ spectra measured with the 
ALICE experiment~\cite{pieta_alice}, the known material budget of the 
apparatus~\cite{koch_alice}, \mt scaling for the \pt spectra and yield of
light vector mesons, heavy quarkonia measurements with the 
ALICE~\cite{jpsi_alice} and CMS experiments~\cite{jpsi_cms,upsilon_cms}, and 
next-to-leading-order pQCD calculations~\cite{vogelsang1,vogelsang2} for
the electron spectra from direct radiation. With the ALICE ITS, electrons
originating from displaced secondary bottom-hadron decay vertices can be
separated from decay vertices of shorter lived charm hadrons and from the 
primary collision vertex. This technique allows the measurement of electrons
from bottom-hadron decays~\cite{alice_hfe_beauty}.

Muons with momenta above 4~\gevc are identified and their momenta are measured
in the pseudorapidity range $-4 < \eta < -2.5$ with the ALICE forward muon 
spectrometer, which consists of a 10 interaction lengths passive front 
absorber, a small angle beam shield, a set of tracking chambers in front of,
in, and behind a 3~Tm dipole magnet, and trigger chambers located behind an
iron wall with a thickness of 7.2 interaction lengths. From the inclusive 
muon candidate spectrum three main background sources have to be removed: 
muons from the decay of primary light hadrons, muons from the decay of hadrons 
produced in secondary interactions in the front absorber, and hadrons that 
punch through the front absorber. Contributions from the decay of W$^\pm$ and
$Z^0$ vector bosons are negligible in the muon \pt range $2 < \pt < 12$~\gevc. 
Punch through hadrons are rejected efficiently requiring that muon candidate 
tracks match a track in the muon trigger system. Decay muons are the dominant 
source of background in the relevant \pt region. 
In \pp collisions~\cite{muon_pp} this contribution can be subtracted using 
information from detailed simulations. In \pbpb collisions~\cite{muon_pbpb} 
the presence of unknown medium effects prevents such a MonteCarlo approach. 
Instead, the decay muon contribution is estimated by extrapolating the \pt 
spectra of pions and kaons measured in \pbpb collisions at 
mid-rapidity~\cite{appels11} to forward rapidity and calculating the 
corresponding decay muon spectra.

\subsubsection{The ATLAS approach to open heavy-flavor measurements}
\label{subsubsec:atlas}
The ATLAS collaboration operates one of the two large general purpose detectors
at the LHC designed to extend the frontiers of particle physics by exploiting
the unprecedented high collision energy and luminosity delivered by the LHC.
The main goals of ATLAS are related to the search for the Standard Model
Higgs boson, precision tests of QCD, electroweak interactions, flavor physics,
and the search for physics beyond the Standard Model in \pp collisions.
Utilizing mainly hard probes such as jets, ATLAS also studies the \pbpb 
collisions provided by the LHC. In the following, a brief characterization
of those subsystems of the ATLAS apparatus is given which are relevant for the 
results published on open heavy-flavor production in \pp collisions up to now. 
ATLAS has not yet published open heavy-flavor data from \pbpb collisions.

The ATLAS apparatus covers almost the full solid angle around the interaction
point. An inner detector (ID) tracking system consisting of a 3 layer silicon 
pixel detector, a 4 layer double-sided silicon strip detector, and a 73 layer
transition radiation straw tube tracker is located in a 2~T solenoidal magnetic
field, and it covers the pseudorapidity range $|\eta| < 2.5$. The ID is
surrounded by a high-granularity liquid-argon electromagnetic calorimeter (EM),
also covering $|\eta| < 2.5$, and hadron calorimetry provided by an 
iron-scintillator tile calorimeter in the mid-rapidity region and 
copper/liquid-argon endcap calorimeters towards forward and backward rapidity.
The outer muon spectrometer (MS) is located inside a toroidal magnetic field
providing a bending power of 2.5~Tm in the barrel and 5~Tm in the end caps.
Muons are tracked in the region $|\eta| < 2.7$ using monitored drift tube
chambers and higher granularity cathode strip chambers in the end caps. Muon
trigger chambers employ resistive plate chambers for $|\eta| < 1.05$ and thin
gap chambers in the region $1.05 < |\eta| < 2.4$. This general purpose setup 
provides ample opportunities for the measurement of open heavy-flavor 
observables. A detailed descriptions of the detector layout of the ATLAS 
experiment can be found elsewhere~\cite{atlas}.

\subsubsection{The CMS approach to open heavy-flavor measurements}
\label{subsubsec:cms}
The CMS collaboration operates the second large general purpose detector at the
LHC. The main goals of CMS are identical to the ones of ATLAS (see 
Sec.~\ref{subsubsec:atlas}). Open heavy-flavor measurement in \pp and \pbpb
collisions with the CMS experiment at the LHC have used the inner tracker and 
the muon system of the apparatus up to now. Therefore, these detector 
components are briefly characterized below. A detailed descriptions of the 
detector layout can be found elsewhere~\cite{cms}.

The central component of the CMS apparatus is a superconducting solenoid with
an internal diameter of 6~m, providing at 3.8~T magnetic field. Immersed in
the magnetic field are the inner tracker surrounded by a lead-tungstate crystal
electromagnetic calorimeter and a brass/scintillator hadron calorimeter. The 
silicon pixel tracker comprises three barrel layers and two endcap disks at 
each barrel end. This device is surrounded by a silicon strip tracker 
consisting of 10 barrel layers and 12 endcap disks at each barrel end. 
Muons are detected in the pseudorapidity interval $|\eta| < 2.4$ with three
types of gas-ionization detectors embedded in the magnet's steel return yoke:
drift tubes, cathode strip chambers, and resistive plate chambers. Due to the
strong magnetic field and the silicon tracker's high granularity a \pt
resolution of about 1\% is reached for muon candidates which are matched 
with reconstructed tracks. The silicon tracker also provides the position
of the primary collision vertex with an accuracy of about 20~$\mu$m allowing
the separation of primary tracks from those originating from displaced 
secondary decay vertices. Weak decays of hadrons, 
\eg $K^0_S \rightarrow \pi^+ \pi^-$, can be reconstructed via an invariant
mass analysis of charged particle tracks pointing back to a common decay
vertex and fulfilling decay topology cuts, even without the identification 
of the decay hadron species. In addition, the CMS experiment includes 
substantial forward calorimetry in the pseudorapidity range $2.9 < \eta < 5.2$
as well as auxiliary forward detectors for triggering purposes and beam-halo
rejection. In \pbpb collisions information provided by the forward detectors
is used for event selection and the event-by-event centrality determination.
As in the case of ATLAS, the CMS experiment gives access to open heavy-flavor
observables in a multitude of channels, not all of which have been addressed
in publications yet. 

\subsubsection{The LHCb approach to open heavy-flavor measurements}
\label{subsubsec:lhcb}
The LHCb experiment was designed as a single-arm forward 
spectrometer~\cite{lhcb} aiming at the measurement of CP violating and rare 
decays of heavy-flavor hadrons in \pp collisions at the LHC. \pbpb collisions 
are not investigated with the LHCb experiment. Nevertheless, in the context of 
this review LHCb provides an important testing ground for pQCD calculations 
regarding the rapidity dependence of heavy-flavor production cross sections 
in \pp collisions at the LHC.

The LHCb apparatus covers the pseudorapidity range $2 < \eta < 5$. The
detector includes a high precision tracking system comprising a silicon 
strip vertex detector surrounding the nominal \pp interaction point, a 
large-area silicon strip detector located upstream of a dipole magnet
providing a bending power of 4~Tm, and three stations of silicon strip
detectors and straw drift tubes located downstream. With this tracking system
a momentum resolution $\Delta p/p$ is achieved between 0.4\% at 5~\gevc and
0.6\% at 100~\gevc. For high \pt tracks, the impact parameter is measured
with a resolution of 20~$\mu$m. Two ring imaging Cherenkov detectors are used
for hadron identification. Photon, electron, and hadron candidates are selected
with a calorimeter system including scintillating pad and pre shower detectors,
an electromagnetic and a hadronic calorimeter. Alternating layers of iron 
absorbers and multiwire proportional chambers compose the muon detection
system.

In addition to a variety of results concerning CP violating and rare 
heavy-flavor hadron decays, the LHCb collaboration has also published 
baseline cross sections of bottom production in \pp collisions at
$\sqrt{s} = 7$~TeV, which can be compared with predictions from pQCD
calculations. 

\subsection{Reference measurements in \pp collisions}
\label{subsec:lhc_pp}
As for the RHIC case, open heavy-flavor measurements in \pp collisions at 
$\sqrt{s} = 7$~TeV at the LHC provide a crucial testing ground for higher order
pQCD calculations in a yet unexplored high energy regime and, at the same time,
they deliver a baseline for corresponding measurements in \pbpb collisions,
where the interaction of heavy-flavor particles with the produced hot and 
dense QCD medium is expected to leave its footprint on open heavy-flavor 
observables.

\subsubsection{Hadronic heavy-flavor hadron decays}
\label{subsubsec:lhc_d_pp}
\begin{figure}
\begin{center}
\includegraphics[width=0.32\textwidth]{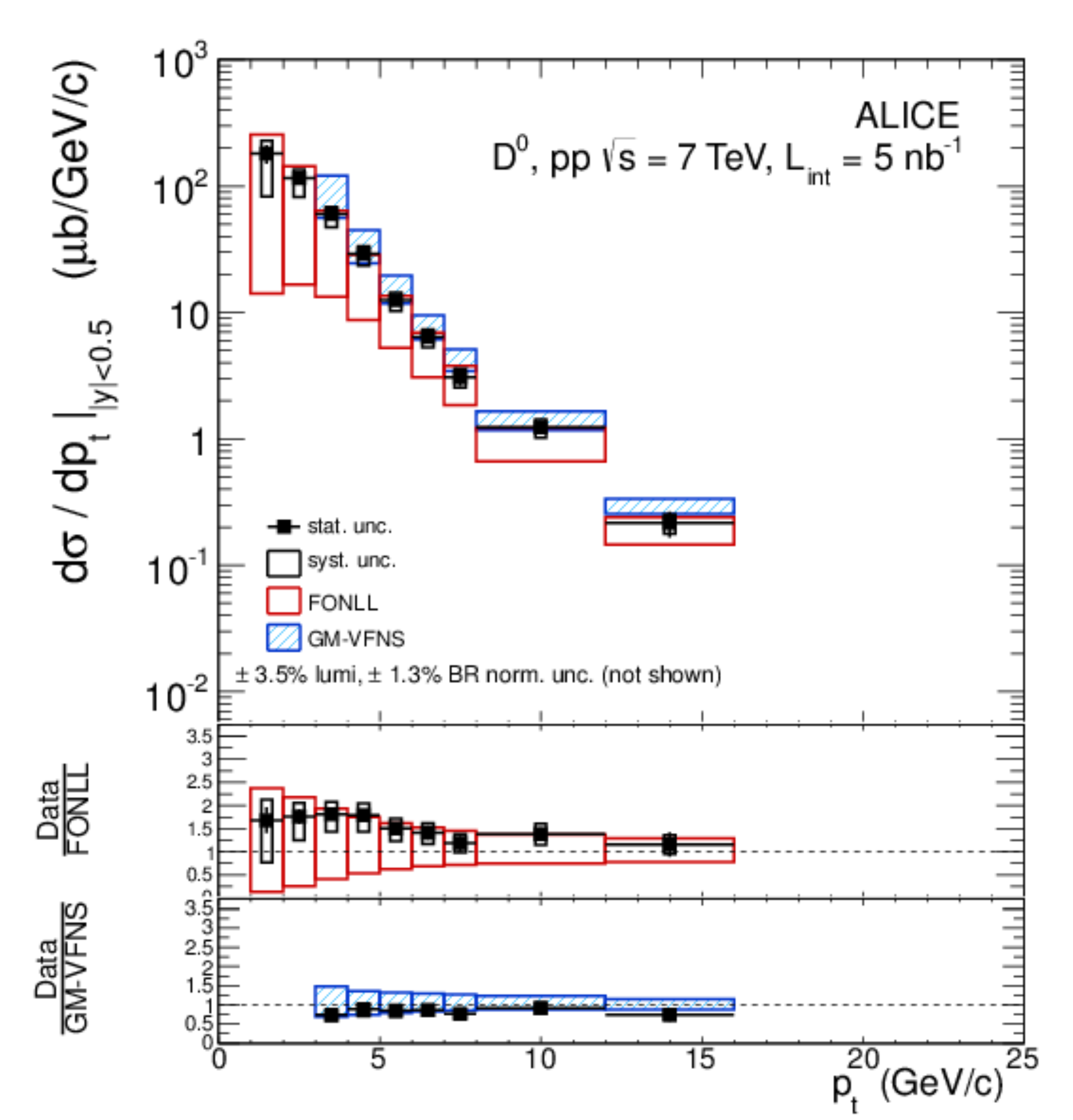}
\includegraphics[width=0.32\textwidth]{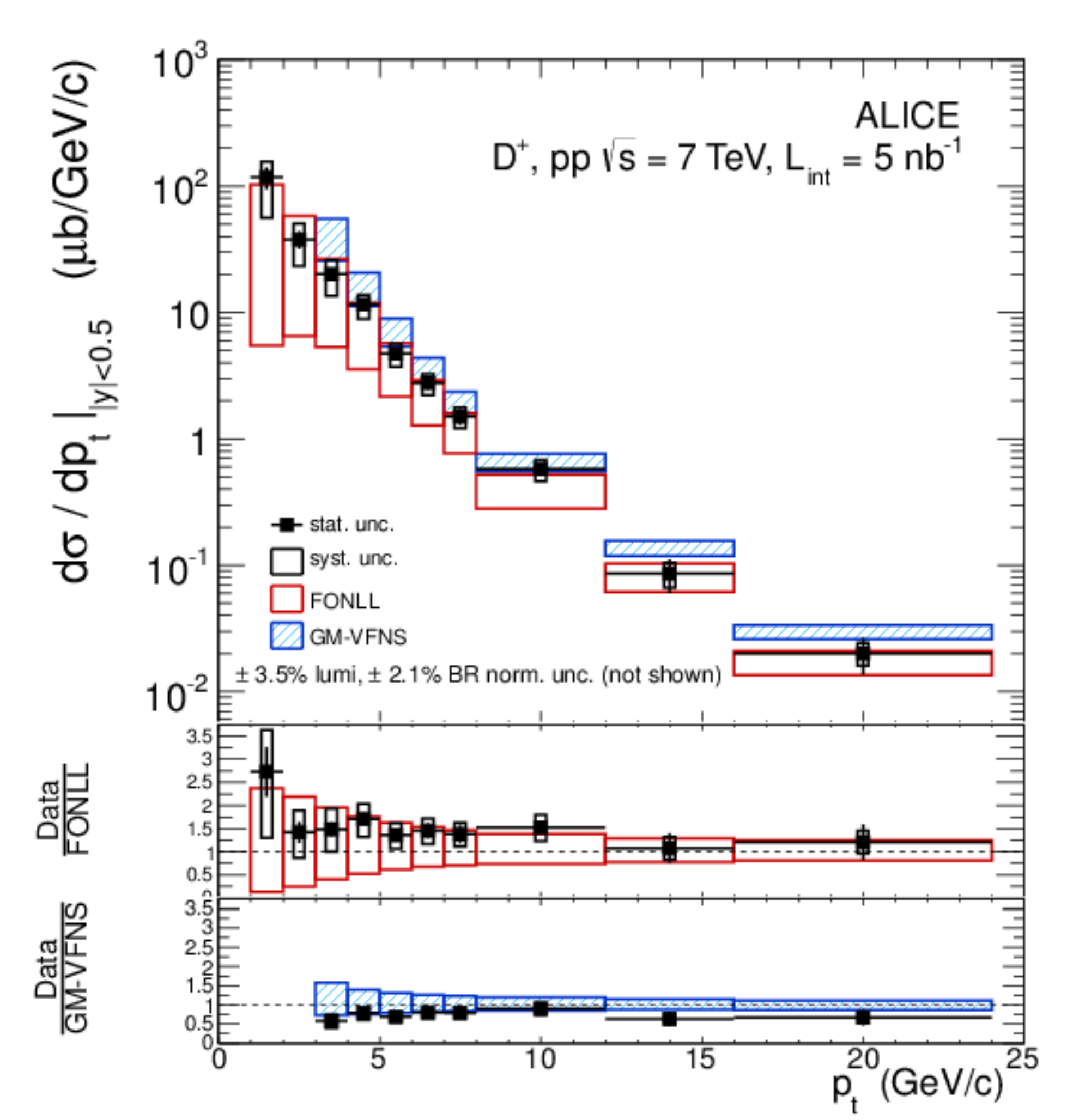}
\includegraphics[width=0.32\textwidth]{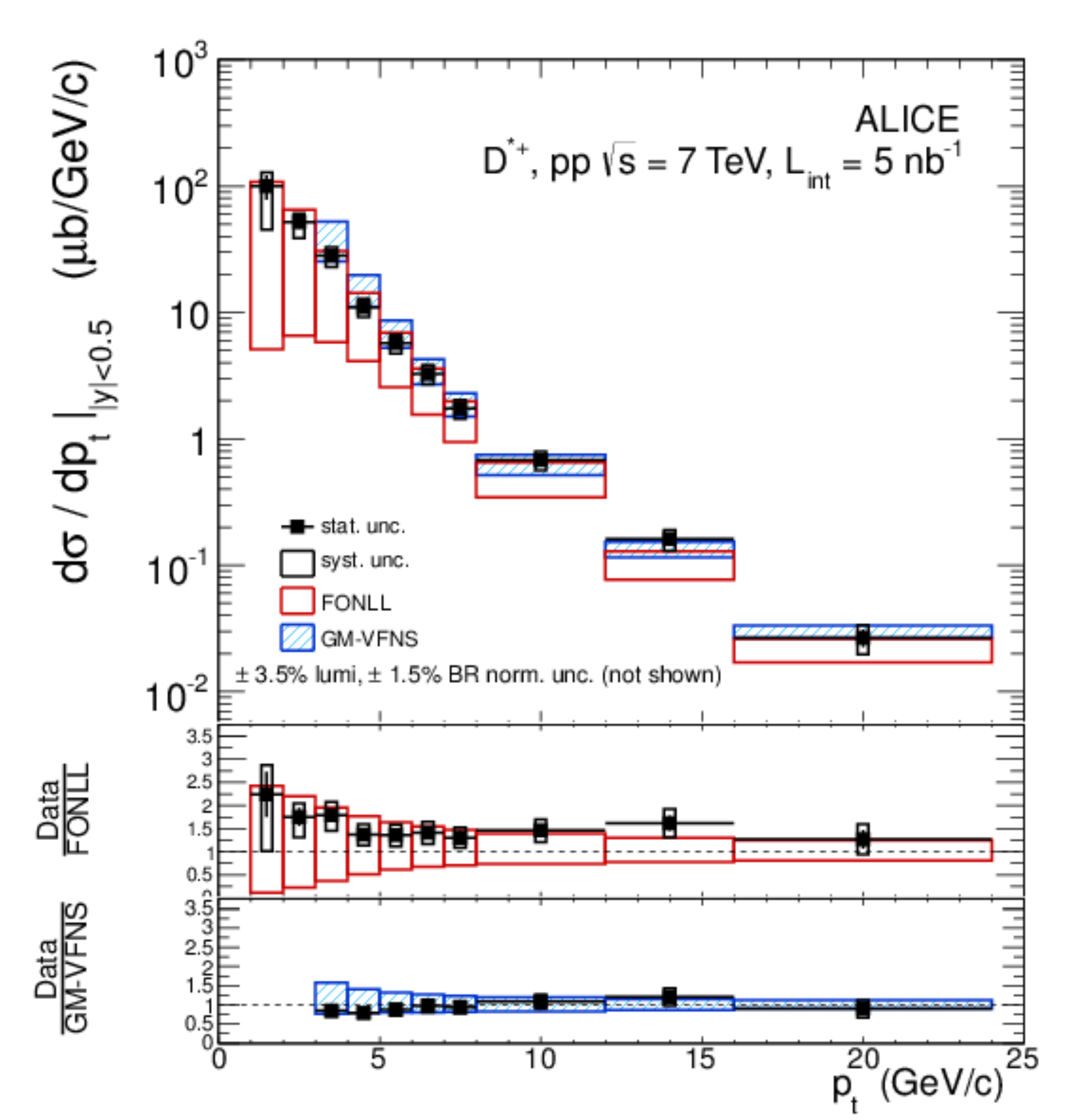}
\end{center}
\caption{\pt-differential production cross sections of \dzero (left panel), 
         \dplus (middle panel), and \dstarplus (right panel) measured at
         mid-rapidity ($|y| < 0.5$ ) with the ALICE experiment in \pp 
         collisions at $\sqrt{s} = 7$~TeV~\cite{d_in_pp} in comparison with 
         pQCD calculations~\cite{fonll3,gmvfns}. The lower panels of the 
         individual plots show the ratio of the measured cross sections and 
         the corresponding predictions from higher order FONLL and GM-VFNS 
         pQCD calculations (reprinted from Ref.~\cite{d_in_pp} with kind 
         permission from Springer Science and Business Media).}
\label{fig:alice_d_in_pp}
\end{figure}

\begin{figure}
\begin{center}
\includegraphics[width=0.32\textwidth]{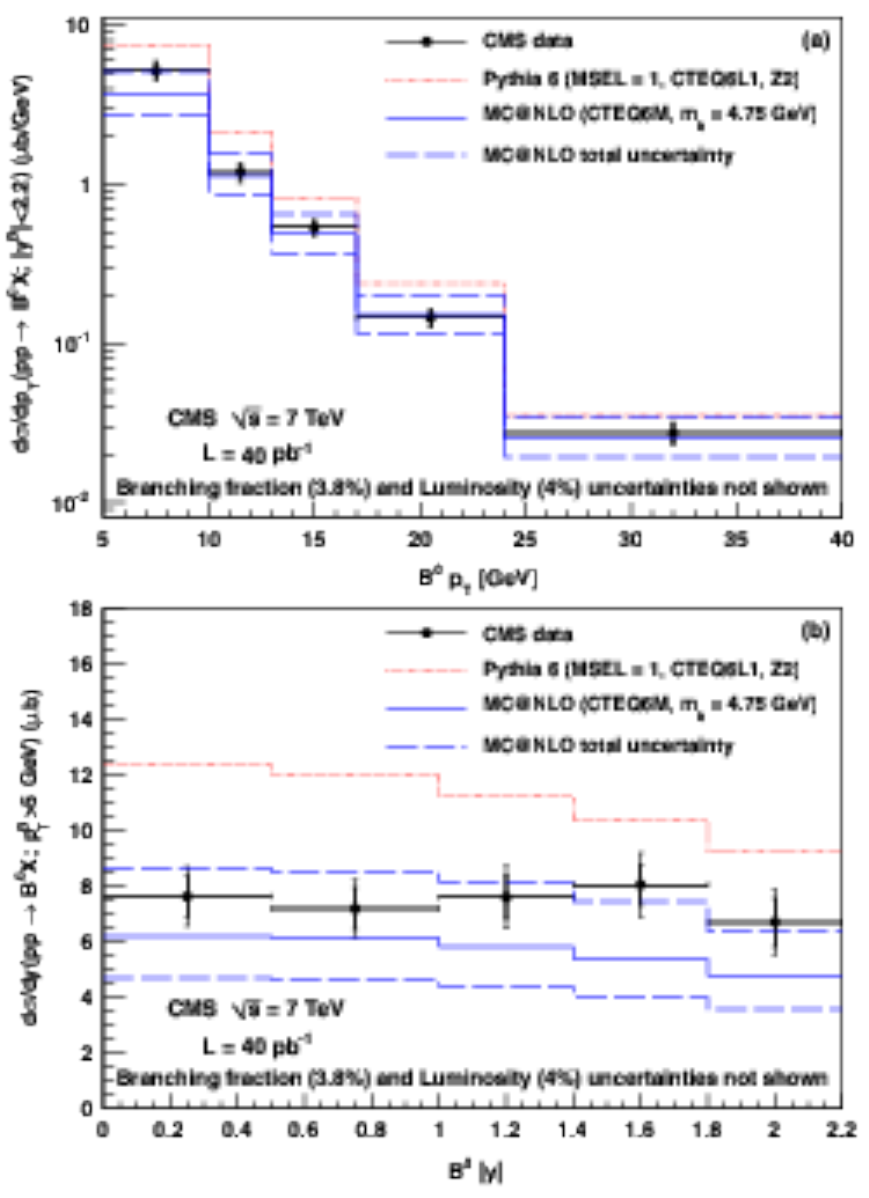}
\includegraphics[width=0.30\textwidth]{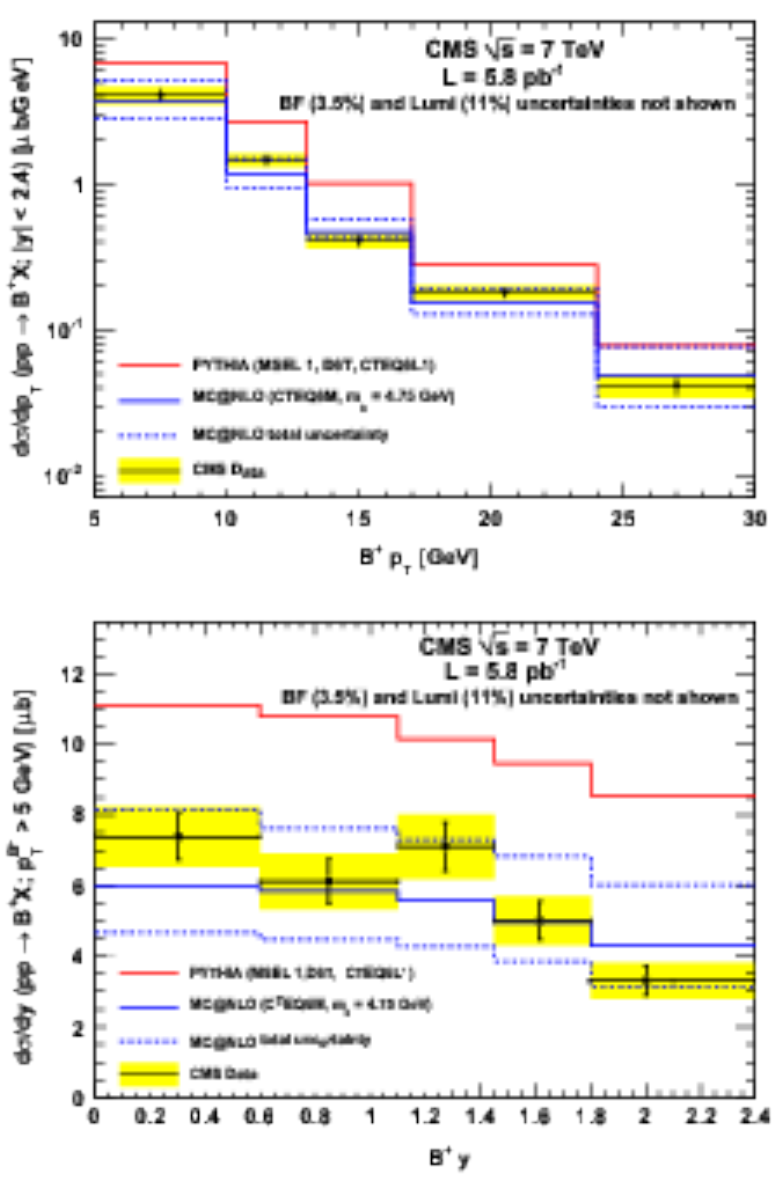}
\includegraphics[width=0.32\textwidth]{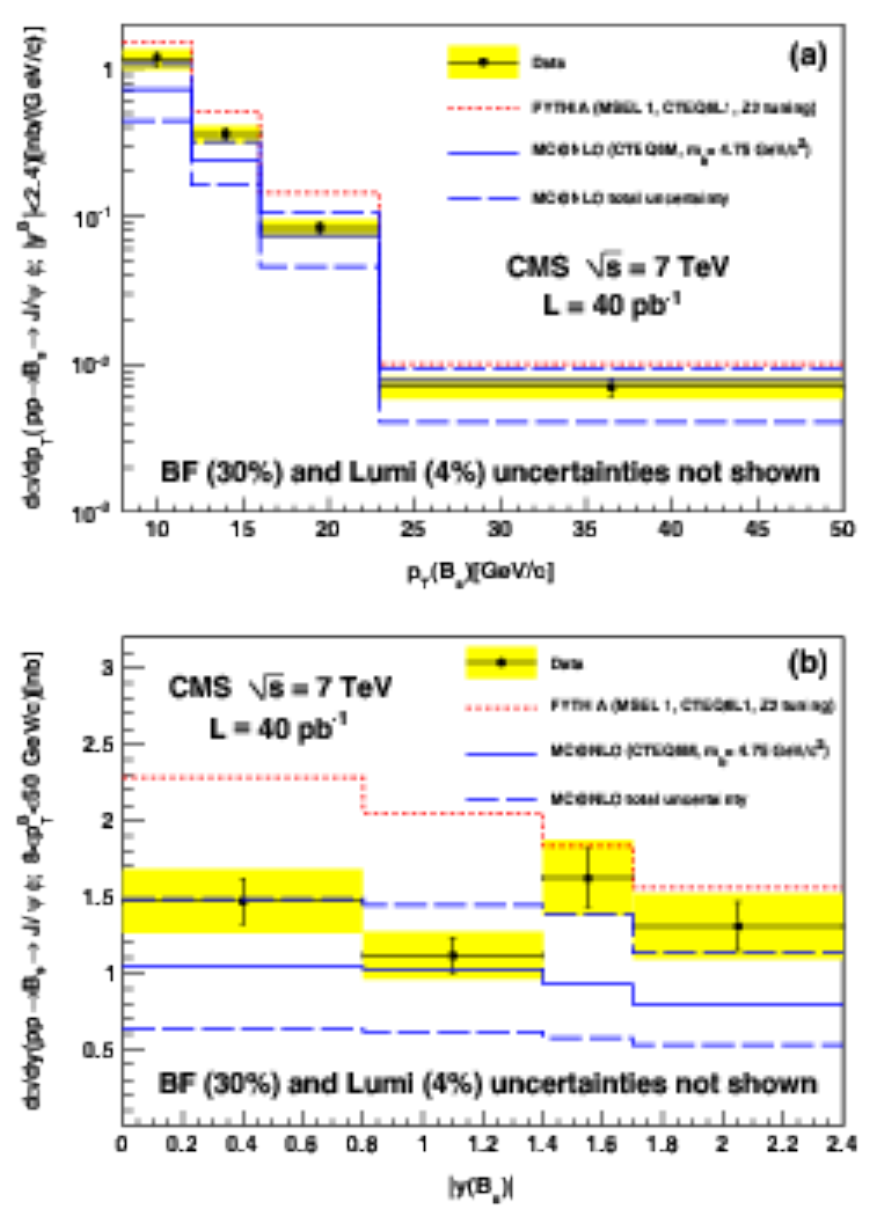}
\end{center}
\caption{\pt- (upper panels) and rapidity-differential (lower 
         panels) production cross sections of \bzero (left
         panels), \bplus (middle panels), and \bszero (right
         panels) mesons measured with the CMS experiment in
         \pp collisions at $\sqrt{s} = 7$~TeV. Predictions from
         higher order pQCD calculations (MC@NLO~\cite{mcnlo}) 
         are compared with the data (reprinted with permission 
         from Refs.~\cite{cms_bzero,cms_bplus,cms_bszero}; 
         Copyright (2011) by the American Physical Society).}
\label{fig:cms_b_in_pp}
\end{figure}

The only measurement of mid-rapidity charm hadron production cross sections 
in \pp collisions at the LHC via the full reconstruction of hadronic decay 
channels has been conducted with the ALICE experiment~\cite{d_in_pp}. 
\dzero, \dplus, and \dstarplus \pt-differential production cross sections 
measured in the range $|y| < 0.5$~\cite{d_in_pp} are shown in 
Fig.~\ref{fig:alice_d_in_pp}. Theoretical calculations based on higher-order 
pQCD calculations (FONLL~\cite{fonll3} and GM-VFNS~\cite{gmvfns}) are in 
reasonable agreement with the data within substantial experimental and 
theoretical systematic uncertainties.

The situation is similar for the production of bottom hadrons in \pp collisions
at the LHC as demonstrated in Fig.~\ref{fig:cms_b_in_pp} which shows the 
production cross sections of \bzero, \bplus, and \bszero mesons as a function
of \pt and, for high \pt mesons, as a function of 
rapidity~\cite{cms_bzero,cms_bplus,cms_bszero}. Calculations within 
the framework of the higher-order MC@NLO pQCD model~\cite{mcnlo} are in 
reasonable agreement with the data although both the experimental and the 
theoretical systematic uncertainties are substantial again.

\bplus meson production in \pp collisions was also investigated with the LHCb 
experiment~\cite{lhcb_bplus} in the kinematic range $0 < \pt < 40$~\gevc and 
$2.0 < y < 4.5$.
The measured \bplus cross sections are in good agreement with FONLL pQCD 
calculations~\cite{fonll3}. Furthermore, the pseudorapidity dependence of 
bottom hadron production cross sections was measured with LHCb~\cite{lhcb_bbar}
in the range $2 < \eta < 6$ without an exclusive identification of the various 
hadron species. Also in this case, higher order pQCD calculations in the 
FONLL~\cite{fonll3} and MCFM~\cite{mcfm} frameworks are able to describe the 
measured data.

\begin{figure}[p]
\begin{center}
\includegraphics[width=0.45\textwidth]{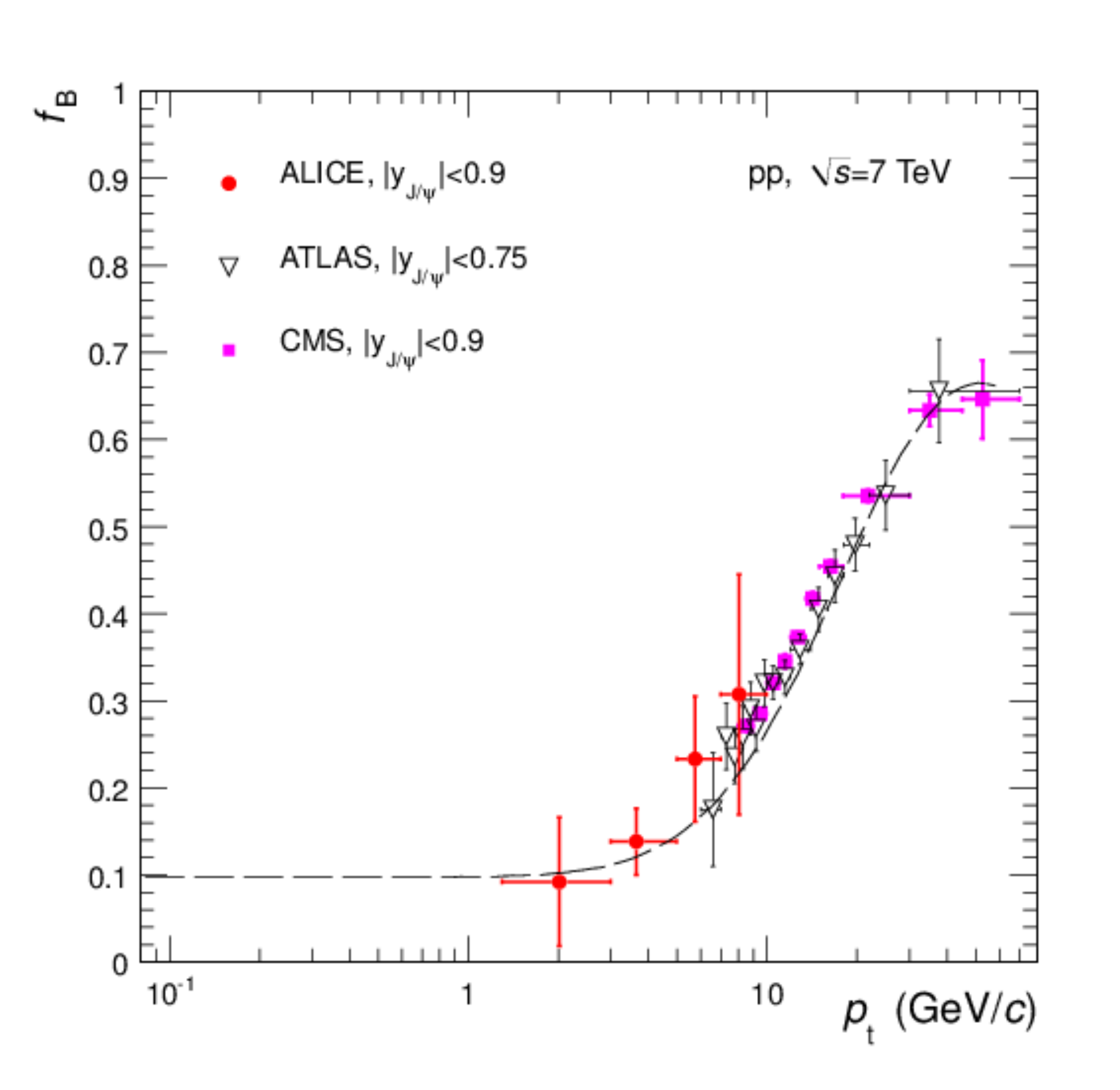}
\includegraphics[width=0.45\textwidth]{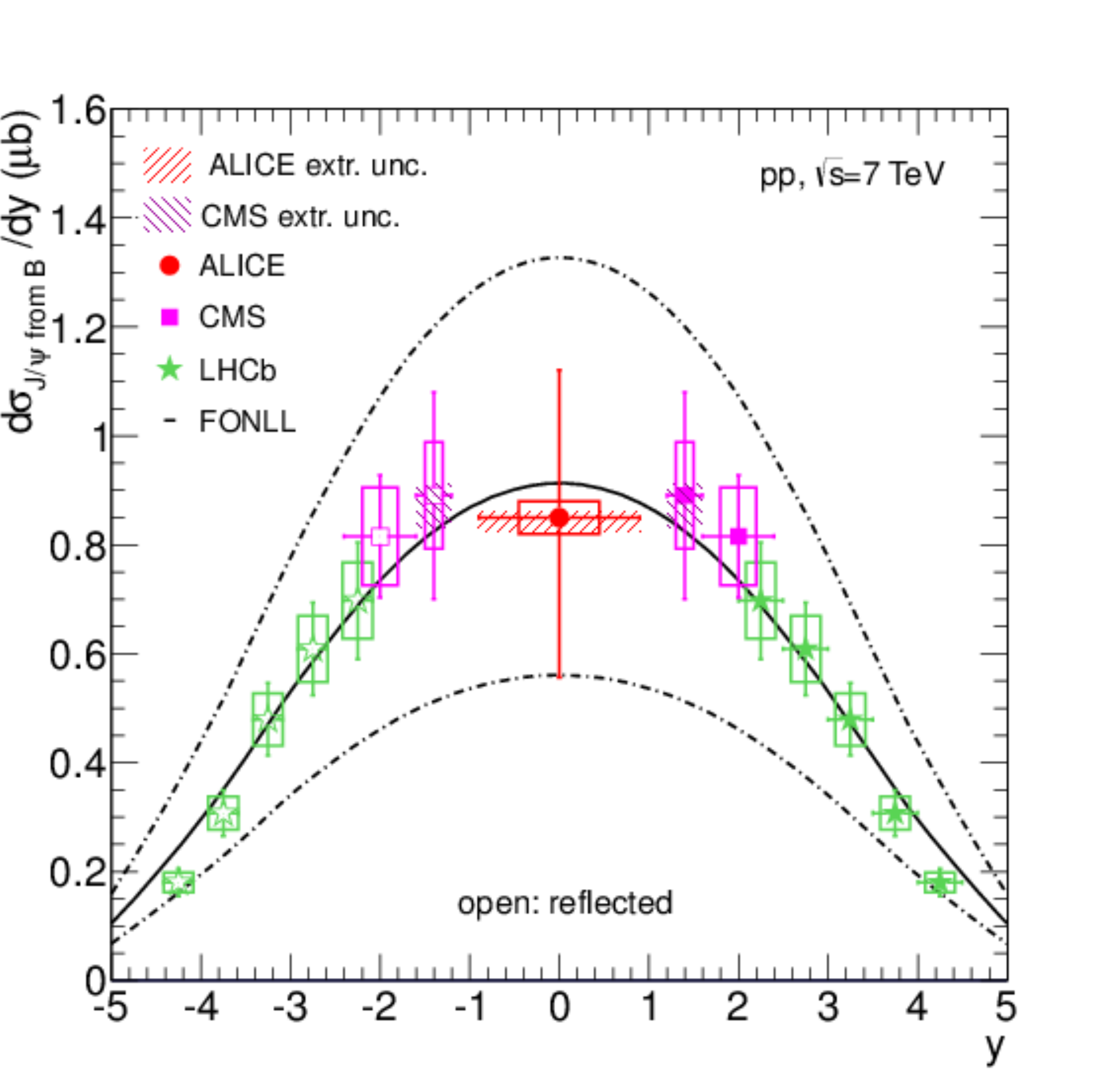}
\includegraphics[width=0.45\textwidth]{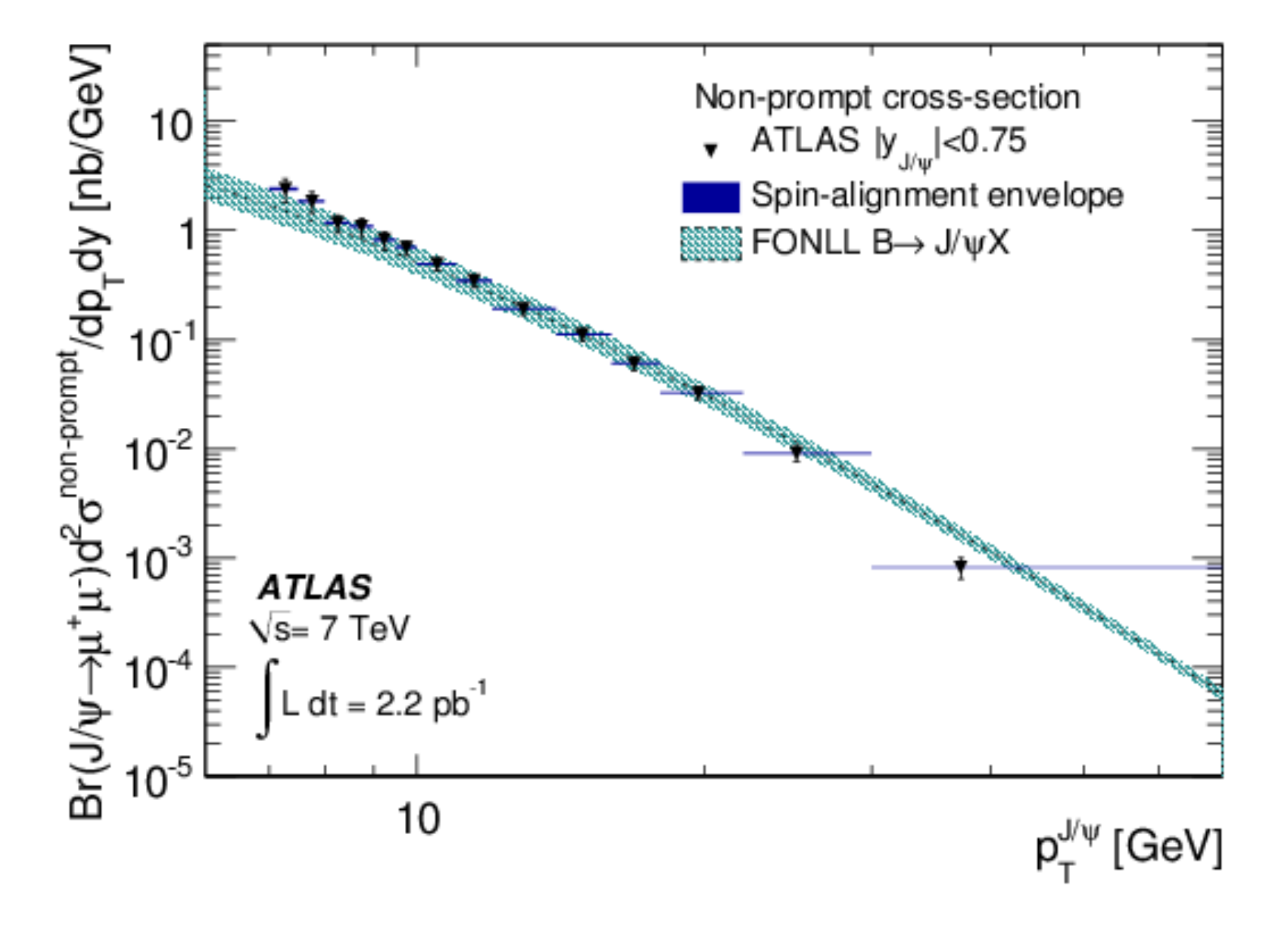}
\includegraphics[width=0.45\textwidth]{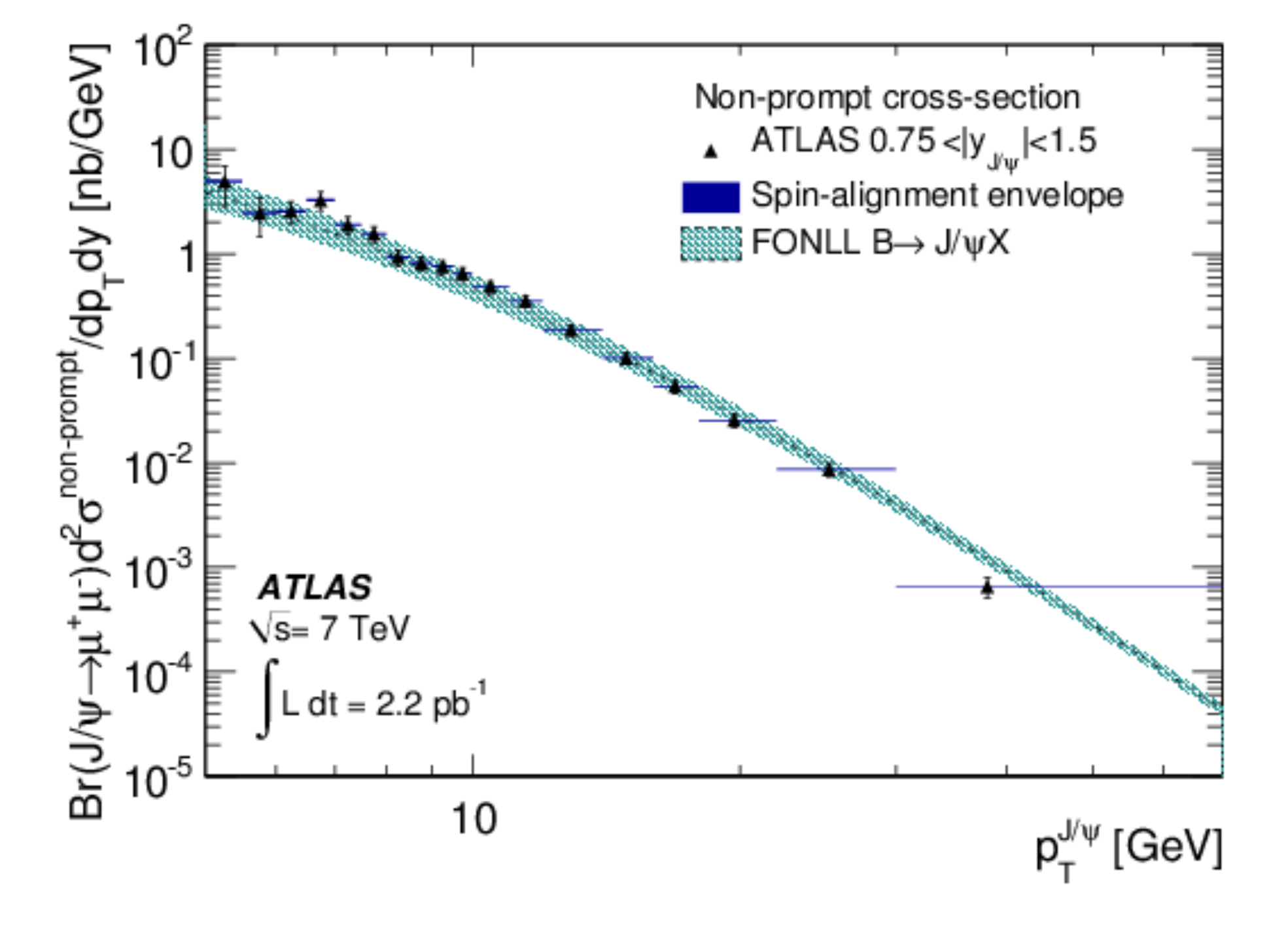}
\includegraphics[width=0.45\textwidth]{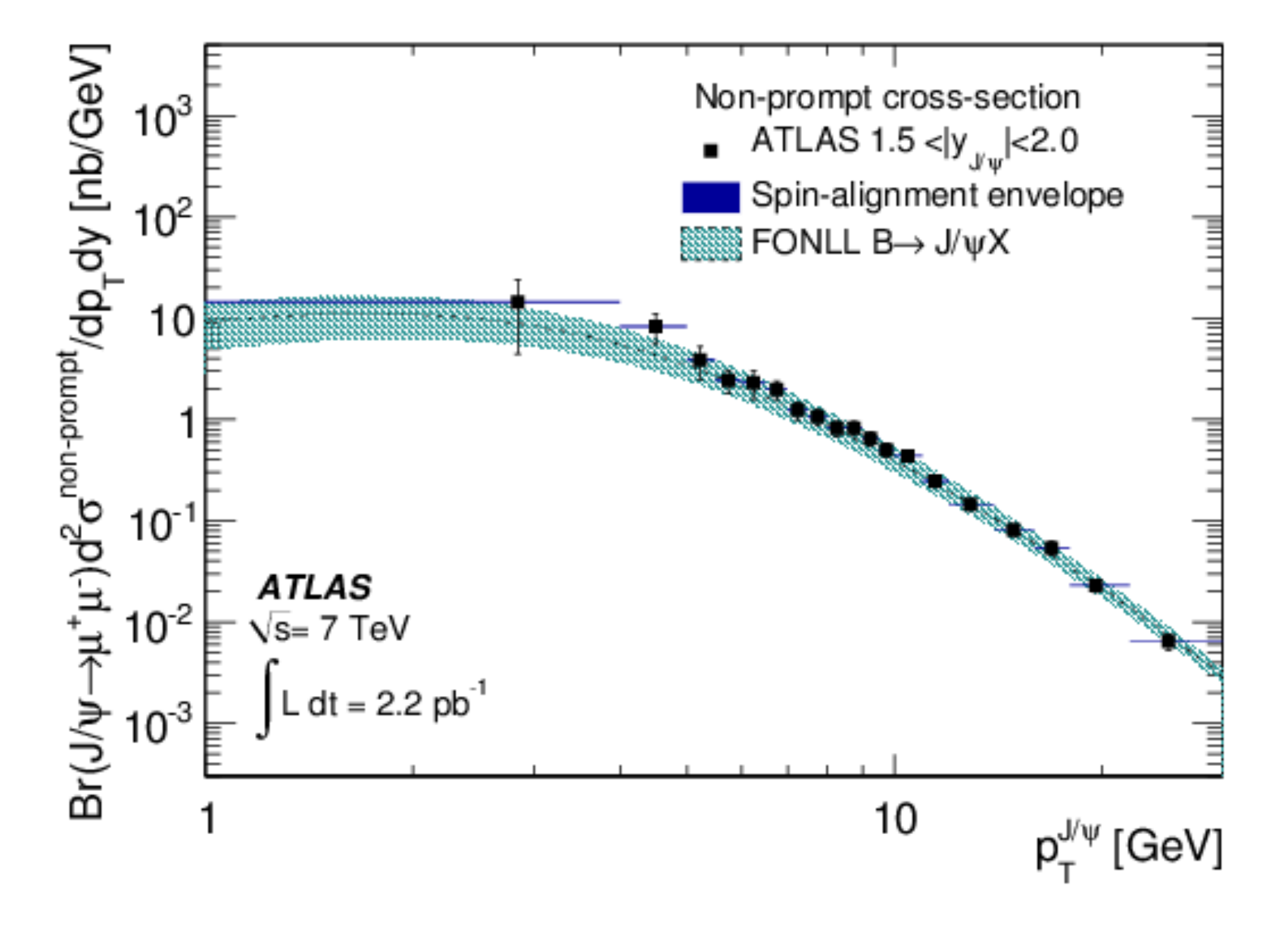}
\includegraphics[width=0.45\textwidth]{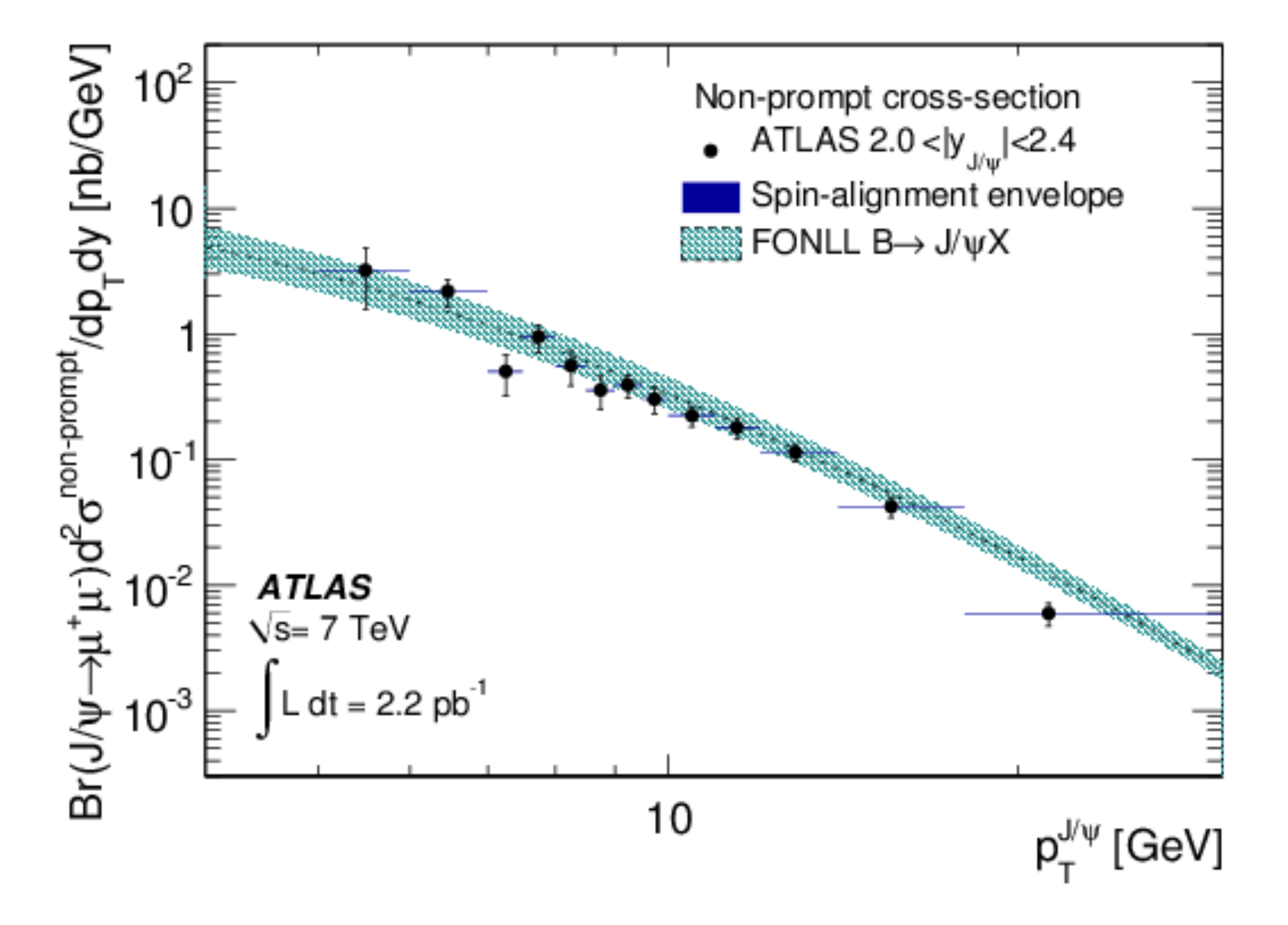}
\end{center}
\caption{Fraction of \jpsi from the decay of bottom hadrons in the inclusive
         \jpsi sample as measured with ALICE~\cite{alice_jpsi_b}, 
         ATLAS~\cite{atlas_jpsi_b}, and CMS~\cite{cms_jpsi_b} in 7~TeV \pp 
         collisions at mid-rapidity. Rapidity-differential production cross 
         section of \jpsi from bottom-hadron decays measured with 
         ALICE~\cite{alice_jpsi_b}, CMS~\cite{cms_jpsi_b}, and 
         LHCb~\cite{lhcb_jpsi_b} (upper two panels, reprinted from 
         Ref.~\cite{alice_jpsi_b} with kind permission from Springer Science
         and Business Media). \pt-differential non-prompt \jpsi
         production cross sections in four rapidity intervals as measured
         with the ATLAS experiment (lower four panels, reprinted from 
         Ref.~\cite{atlas_jpsi_b}). FONLL pQCD predictions of \jpsi cross 
         sections from bottom-hadron decays are compared with the measured 
         differential \jpsi cross sections in all but the upper left panel. 
         (see text for further details.)}
\label{fig:nonprompt_jpsi_pp}
\end{figure}

An alternative approach to bottom hadron production is the measurement of 
non-prompt \jpsi mesons originating from displaced bottom hadron decay 
vertices. All four large experiments at the LHC include vertex spectrometers 
with sufficient resolution for such measurements. The relative contribution
$f_{\rm B}$ of bottom hadron decays to the inclusive \jpsi yield increases at
mid-rapidity from $\approx$10\% at low \jpsi \pt to $\approx$2/3 at high \pt 
as demonstrated in the upper left panel of Fig.~\ref{fig:nonprompt_jpsi_pp} 
showing data from the ALICE~\cite{alice_jpsi_b}, ATLAS~\cite{atlas_jpsi_b}, and 
CMS~\cite{cms_jpsi_b} experiments measured in \pp collisions at 
$\sqrt{s} = 7$~TeV. The dashed line superimposed to the data 
indicates a semi-phenomenological function which is used to extrapolate the 
measured fraction down to zero \pt. As an example, the ATLAS measurement of the
\pt-differential production cross section on non-prompt \jpsi 
mesons~\cite{atlas_jpsi_b} is shown for four different rapidity intervals 
between mid-rapidity and $2.0 < y_{{\rm J}/\psi} < 2.4$ in the lower four panels 
of Fig.~\ref{fig:nonprompt_jpsi_pp}. Only in the rapidity interval 
$1.5 < y_{{\rm J}/\psi} < 2.0$ the ATLAS measurement reaches down to a \jpsi
meson \pt of 1~\gevc, whereas the minimum \pt in the other rapidity intervals
is 5~\gevc or even higher. FONLL predictions are in agreement with the measured
\pt-differential production cross section for all rapidity selections.
Furthermore, the measured rapidity dependence of the \jpsi cross section
from bottom hadron decays is reproduced by FONLL within experimental and
theoretical uncertainties as is demonstrated in the upper right panel of
Fig.~\ref{fig:nonprompt_jpsi_pp}, which includes data from 
ALICE~\cite{alice_jpsi_b}, CMS~\cite{cms_jpsi_b}, and LHCb~\cite{lhcb_jpsi_b}. 
ALICE has measured the \pt-differential cross section of non-prompt 
\jpsi meson production in the transverse momentum range $\pt > 1.3$~\gevc 
at mid-rapidity ($|y| < 0.9$). The shape of the corresponding FONLL prediction
was used to extrapolate the measurement to zero \pt. A corresponding procedure
was applied for the CMS measurement in the range $1.2 < |y| < 1.6$. The 
systematic uncertainties related to this extrapolation procedure are indicated
by the slashed boxes in the upper right panel of 
Fig.~\ref{fig:nonprompt_jpsi_pp}. Neither the CMS data in the interval 
$1.6 < |y| < 2.4$ nor the LHCb data had to be extrapolated since non-prompt 
\jpsi mesons have been measured down to zero \pt in these cases. The cross
section measured with the LHCb experiment at forward rapidity (closed symbols)
have been reflected around $y = 0$ (open symbols). The corresponding FONLL
calculation~\cite{fonll3} and its uncertainty are indicated by the solid and 
dashed lines, respectively.

\subsubsection{Semielectronic heavy-flavor hadron decays}
\label{subsubsec:lhc_e_pp}
\begin{figure}[t]
\begin{center}
\includegraphics[width=0.49\linewidth]{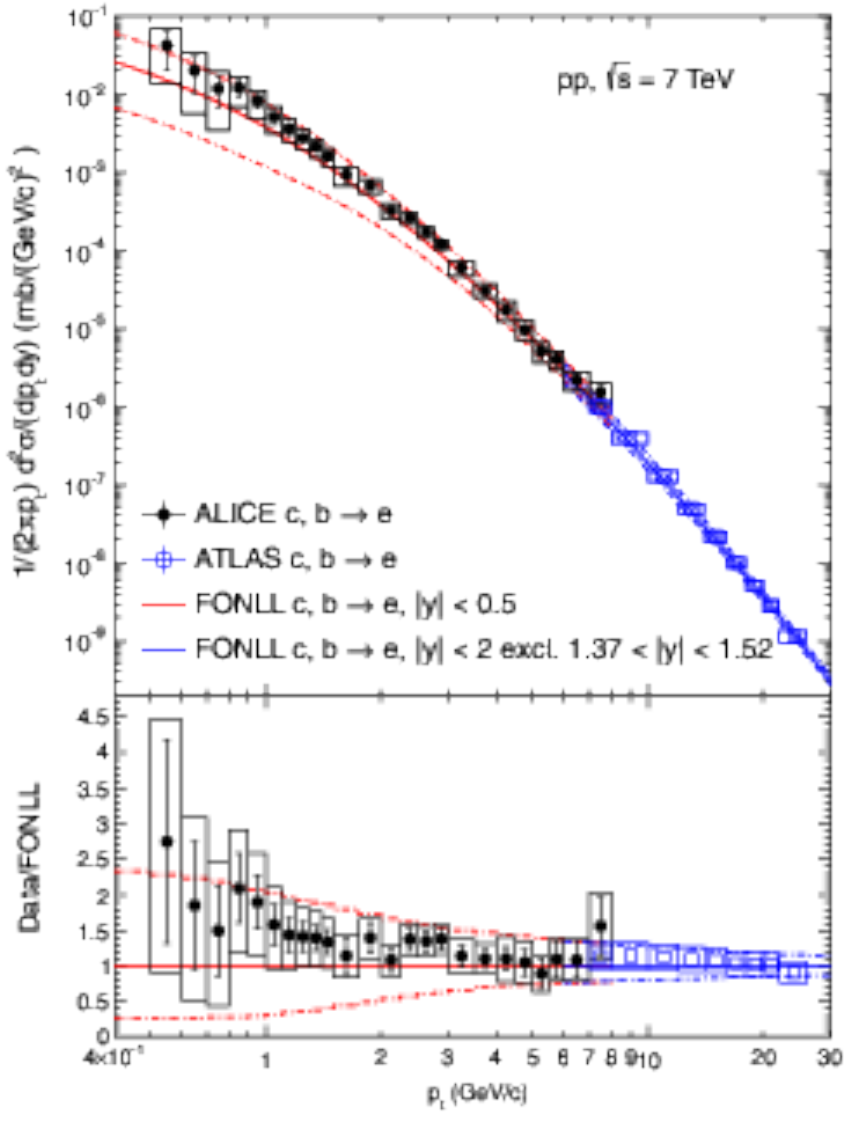}
\includegraphics[width=0.49\linewidth]{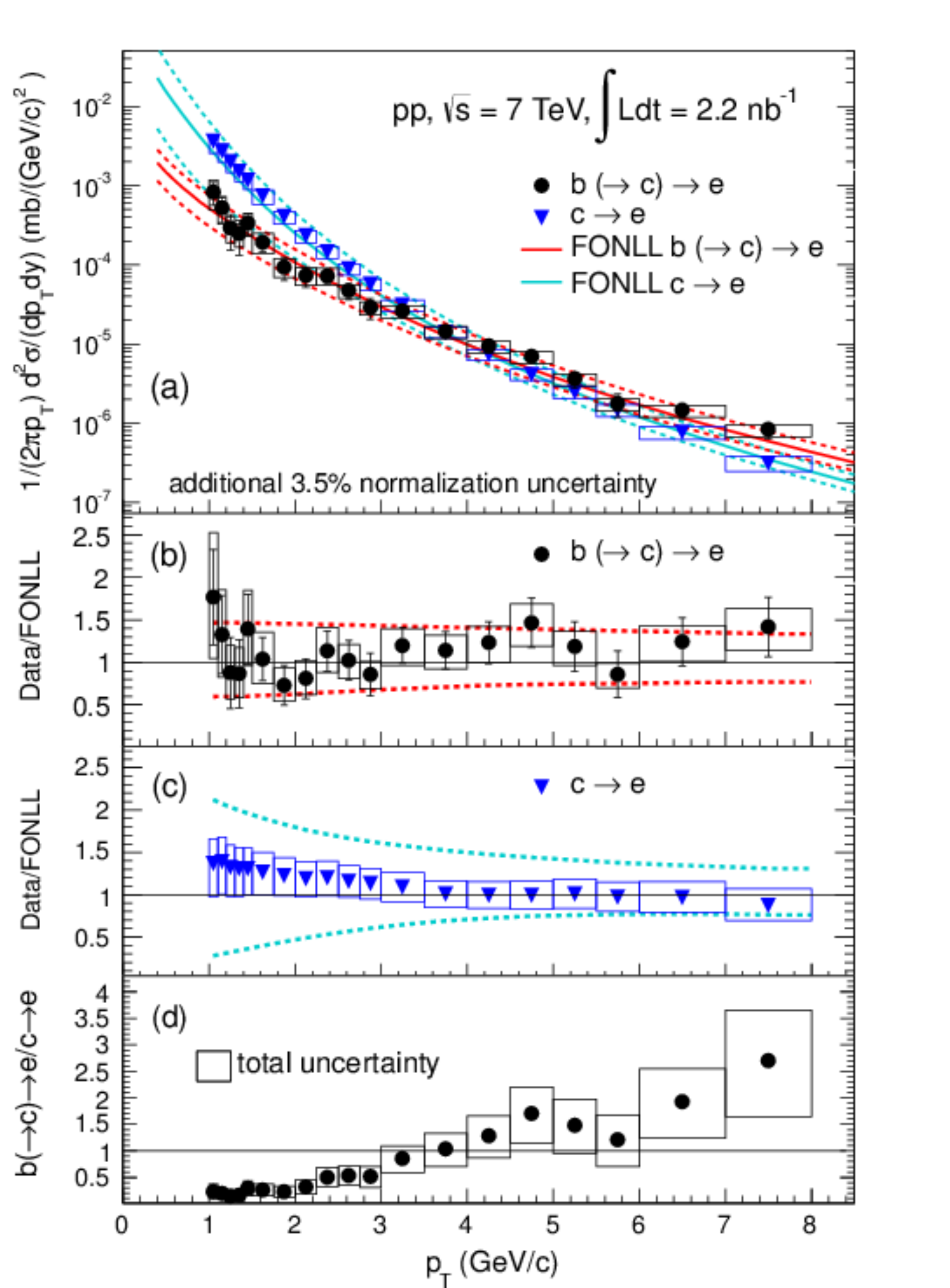}
\end{center}
\caption{\pt-differential production cross sections of electrons from 
         heavy-flavor hadron decays measured around mid-rapidity with the 
         ALICE~\cite{alice_e_pp} and ATLAS experiments~\cite{atlas_e_mu} in 
         \pp collisions at $\sqrt{s} = 7$~TeV (left panel, reprinted with 
         permission from Ref.~\cite{alice_e_pp}; Copyright (2012) by the 
         American Physical Society). \pt-differential production cross sections 
         of electrons from charm- and from bottom-hadron decays measured with 
         ALICE at mid-rapidity in \pp collisions at $\sqrt{s} = 7$~TeV (right
         panel, reprinted from Ref.~\cite{alice_hfe_beauty}). In both panels
         corresponding predictions from a FONLL pQCD calculation~\cite{fonll3} 
         are shown for comparison.}
\label{fig:e_in_pp}
\end{figure}

Both ALICE and ATLAS have published first cross section measurements of
electrons from semileptonic decays of heavy-flavor hadrons in \pp collisions
at $\sqrt{s} = 7$~TeV at the LHC. The ALICE measurement~\cite{alice_e_pp} is 
restricted to a narrow interval around mid-rapidity ($|y| < 0.5$). With ATLAS, 
electrons were reconstructed~\cite{atlas_e_mu} in the larger range $|y| < 2$, 
excluding the interval $1.37 < |y| < 1.52$ in which the electron identification
and the resolution of the energy measurement is degraded by a large amount of 
material in front of the first detector elements. The resulting 
\pt-differential production cross sections, normalized per unit rapidity, are 
depicted in the left panel of Fig.~\ref{fig:e_in_pp}. While the ALICE data 
include most of the total electron cross section from heavy-flavor hadron 
decays, the ATLAS data extend the measurement to higher \pt. Exploiting
the excellent vertexing capabilities of the ALICE ITS, electrons from charm
and bottom hadron decays can be separated from each other in \pp collisions
at $\sqrt{s} = 7$~TeV~\cite{alice_hfe_beauty}. The resulting \pt-differential 
production cross sections are shown in the right panel of 
Fig.~\ref{fig:e_in_pp}. Predictions from a FONLL pQCD calculation~\cite{fonll3}
are in good agreement both with the inclusive electrons cross sections from 
heavy-flavor hadron decays as well as with the individual cross sections of 
electrons from charm- and bottom-hadron decays, respectively, within 
experimental and theoretical uncertainties.

\begin{figure}[t]
\begin{center}
\includegraphics[width=0.49\textwidth]{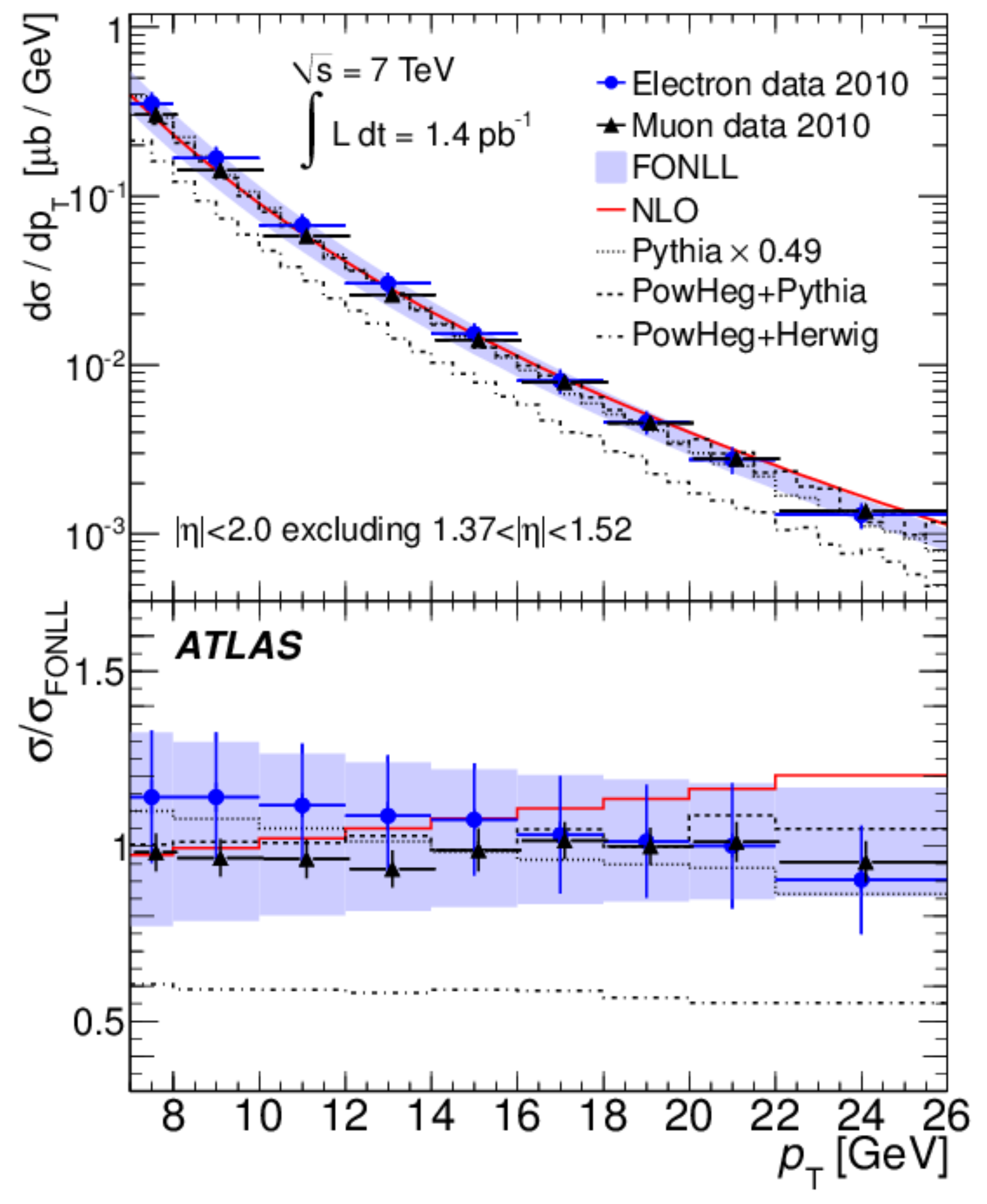}
\includegraphics[width=0.49\textwidth]{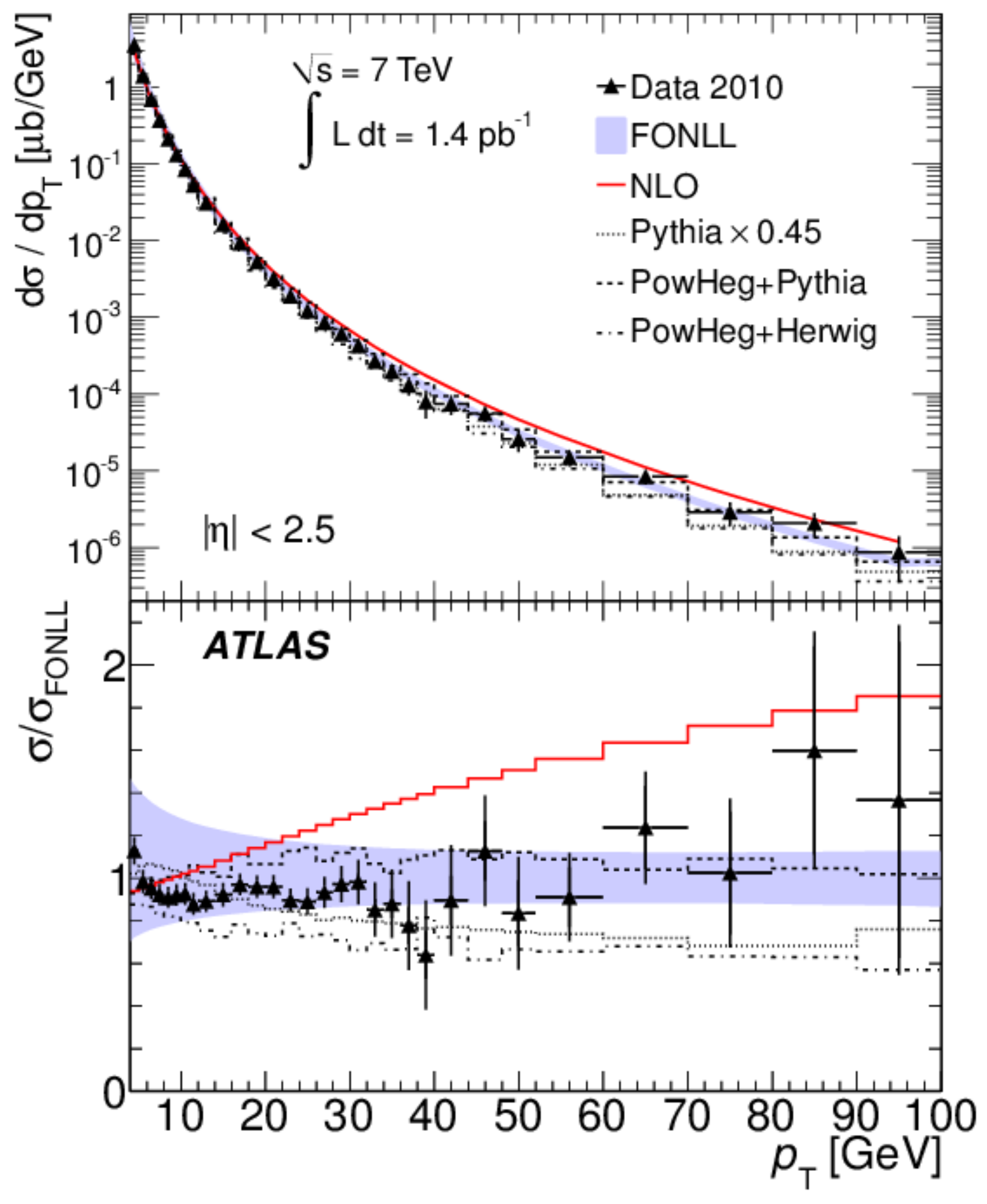}
\end{center}
\caption{\pt-differential production cross sections of leptons from heavy-flavor
         hadron decays measured with the ATLAS experiment in \pp collisions at
         $\sqrt{S} = 7$~TeV at the LHC. Electron and muon production cross
         sections for $|\eta| < 2.0$, excluding the region 
         $1.37 < |\eta| < 1.52$ (left). Muon production cross section for
         $|\eta| < 2.5$ (right). Predictions from FONLL and other pQCD 
         calculations are compared with the data. Ratios of the measured
         cross sections and of other predictions to the FONLL calculation
         are shown in the lower panels (see text for further details)
         (reprinted from Ref.~\cite{atlas_e_mu}).}
\label{fig:atlas_l_vs_fonll}
\end{figure}

\subsubsection{Semimuonic heavy-flavor hadron decays}
\label{subsubsec:lhc_mu_pp}
Both with the ALICE~\cite{muon_pp} and ATLAS~\cite{atlas_e_mu} experiments 
not only electrons but also muons from open heavy-flavor hadron decays have 
been measured at the LHC. The left panel of Fig.~\ref{fig:atlas_l_vs_fonll} 
shows a comparison of the muon \pt-differential production cross section from
ATLAS with the electron data (also shown in Fig.~\ref{fig:e_in_pp}) for 
$\pt > 7$~\gevc in the pseudorapidity range $|\eta| < 2.0$, excluding the 
interval $1.37 < |\eta| < 1.52$ in which no electron measurement is available. 
In this high \pt range the mass difference between electrons and muons is 
irrelevant. Therefore, one would expect the cross sections of electrons and 
muons from semileptonic heavy-flavor hadron decays to be equal which is 
confirmed within experimental uncertainties by the ATLAS data~\cite{atlas_e_mu}.
The data are confronted with various pQCD-based model calculations. The 
state-of-the are FONLL pQCD prediction~\cite{fonll3}, shown as a solid
band in Fig.~\ref{fig:atlas_l_vs_fonll}, agrees well with the data within 
experimental and theoretical uncertainties. In order to simplify the
quantitative comparison of model predictions with the data the lower panels
in Fig.~\ref{fig:atlas_l_vs_fonll} show the ratio of the data and the other
model calculations to the FONLL prediction on a linear scale. The NLO 
calculation POWHEG~\cite{powheg1,powheg2} stops on the parton level and, 
therefore, has to be interfaced with parton shower generators to allow a 
comparison with decay lepton data. Here, PYTHIA~\cite{pythia} and 
HERWIG~\cite{herwig} have been used for the parton shower 
simulation. Whereas POWHEG+PYTHIA agree reasonably well with the data and 
with FONLL, this is different for POWHEG+HERWIG which predicts a significantly 
reduced lepton cross section. It was shown that this discrepancy can not be 
accounted for by the different hadron decay models. In addition, a LO PYTHIA 
pQCD calculation~\cite{pythia}, which includes the parton shower simulation, 
predicts total cross sections approximately a factor two larger than the 
measurement, but with the proper shape of the decay lepton \pt distribution.

It is an interesting question whether the next-to-leading-log (NLL) resummation
terms included in the FONLL pQCD calculation are actually important at the LHC 
or whether an NLO calculation without this resummation would be sufficient for 
the description of open heavy-flavor data. In order to address this issue, an 
NLO central value prediction was calculated with the FONLL code excluding the 
NLL resummation~\cite{fonll3}. An indication that this approach yields too hard
lepton \pt spectra is visible in the left panel of 
Fig.~\ref{fig:atlas_l_vs_fonll} already, which shows the electron and muon \pt 
spectra up to 26~\gevc. The right panel of Fig.~\ref{fig:atlas_l_vs_fonll} 
displays the \pt-differential production cross section of muons from 
heavy-flavor hadron decays in the extended transverse momentum range up to 
100~\gevc~\cite{atlas_e_mu}. While the FONLL pQCD prediction and, to some 
extent, also the other pQCD calculations are in good agreement with the 
measurement, the NLO FONLL calculation ignoring the NLL resummation is clearly 
not consistent with data, demonstrating for the first time at a hadron collider
the importance of the resummation of high order terms in the pQCD 
expansion.

\begin{figure}[t]
\begin{center}
\includegraphics[width=0.62\linewidth]{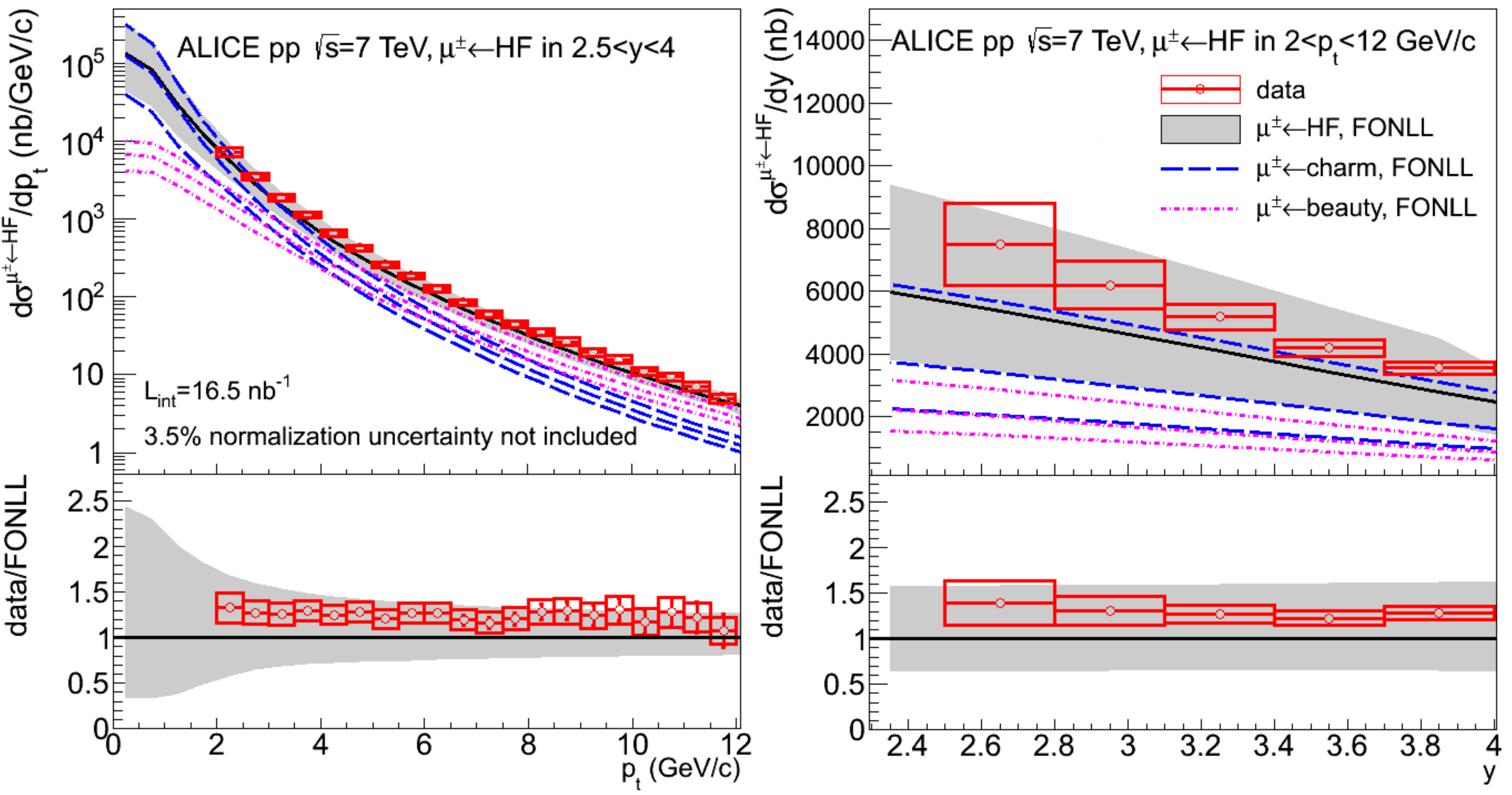}
\includegraphics[width=0.35\linewidth]{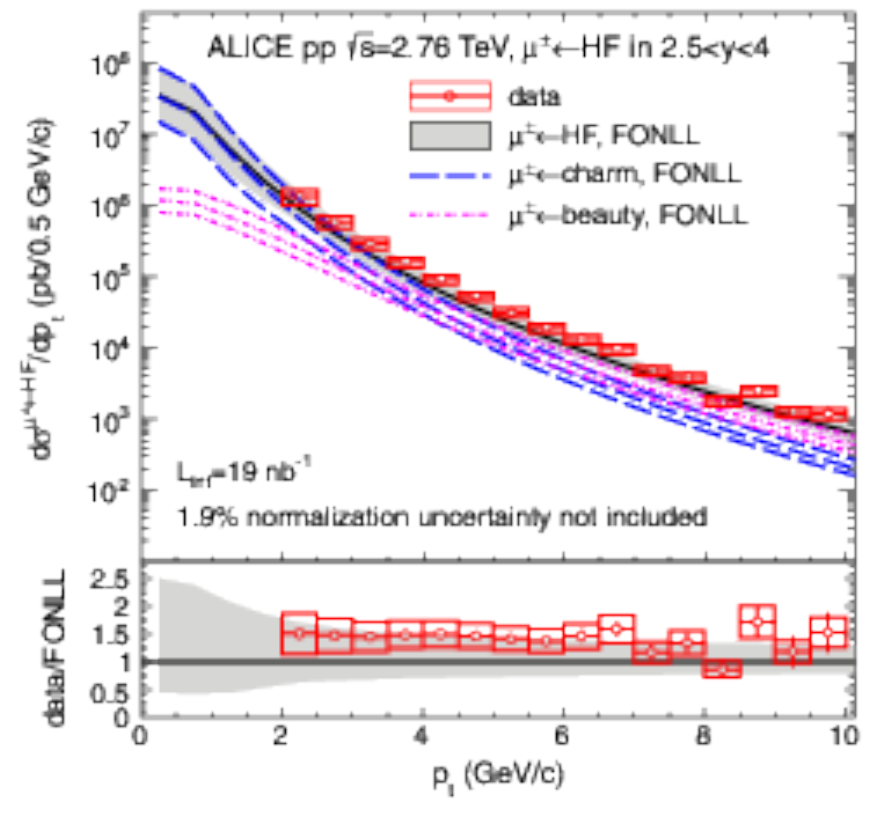}
\end{center}
\caption{\pt- and rapidity-differential production cross sections of muons 
         from heavy-flavor hadron decays measured with the ALICE experiment
         in the ranges $2 < \pt < 12$~\gevc and $-4 < y < -2.5$ in \pp 
         collisions at $\sqrt{s} = 7$~TeV (left and middle panel, reprinted 
         from Ref.~\cite{muon_pp}), and \pt-differential cross section at 
         $\sqrt{s} = 2.76$~TeV (right panel, reprinted with permission from 
         Ref.~\cite{muon_pbpb}; Copyright (2012) by the American Physical 
         Society) in comparison with predictions from FONLL pQCD calculations. 
         Error bars (boxes) represent the statistical (systematic) 
         uncertainties, where an additional 3.5\% absolute normalization 
         uncertainty is not shown. Grey bands indicate the FONLL 
         prediction~\cite{fonll3} for muons from heavy-flavor hadron decays 
         and the long dashed (short dashed) curves represent the FONLL 
         predictions for charm (bottom) hadron decays separately. The ratios 
         of the data to the FONLL prediction are shown in the lower panels 
         on a linear scale.}
\label{fig:alice_mu_in_pp}
\end{figure}

At forward rapidity ($-4 < y < -2.5$) the \pt- and rapidity-differential 
production cross sections of muons from heavy-flavor hadron decays have been 
measured in \pp collisions at $\sqrt{s} = 7$~TeV~\cite{muon_pp} and 
2.76~TeV~\cite{muon_pbpb} with the ALICE experiment as shown in 
Fig.~\ref{fig:alice_mu_in_pp}. Corresponding FONLL pQCD calculations are in 
good agreement with the measurement within the systematic uncertainties. In 
both semileptonic decay channels, \ie electrons and muons, the FONLL pQCD 
calculations suggest that for $\pt > 5-6$~\gevc the dominant contribution is 
from bottom hadron decays while at lower \pt charm hadron decays represent the 
relevant lepton source.

The correlation of single high-\pt muons with jets was measured with the
CMS experiment in \pp collisions at $\sqrt{s} = 7$~TeV~\cite{cms_mu_b}. Events 
with high-\pt muons ($\pt^\mu > 6$~\gevc) have been selected in the range 
$|\eta^\mu| < 2.1$. Events containing bottom hadrons were discriminated from 
background via the measurement of the muon transverse momentum with respect 
to the closest jet.
The resulting \pt- and pseudorapidity-differential cross sections of muons
from bottom-hadron decays have been confronted with pQCD calculations executed
in the MC@NLO framework~\cite{mcnlo}. While the measured cross section are 
slightly larger than theoretically predicted, the data still agree with the 
model within the combined systematic uncertainties.

\subsubsection{FONLL-driven $\sqrt{s}$ scaling}
\label{subsubsec:pp_scaling}
The heavy-flavor production cross sections measured in \pp collisions at 
$\sqrt{s} = 7$~TeV can be scaled to $\sqrt{s} = 2.76$~TeV to provide 
a reference for the \pbpb data at the same energy per nucleon-nucleon
pair for cases in which a direct measurement of a \pp reference at 2.76~TeV
is not available. This was demonstrated for the ALICE case in 
Ref.~\cite{reference}. Since FONLL pQCD calculations~\cite{fonll3} have been 
shown to be in reasonable agreement with all open heavy-flavor observables 
measured in \pp  collisions at $\sqrt{s} = 7$~TeV, FONLL was used for the 
necessary $\sqrt{s}$ scaling in this case. The scaling factors for D mesons, 
electrons, and muons were defined as the ratios of the corresponding cross 
sections from FONLL calculations at 2.76 and 7~TeV. The assumption was used 
that neither the factorization and renormalization scales nor the heavy-quark 
masses in the FONLL calculation vary with $\sqrt{s}$. To evaluate the 
uncertainties of the scaling factors, the scales and heavy-quark masses were 
varied over reasonable ranges and the envelopes of the resulting scaling 
factors were determined. For all open heavy-flavor observables the resulting 
uncertainties of the scaling functions are similar. They range from 
$\sim 25$\% at $\pt = 2$~\gevc to less than 10\% for $\pt > 10$~\gevc.

As mentioned above, with all LHC experiments \pp data have been recorded
at $\sqrt{s} = 2.76$~TeV, albeit with limited statistics only. \pt-differential
muon spectra from heavy-flavor hadron decays have been measured with the ALICE
experiment~\cite{muon_pbpb} as shown in the right panel of 
Fig.~\ref{fig:alice_mu_in_pp}. 
Furthermore, D-meson \pt- differential production cross 
sections have been measured with ALICE at the same energy~\cite{d_pp_low}. 
Both the muon and the D-meson data are in good agreement within uncertainties 
with the reference obtained via the FONLL-driven $\sqrt{s}$ scaling procedure 
from the 7~TeV \pp measurements to 2.76~TeV.

\subsection{Heavy flavor in \pbpb collisions}
\label{subsec:lhc_pbpb}
\subsubsection{Hadronic heavy-flavor hadron decays}
\label{subsubsec:lhc_d_pbpb}
\begin{figure}[p]
\begin{center}
\includegraphics[width=0.60\linewidth]{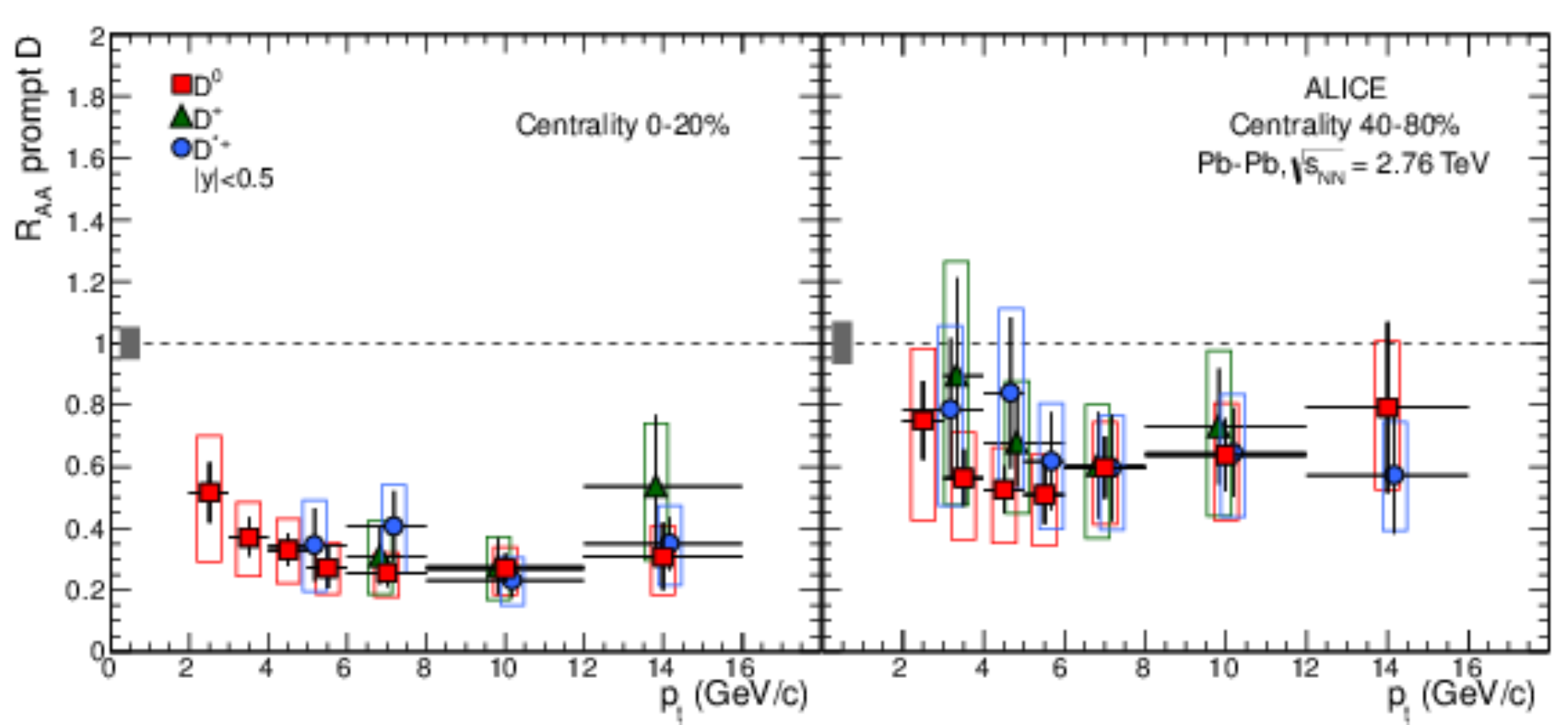}
\includegraphics[width=0.60\textwidth]{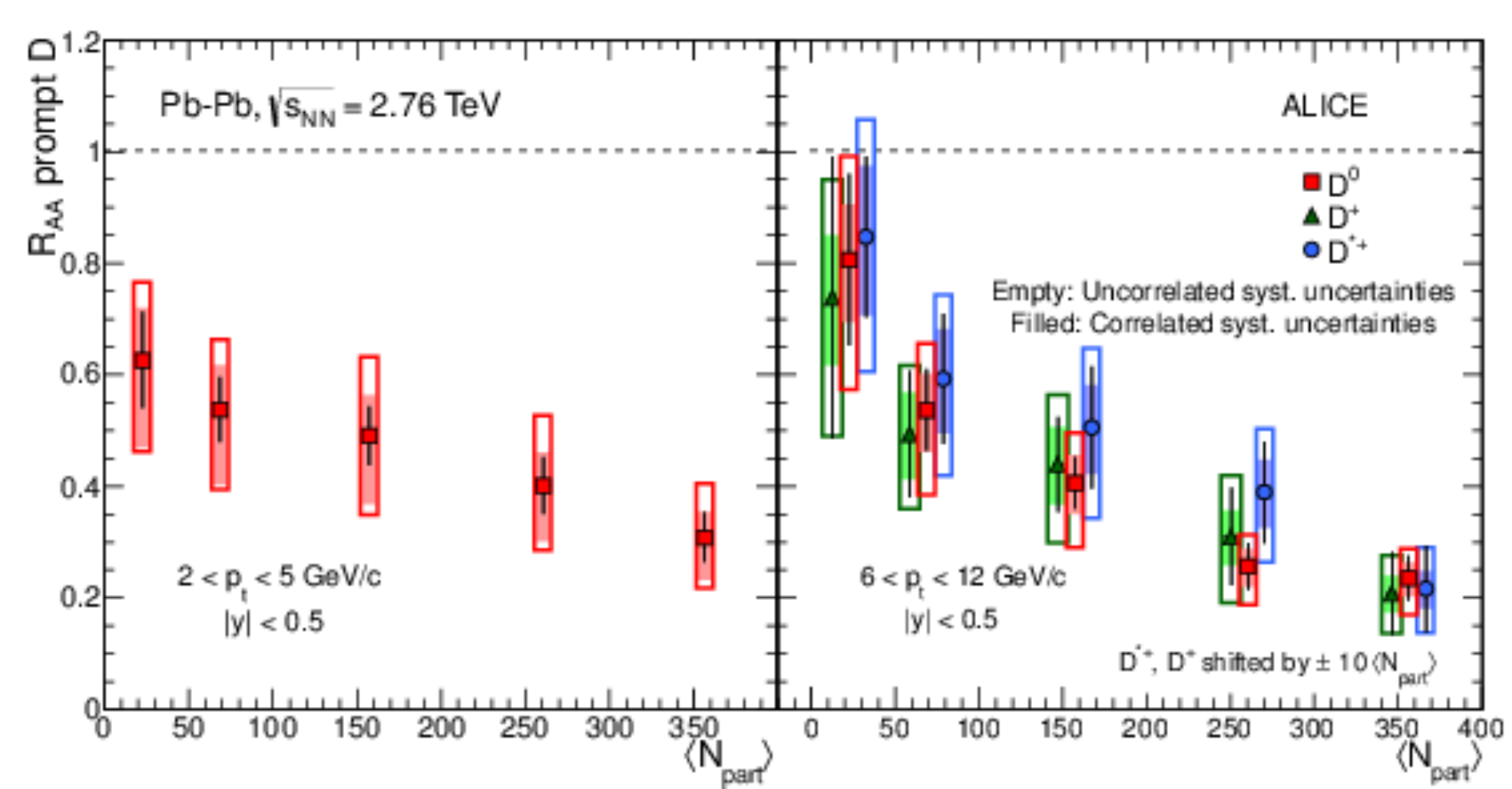}
\includegraphics[width=0.40\linewidth]{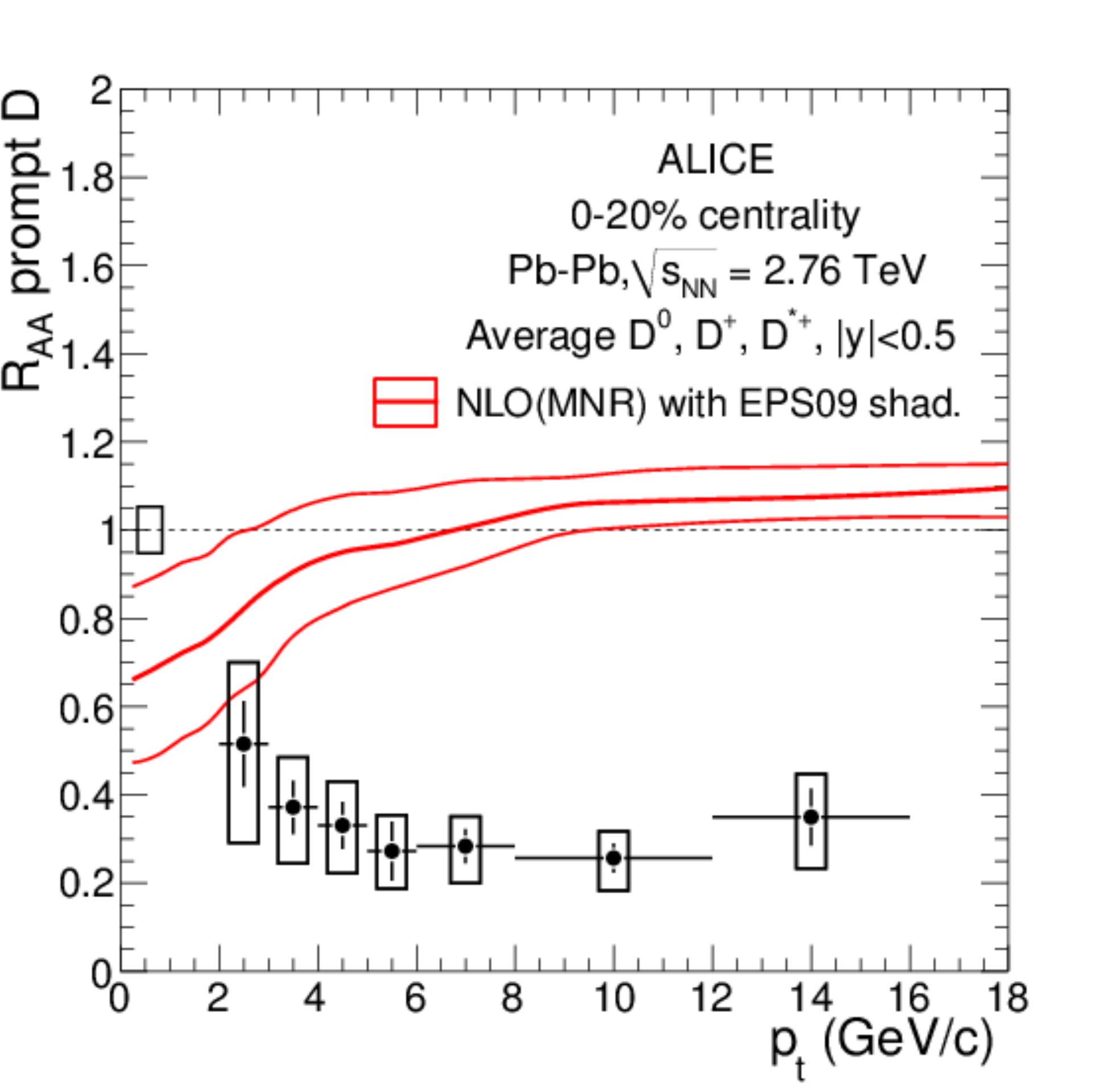}
\includegraphics[width=0.40\linewidth]{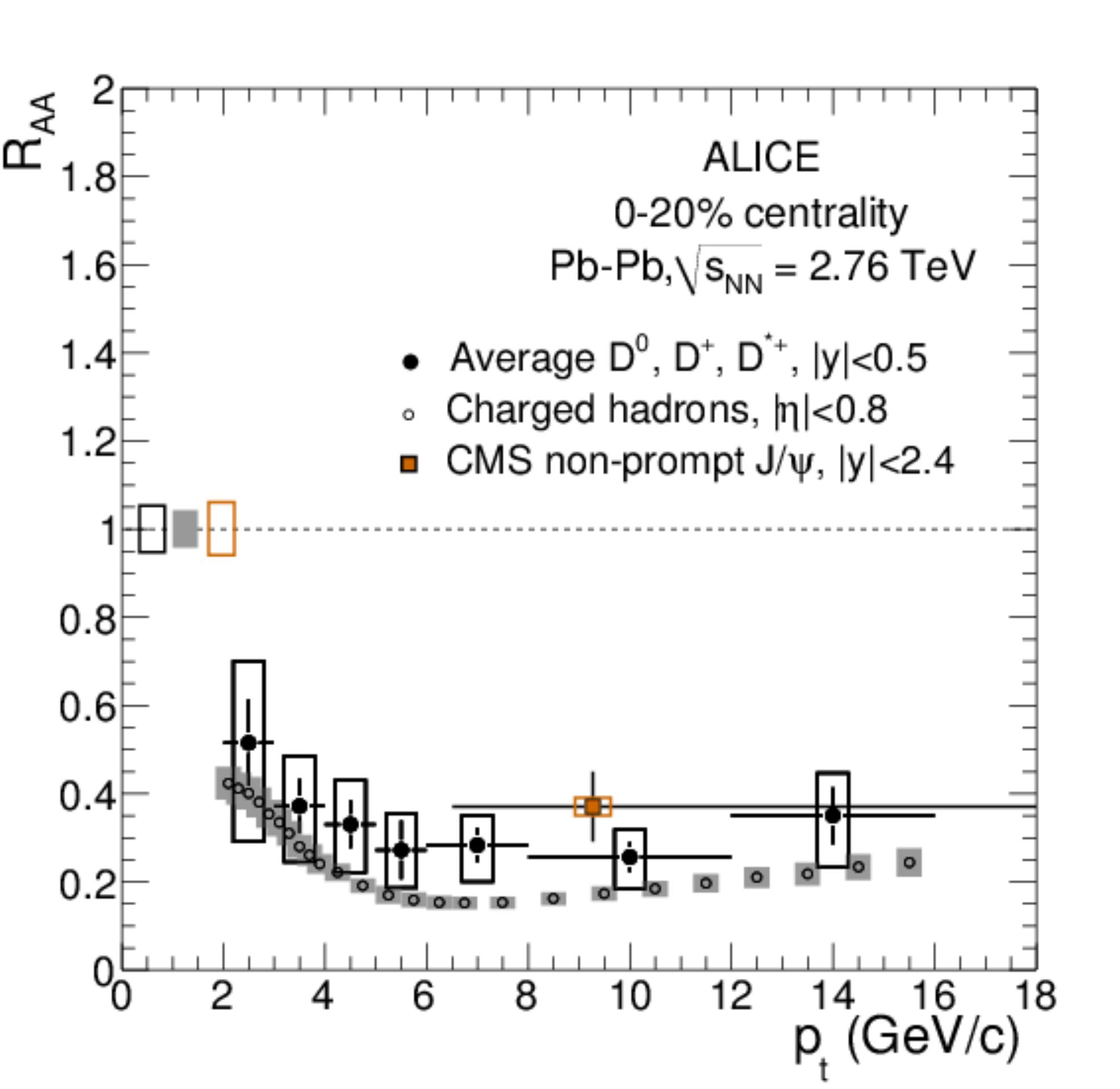}
\end{center}
\caption{Nuclear modification factor $\raa$ of prompt \dzero, \dplus, and
         \dsplus mesons measured with the ALICE experiment in the 20\% most 
         central (upper left) and in 40-80\% central (upper right) \pbpb 
         collisions at $\sqrt{s_{\rm NN}} = 2.76$~TeV. Error bars (boxes)
         show the statistical (systematic) uncertainties. The normalization
         uncertainties are indicated by the full boxes at one. Centrality
         dependence of $\raa$ for \dzero mesons with $2 < \pt < 5$~\gevc
         (middle left) and all D-meson species with $6 < \pt < 12$~\gevc
         (middle right). Average D-meson $\raa$ in the 20\% most central 
         \pbpb collisions in comparison with results from an NLO pQCD 
         calculation including EPS09 nuclear shadowing~\cite{eps09} (lower 
         left). D-meson $\raa$ compared with corresponding values of the 
         nuclear modification factors of charged hadrons as measured with 
         ALICE~\cite{raa_charged} and of non-prompt \jpsi mesons from 
         bottom-hadron decays as measured with the CMS 
         experiment~\cite{cms_jpsi_pbpb} (lower right) (reprinted from 
         Ref.~\cite{d_in_pbpb} with kind permission from Springer Science
         and Business Media).}
\label{fig:d_in_pbpb}
\end{figure}

First results on open heavy-flavor production in \pbpb collisions at
$\sqrt{s_{\rm NN}} = 2.76$~TeV at the LHC have been published recently.
One of the major deficiencies of the measurements at RHIC has been overcome
at the LHC right away. With ALICE, D mesons could be reconstructed even in 
central \pbpb collisions~\cite{d_in_pbpb} due to the excellent performance
of the apparatus. Charged particle tracks are reconstructed in the ALICE TPC
and ITS with a typical efficiency of 90\%. Furthermore, the ITS provides 
excellent vertexing precision, \eg the track impact parameter resolution in the
transverse plane is better than 50~$\mu$m for $\pt \ge 2$~\gevc. With the TOF
system, which is instrumental for hadron identification, a time resolution in 
the order of 100~ps can be achieved.
The D-meson transverse momentum spectra measured with ALICE in \pp and \pbpb 
collisions at the LHC contain feed down from bottom-hadron decays. This 
contribution was estimated based on the bottom production cross sections 
predicted from the FONLL calculation, which in \pp collisions is in good 
agreement with the ALICE measurements of \jpsi and electrons from 
bottom-hadron decays as demonstrated in Figs.~\ref{fig:nonprompt_jpsi_pp} and 
\ref{fig:e_in_pp}, respectively. The feed down from bottom hadron 
decays was subtracted from the inclusive D-meson spectra resulting in the 
so-called prompt D-meson yields. For \pbpb 
collisions, the unknown nuclear modification factor of bottom hadrons 
introduces an additional systematic uncertainty on the prompt D-meson yields 
in \pbpb collisions. The conservative assumption was employed that the nuclear 
modification factor of bottom hadrons relative to the prompt D-meson $\raa$ 
fulfills the condition $0.3 < R^B_{\rm AA}/R^D_{\rm AA} < 3$. The resulting nuclear
modification factors of prompt \dzero, \dplus, and \dsplus mesons in the 
20\% most central \pbpb collisions are shown in the upper left panel of 
Fig.~\ref{fig:d_in_pbpb}. A strong suppression is observed for all species 
reaching a factor 3-5 for $\pt > 8$~\gevc. Towards more peripheral \pbpb 
collisions nuclear modification factors closer to one are measured for prompt 
D mesons as demonstrated for 40-80\% central \pbpb collisions in the upper 
right panel of Fig.~\ref{fig:d_in_pbpb}. The centrality dependence of the 
D-meson $\raa$ is quantified in the middle panel of Fig.~\ref{fig:d_in_pbpb}, 
which shows the nuclear modification factor of \dzero mesons in the transverse 
momentum range $2 < \pt < 5$~\gevc as a function of the average number of 
participants $\langle N_{\rm part} \rangle$ on the left and the same dependence 
for all D-meson species in the range $6 < \pt < 12$~\gevc on the right. Clearly
the D-meson suppression increases towards more central \pbpb collisions.

It is an important issue whether the observed suppression of prompt D mesons
at intermediate and high \pt might be related to shadowing of the parton 
distribution functions at the LHC. The comparison of an NLO pQCD calculation 
including the state of the art EPS09 parametrization of parton 
shadowing~\cite{eps09} with the measured D-meson $\raa$ as shown in the lower
left panel of Fig.~\ref{fig:d_in_pbpb} provides evidence that this is not the 
case, \ie the observed suppression clearly is a final state effect related to 
the hot and dense medium produced in \pbpb collisions at the LHC. While parton 
shadowing seems not to be relevant for the interpretation of the prompt D-meson
$\raa$, it is obvious that a highly desirable experimental confirmation of 
this conclusion will require data from proton-nucleus collisions at the LHC.

One of the most important experimental question left open by heavy-flavor 
measurements at RHIC is related to the expected mass hierarchy of hadron 
suppression. Radiative in-medium energy loss should be larger for light quark 
flavors compared to charm and, in particular, bottom quarks. However, the 
measured nuclear modification factor of electrons from heavy-flavor hadron 
decays at high \pt was comparable in magnitude to the light-flavor dominated 
neutral pion $\raa$ in \auau collisions at RHIC, giving no hint of a quark 
mass hierarchy. The more exclusive measurements from \pbpb collisions at the 
LHC might provide new insight into this issue. The nuclear modification factor 
of prompt D mesons measured~\cite{d_in_pbpb} as a function of \pt with ALICE 
in the 20\% most central \pbpb collisions at mid-rapidity at the LHC is 
compared with corresponding ALICE measurements of charged 
hadrons~\cite{raa_charged} and non-prompt \jpsi mesons from bottom-hadron 
decays as measured with the CMS experiment~\cite{cms_jpsi_pbpb} in the lower 
right panel of Fig.~\ref{fig:d_in_pbpb}. While with the current statistical and 
systematic uncertainties a definite conclusion can not be drawn, the ALICE
data hint towards a stronger suppression of charged hadrons relative to 
D mesons, as one would expect from a scenario in which induced gluon radiation 
is the dominant energy loss mechanism for partons propagating through a dense 
and color-charged medium. The nuclear modification factor of non-prompt 
\jpsi mesons from bottom-hadron decays measured with the CMS experiment, 
clearly supports this picture because it is larger than the $\raa$ measured 
for charged hadrons. With CMS the $\raa$ of non-prompt \jpsi mesons from 
bottom-hadron decays was also measured in the 80\% most peripheral collisions 
at $\sqrt{s_{\rm NN}} = 2.76$~TeV~\cite{cms_jpsi_pbpb}. Within uncertainties the
measured value is consistent with the measurement in central \pbpb collisions. 
In addition to the obvious necessity to collect statistically more significant 
data samples it is also mandatory to measure the nuclear modification factor 
of heavy- and light-flavor observables in proton-nucleus collisions at the LHC,
such that cold nuclear matter effects can be separated from medium 
modifications due to the interaction of heavy quarks with the hot QCD matter 
produced in nuclear collisions at the LHC. 

\begin{figure}[t]
\begin{center}
\includegraphics[width=0.70\linewidth]{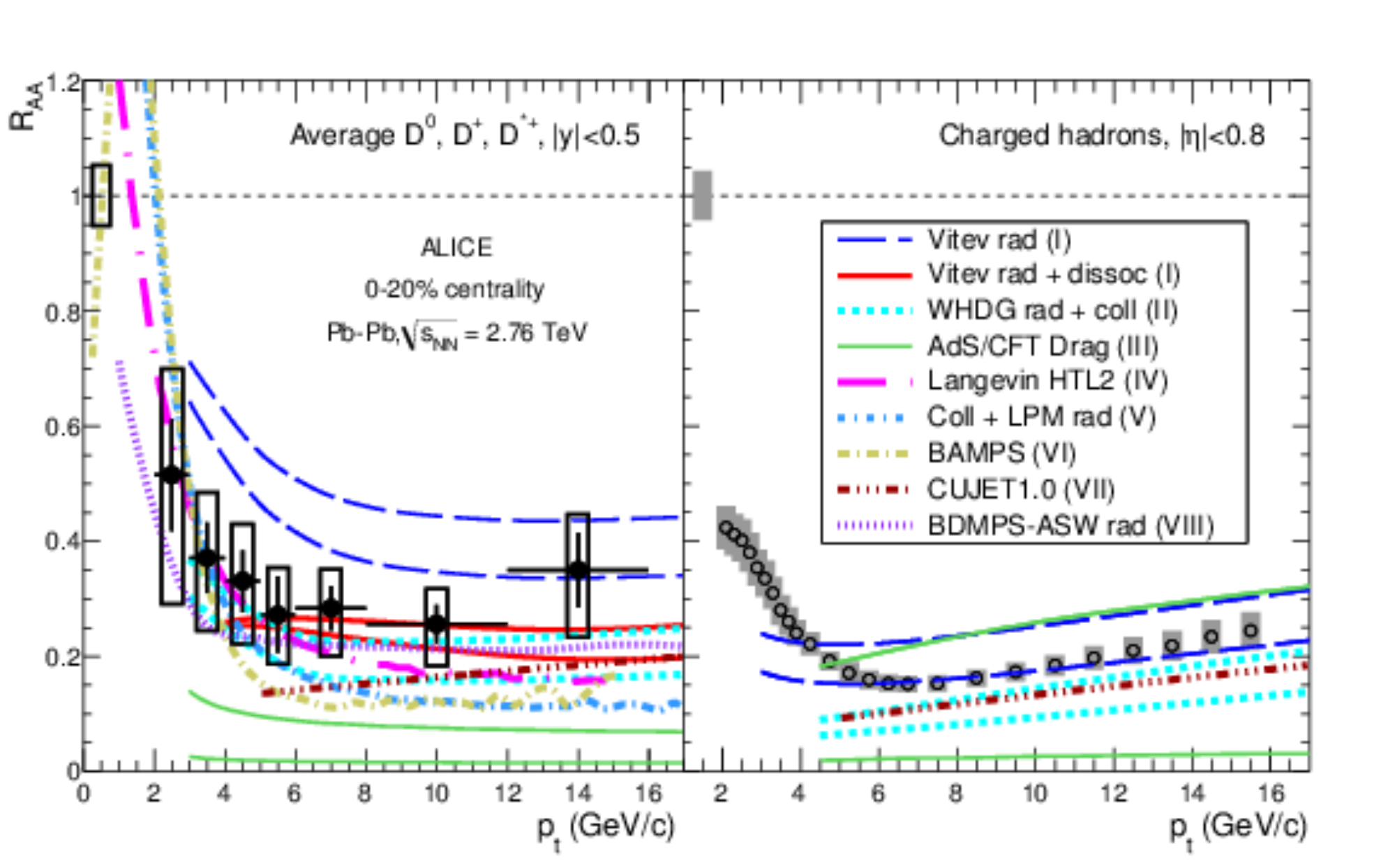}
\end{center}
\caption{Average nuclear modification factor $\raa$ of various D-meson species
         measured as a function of \pt with the ALICE experiment at mid-rapidity
         in the 20\% most central \pbpb collisions at 
         $\sqrt{s_{\rm NN}} = 2.76$~TeV~\cite{d_in_pbpb} (left panel) and the 
         corresponding $\raa$ of charged hadrons in a slightly wider 
         pseudoradidity interval~\cite{raa_charged} (right panel). Error bars 
         (boxes) correspond to the statistical (systematic) uncertainties of 
         the measurements. The normalization uncertainties, which are almost 
         fully correlated for the two measurements, are indicated by the boxes 
         around one. Predictions from various model calculations are compared 
         to the data (see text for further details) (reprinted from 
         Ref.~\cite{d_in_pbpb} with kind permission from Springer Science
         and Business Media).}
\label{fig:d_in_pbpb_models}
\end{figure}

In the same context, it is important to confront theoretical predictions with 
the nuclear modification factors as measured in central \pbpb collisions at 
the LHC. A comparison of various model predictions with the average D meson 
$\raa$~\cite{d_in_pbpb} (left panel) and with the $\raa$ of charged 
hadrons~\cite{raa_charged} (right panel) measured with the ALICE experiment in 
the 20\% most central \pbpb collisions is shown in 
Fig.~\ref{fig:d_in_pbpb_models}. In several models the nuclear modification 
factors are calculated both for D mesons and for charged hadrons. 
Among those, calculations which supplement radiative energy loss with the 
in-medium dissociation of D meson (curves I~\cite{sharma09,he11} in 
Fig.~\ref{fig:d_in_pbpb_models}) or with collisional energy loss via the WHDG 
(curves II~\cite{whdg}) or CUJET1.0 (curves VII~\cite{cujet}) implementations 
describe $\raa$ of D mesons and hadrons reasonably well at the same time. In 
the model incorporating in-medium formation and dissociation of D mesons the 
medium density is tuned such that it describes the inclusive jet suppression 
measured at the LHC. This is different for the WHDG and CUJET1.0 calculations 
which extrapolate the medium density required to describe the pion suppression 
observed at RHIC to LHC conditions. This extrapolation might contribute to the 
fact that the nuclear modification factors calculated for charged hadrons in 
these two models are slightly smaller than measured. The average D-meson 
$\raa$ calculated in a completely different model approach based on AdS/CFT 
drag coefficient (curves III~\cite{horowitz11}) underestimates the measurement 
significantly. The predictive power for the charged hadron $\raa$ is only 
limited in this model. Other models, including a Langevin transport calculation
(curve IV~\cite{alberico11a,alberico11b}), a calculation including collisional 
and radiative energy loss in the medium (curve V~\cite{gossiaux09a,gossiaux10}),
a pQCD based partonic transport calculation (curve VI~\cite{fochler11}), and a 
calculation employing radiative energy loss only (curve VIII~\cite{asw05}), 
provide the average D-meson $\raa$ only. In general, these model 
calculations are in reasonable agreement with the data with a tendency to 
underpredict the measured D-meson $\raa$.

\begin{figure}[t]
\begin{center}
\includegraphics[width=0.5\linewidth]{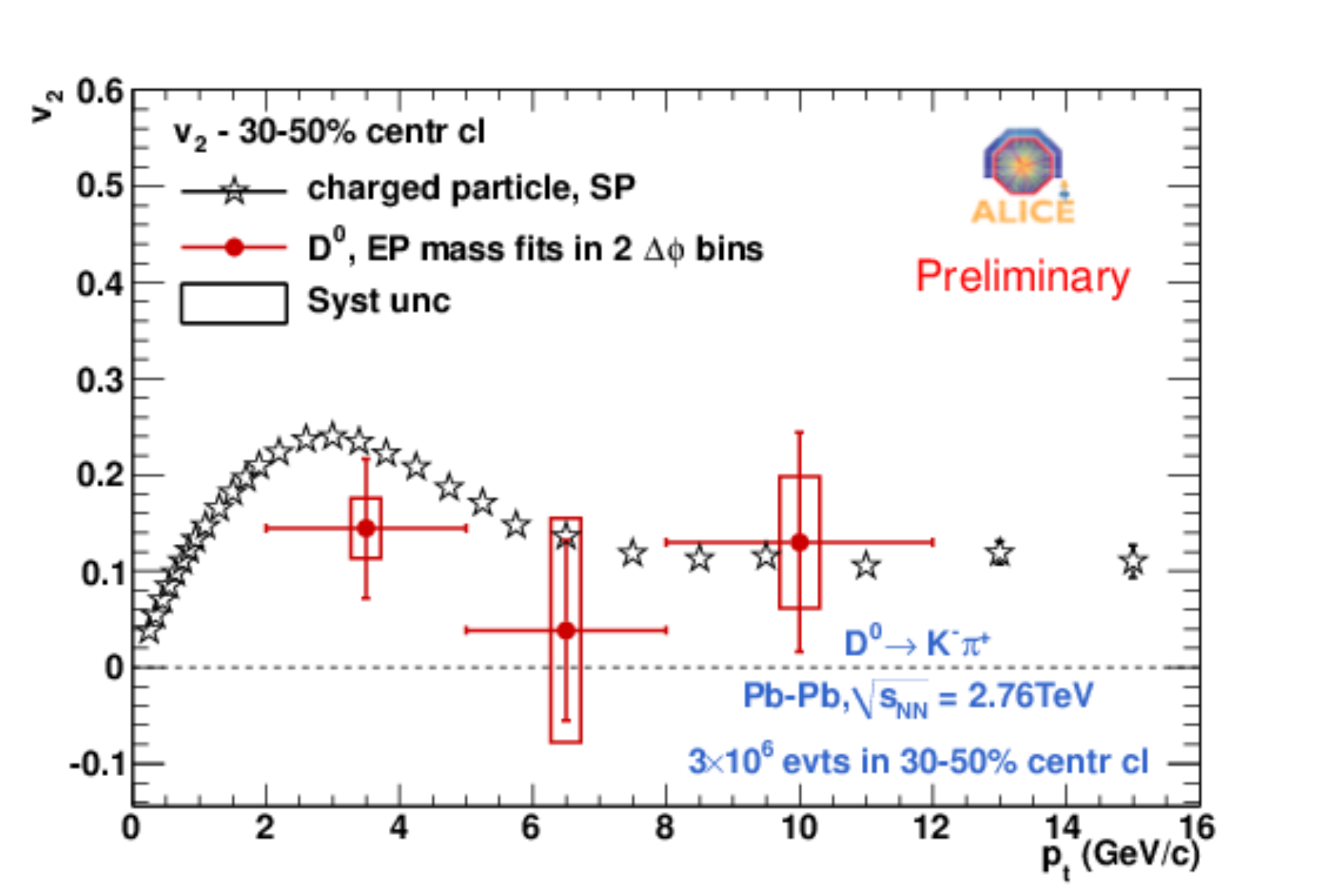}
\end{center}
\caption{Elliptic-flow strength $v_2$ of \dzero mesons measured with the ALICE
         experiment~\cite{d0_flow} as a function of \pt in mid-central \pbpb 
         collisions in comparison with the corresponding $v_2$ of unidentified 
         charged particles~\cite{alice_v2} (reprinted from 
         Ref.~\cite{d0_flow}).}
\label{fig:d_flow}
\end{figure}

The measurement of nuclear modification factors of D mesons demonstrates the 
strong interaction of heavy quarks produced in the earliest phase of \pbpb 
collisions at the LHC with the hot and dense partonic medium that is formed 
afterwards. It is an important question whether heavy quarks participate in
the collective expansion of this medium and whether they might thermalize. 
This question can be addressed via the measurement of the elliptic flow 
strength $v_2$ of particles carrying heavy quarks. $v_2$ was measured
for neutral D mesons~\cite{d0_flow} in mid-central \pbpb collisions with 
the ALICE experiment as shown in Fig.~\ref{fig:d_flow}. A correction for 
feed down from B-meson decays was not applied yet. While statistical and 
systematic uncertainties are still large, an indication for a non-zero 
$v_2$ of D mesons is observed. Clearly, more statistics is necessary to
be able to evaluate the centrality dependence of D-meson flow, and it will
be important to reduce the systematic uncertainties before any strong
conclusion can be drawn.

\subsubsection{Semileptonic heavy-flavor hadron decays}
\label{subsubsec:lhc_emu_pbpb}
\begin{figure}[t]
\begin{center}
\includegraphics[width=0.58\linewidth]{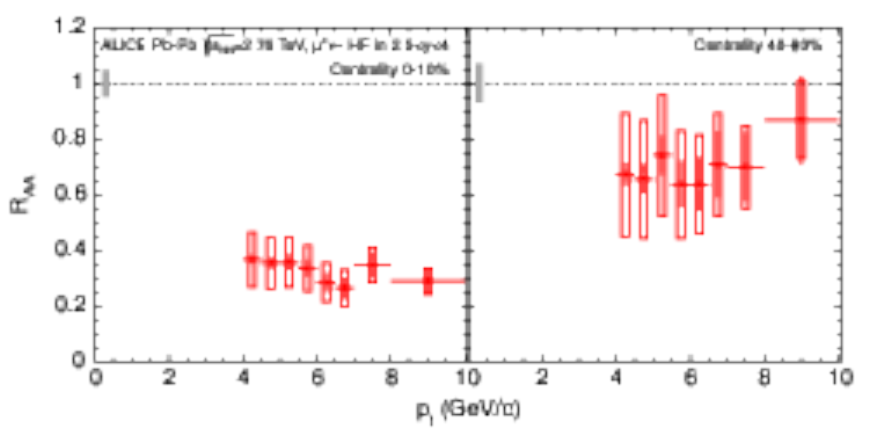}
\includegraphics[width=0.41\textwidth]{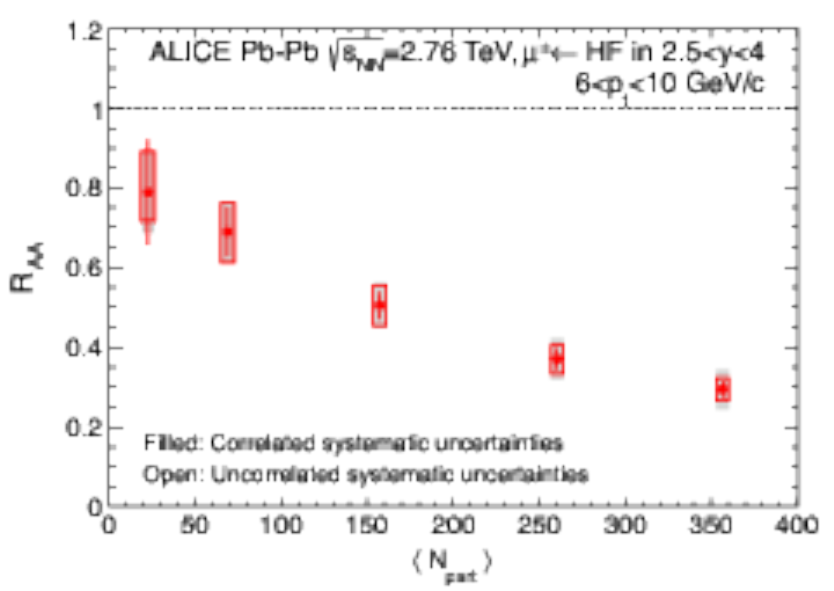}
\end{center}
\caption{Nuclear modification factor of muons from heavy-flavor hadron decays
         measured as a function of \pt with the ALICE experiment in the 
         rapidity range $2.5 < y < 4.0$ in the 10\% most central and in 40-80\%
         central \pbpb collisions (left panel). Centrality dependence of the 
         muon $\raa$ as a function of the average number of participating 
         nucleons in the ranges $2.5 < y < 4.0$ and $4 < \pt < 10$~\gevc 
         (right panel) (reprinted with permission from Ref.~\cite{muon_pbpb}; 
         Copyright (2012) by the American Physical Society).}
\label{fig:muons_in_pbpb}
\end{figure}

Both electrons and muons from heavy-flavor hadron decays have been measured
with the ALICE experiment in \pbpb collisions at $\sqrt{s_{\rm NN}} = 2.76$~TeV.
The electron analysis in the \pt range up to 6\~gevc at mid-rapidity 
($|y| < 0.5$) is currently being finalized. Preliminary 
results~\cite{alice_e_pbpb} indicate a strong suppression in central 
collisions consistent with the suppression observed for muons from heavy-flavor
hadron decays (see below) although the systematic uncertainties are still 
large. Towards more peripheral collisions the high-\pt heavy-flavor electron 
suppression decreases, again consistent with the muon case.

For muons from heavy-flavor hadron decays the nuclear modification factors
as measured with the ALICE experiment~\cite{muon_pbpb} at forward rapidity 
($2.5 < y < 4.0$) in the 10\% most central and 40-80\% central collisions are 
presented as a function of \pt in the left panel of 
Fig.~\ref{fig:muons_in_pbpb}.
In contrast to the D-meson $\raa$ measurement the muon \pp reference could be 
taken directly from the ALICE \pp measurement at $\sqrt{s} = 2.76$~TeV and not 
from the FONLL-driven $\sqrt{s}$ scaling~\cite{reference} of the \pp data at 
7~TeV. This was possible only because a hardware muon candidate trigger allowed
a significant muon sample to be recorded in the relatively short \pp run 
period of the LHC at 2.76~TeV. In the transverse momentum range 
$4 < \pt < 10$~\gevc a strong suppression of muons from heavy-flavor hadron 
decays is observed in the most central \pbpb collisions without a pronounced 
\pt dependence. In 40-80\% central \pbpb collisions a suppression is observed 
as well, although to a much lesser extent, again without any significant \pt 
dependence. The pronounced centrality dependence of the muon $\raa$ from 
heavy-flavor hadron decays is quantified in the right panel of 
Fig.~\ref{fig:muons_in_pbpb} which presents the nuclear modification factor of 
such muons measured with ALICE in the range $6 < \pt < 10$~\gevc at forward 
rapidity as a function of the average number of participating nucleons in 
various \pbpb centrality classes. 
While in peripheral \pbpb collisions only a small 
suppression is observed, it increases strongly towards central collisions. 
However, it is important to note that for the decay muons in the \pt range 
investigated here, a significant fraction originates from the decay of bottom 
hadrons. Therefore, these data emphasize again (see also the lower right panel 
of Fig.~\ref{fig:d_in_pbpb}) that not only charm but also bottom quarks are 
coupled strongly to the hot and dense QCD medium produced in high-energy 
heavy-ion collisions.

\section{Summary and outlook}
\label{sec:summary}
Hadrons carrying heavy flavor, \ie charm or bottom quarks, have been shown 
to be among the most interesting probes in modern particle and nuclear physics.
Because of the large quark masses heavy-flavor production proceeds mainly 
through hard parton-parton scattering processes in the earliest stage of 
hadronic collisions. As such the measurement of heavy-flavor production
in \pp collisions provides a crucial testing ground for perturbative QCD
calculations. State of the art FONLL pQCD calculations are in agreement with 
all of such measurements within statistical and systematic uncertainties
at the Fermilab Tevatron, the BNL RHIC, and at the CERN LHC. Furthermore,
these measurements serve as a baseline for heavy-flavor studies in 
nucleus-nucleus collisions, in which the heavy quarks propagate through and
interact with the hot and dense medium produced in the nuclear collisions.
The investigation of in-medium modifications of heavy-flavor observables
can shed light on the properties of the QCD medium and the nature of 
parton-medium interactions. This is the case both for heavy-quarkonia, 
which have not been discussed in this review (see Ref.~\cite{rapp10} instead) 
and for which open heavy-flavor measurements can provide a natural baseline, 
and for open heavy-flavor observables, which have been in the focus of this 
review.

While extensive systematic studies of the production of heavy quarkonia have
been conducted in the nucleus-nucleus collision program at the CERN SPS, the
production of open heavy-flavor hadrons does not play a role in the collision
dynamics in this rather low energy regime. 

First systematic studies of open heavy-flavor production in nucleus-nucleus
collisions have been conducted at RHIC, mainly in the semi-electronic decay 
channel, where it was verified that the total heavy-flavor yields scale with 
the number of binary collisions. Measurements of the nuclear modification factor
and of the elliptic flow strength of electrons from heavy-flavor hadron decays
demonstrate the strong coupling of heavy quarks with the hot and dense
QCD medium produced in \auau collisions at RHIC. To further elucidate 
the properties of hot QCD matter it is important to shed light on the
relevant interaction mechanism of heavy quarks with the medium. It is still
not clear which mechanisms are finally responsible for the strong suppression
observed for high \pt electrons in central \auau collisions at RHIC.

At the CERN LHC, the systematic study of open heavy-flavor production in
nucleus-nucleus collisions is continued in an unprecedented high-energy
regime. In addition, for the first time, exclusive charm and bottom 
measurements have become possible in nucleus-nucleus collisions thanks to 
the availability of silicon vertex spectrometers that allow the separation 
of secondary, displaced heavy-flavor hadron decay vertices from the primary 
collision vertex. The general picture established at RHIC was confirmed by the
first measurements at the LHC already, but major additional progress is 
expected in the near future through the systematic study of the centrality,
\pt, and rapidity dependence of exclusive charm and bottom observables both 
in \pbpb and soon in \ppb collisions. Table~\ref{tab:summary} gives an
overview over the measured open heavy-flavor channels and their kinematic
coverage at RHIC and at the LHC until July 2012. It should be noted
that since then preliminary results have become available, in particular from 
the LHC, which significantly extend the kinematic coverage in most channels.

\begin{table}[p]
\begin{center}
\caption{Overview over measurements related to open heavy-flavor production 
         conducted at RHIC ($\snn \le 0.2$~TeV) and at the LHC 
         ($\snn \le 7$~TeV) until July 2012. For each collision system,
         center-of-mass energy $\snn$, and channel investigated the phase space 
         coverage is given as well.}
\label{tab:summary}
\begin{tabular}{llllll}
\\
\hline
system & $\snn$ [TeV] & channel & \pt coverage [\gevc] & $y$ coverage & ref. \\ 
\hline
\pp & 0.2 & \dzero  & $0.6 < \pt < 2.0$ & $|y| < 1$ & \cite{star_d_pp} \\
    &     & \dstar & $2 < \pt < 6$ & $|y| < 1$ & \cite{star_d_pp} \\
    &     & $c,b \rightarrow e^\pm$ & $0.3 < \pt < 9.0$ & $|y| < 0.35$ & \cite{ppg077} \\
    &     & $c,b \rightarrow e^\pm$ & $1.2 < \pt < 10.0$ & $|y| < 1$ & \cite{star_e_pp} \\
    &     & $c \rightarrow e^\pm$ & $2 < \pt < 7$ & $|y| < 0.35$ & \cite{ppg094} \\
    &     & $b \rightarrow e^\pm$ & $2 < \pt < 7$ & $|y| < 0.35$ & \cite{ppg094} \\
    &     & $c \rightarrow e^\pm$ & $2.5 < \pt < 9.5$ & $|y| < 0.7$ & \cite{star_e_b_d} \\
    &     & $b \rightarrow e^\pm$ & $2.5 < \pt < 9.5$ & $|y| < 0.7$ & \cite{star_e_b_d} \\
    &     & $c,b \rightarrow e^+e^-$  & $\pte > 0.2$ & $|y| < 0.35$ & \cite{ppg085,ppg088}\\
    &     & $c,b \rightarrow \mu^-$  & $1 < \pt < 7$ & $1.4 < |y| < 1.9$ & \cite{ppg117} \\
\hline
\dau & 0.2 & \dzero  & $0.1 < \pt < 3.0$ & $|y| < 1$ & \cite{star_d_dau} \\
     &     & $c,b \rightarrow e^\pm$ & $0.85 < \pt < 8.5$ & $|y| < 0.35$ & \cite{ppg131} \\
     &     & $c,b \rightarrow e^\pm$ & $1 < \pt < 10$ & $|y| < 1$ & \cite{star_e_auau} \\
\hline
\cucu & 0.2 & $c,b \rightarrow \mu^-$  & $1 < \pt < 4$ & $1.4 < |y| < 1.9$ & \cite{ppg117} \\
\hline
\auau & 0.13 & $c,b \rightarrow e^\pm$ & $0.5 < \pt < 3.0$ & $|y| < 0.35$ & \cite{ppg011} \\
\hline
\auau & 0.2 & \dzero  & $\pt < 2.0$ & $|y| < 1$ & \cite{star_d_auau} \\
      &     & $c,b \rightarrow e^\pm$ & $0.3 < \pt < 9.0$ & $|y| < 0.35$ & \cite{ppg077} \\
      &     & $c,b \rightarrow e^\pm$ & $1.2 < \pt < 10.0$ & $|y| < 1$ & \cite{star_e_auau} \\
      &     & $c,b \rightarrow e^+e^-$  & $\pte > 0.2$ & $|y| < 0.35$ & \cite{ppg088} \\
\hline
\hline
\pp & 7.0 & \dzero, \dplus, \dstarplus & \dplus, \dstarplus: $1 < \pt < 24$ & $|y| < 0.5$ & \cite{d_in_pp}\\
    &     &                         & \dzero: $1 < \pt < 16$ & & \\
    &     & \bzero, \bplus, \bszero & \bplus, \bzero: $\pt > 5$ & \bplus, \bszero: $|y| < 2.4$ & \cite{cms_bzero,cms_bplus,cms_bszero}\\
    &     &                         & \bszero: $\pt > 5$ & \bzero: $|y| < 2.2$ & \\
    &     & \bplus & $0 < \pt < 40$ & $2.0 < y < 4.5$ & \cite{lhcb_bplus}\\
    &     & b hadron $\rightarrow$ \jpsi & $\pt > 0$ & $|y| < 0.9$ & \cite{alice_jpsi_b}\\
    &     & b hadron $\rightarrow$ \jpsi & $7 < \pt < 70$ & $|y| < 0.75$ & \cite{atlas_jpsi_b}\\
    &     & b hadron $\rightarrow$ \jpsi & $1 < \pt < 30$ & $1.5 < |y| < 2.0$ & \cite{atlas_jpsi_b}\\
    &     & b hadron $\rightarrow$ \jpsi & $\pt > 6.5$ & $|y| < 1.2$ & \cite{cms_jpsi_b}\\
    &     & b hadron $\rightarrow$ \jpsi & $\pt > 0$ & $1.6 < |y| < 2.4$ & \cite{cms_jpsi_b}\\
    &     & $c,b \rightarrow e^\pm$ & $0.5 < \pt < 8.0$ & $|y| < 0.5$ & \cite{alice_e_pp}\\
    &     & $c,b \rightarrow e^\pm$ & $7 < \pt < 26$ & $|y| < 2.0$ & \cite{atlas_e_mu}\\
    &     & $c \rightarrow e^\pm$ & $1 < \pt < 8$ & $|y| < 0.8$ & \cite{alice_hfe_beauty}\\
    &     & $b \rightarrow e^\pm$ & $1 < \pt < 8$ & $|y| < 0.8$ & \cite{alice_hfe_beauty}\\
    &     & $c,b \rightarrow \mu^-$  & $2 < \pt < 12$ & $2.5 < y < 4.0$ & \cite{muon_pp}\\
    &     & $c,b \rightarrow \mu^-$  & $7 < \pt < 24$ & $|y| < 2.0$ & \cite{atlas_e_mu}\\
pp  & 2.76 & \dzero, \dplus, \dstarplus & $1 < \pt < 12$ & $|y| < 0.5$ & \cite{d_pp_low}\\
    &     & $c,b \rightarrow \mu^\pm$  & $2 < \pt < 10$ & $2.5 < y < 4.0$ & \cite{muon_pbpb}\\
\pbpb & 2.76 & \dzero, \dplus, \dstarplus & $2 < \pt < 16$ & $|y| < 0.5$ & \cite{d_pp_low}\\
      &      & b hadron $\rightarrow$ \jpsi & $6.5 < \pt < 30$ & $|y| < 2.4$ & \cite{cms_jpsi_pbpb}\\
      &      & $c,b \rightarrow e^\pm$ & $3.5 < \pt < 6.0$ & $|y| < 0.5$ & \cite{alice_e_pbpb}\\
      &      & $c,b \rightarrow \mu^\pm$ & $4 < \pt < 10$ & $2.5 < |y| < 4.0$ & \cite{muon_pbpb}\\
\hline
\end{tabular}
\end{center}
\end{table}

The future is bright for open heavy-flavor measurements in high-energy
nucleus-nucleus collisions. The experimental program at the LHC has just
begun. At RHIC, after the very successful first decade of operations at the
world's first dedicated heavy-ion collider, the accelerator is upgraded to
deliver higher luminosities and concurrently the PHENIX and STAR detectors
are undergoing major upgrade programs. Key elements of these upgrades are
silicon vertex detectors aimed towards the separation of displaced heavy-flavor
hadron decay vertices from the primary collision vertex. With these upgrades 
in place it will be possible both at RHIC~\cite{frawley08} as well as at the 
LHC to measure the yields and phase space distributions of charm and bottom 
hadrons in \pp, p/d-nucleus, and nucleus-nucleus collisions via exclusive 
channels, providing unique hard probes for hot QCD matter even at low \pt. 
In particular the measurement of heavy-flavor baryons, notably the \lambdac,
would be important even though this is very challenging, markedly in
nucleus-nucleus collisions. 

Precision measurements of the nuclear modification factors of heavy-flavor 
hadrons will address the issue whether the expected mass ordering of quark 
energy loss in the hot and dense medium is realized in nature or not.
This is one of the crucial questions related to the in-medium parton dynamics 
and, in particular, to the mechanism of parton energy loss in the medium.
The latter is up to now not understood on a fundamental level. Tightly
connected to the energy loss of a hard probe is the reaction of the medium
to the resulting energy deposit. Regarding these issues, heavy-flavor 
measurements can contribute a unique piece of information. With the high 
resolution vertex detectors at RHIC and at the LHC it should be possible to 
``tag'' charm and, in particular, bottom jets on an event-by-event 
basis. In contrast to high-\pt particle production, which at RHIC and at the 
LHC is dominated by gluon jets, such heavy-flavor tags give clean access to 
quark jets. Furthermore, momentum conservation in the hard scattering process 
gives rise to a back-to-back correlation in azimuth of the quark and antiquark 
jets produced associately. Therefore, tagged heavy-quark jets constitute a new 
quality of hard probes as they give much better access to the kinematics of
the initially scattered partons and allow the investigation of the mass
dependence of parton energy loss and the medium response. Also, the 
precision measurement of heavy-flavor hadron flow will be possible at RHIC
and at the LHC, such that the question can be addressed to what extent 
heavy quarks reach equilibrum with the hot and dense medium. This is a crucial 
issue as it is directly related to the transport properties of the medium. 

With precision vertexing it should furthermore be possible to measure at RHIC 
and at the LHC the contribution of correlated semileptonic heavy-flavor hadron 
decays to the dilepton continuum in the intermediate mass region, which at RHIC
is dominated by charm hadron decays while at the LHC bottom hadron decays
contribute significantly. In nucleus-nucleus collisions, these dileptons carry
information about the interaction of heavy quarks with the medium in a twofold 
way. First, the quarks lose energy in the medium and second the opening
angle between the quark and the antiquark can be modified. In the extreme case,
the heavy quarks thermalize with the medium such that the angular correlation
between the quark and antiquark is completely lost. Again, a comparison of
charm and bottom measurements would be useful regarding the issue of the
parton energy loss mechanism and it could help to better understand the 
transport properties of the medium. Furthermore, it should be noted that
only two sources are expected to contribute significantly to the intermediate 
mass dilepton continuum, \ie correlated heavy-flavor hadron decays and thermal
radiation from the hot medium. If these two contributions can be disentangled
with the help of precision vertexing the temperature of the emitting source
and its evolution with time can be investigated in a unique way.

In the longer term future, open charm production will be addressed at the FAIR 
facility currently under construction at GSI. At this fixed-target machine 
charm can be investigated close to the production threshold in proton-nucleus
and nucleus-nucleus collisions, where models predict in-medium effects which 
in a unique way could shed light on the properties of the baryon-rich form of 
dense QCD matter produced at FAIR~\cite{cbm_book}.

\end{document}